\documentclass[12pt, a4paper, twoside]{mythesis}

\usepackage[english]{babel}
\usepackage[latin1]{inputenc}
\usepackage{pdfpages}
\usepackage{moreverb}
\usepackage{fancyhdr}
\usepackage{setspace}
\usepackage{listings}
\usepackage{amssymb}
\usepackage{amsfonts}
\usepackage{amsmath}
\usepackage{amsthm}
\usepackage{thmtools}
\usepackage{relsize}
\usepackage{float}
\usepackage[english]{varioref}
\usepackage[a4paper,top=3cm,bottom=3cm,left=3cm,right=3cm,bindingoffset=5mm,twoside=true]{geometry}
\usepackage{indentfirst}
\usepackage{multirow}
\usepackage{colortbl}
\usepackage{chngpage}
\usepackage{latexsym}
\usepackage{siunitx}
\usepackage{circuitikz}
\usepackage{tikz}
\usepackage{enumerate}
\usepackage{appendix}
\usepackage{subfloat}
\usepackage{braket}
\usepackage{bbm}
\usepackage{cite}
\usepackage[Lenny]{fncychap}
\usepackage[colorlinks=false]{hyperref}
\usepackage{color}
\definecolor{dkgreen}{RGB}{47, 152, 92}
\usepackage{algpseudocode}
\usepackage{algorithm}

\begin{document}

\begin{titlepage}

\begin{figure}
\begin{center}
\includegraphics[width=9cm]{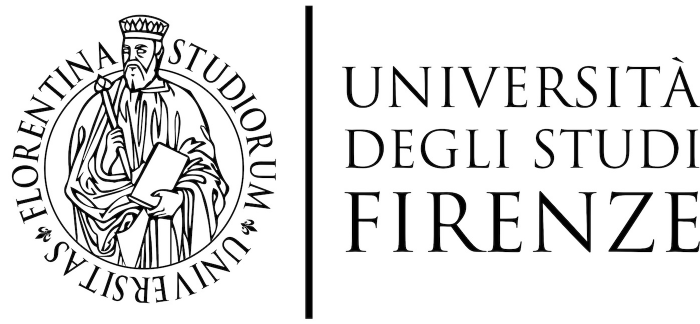}
\end{center}
\end{figure}

\begin{center}
{\Large{\textsc{Università degli Studi di Firenze}}} \\
%\rule[0.1cm]{13.8cm}{0.1mm}
{\textsc{Dipartimento di Ingegneria dell'Informazione (DINFO)}}\\
{\textsc{Corso di dottorato in Ingegneria dell'Informazione}}\\
{\textsc{Curriculum: Dinamica non lineare e sistemi complessi}}\\
{\textsc{Settore scientifico-disciplinare: FIS/03}}\\
\vspace{5mm}
\end{center}

\begin{center}
{\LARGE{\bf NOISE AS A RESOURCE}}\\
{\large{Probing and manipulating classical and quantum dynamical systems via stochastic measurements}}
\end{center}

\par
\noindent

\begin{center}
{{\small{\textsc{Candidate}}}}\\
\vspace{5mm}
{\normalsize{Stefano Gherardini}}\dotfill\\
\vspace{5mm}
{{\small{\textsc{Supervisors:}}}}\\
{\normalsize{Prof. Filippo Caruso}}\dotfill\\
\vspace{5mm}
{\normalsize{Prof. Stefano Ruffo}}\dotfill\\
\vspace{5mm}
{\normalsize{Prof. Giorgio Battistelli}}\dotfill\\
\vspace{5mm}
{\normalsize{Prof. Luigi Chisci}}\dotfill
\end{center}

\begin{center}
{\textsc{Ciclo XXX, 2014-2017}}
\end{center}

\end{titlepage}

\newpage{
\thispagestyle{empty}
{\textsc{PhD Thesis Evaluators:}}\\
{\normalsize{Prof. Mauro Paternostro, Queen's University of Belfast (Northern Ireland).}}\\
{\normalsize{Prof. Claudio Altafini, Link\"{o}ping University (Sweden).}}\\ \\ \\

{\textsc{PhD Thesis Committee:}}\\
{\normalsize{Prof. Francesco Ticozzi, University of Padua (Italy).}}\\
{\normalsize{Prof. Daniel Remondini, University of Bologna (Italy).}}\\
{\normalsize{Prof. Luca Munaron, University of Turin (Italy).}}
}

\newpage{
\thispagestyle{empty}
\clearpage
\chapter*{Abstract}
%\label{abstract}
%\addcontentsline{toc}{chapter}{Abstract}
%\fancyhead{}
%\fancyhead[LEH]{ABSTRACT}
%\fancyhead[RO]{ABSTRACT}

In this thesis, common features from the theories of open quantum systems, estimation of state dynamics and statistical mechanics have been integrated in a comprehensive framework, with the aim to analyze and quantify the energetic and information contents that can be extracted from a dynamical system subject to the external environment. The latter is usually assumed to be deleterious for the feasibility of specific control tasks, since it can be responsible for uncontrolled time-dependent (and even discontinuous) changes of the system.

However, if the effects of the random interaction with a noisy environment are properly modeled by the introduction of a given stochasticity within the dynamics of the system, then even noise contributions might be seen as control knobs. As a matter of fact, even a partial knowledge of the environment can allow to set the system in a dynamical condition in which the response is optimized by the presence of noise sources. In particular, we have investigated what kind of measurement devices can work better in noisy dynamical regimes and studied how to maximize the resultant information via the adoption of estimation algorithms. Moreover, we have shown the optimal interplay between quantum dynamics, environmental noise and complex network topology in maximizing the energy transport efficiency. Then, foundational scientific aspects, such as the occurrence of an ergodic property for the system-environment interaction modes of a randomly perturbed quantum system or the characterization of the stochastic quantum Zeno phenomena, have been analyzed by using the predictions of the large deviation theory. Finally, the energy cost in maintaining the system in the non-equilibrium regime due to the presence of the environment is evaluated by reconstructing the corresponding thermodynamics entropy production.

In conclusion, the present thesis can constitute the basis for an effective \textbf{resource theory of noise}, which is given by properly engineering the interaction between a dynamical (quantum or classical) system and its external environment.

}

\clearpage\null\thispagestyle{empty}\newpage
\pagenumbering{Roman}
\tableofcontents

\mainmatter

\phantomsection
\chapter*{Introduction}
\label{introduction}
\addcontentsline{toc}{chapter}{Introduction}
\fancyhead{}
\fancyhead[LEH]{INTRODUCTION}
\fancyhead[RO]{INTRODUCTION}

The term \textbf{stochasticity} quantifies the lack of predictability of a sequence of events. Predictability is ensured only from a probabilistic point of view in the distribution of the random variables' outcomes, computed over a very large number of replicas of the same sequence of events~\cite{Papoulis1984}. For this reason, the introduction of stochastic processes in the mathematical formulation of real systems dynamics at any dimensional scale has allowed the modeling of a wide class of static and dynamical uncertainties, as, for example, the unavoidable presence of noise on a measuring device, or the occurrence of spontaneous transitions of a quantum mechanical system (such as atoms, molecules or subatomic particles) from an excited energy state to a lower energy one with the resulting emission of a quantum of light (photon).

On the other side, \textbf{information} is an actually universal term too: it quantifies the relevance of a given amount of data in relation to our knowledge of their (real or abstract) content~\cite{Cover2006}. This concept hides inside the existence of a cognitive process given by the presence of an observer, providing an uncertain estimate of the issues under investigation after a repeated sequence of observations. In this regard, especially in quantum mechanics, the measurement process is to be considered as random, so as to describe the predicted outcome of the measurements within the same probabilistic framework, which is used to model the evolution of the observed phenomena. In particular, the postulates of the quantum measurement theory enable to quantitatively calculate the probability distributions of the measurement outcomes and determine the corresponding back-actions on the quantum system dynamics. Accordingly, the interplay between the application of sequences of quantum measurements and the presence of stochastic contributions within the system dynamics, due to the random interaction with the environment, is actually a topic which is worth exploring.

Finally, in order to design nanoscale machines and engines, which can be characterized by automatic functionalities and computing capability, it can be required to evaluate in routine transformations their \textbf{energy} consumption, in comparison with the efficiency of the device. A larger efficiency-energy rate unavoidably translates into novel challenges on how to fully exploit (by means of new methods) the minimum information content coming from the measurements or proper interactions with the environment.

In this thesis, all of these concepts are contextualized in the framework of open dynamical systems, i.e. systems that interact with an external environment, leading to the introduction of stochasticity contributions in the form of disorder or noise. Generally, both disorder and noise are considered deleterious to accurately manipulate/control the system; however, with the present thesis, we want to provide the basic points, which define a \textbf{resource theory of noise}, i.e.
\begin{itemize}
\item
The capability to exploit measurement devices, that can work \textit{better} in noisy dynamical regimes by using estimation algorithms able to extract information from fictitious variations of the device's outcomes due to a measurement noise source. As it will be explained in Chapter~1, an example of such a device is given by binary (threshold) sensors.
\item
The knowledge of physical phenomena, which are well optimized by nature (after thousand of years of continuous natural adaptation to external boundary conditions) in their dynamical behaviour thanks to the optimal interaction with the environmental degrees of freedom. In this context, recent studies~\cite{Lee2007SCI316,Engel2007NAT446} in the novel research field of \textit{quantum biology}~\cite{qbiobook} have shown that a remarkably high efficiency in the excitation energy transfer over light-harvesting complexes can be achieved only when noise sources affects each site of such a biological systems. The presence of noise, indeed, allows the electron excitations not to remain in any local minimum of the path potential, but to efficiently proceed to the reaction site, where they are chemically processed. The optimal interplay between quantum coherence and environmental noise to realize efficient energy transport phenomena is also called \textit{Noise-Assisted Transport (NAT)}. In Chapter~2, a scalable transport emulator based on optical fiber cavity network, which has experimentally reproduced NAT, will be shown.
\item
The identification of mathematical tools (especially from statistical mechanics), which allow to make predictions about the behaviour of a dynamical system subject to random interactions with the environment. In this regard, by modelling how the system is externally influenced in a repeated sequence of events, the \textit{Large Deviation (LD) theory}~\cite{Touchette1,Dembo1} has turned out to be the most appropriate method with such a predictive feature. The application of the LD theory to open quantum systems will be introduced in Chapter~3.
\item
The design of \textit{noise sensing} algorithms. As a matter of fact, in order to consider the noise entering into the system as an effective resource, it can be required to be able to infer the noise fluctuation profiles, which are uniquely determined by the corresponding power spectral density in the frequency domain~\cite{Degen2017}. Moreover, as shown again in Chapter~3, an a-priori modeling of the noise occurrence within the system dynamics is essential to enhance the predictions from LD theory, and, then, drive the system in a target non-equilibrium regime of the system-environment configurations space.
\item
The introduction of a figure of merit, which can measure the degree of \textit{energy dispersion} within the system due to the presence of external noise contributions. The latter is given by the \textit{thermodynamic entropy}, which quantifies how much the current state of the system, after repeated system-environment interactions, differs from a configuration corresponding to states at minimal energy~\cite{Groot1984}. As a matter of fact, noise terms can drive the system towards novel dynamical regimes that could not be otherwise achieved, and it is worth asking questions about the energetic cost needed to maintain the system in such non-equilibrium condition. In this specific framework, in Chapter~4 the most important result is given by the characterization of the thermodynamic irreversibility for an arbitrary open quantum system subject to external environments.
\end{itemize}
In this way, the presence of noise can be effectively seen as a \textit{control knob}, which would allow one to set-up a given dynamical system in a suitable configuration, in which the responses of the system itself are optimized by the presence of an external environment.

%-------------------------------------------------------------------------
\subsection*{Thesis outline}

Specifically, the following macro-themes will be addressed in detail within the thesis:
\begin{itemize}
\item
\textbf{Chapter 1: State estimation via networks of binary sensors.} We will address state estimation for complex discrete-time systems with binary (threshold) measurements by following both deterministic and probabilistic \textit{Moving Horizon Estimation} (MHE) approaches. The outputs of binary sensors (probes) can take only two possible values according to whether the sensed variable exceed or not a given threshold. For both classical and quantum systems, the solution of state estimation problems with binary sensors is of absolute scientific and technological relevance, because such devices provide the least amount of information as possible. As a matter of fact, especially in the continuous-time case, the information coming from a binary sensor is strictly related to the threshold-crossing instants (by the sensed variable), and \textit{system-observability} can be ensured only when the number of threshold-crossing instants is sufficiently large, as well as for irregularly sampled systems. In this regard, we will show that by using the probabilistic approach to state estimation the proposed estimators exhibit \textit{noise-assisted} features, so that the estimation accuracy is improved under the presence of measurement noise.
\item
\textbf{Chapter 2: Noise-assisted quantum transport.} We will address how excitations energy transport over complex networks can be performed with remarkably high efficiency only via the optimal interplay between quantum coherence and environmental noise. The presence of coherence, indeed, leads to a very fast delocalization of excitations, that in this way can simultaneously exploit several paths to the target site. However, the transmission of energy can be prevented by the occurrence of destructive interference between different pathways and by the presence of energy gaps between the network sites. In particular, we have experimentally shown that, in specific dynamical conditions, the additional and unavoidable presence of static disorder and noise positively affects the transmission efficiency, thus leading to the evidence of a \textit{noise-assisted quantum transport} paradigm.
\item
\textbf{Chapter 3: Large deviations and stochastic quantum Zeno phenomena.} We will address how to model the \textit{stochastic interaction between a quantum (many-body) system and the external environment} by using the tools of the non-equilibrium statistical mechanics. In particular, we will analyze through the LD theory the effects on quantum system dynamics given by the presence of some noise sources and the application of sequences of quantum measurement within the framework of the \textit{stochastic quantum Zeno effect (QZE)}. The quantum Zeno effect states that in case of a frequent enough series of measurements, projecting back the quantum system to the initial state, its dynamical evolution gets completely frozen, while the LD theory concerns the asymptotic exponential decay of a given system probability function due to large fluctuations of some stochastic variables entering into its dynamics. We will show how to derive the typical value (not necessarily equal to the mean value) of the probability that the system remains in its initial state (survival probability) after a randomly-distributed sequence of quantum measurements, so that it can be used as a control knob to protect and manipulate information contents within open quantum systems. The chapter ends with the introduction of a novel (quantum Zeno-based) noise filtering scheme for the detection of time correlations in random classical fields coupled to the quantum system used as a probe. Indeed, time correlations in the noise field determine whether and how fast the typical value of the survival probability converges to its statistical mean, and, consequently, how the standard deviation of the survival probability over many realisations can reveal information on the noise field.
\item
\textbf{Chapter 4: Quantum thermodynamics.} In this chapter, we will finally address the characterization and reconstruction of general thermodynamical quantities such as work, heat and entropy. In the quantum regime, the dynamics of systems is highly stochastic, in the sense that thermal and quantum fluctuations become of the same order of magnitude as the averages of the physical quantities defining the system Hamiltonian. Therefore, the characterization of such fluctuations with the tools of the non-equilibrium statistical mechanics is crucial to understand both the dynamics of an open system and the ways whereby the environmental stochasticity affects the system itself. In particular, by starting from the analysis of the fluctuation theorem for open quantum systems, we will introduce an efficient protocol (relying on the two-time quantum measurements scheme) to determine the characteristic functions of the stochastic entropy production of an arbitrary quantum many-body system. It is worth noting that the concept of \textit{entropy} is important not only in thermodynamics, where it allows to characterize the irreversibility of a dynamical system, but also in information theory to measure the amount of lost information within a communication channel~\cite{Cover2006}.
\end{itemize}
For the sake of clarity, the notation of each chapter is introduced before being adopted. However, the same symbols with a different meaning can be found in various chapters of the thesis. Furthermore, throughout the thesis the Dirac notation (or bra-ket notation) will be largely used. The bra-ket notation was initially introduced by Dirac to represent in a compact way a vector (linear) space, and, in particular, in quantum mechanics is the standard notation to describe quantum states. In general, a collection of physical quantities is represented by a row or column vector. In this regard, the Dirac notation uses two distinct symbols: given the generic vector $v$, the \textit{ket}
\begin{equation}\label{ket}
|v\rangle\Leftrightarrow\begin{pmatrix}v_{1} \\\vdots \\ v_{n} \end{pmatrix}
\end{equation}
corresponds to the column vectors, while the \textit{bra}
\begin{equation}\label{bra}
\langle v|\Leftrightarrow(v_{1}^{\ast},\cdots,v_{n}^{\ast})
\end{equation}
to a row vector. Observe that in (\ref{ket}), $v_{i}$, $i = 1,\ldots,n$, is the $i-$th element of the vector $v$, while in (\ref{bra}) the superscript $\ast$ denotes complex conjugation. Finally, in the thesis ${\rm col}(\cdot)$ will denote the matrix obtained by stacking its arguments one on top of the other, and ${\rm diag}(m_{1},\ldots,m_{q})$ will be the diagonal matrix whose diagonal elements are the scalars $m_{1},\ldots,m_{q}$. Further, given a matrix $M$, ${\rm vec}(M)$ is the linear transformation which converts the matrix $M$ into a column vector and $\||v\rangle\|_{M} \equiv \langle v|M|v\rangle$.

\chapter{State estimation with binary sensors}
\label{chapter1}
\fancyhead{}
\fancyhead[LEH]{\leftmark}
\fancyhead[RO]{\rightmark}

\textit{
In this chapter, we will address how to solve the problem to accurately infer a given system dynamics via the adoption of measurement devices (sensors) providing a minimal amount of information. Such devices are modelled as binary sensors, whose output can take only two possible values according to whether the sensed variable exceed or not a given threshold. This issue is crucial when we want to analyze phenomena in which the exact outcomes coming from the measurement process are fundamentally unpredictable, so that our knowledge of the real world is given only by computing the probabilities of such outcomes.
\footnote{The part of this chapter related to moving-horizon state estimation for discrete-time dynamical systems has been published in the following scientific papers: ``Moving horizon state estimation for discrete-time linear systems with binary sensors'' in \textit{54th International Conference on Decision and Control (CDC)}, December 15-18, 2015, Osaka (Japan)~\cite{GherardiniCDC}; ``Moving horizon estimation for discrete-time linear systems with binary sensors: algorithms and stability results'' in \textit{Automatica} \textbf{85}, 374-385 (2017)~\cite{GherardiniAutomatica}; ``MAP moving horizon state estimation with binary measurements'' in \textit{The 2016 American Control Conference (ACC)}, July 6-8, 2016, Boston (USA)~\cite{GherardiniMAP}; ``MAP moving horizon field estimation with threshold measurements for large-scale systems'' in preparation, 2018~\cite{GherardiniFIELDmap}.}
}

\section*{Introduction}

Every measurement of a given physical (classical or quantum) quantity is \textit{uncertain}. About classical systems, the unavoidable presence of external noise sources (especially in the measurement device) introduces systematic errors, which makes our knowledge of the process partial and uncertain. About quantum systems, indeed, uncertainty relations are consistently present in the physical behaviour of systems such as electrons and light, which behave sometimes like waves and sometimes like particles, in accordance with the Heisenberg's \textit{uncertainty principle}. The latter, as given in Ref.~\cite{feynman93}, literally states that \textit{any determination of the alternatives taken by a process capable of following more than one alternative destroys the interference between them}. As a consequence, if we introduce an additional (macroscopic) system (i.e. an \textit{observer}), which effectively measures the expectation value of the position (momentum) operator along one or more of the path followed by the particle, then our knowledge of the momentum (position) is prevented by the presence of quantum fluctuations introduced by the observer. More formally, the uncertainty principle states that
\begin{equation}
\Delta X \Delta P \geq \frac{\hbar}{2},
\end{equation}
where $\Delta X$ and $\Delta P$ denotes, respectively, the standard deviations of the position and momentum operators, and $\hbar$ is the reduced Planck's constant. In the same way, as dual definition, the principle states that it is not possible to \textit{prepare} a quantum system state, which admits simultaneously well-defined values of the position and momentum observables, $X$ and $P$ respectively, after being measured. Position and momentum, indeed, are non-commutating operators, satisfying the relation $[X,P]=i\hbar$. In the present form, as given for example by the well-known double-slit experiment, originally performed by Davisson and Germer in $1927$~\cite{davisson1928}), the Heisenberg's uncertainty principle sets a lower bound to the accuracy that can be reached in performing a measurement on a given system observable. As clearly shown by Feynman in \cite{feynman93}, the connection between uncertainty in classical and quantum systems is mainly given by the following two observations: (i) the quantum mechanical laws of the physical world approach very closely to the classical ones when the size of the dynamical systems involved in the experiment increases; (ii) the concept of probability is not altered in quantum mechanics: what radically changes are the methods of determining the outcome probabilities, which are provided by the postulates of quantum mechanics.

In this chapter, we will analyze, for a given dynamical system, what is the ultimate limit in estimating its state by using \textit{binary sensors}, which provide the minimal amount of information from the measurement process. In particular, we will evaluate when the system is \textit{observable} as a function of the number of sensors and their placement within the system domain, and proper mathematical estimators will be introduced with the aim to increase the information that can be extracted from the system. Furthermore, as it will be shown in the second part of this chapter, the presence of measurement noise can be a helpful source of information when a probabilistic approach to estimation is adopted. The above stated paradigm will be recast in the branch of \textit{noise-assisted estimation/metrology}.

\subsection*{Binary sensors}

Many examples, requiring the use of a binary sensor as measurement device, can be found both from classical and quantum systems. Binary (threshold) sensors are measurement devices, which are nowadays commonly exploited for monitoring/control aims in a wide range of application domains.

A non-exhaustive list of existing binary sensors, involving classical systems, includes: industrial sensors for brushless dc motors, liquid levels, pressure switches; chemical process sensors for vacuum, pressure, gas concentration and power levels; switching sensors for exhaust gas oxygen (EGO or lambda sensors), ABS, shift-by-wire in automotive applications; gas content sensors ($CO$, $CO_2$, $H_2$, etc.) for gas \& oil industry; traffic condition indicators for \textit{asynchronous transmission mode} (ATM) networks; and medical sensors/analyses with dichotomous outcomes.

Regarding nanoscale systems, instead, any ideal detector of quantum system dynamics performs projective measurements. In case the system is a qubit, the measurements have only two possible values, $0$ and $1$, corresponding to the two qubit states. Accordingly, the probabilities of these outcomes are equal to the matrix elements of the qubit's density matrix, which describes the statistical ensemble of the corresponding quantum state~\cite{KorotkovPRB}. After the measurement, the quantum state is projected onto the subspace corresponding to the measurement outcome. More practically, quantum binary sensors can be modelled and realized in several ways. (i) One can use the model of \textit{indirect projective measurements}, for which the quantum system (whose state has to be estimated) interacts with an ancillary qubit, that is later measured by means of standard projective measurements. (ii) We can adopt a \textit{linear detector in binary-outcome mode}: the output of a linear detector is compared with a certain threshold, so as to determine if the output falls into the region given by result $0$ or the region corresponding to result $1$. (iii) Binary detectors from solid-state qubits: the qubit used as the quantum binary sensor is realized by a superconducting loop, which is interrupted by a Josephson junction~\cite{Katz2006}. The qubit, then, is measured by changing the magnetic flux through the loop, so that only one of the two states of the qubit can tunnel outside the potential profile given by the magnetic field. Finally, the tunnelling event or its absence, which is a binary measurement, is recorded by using another superconducting quantum interference device~\cite{KorotkovPRB}.

In all of these applications, binary sensors represent the only viable solution for \textit{real-time monitoring}. In any case, especially if used for macroscopic systems, they provide a remarkably more cost-effective alternative to continuous-valued sensors at the price of an accuracy deterioration which can, however, be compensated by using many binary sensors (for different variables and/or thresholds) in place of a single one or few linear sensors. In other words, the idea is that by a large number of low-resolution sensing devices it is possible to achieve the same estimation accuracy that a few (possibly a single one) high-resolution sensors could provide.

Moreover, binary (threshold) measurements arise naturally in the context of networked state estimation when, in order to save bandwidth and reduce the energy consumption due to data transmission, the measurements collected by each remote sensor are compared locally with a time-varying threshold and only the information pertaining to the threshold-crossing instants is transmitted to the computing center. This latter setting falls within the framework of event-based or event-triggered state estimation~\cite{BaBeCh,Lazar,likelihood}, and is more challenging as compared to the usually addressed settings due to the minimal information exchange. A binary measurement just conveys a minimal amount (i.e. a single bit) of information, implying possible communication bandwidth savings and consequently a greater energy efficiency. Thus, it is of paramount importance to fully exploit the little available information by means of smart estimation algorithms.

In the existing literature \cite{state_reconstruction,Irr-sampling} investigated observability and observer design for linear time-invariant (LTI) continuous-time systems under binary-valued output observations, while the work in \cite{Wang1,Wang2} addressed system identification using binary sensors. A possible solution for coping with the high nonlinearity associated with binary measurements within a stochastic framework is particle filtering~\cite{Djuric_2,Ristic}. Such techniques, however, suffer from the so-called curse of dimensionality (i.e., the exponential growth of the computational complexity as the state dimension increases) and from the lack of guaranteed stability and performance, being based on Monte Carlo integration.

Both limitations are discussed and solved for wide classes of dynamical systems. In particular, state estimation with binary (threshold) output measurements will be addressed by following a \textit{moving horizon estimation} (MHE) approach. MHE techniques were originally introduced to deal with uncertainties in the system knowledge~\cite{Jazwinski} and, in recent years, have gathered an increasing interest thanks to their capability of taking explicitly into account constraints on state and disturbances in the filter design~\cite{RaoRawLee01}, and on the possibility of having guaranteed stability and performance even in the nonlinear case~\cite{RaRaMa03,NLMHE,AlBaBaZavCDC10}. Moreover, MHE has been successfully applied in many different contexts, ranging from switching and large-scale systems~\cite{AlBaBaTAC05,GuoHuang13,FaFerrSca10,HabVerh13,SchnHannMarq15} to networked systems~\cite{Farina1,Farina2,quantized_measurement}.

The novel contributions here introduced in solving state estimation problems by using binary measurements can be split into two approaches:
\begin{itemize}
\item
{\bf Deterministic approach}: No probabilistic description of the system disturbance and measurement noise is supposed to be available. The estimates are computed by minimizing suitable cost functions, which are defined over a given time-horizon (advancing in time) of finite length, possibly subject to linear inequality constraints accounting for the threshold measurements. Specifically, for such a case two different cost functions are proposed and analyzed. About the first cost function, only the threshold-crossing instants are taken into account, so as to penalize the distance of the expected continuous outputs (based on the state estimates) from the threshold at those instants. The main advantage of this solution is that the resulting cost function is quadratic. The second cost function, instead, exploits all the available information by defining a piece-wise quadratic term which accounts for all the available binary measurements, but requires the solution of a convex optimization problem at each time instant. Stability results will be proved for the two different choices of the cost function.
\item
{\bf Probabilistic approach}: According to the deterministic approach, information contributions from binary measurements are given only in correspondence of the sampling instants in which some outcomes change their values. As a consequence, there is no or very little information available for estimation purposes whenever no or very few binary sensor switchings occur. Therefore, a probabilistic approach is recommended. In this regard, we can exploit binary sensor readings to infer information about the probability distribution of the variable of interest. To clarify this point, let us assume that a very large number of binary sensors of the same type (i.e. measuring the same variable with the same threshold) is available and that the distribution of their measurement noise (e.g. Gaussian with zero mean and given standard deviation) is known. Then, thanks to the high number of measurements, the relative frequency of $1$ (or $0$) values occurring in the sensor readings could be considered as a reasonable estimate of the probability that the sensed variable is above (or below) the threshold and this, in turn, by exploiting the knowledge of the measurement noise distribution, allows to extract information about the location of the value of the sensed variable with respect to the threshold. The above arguments suggest that, adopting a probabilistic approach to estimation using binary measurements, the presence of measurement noise can be a helpful source of information. Accordingly, a \textit{noise-assisted paradigm} for state estimation with binary measurements can be stated by taking advantage of the fact that the measurement noise randomly shifts the analog measurement, thus making possible to infer statistical information on the sensed variable.
\end{itemize}

\section{Problem formulation}

Now, let us consider the problem of recursively estimating the state $|x_{t}\rangle$ of the following discrete-time nonlinear dynamical system:
\begin{equation}\label{1}
\begin{array}{rcl}
|x_{t+1}\rangle  & = &  f(|x_{t}\rangle,|u_{t}\rangle) + |w_{t}\rangle \\
z_{t}^{i} & = &  g^{i}(|x_{t}\rangle) + v_{t}^{i},\hspace{3mm}i=1,\ldots,p
\end{array}
\end{equation}
from \textit{binary (threshold)} measurements
\begin{equation} \label{2}
\begin{array}{rclcl}
y_{t}^{i} & = & h^i(z_{t}^{i}) & = &  \left\{
\begin{array}{ll} +1, & \mbox{if }  z_{t}^{i} \geq \tau^{i} \\
                         -1,   & \mbox{if }  z_{t}^{i}  < \tau^{i}
\end{array}
\right.
\end{array}.
\end{equation}
In (\ref{1})-(\ref{2}): $|x_{t}\rangle\in \mathbb{R}^n$ is the state to be estimated, $|u_{t}\rangle \in \mathbb{R}^m$ is a known input, $|z_{t}\rangle = {\rm col} \left( z_t^i \right)_{i=1}^p \in \mathbb{R}^p$, and $\tau^i$ is the threshold of the $i-$th binary sensor. Instead, $|w_{t}\rangle$ and $|v_{t}\rangle = {\rm col} \left( v_t^i \right)_{i=1}^p$ are the process and, respectively, measurement noises assumed \textit{unknown but bounded}. The process noise is an additional disturbance affecting the system dynamics, which accounts for uncertainties in the mathematical model, while the measurement noise models the effects of the environment on the measurement devices. Notice from (\ref{1})-(\ref{2}) that sensor $i$ provides a binary measurement $y_t^i \in \{-1,+1\}$ (two-level measurement quantization) according to whether the noisy function of the state $z_{t}^{i} = g^{i}(|x_{t}\rangle) + v_{t}^{i}$ falls below or above the threshold $\tau^i$. Hereafter, for the sake of simplicity, we will use in the next sections of this chapter $x_{t}$, $u_{t}$, $w_{t}$ and $v_{t}$ instead of $|x_{t}\rangle$, $|u_{t}\rangle$, $|w_{t}\rangle$ and $|v_{t}\rangle$. The bra-ket notation will be resumed when quantum dynamical systems will be taken into account.

Let us observe that the aforementioned problem includes, as a special instance, the case of quantized sensors with an arbitrary number of levels. Indeed, a $d$-level quantizer, for generic $d  \geq 2$, can be easily realized by using $d-1$  binary (threshold) sensors for the same physical variable but with appropriate different thresholds. The considered setting with multiple binary sensors (which can measure the same physical variable with different thresholds, but also different physical variables) is clearly more general. Moreover, it is worth to point out how the problem of estimating the state of a dynamical system via the adoption of binary sensors reveals a very deep connection with the observability properties of the system. In this regard, let us recall from control theory the definition of \textit{observability}: observability is a measure of the observer capability to infer the state of a dynamical system from the knowledge of some external outputs coming from the measurement devices. For the analyzed case, at least for the deterministic approach, the measurement outcomes are obtained by sampling the outputs of the system in correspondence of a set of non-periodic and irregularly-spaced time instants. As a matter of fact, the available information from binary sensors is set in correspondence of the threshold-crossing instants. Thus, under these hypotheses, the state observability may be lost, and only the moving horizon approach will guarantee an asymptotically bounded estimation error.

\section{Deterministic approach}

In this section, the results of \cite{GherardiniCDC,GherardiniAutomatica} are discussed. In a deterministic context, the available information from a binary sensor is essentially concentrated at the sampling instants in which the measurement outcomes have switched value~\cite{Irr-sampling}. However, as shown in \cite{GherardiniCDC,GherardiniAutomatica}, some additional information from the measurements can be exploited also in the non-switching sampling instants by penalizing the values of the estimated quantities, whose predicted measurement is on the opposite side with respect to the binary sensors reading.

In this regard, let us assume that the discrete-time dynamical system of (\ref{1}) is \textit{linear}, i.e.
\begin{equation}\label{3}
\begin{array}{rcl}
x_{t+1}  & = &  Ax_{t}+Bu_{t}+ w_{t} \\
z_{t}^{i} & = &  C^{i}x_{t}+ v_{t}^{i},\hspace{3mm}i=1,\ldots,p
\end{array}
\end{equation}
where $A, B, C = {\rm col} \left( C^i \right)_{i=1}^p$ are matrices of compatible dimensions. The binary measurement equation, instead, remains unchanged and is given again by (\ref{2}). The system (\ref{2})-(\ref{3}) represents a very special instance of a linear system with output nonlinearity, generally called \textit{Wiener} system~\cite{Wiener1}. However, due to the discontinuous nature of the measurement function of (\ref{2}), all the standard state estimation techniques for Wiener systems that require a certain smoothness of the output nonlinearity (see for example~\cite{Wiener2} and the references therein) cannot be applied. In fact, while general-purpose nonlinear estimators accounting for such a discontinuity (e.g. the particle filter) could be used, the peculiar nature of the considered output nonlinearity deserves special attention and, for optimal exploitation of the poor available information, the development of ad-hoc receding-horizon estimators, that will be presented in the sequel, is required.

Before addressing the estimation problem, some preliminary considerations on the information provided by binary multi-sensor observations are useful. With this respect, it has been pointed out in~\cite{Irr-sampling} that, in the continuous-time case, the information provided by a binary sensor is strictly related to the threshold-crossing instants. In this case, indeed, at every instant corresponding to a discontinuity of the binary signal $y^i$, it is known that the signal $z^i$ is equal to the threshold value $\tau^{i}$, implying that the \textit{linear} measurement $z^i = \tau^i$ is available. Hence, observability with binary sensors for continuous-time linear systems can be analyzed within the more general framework of observability for \textit{irregularly sampled systems}~\cite{Irr-sampling}. In particular, observability can be ensured when the number of threshold-crossing instants, which corresponds to the number of available irregularly sampled linear measurements, is sufficiently large.

The situation is, however, different for \textit{discrete-time systems}. To see this, let us consider a generic time instant $k$ in which the binary signal $y_k^i$ changes sign, i.e., $y_{k}^{i}y_{k+1}^{i}<0$. Then, it is not possible to state, as in the continuous-time case, that $z_{k}^{i}$ coincides with the threshold $\tau^i$. Conversely, it can be simply concluded that there exists $\alpha \in [0,1]$ such that
\begin{equation}\label{4}
\alpha \, z_{k}^{i}+(1-\alpha) \, z_{k+1}^{i}=\tau^{i} \, ,
\end{equation}
where the exact value of $\alpha$ is clearly unknown and unobservable from the binary measurements. Notice that (\ref{4}) simply states that if the binary output $y_k^i$ switches from discrete time $k$ to $k+1$, then the threshold $\tau^i$ must lie in the interval between $z_k^i$ and $z_{k+1}^i$. In view of (\ref{4}), such discrete time instants $k$, at which the output of some binary sensor changes value, will be more appropriately referred to as \textit{output switching} or simply \textit{switching} instants, instead of \textit{threshold-crossing} instants like in the continuous-time case considered in \cite{Irr-sampling}. It is easy to see that (\ref{4}) corresponds to an \textit{uncertain linear measurement}, i.e.
\begin{equation}\label{5}
\alpha \, z_{k}^{i}+(1-\alpha) \, z_{k+1}^{i} = C^i x_{k} + \delta^i_{k} + \zeta^i_{k},
\end{equation}
where $\delta^i_{k} $ is the uncertainty and $\zeta^i_{k}$ the measurement noise given by the following relations:
\begin{eqnarray}
\delta^i_{k} &=& (1-\alpha)C^{i}(A-\mathbbm{1})x_{k}+(1-\alpha)C^{i}Bu_{k},\label{6} \\
\zeta^i_{k}   &=& \alpha\, v_{k}^{i}+(1-\alpha) \, v_{k+1}^{i} + (1-\alpha) \, C^{i} \, w_{k}\label{7} \, ,
\end{eqnarray}
with $\mathbbm{1}$ equal to the identity operator. As a consequence, even in presence of bounded disturbances, the uncertainty associated with the measurement (\ref{4}) depends on $x_k$ and $u_k$. Recalling that, in general in the context of state estimation for uncertain systems, boundedness of the state trajectories is a prerequisite for the boundedness of the estimation error - see, for instance, the discussion in Section 2.1 of \cite{BlMi} - our attention will be restricted to the case of bounded state and input trajectories by making the following assumption:\vspace{.2cm}
\begin{enumerate}[\bf {A}1]
\item At any time $t$, the vectors $x_t$, $u_t$, $w_t$, $v^i_t, \, i=1, \ldots, p$, belong to the compact sets $X$, $U$, $W$, and $V^i, \, i=1, \ldots,p$, respectively.
\end{enumerate}\vspace{.2cm}
In practice, the compact sets $X$, $U$, $W$, $V^i$ need not be known by the estimator; they will only be used for stability analysis purposes.

\subsection{Moving horizon estimation}

In order to estimate the state $x_t$ of the linear system $(\ref{3})$ given the binary measurements (\ref{2}), a MHE approach is adopted. Then, by considering a sliding window $\mathfrak W_t = \{t-N, t-N+1, \ldots, t\}$, the goal is to find estimates of the state vectors $x_{t-N},\ldots,x_{t}$ on the basis of the information available in  $\mathfrak W_t$ and of the state prediction $\overline{x}_{t-N}$ at the beginning of $\mathfrak W_t$.  Let us denote by $\hat{x}_{t-N|t},\ldots,\hat{x}_{t|t}$ the estimates of $x_{t-N},\ldots,x_{t}$, respectively, to be obtained at any stage $t$.

Following the discussion at the end of the previous section, a first natural approach for constructing a MH estimator would amount to considering the information provided by the switching instants inside the sliding window $\mathfrak W_t$, in order to define the cost-function to be minimized. Accordingly, for any time instant $t \ge N$ and for any sensor index $i$, let us define the set $\mathfrak{I}_{t}^i$ of switching instants as
\begin{equation}\label{8}
\mathfrak{I}^{i}_t = \{ k \in \mathfrak{W}_t : k+1  \in \mathfrak{W}_t \mbox{ and } y_k^i \, y_{k+1}^i < 0 \} .
\end{equation}
Then, the following least-squares cost function can be defined:
\begin{eqnarray}\label{9}
J^A_{t} &=& \|\hat{x}_{t-N|t}-\overline{x}_{t-N}\|^{2}_{P}+ \sum_{k=t-N}^{t-1}\|\hat{x}_{k+1|t}-A\hat{x}_{k|t}-Bu_{k}\|^{2}_{Q}\nonumber \\
&+& \sum_{i=1}^{p}\sum_{k\in\mathfrak{I}^{i}_t} \|C^{i} \, \hat{x}_{k|t}-\tau^{i}\|^{2}_{R^{i}},
\end{eqnarray}
where the positive definite matrices $P\in\mathbb{R}^{n\times n}$, $Q\in\mathbb{R}^{n\times n}$ and the positive scalars $R^i , \, i = 1, \ldots p$, are design parameters to be suitably chosen. The first term, weighted by the matrix $P$, penalizes the distance of the state estimate at the beginning of the sliding window from the prediction $\overline{x}_{t-N}$. The second contribution, weighted by the matrix $Q$, takes into account the evolution of the state in terms of the state equation (\ref{3}). Finally, for each sensor $i$ the third term weighted by the scalar $R^i$ penalizes the distances of the expected output (based on the state estimates) $C^{i} \, \hat{x}_{k|t}$ from the threshold $\tau^i$ at the switching instants. Let us note that considering the distance from the threshold at the switching instant is equivalent, for sampled-data systems, to considering the beginning of the time interval $[k T_s, (k+1) T_s]$ in which the threshold crossing happens. As a matter of fact, since for a sampled-data system a binary sensor does not provide a precise information on the threshold crossing instant in the interval $[k T_s, (k+1) T_s]$, considering the distance from the threshold at the beginning of the time interval is just a choice, not necessarily optimal. As an alternative, with little modifications, one could consider for instance the middle point of the interval. Such modifications would not affect the properties (e.g. stability) of the estimator. Thus, at each time $t \ge N$, the estimates in the window $\mathfrak W_t$ can be obtained by solving the following optimization problem. \\ \\
\textbf{Problem $E_{t}^{A}$:} Given the prediction $\overline{x}_{t-N}$, the input sequence $\{ u_{t-N}, \ldots, u_{t-1} \}$, and the sets $\mathfrak{I}^{i}_t , \, i = 1, \ldots, p$,  find the optimal estimates $\hat{x}^{\circ}_{t-N|t},\ldots,\hat{x}^{\circ}_{t|t}$ that minimize the cost function (\ref{9}). \vspace{.2cm}
Concerning the propagation of the estimation procedure from Problem $E_{t}^{A}$ to Problem $E_{t+1}^{A}$, different prediction strategies may be adopted.
For instance, a first possibility consists of assigning to $\overline{x}_{t-N+1}$ the value of the estimate of $x_{t-N+1}$ made at time instant $t$, i.e. $\bar{x}_{t-N+1} = \hat{x}^{\circ}_{t-N+1|t}$. As an alternative, following \cite{NLMHE}, the state equation of the noise-free system can be applied to the estimate $\hat{x}^{\circ}_{t-N|t}$. In this case, the predictions are recursively obtained by
\begin{equation}\label{10}
\overline{x}_{t-N+1}=A\hat{x}_{t-N|t}^{\circ}+Bu_{t-N},\hspace{3mm}t=N,N+1,\ldots \, .
\end{equation}
Such a recursion is initialized with some a priori prediction $\overline x_0$ of the initial state vector. Hereby, this latter possibility will be adopted as it will facilitate the derivation of the stability results (see the next subsection).

The main positive feature of Problem $E_{t}^{A}$ is that it admits a closed-form solution since the cost function (\ref{9}) depends quadratically on the estimates $\hat{x}_{t-N|t},\ldots,\hat{x}_{t|t}$ (for the readers' convenience an explicit expression for the solution is reported in the Appendix~\ref{chapter:appA}). On the other hand, such a cost takes into account \textit{only} the information pertaining to the switching instants, which, however, is intrinsically uncertain. In order to overcome such a limitation, a different cost function can be considered by taking into account all the time instants in the sliding window $\mathfrak W_t$. To this end, for any sensor $i = 1 , \ldots, p$, let us define the functions
\begin{equation}\label{11}
\omega^i(z^{i},y^i)= \left\{ \begin{array}{ll}
1, & \mbox{if} ~ \left( z^i - \tau^i \right) y^i < 0 \\
0, & \mbox{otherwise}
\end{array} \right.
\end{equation}
Suppose now that at time $k$ the sensor $i$ provides a measurement $y^i_k = 1$. Then, the information provided by such a measurement is that the linear measurement $z^i_k$ is above the threshold $\tau^i$, i.e. belongs to the semi-interval $[\tau^i, +\infty)$. Such information can be included in the cost function by means of a term of the form $\omega^{i}(C^{i}\hat{x}_{k|t},1)~\|C^{i}\hat{x}_{k|t}-\tau^{i}\|^{2}_{R^{i}}$, which penalizes the distance of the expected output $C^{i}\hat{x}_{k|t}$ from $[\tau^i, +\infty)$. Similarly, in the case $y^i_k = -1$, a term of the form $\omega^{i}(C^{i}\hat{x}_{k|t},-1)~\|C^{i}\hat{x}_{k|t}-\tau^{i}\|^{2}_{R^{i}}$ can be used to penalize the distance of the expected output $C^{i}\hat{x}_{k|t}$ from $(-\infty,\tau^i]$. Summing up, the inclusion of such terms gives rise to a cost function of the following form:
\begin{eqnarray}\label{12}
J^B_{t} &=& \|\hat{x}_{t-N|t}-\overline{x}_{t-N}\|^{2}_{P}  + \sum_{k=t-N}^{t-1}\|\hat{x}_{k+1|t}-A\hat{x}_{k|t}-B u_{k}\|^{2}_{Q}\nonumber \\
&+&\sum_{i=1}^{p}\sum_{k=t-N}^{t} \omega^i (C^{i}\hat{x}_{k|t},y_{k}^{i}) \| C^{i}\hat{x}_{k|t}-\tau^{i}\|^{2}_{R^{i}} \, .
\end{eqnarray}
While a closed-form expression for the global minimum of (\ref{12}) does not exist, since $J_t^B$ is piece-wise quadratic, it is easy to see that the cost $J_t^B$ enjoys some nice properties. In fact, while each function $\omega^i \left( C^i \hat{x}_{k|t}, y_{k}^{i}\right)$ per se is discontinuous, the product $\omega^i \left( C^i \hat{x}_{k|t}, y_{k}^{i}\right)\| C^i \hat{x}_{k|t} - \tau^i \|^2_{R^i}$ is continuous since at the points of discontinuity of
$\omega^i \left( C^i \hat{x}_{k|t}, y_{k}^{i}\right)$, i.e. for $C^i \hat{x}_{k|t} = \tau^i$, the product vanishes. Further, for similar reasons, also the derivative $2 \omega^i \left( C^i \hat{x}_{k|t}, y_{k}^{i}\right) R^i (C^i )' ( C^i \hat{x}_{k|t} - \tau^i )$ of the product turns out to be continuous even at $C^i \hat{x}_{k|t} = \tau^i$. Thus, the product $\omega^i \left( C^i \hat{x}_{k|t}, y_{k}^{i}\right)\| C^i \hat{x}_{k|t} - \tau^i \|^2_{R^i}$ is continuously differentiable on $\mathbb R^n$, such that the overall cost function $J_t^B$ is continuously differentiable with respect to the estimates $\hat{x}_{t-N|t},\ldots,\hat{x}_{t|t}$ and also strictly convex (since $P>0$ and $Q>0$). Hence, standard optimization routines can be used in order to find its global minimum. Clearly, since an optimization has to be performed, it is also reasonable to include constraints accounting for the available information on the state trajectory so that the solver can work on a bounded solution set. In particular, in order to preserve convexity, it is advisable to consider a convex set $\mathcal X$ containing $X$ (if $X$ is convex, one can simply set $\mathcal X = X$; in general, choosing $\mathcal X$ as a convex polyhedron is preferable so that only linear constraints come into play). Then, at any stage $t=N,N+1,\ldots$, the following optimization problem has to be solved. \\ \\
\textbf{Problem $E_{t}^{B}$:} Given the prediction $\overline{x}_{t-N}$, the input sequence $\{ u_{t-N}, \ldots, u_{t-1} \}$, the measurement sequences $\{ y^i_{t-N}, \ldots, y^i_t , \, i = 1, \ldots, p \}$, find the optimal estimates $\hat{x}^{\circ}_{t-N|t},\ldots,\hat{x}^{\circ}_{t|t}$ that minimize the cost function (\ref{12}) under the constraints $\hat{x}^{\circ}_{k|t} \in \mathcal X$ for $k= t-N , \ldots,t$.
\vspace{.3cm}

Also in this case, the predictions $\overline{x}_{t-N}$ are supposed to be recursively obtained via equation (\ref{10}) starting from a prior prediction $\overline x_0$. Of course, if no information on the set $X$ is available or if it is preferable to resort to an unconstrained optimization routine, one can simply let $\mathcal X = \mathbb R^n$. As a final remark, it is worth pointing out that for the two previously presented optimization problems there is a \textit{trade-off} between estimation accuracy and computational cost. As a matter of fact, the cost in Problem $E_{t}^{A}$ is quadratic but accounts only for part of the information provided by the sensors, while Problem $E_{t}^{B}$ accounts for all the available information but requires a convex optimization program to be solved. To summarize
\begin{itemize}
\item
The solution of Problem $E_{t}^{A}$ requires simply the minimization of a strictly convex quadratic form in $(n+1) N$ variables, where $n$ is the plant order. Standard techniques like Gaussian elimination can solve this kind of problems with complexity $O(n^3 N^3)$, but faster algorithms are available. This means that this approach is much computationally cheaper as compared to particle filtering algorithm, which usually require in the order of $O(10^n)$ particles to provide satisfactory performance.
\item
As for the solution of Problem $E_{t}^{B}$, it entails the minimization of a convex and continuously differentiable piecewise quadratic cost function. It is known that this kind of problems can be solved in finite time by means of sequential quadratic programming~\cite{PWCP}. Further, many computationally efficient algorithms are available which are able to handle problems with hundreds of optimization variables~\cite{OPT2} and enjoys super-linear convergence~\cite{bounds}. Nevertheless, application of Problem $E_{t}^{B}$ is possible only when the number $n$ of state variables is not too large and the sampling interval is sufficiently large so as to allow the optimization to terminate. In the other cases, one must resort to Problem $E_{t}^{A}$.
\end{itemize}

\subsubsection{Accounting for additional constraints}

Provided that some information on the bounds of the process disturbance $w_t$ and measurement noises $v_t^i $ is available, additional constraints can be considered in the determination of the state estimates. For instance, considering a convex (usually polyhedral) set $\mathcal W$ containing $W$, one can impose the constraints
\begin{equation}\label{13}
\hat{x}_{k+1|t}-A\hat{x}_{k|t}-B u_{k} \in \mathcal W \, , \quad k = t-N, \ldots, t-1
\end{equation}
in the solution of the optimization problem. Moreover, assuming the knowledge of upper bounds $\rho_{V}^{i}$ on the amplitudes $|v_t^i | , \, i=1, \ldots, p$, of the measurement noises, for each $k$ and each $i$, the constraints
\begin{equation}\label{14}
\begin{cases}
C^{i}\hat{x}_{k|t}<\tau^{i} +\rho_{V}^{i},\hspace{3mm}\text{if}\hspace{2mm}y_{k}^{i}=-1\\
C^{i}\hat{x}_{k|t}>\tau^{i} - \rho_{V}^{i},\hspace{2mm}\text{if}\hspace{2mm}y_{k}^{i}=1
\end{cases}
\end{equation}
can be imposed. With this respect, it is an easy matter to see that the constraints in (\ref{14}) define a polyhedron in the state space as summarized in the following proposition (the proof is reported in the Appendix~\ref{chapter:appA}). \\ \\
{\bf Proposition 1.1:}
Given the vector $\hat{\chi}_{t}= vec \left ( [ \hat x_{t-N|t} \cdots \hat x_{t|t}  ]' \right )$ of the estimates in the observation window,
the constraints in (\ref{14}), for $k = 0, \ldots, N$ and $i=\, \ldots, p$, can be written in compact form as
\begin{equation}\label{15}
\Gamma_{t}\hat{\chi}_{t}<\gamma_{t},
\end{equation}
where
\begin{equation}\label{16}
\begin{split}
&\Gamma_{t} = \left[\Phi_{t}(C\otimes I_{N})\right]\in\mathbb{R}^{pN\times nN},\hspace{.5cm}\gamma_{t} = \left[\Phi_{t}vec(\mathcal{T}')+vec(\mathcal{V})\right]\in\mathbb{R}^{pN}, \\
&\Phi_{t}= -diag(y_{t-N}^{1},\ldots,y_{t}^{1},y_{t-N}^{2},\ldots,y_{t}^{2},\ldots,y_{t-N}^{p},\ldots,y_{t}^{p})\in\mathbb{R}^{pN\times pN}, \\
\end{split}
\end{equation}
and
\begin{equation}
\mathcal{T} = \begin{bmatrix} \tau^{1} &  \cdots & \tau^{1} \\
\vdots & \vdots & \vdots \\
\tau^{p} & \cdots & \tau^{p}
\end{bmatrix}\in\mathbb{R}^{p\times N}, \; \mathcal{V}= \begin{bmatrix} \rho_{V}^{1} & \cdots & \rho_{V}^{1} \\
\vdots & \vdots & \vdots \\
\rho_{V}^{p} & \cdots & \rho_{V}^{p} \end{bmatrix}\in\mathbb{R}^{p\times N}.
\end{equation}
%\vspace{.3cm}
While the inclusion of the constraints (\ref{13}) and (\ref{15}) in the convex optimization problem $E_{t}^{B}$ is natural, in some circumstances it may be interesting to combine them also with the quadratic cost $J_{t}^{A}$. For example, minimizing $J_{t}^{A}$ under the linear constraints (\ref{15}) can be a way to account for the information concerning the non switching instants without the necessity of considering the piece-wise quadratic cost. In fact, this would result in a quadratic programming problem (being the cost quadratic and the constraints linear) for which many efficient solvers are available. It is worth to point out that what is, among the above mentioned options, the best choice clearly depends on the situation under consideration and, in particular, on the available computational resources, on the available information (the bounds $\rho_{V}^{i}$ may be unknown), and on the necessity (or not) of having estimates satisfying the constraints (since clearly this property is guaranteed only if the constraints are taken into account in the estimator design). Nevertheless, in the next section it will be shown that both costs $J_t^A$ and $J_t^B$ imply some nice stability properties of the resulting MH estimator.

Furthermore, as final remark, let us observe that while the considered system dynamics is linear, we do not have access to the linear measurements $z_t = C x_t + v_t$ but rather to the nonlinear (binary) measurements $y_t^i = h^i(z_t^i)$, for which we cannot apply neither the Kalman filter due to nonlinearity of $h^i(\cdot)$ nor the extended Kalman filter due to the discontinuous nature of $h^i(\cdot)$. It is however worth noting that the simplified quadratic cost $J^A_t$ amounts to considering  a \textit{fictitious} linear measurement of the form $C^i x_{k} = \tau^i + \zeta^i_k$ for each switching instant $k$ in the observation window. In this case and supposing that no constraints are imposed, the estimates could be computed also via a Kalman-like filter.
In all the other cases, i.e. when the piecewise quadratic cost $J^B_t$ is used or constraints are imposed in the optimization, this is no longer possible.

\subsection{Stability analysis}

Here, we analyze the stability properties of the state estimators obtained by solving, at each time instant, either Problem $E^A_t$ or $E^B_t$. Specifically, a complete analysis is first provided in the more involved case of Problem $E^B_t$. This will be followed by a short discussion on the main differences in the analysis with respect to Problem $E^A_t$. Notice that the analysis carried out in \cite{NLMHE} for the nonlinear case cannot be directly applied in the present context, since the binary sensors do not satisfy the observability requirement of \cite{NLMHE}. The proofs of all results can be found in the Appendix~\ref{chapter:appA}.

For each sensor $i$ and for each time instant $t\geq N$, let us denote by $\Theta_{t}^{i}$ the observability matrix concerning the set $\mathfrak{I}_{t}^{i}$ of the switching instants in the observation window $\mathfrak{W}_t$, i.e,
\begin{equation}\label{17}
\Theta_{t}^{i}={\rm col}(C^{i}A^{k-t+N})_{k\in\mathfrak{I}_{t}^{i}}.
\end{equation}
Then, the observability matrix related to the switchings in $\mathfrak{W}_t$ of all binary sensors is
\begin{equation}\label{18}
\Theta_{t}={\rm col}(\Theta_{t}^{i})_{i=1}^{p}.
\end{equation}
Please notice that the observability matrix defined in (\ref{17})-(\ref{18}) is actually related to the linear subsystem (\ref{3}), with output $z_t$, of the overall system (\ref{2})-(\ref{3}) considering only those discrete-time instants at which some binary sensor output switches. Thus, the following \textit{uniform observability} assumption is needed in order to ensure that enough information is provided by the binary sensors in each window $\mathfrak{W}_t$.\vspace{.2cm}
\begin{enumerate}[\bf {A}1]
\setcounter{enumi}{1}
\item \label{19}
For any $ t\geq N$, ${\rm rank}(\Theta_{t})=n$, with $n=\dim(x_{t})$.
\end{enumerate}\vspace{.2cm}
The above uniform observability assumption is made in accordance with the observation that each output switching can be associated with a linear (albeit uncertain) measurement of the form (\ref{4}). Hence, each switching instant $k$ can be thought of as a sampling instant for the linear output $z^i_k$.
This means that observability of the system depends crucially on the output switching instants in each observation window which, in turn, clearly depend on the thresholds and of the time window length $N$. In practice, the threshold (or the thresholds when multiple sensors are available) and the time window length $N$ must be chosen taking into account the system dynamics so as to ensure that such an irregular sampling preserves observability. For instance, when only one binary sensor is available, clearly $N$ should be substantially greater than $2n-1$, with $n = \dim(x_t)$, so as to ensure that at least $n$ output switching instants are present in each observation window. While some analytical results on observability under irregular sampling are available~\cite{Irr-sampling}, the simplest approach amounts to studying, for instance by numerical simulations, how the observability measure $\delta$ varies as a function of the thresholds and of the time window length $N$. Of course, depending on the system dynamics, time-invariant thresholds may not be sufficient to always ensure uniform observability (think for example to the case of a constant linear output). In these cases, observability can be recovered by making each threshold oscillate in the range of variability of the corresponding continuous output $z^i_t$ with a sufficiently high frequency and by choosing $N$ so that each observation window contains a sufficient number of threshold oscillation periods. This latter solution is particularly convenient in case $z^i_t$ is a measurement collected by a remote sensor and a time-varying threshold $\tau^i_t$ is used for transmission scheduling.

Before stating the main stability results, some preliminary definitions are needed. Given a symmetric matrix $S$, let us denote by $\underline{\lambda}(S)$ and $\overline{\lambda}(S)$ the minimum and maximum eigenvalues of $S$, respectively. Further, given a matrix $M$, we define by $\|M\|\equiv\overline{\lambda}(M'M)^{1/2}$ its norm. Given a generic subset $\Psi$ of an Euclidean space, $ \rho_{\Psi}\equiv\overline  \rm sup_{v \in \Psi } \| v \|$. Then, given a generic quantity $G^{i}$ related to the $i-$th binary sensor, let us denote $\overline{G}\equiv\max_{i}\|G^{i}\|$ and $\underline{G}\equiv\min_{i}\|G^{i}\|$. Finally, the \textit{uniform observability measure} associated to the matrices $\Theta_t$ is given by
\[
\delta = \inf_{t \ge N}  \left \| \Theta_t \right\| =  \inf_{t \ge N} {\, \underline{\lambda}(\Theta_{t}'\Theta_{t})^{1/2}} \, .
\]
Recalling that, under assumption {\bf A2}, $\delta >0$, we can state the following result. \\ \\
{\bf Theorem 1.1:}
Let assumptions {\bf A1} and {\bf A2} hold. For each $t \ge N$, let the estimate $\hat x_{t-N,t}^\circ$ be generated by solving Problem $E^B_t$, with $\overline x_{t-N}$ recursively obtained via equation (\ref{10}), and consider the estimation error $e_{t-N} \equiv x_{t-N} - \hat x_{t-N|t}^\circ$. Then, the weighted norm of the estimation error can be recursively bounded as
\begin{equation}\label{19}
\|e_{t-N}\|^{2}_{P}\leq a_{1}\|e_{t-N-1}\|^{2}_{P}+a_{2},\hspace{3mm}t=N,N+1,\ldots
\end{equation}
where
\begin{equation}\label{20}
\begin{split}
& a_{1}=\frac{b_{1}\|A\|^{2}}{b_{2}} , \\
& a_2 = \frac{c_1 \, \| A - \mathbbm{1} \|^2 \, \rho^2_{\mathcal X}+ c_2 \, \| B \|^2 \, \rho_U^2 +  c_3 \, \rho_W^2 + c_ 4 \, \overline{\rho}_V^2 }{b_2}, \\
&b_{1}= \frac{\overline{\lambda}(P)}{\underline{\lambda}(P)} \left[4+\frac{d_1}{\underline{\lambda}(Q)}\left(d_2+\overline{R} \right)\right],\hspace{3mm}b_{2}=\left(\frac{1}{2}+\frac{\delta^{2}\underline{R}}{4\overline{\lambda}(P)}\right) \\
\end{split}
\end{equation}
and $c_1$, $c_2$, $c_3$, $c_4$, $d_1$, $d_2$ are suitable constants (given in the proof). In addition, if the weights $Q$ and $R^{i}$, $i=1,\ldots,p$, are selected such that $a_{1}<1$, the norm of the estimation error turns out to be asymptotically bounded in that
\[
\limsup_{t \rightarrow + \infty} \|e_{t-N}\| \le e^{\circ}_{\infty}\equiv \left ( \frac{a_{2}}{1-a_{1}} \right )^{1/2} \, .
\]
\mbox{   }\hfill $\square$

The reason for analyzing the estimate at the beginning of the observation window is that, due to the nature of the MHE estimation scheme, the estimate $\hat x^\circ_{t-N|t}$ is used to generate the prediction $\bar x_{t-N+1}$ used at time $t+1$. This makes it possible to recursively write $e_{t-N+1} = x_{t-N+1} - \hat x^\circ_{t-N+1|t+1}$ as a function of $e_{t-N} = x_{t-N} - \hat x^\circ_{t-N|t}$. Let us note that even in the noise-free case, i.e., when the process disturbance and the measurement noise are zero and hence $\rho_W = \rho_V = 0$, the asymptotic bound $e_\infty^{\circ}$ on the estimation error does not go to zero due to the presence of the term $c_1 \, \| A - \mathbbm{1} \|^2 \, \rho^2_{\mathcal X}  + c_2 \, \| B \|^2 \, \rho_U^2$ in $a_2$. Indeed, such a term accounts for the intrinsic uncertainty associated with the switching instants in discrete-time. With this respect, it is worth recalling that, when the discrete-time system under consideration is obtained by sampling a continuous-time system, the quantities $\| A - \mathbbm{1} \|$ and $\| B \|$ vanish as the sampling interval $T_s$ goes to zero. This means that the smaller is the sampling interval, the smaller turns out to be the asymptotic bound on the estimation error since the information concerning the switching instants becomes more precise.

Another important issue concerns the solvability of the stability condition $a_1 <1$. In particular, the following result can be readily proved. \\ \\
{\bf Proposition 1.2:} Let assumption {\bf A2} hold. Then, when $\delta >0$,  it is always possible to select the weights $P$, $Q$ and $R^{i}$, $i=1,\ldots,p$, so that $a_{1}<1$. In particular, for given $Q$ and $R^{i}$, $i=1,\ldots,p$, the condition $a_{1}<1$ can be satisfied by letting $P = \varepsilon \overline{P}$, with $\overline{P}$ any positive definite matrix, provided that $\varepsilon$ is suitably small.

Hence, if the observability measure $\delta$ is strictly positive, it is sufficient to choose $P$ sufficiently small in order to ensure the satisfaction of the stability condition $a_1<1$.  This result is in accordance with the well-known results on stability of MHE algorithms which stipulate that stability is ensured provided that the weight on the prediction is sufficiently small \cite{NLMHE}.

For the sake of clarity, the following remarks about the stability results of Theorem~1.1 have to be stated:
\begin{itemize}
\item
Let us consider now the case in which, for each $t \ge N$, the estimate $\hat x_{t-N|t}^\circ$ is generated by solving Problem $E^A_t$, with $\overline x_{t-N}$ recursively obtained via equation (\ref{18}). In particular, the estimates $\hat x^\circ_{t-N} , \ldots, \hat x^\circ_{t}$ are readily obtained as the unique global minimum of the strictly convex quadratic function $J_t^A$. A close inspection of the proof of Theorem~1.1 shows that the same line of reasoning can be applied also for Problem $E^A_t$. The main difference is that, when deriving the lower bound for the optimal cost, each term $\iota(\hat{x}_{k|t}^{\circ},\hat{x}_{k+1|t}^{\circ})$ in the proof of Theorem~1.1 (see Appendix~\ref{chapter:appA}) can be simply replaced with the quantity $\|C^{i}\hat{x}^{\circ}_{k|t}-\tau^{i}\|^{2}$ in accordance with the definition of cost $J^{A}_{t}$. Then an inequality analogous to (\ref{19}) can be derived, with the important difference that, in the definition of the novel $a_2$, $\rho_{\mathcal X}$ can be replaced by $\rho_X$, which is consistent with the fact that the constraint set $\mathcal X$ is not used in the solution of Problem $E^A_t$.
\item
While the foregoing analysis does not account for the possible presence of the additional constraints, analogous results could be easily obtained also when the constraints (\ref{13}) and/or (\ref{15}) are imposed in the determination of the state estimates. In this case, the bound on the estimation error turns out to be smaller thanks to the additional information provided by such constraints.
\item
The extension of the stability results reported here to the case in which the binary measurements are obtained by thresholding nonlinear output maps and/or the system dynamics is nonlinear does not entail particular conceptual difficulties, by combining the analysis of Theorem~1.1 with that of~\cite{NLMHE,AlBaBaZavCDC10}. On the other hand, in this case, establishing a link between the observability properties and the number of threshold crossing instants appears more challenging. Further, for nonlinear output maps, the resulting cost functions need not be convex.
\end{itemize}

\subsection{Numerical examples}

Here, we present some numerical examples in order to show the effectiveness of the proposed MHE algorithms by adopting binary measurements. In particular, two different case-studies will be considered: a first simple example concerns an hydraulic system composed of two tanks, and a second example on networks of $2$-mass $2$-spring oscillators with multiple binary sensors. In both numerical examples, the performance of the estimators has been evaluated in terms of the Root Mean Square Error (RMSE):
\begin{equation}\label{21}
\text{RMSE}(t)=\left(\sum_{l=1}^{L}\frac{\|e_{t,l}\|^{2}}{L}\right)^{\frac{1}{2}},
\end{equation}
where $\|e_{t,l}\|$ is the norm of the estimation error at time $t$ in the $l-$th simulation run, averaged over $L$ Monte Carlo trials. The estimation error is computed at time $t$ on the basis of the estimate $\hat{x}^{\circ}_{t-N+1|t}$.

\subsubsection{Hydraulic systems}

\begin{figure}[h!]
\centering
\includegraphics[scale=4]{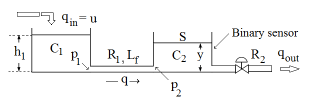}
\caption{Hydraulic system composed of two tanks in cascade.}
\label{fig:serbatoi}
\end{figure}

\begin{figure}[h!]
\centering
\includegraphics[scale=2.6]{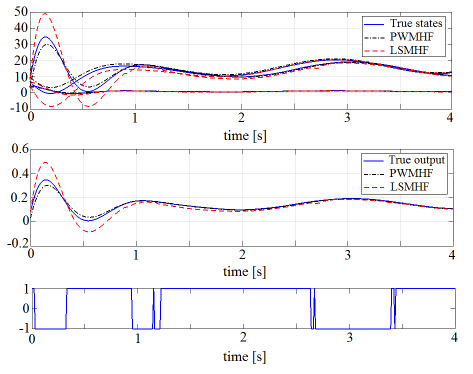}
\caption{True values and estimates of both the state and the output of the system and the binary sensor signal for a randomly chosen simulation.}
\label{fig:estimates}
\end{figure}

\begin{figure}[h!]
\centering
\includegraphics[scale=2.6]{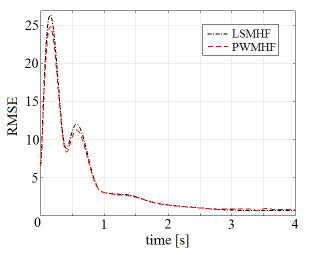}
\caption{RMSEs of the LSMH and PWMH filters.}
\label{fig:rmse_y}
\end{figure}

Let us consider the following continuous-time linear dynamical system:
\begin{equation}
A_{c}=\begin{pmatrix} 0 & 0 & -\frac{1}{C_{1}} \\ 0 & -\frac{1}{R_{2}C_{2}} & \frac{1}{C_{2}} \\ \frac{1}{L_{f}} & -\frac{1}{L_{f}} & -\frac{R_{1}}{L_{f}}
\end{pmatrix}\hspace{5mm}B_{c}=\begin{pmatrix} \frac{1}{C_{1}} \\ 0 \\ 0 \end{pmatrix}\hspace{5mm}
C=\begin{pmatrix} 0 & \frac{C_{2}}{S} & 0 \end{pmatrix},
\end{equation}
that models an hydraulic system composed of two tanks in cascade (see Fig.~\ref{fig:serbatoi}), in which $C_{1}$ and $C_{2}$ are the hydraulic capacities of the two tanks, $R_{1}$ and $R_{2}$ the hydraulic resistances of the connection pipe between the tanks and the output conduit, respectively, $L_{f}$ is the inertance of the connection pipe and $S$ is the output tank area. In correspondence of the second tank a binary sensor is placed, whose threshold value has been chosen equal to $0.17$ $[m]$.
\\
The state $x$ of the system is represented by the vector $(p_{1},p_{2},q)'$, where $p_{1}$ and $p_{2}$ are the pressures in the connection pipe and in the output conduit and $q$ is the flow-rate of liquid in the connection pipe. The values of $C_{1}$, $C_{2}$, $R_{1}$, $R_{2}$, $L_{f}$ and $S$ have been taken equal to $0.05$ $[Pa^{-1}\hspace{0.5mm}m^{3}]$, $0.01$ $[Pa^{-1}\hspace{0.5mm}m^{3}]$, $2$ $[Pa\hspace{0.5mm}s\hspace{0.5mm}kg^{-1}]$, $15$ $[Pa\hspace{0.5mm}s\hspace{0.5mm}kg^{-1}]$, $2$ $[Pa\hspace{0.5mm}s^{2}\hspace{0.5mm}m^{-3}]$ and $1$ $[m^{2}]$ respectively. Finally, the input signal $u$ is supposed to be characterized by a periodic behavior, i.e. $u=a~\sin(2\pi ft)+u_{0}$, with $a=0.75$ $[m^{3}s^{-1}]$, $f=0.5$ $[Hz]$ and $u_{0}=1$ $[m^{3}s^{-1}]$. The components of the initial state $x_{0}$ and the noises $w_{t}$ and $v_{t}$ are supposed to be mutually independent random variables uniformly distributed in the intervals $[0,10]$, $[-10^{-2},10^{-2}]$ and $[-10^{-2},10^{-2}]$ and the weight matrices $P$, $Q$ and $R$ are taken equal to $10^{6}\mathbbm{1}_{3}$, $10^{-8}\mathbbm{1}_{3}$ and $10^{6}$, respectively. The duration of each simulation experiment is fixed to $800\hspace{0.5mm}T_{s}$, where the sampling time $T_{s}$ is equal to $0.01$ $[s]$. \\
Now, for the sake of brevity, we shall denote as the \textit{Least-Squares Moving Horizon Filter} (LSMHF) and as the \textit{Piece-Wise Moving Horizon Filter} (PWMHF) the filters obtained by solving, respectively, Problem $E_{t}^{A}$ and Problem $E_{t}^{B}$. The PWMHF has been implemented by means of the Matlab Optimization Toolbox, and in particular by using the routine \textit{fminunc}. Fig. \ref{fig:estimates} illustrates the behaviour of the true values and the estimates of both the state and the output of the system for a randomly chosen simulation, along with the binary sensor signal, where the number of samples $N$ of the estimation sliding window is equal to $5$.
\\
In the considered settings, LSMHF and PWMHF have a similar behavior, as it can been seen especially in Fig.~\ref{fig:rmse_y}, where the RMSEs for the proposed filters are plotted. The PWMHF exhibits better performance in the transient thanks to the additional information taken into account in the definition of cost $J_t^B$.
\\
Moreover, as a final remark, it can be noted that, due to the binary sensor nonlinearity, the presence of a single sensor implies that a certain transient time (i.e., a certain number of threshold crossings) is needed by the filters.

\subsubsection{2-mass 2-spring oscillators}

\begin{figure}[h!]
\centering
\includegraphics[scale = 4]{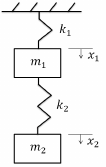}
\caption{$2$-mass $2$-spring mechanical oscillator of example 1.}
\label{fig:oscillators}
\end{figure}

\begin{figure}[h!]
\centering
\includegraphics[scale = 0.625]{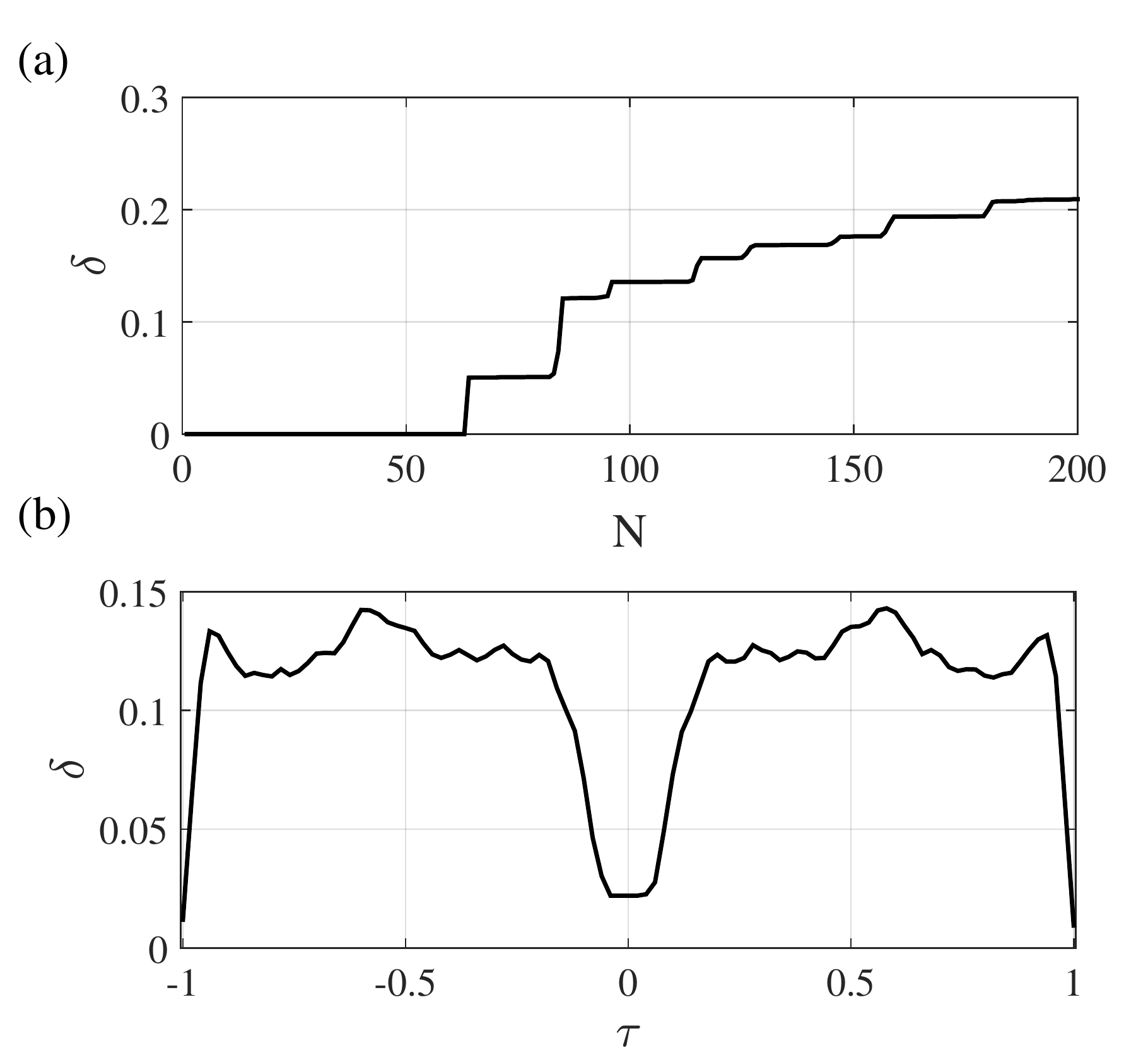}
\caption{Example 1 - (a) Observability measure $\delta$ as a function of the length $N$ of the estimation sliding window (with $\tau = 0.5$). (b) Observability measure $\delta$ as a function of the threshold value $\tau$  (with $N = 100$). The results in (a)-(b) have been evaluated over $100$ Monte Carlo trials.}
\label{fig:observability_measure}
\end{figure}

Let us, initially, consider the $2$-mass $2$-spring mechanical system of Fig.~\ref{fig:oscillators}. The state of the system is defined as $x=\left[x_{1},\dot{x}_{1},x_{2},\dot{x}_{2}\right]'$, where $x_{1}$ and $x_{2}$ are the displacements of the two masses from their static equilibrium positions. Accordingly the system is described by the continuous-time linear state equations $\dot{x}(t) = A_c x(t)$ with
\begin{equation}\label{22}
\begin{split}
&A_{c}=\begin{bmatrix} 0 & 1 & 0 & 0 \\ -\frac{(k_{1}+k_{2})}{m_{1}} & 0 & \frac{k_{2}}{m_{1}} & 0 \\ 0 & 0 & 0 & 1 \\ \frac{k_{2}}{m_{2}} & 0 & -\frac{k_{2}}{m_{2}} & 0 \end{bmatrix}\end{split}
\end{equation}
where $k_{1}, k_{2}$ are the stiffnesses of the springs and $m_{1}, m_{2}$ the corresponding masses.
The parameters are set to $m_{1}=1=m_{2}=1$ [Kg], $k_{1}=k_{2}=10$ [N/m], and the continuous-time model is discretized with sampling interval $T_s = 0.1$ [s]. Further, it is assumed that only the displacement $x_2$ (third state component) is measured by a single threshold sensor so that the output matrix turns out to be $C = \left[ 0, 0, 1, 0 \right]$. In all the simulations, the initial state is chosen so as to impose the harmonic motion condition, i.e. $\overline{x}_{0} = [0.618, 0, 1, 0]'$, making the two masses oscillate with the same frequency but different amplitudes within the interval $[-1,1]$; the initial phase of the oscillations is a uniformly distributed random variable. The process disturbance is taken equal to zero, while the measurement noise is a white sequence with uniform distribution in the interval $[-\rho_V,\rho_V]$. In order to tune the proposed MHE algorithms for appropriate performance, the threshold value $\tau$ of the binary sensor and the length $N$ of the estimation sliding window need to be properly selected. To this end, it has been analyzed by means of numerical simulations how the observability measure $\delta$ varies as a function of $N$ and $\tau$, as shown in Fig.~\ref{fig:observability_measure} with a simulation time interval of $50$ [s] and a noise level $\rho_{V} = 0.05$.
As shown in Fig.~\ref{fig:observability_measure}, observability requires sufficiently large window size ($N \geq  60$ with $\tau = 0.5$). Also notice that the observability measure as a function of $N$ has a monotonically increasing behaviour with some characteristic plateaus. Further, it is perfectly symmetric with respect to $\tau$: if the threshold value is outside the range $[-1,1]$ of the system output, then no information is provided by the binary sensor; $\tau = 0$ also implies poor observability as sampling the sinusoid in proximity of zero provides little information about the sinusoid amplitude. From Fig.~\ref{fig:observability_measure}, we chose $N = 100$ and $\tau = 0.5$ for the forthcoming simulation results, so that assumption {\bf A2} holds. For the weight matrices we selected $Q = \mathbbm{1}_{4}$, $R=1$ and $P = \epsilon \mathbbm{1}_4$ with $\varepsilon<10^{-4}$ in order to satisfy the stability condition $a_1 < 1$ according to Proposition 1.2.

In order to appreciate the accuracy of the proposed algorithms and take into account the timescales of the systems, Monte Carlo simulations have been performed by randomly varying the measurement noise realization, the phase of the oscillations for the true state trajectories, and the a priori prediction $\overline{x}_0$, which is randomly generated with uniform distribution in $[-5,5]^4$.
\begin{figure}[h!]
\centering
\includegraphics[scale = 3.4]{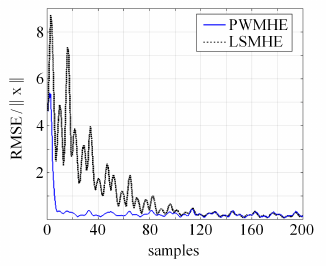}
\caption{Example 1 - Normalized RMSEs of the LSMHE and PWMHE filters, evaluated over $100$ Monte Carlo trials.}
\label{fig:RMSE}
\end{figure}
As performance index, in Fig.~\ref{fig:RMSE} we have plotted the RMSE (as given in Eq.~(\ref{21})) normalized by the Euclidean norm of the true system state, with $L = 100$ Monte Carlo trials. Fig.~\ref{fig:RMSE} confirms the effectiveness of the MHE algorithms for state estimation with binary observations: the estimates resulting from both algorithms converge to the true trajectories of the system state vector. As before, the PWMHE algorithm exhibits much better performance in the transient regime.

The computational burden of solving both Problems $E_{t}^{A}$ and $E_{t}^{B}$, as a function of the length $N$ of the estimation sliding window, has been evaluated by means of the CPU time per iteration step (a notebook with an Intel Core i7-2640M CPU @ 2.80 GHz has been used in simulations).
The results are reported in Table~\ref{table_1}.
\begin{table}[tb]
\centering
\begin{tabular}{|c|c|c|}
\hline
N               & LSMHE                   & PWMHE     \\
\hline
1               & 0.50$\cdot10^{-3}$     & 0.25      \\
5               & 0.56$\cdot10^{-3}$     & 0.42      \\
20              & 1.79$\cdot10^{-3}$     & 1.11      \\
35              & 3.23$\cdot10^{-3}$     & 2.07      \\
50              & 5.30$\cdot10^{-3}$     & 3.19      \\
100             & 22.83$\cdot10^{-3}$    & 7.53      \\
150             & 78.90$\cdot10^{-3}$    & 15.70      \\
\hline
\end{tabular}
\caption{CPU time (in [s]) per iteration step for different values of $N$.}
\label{table_1}
\end{table}
Notice that PWMHE is by far more computationally expensive than LSMHE (computing time three orders of magnitude larger in this specific small-size example). As a matter of fact, the solution of Problem $E_{t}^{A}$ can be found analytically by an explicit matrix formula, while for the solution of $E_{t}^{B}$ a convex mathematical programming problem has to be solved. However, it is worth to point out that the PWMHE algorithm has been implemented by using standard functions of the Matlab Optimization Toolbox, without resorting to ad-hoc optimization routines. Hence, we are confident that much faster computing times can be achieved. The analysis of the dependence of the performance on the threshold $\tau$ and the noise level $\rho_V$ can be found in Ref.~\cite{GherardiniAutomatica}.

\begin{figure}[h!]
\centering
\includegraphics[scale = 6]{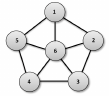}
\caption{Network of six coupled $2$-mass $2$-spring oscillators. Each node of the network has a binary sensor, monitoring the corresponding third state component.}
\label{fig:network}
\end{figure}

\begin{figure}[h!]
\centering
\includegraphics[scale = 3.15]{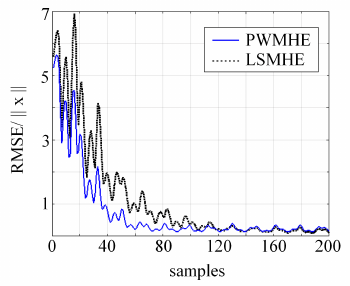}
\caption{Normalized RMSEs of the LSMHE and PWMHE filters, evaluated over $100$ Monte Carlo trials, for a network of six $2$-mass $2$-spring oscillators.}
\label{fig:RMSE_network}
\end{figure}

Finally, in order to numerically assess the performance of the proposed MHE algorithms when the dimensionality of the system state and the number of binary sensors increase, the network in Fig.~\ref{fig:network} of six coupled $2$-mass $2$-spring oscillators (like the one in Fig.~\ref{fig:oscillators}) is considered. It is assumed that each node is equipped with a binary sensor measuring the third component of the local state vector, with threshold belonging to the range $[-1,1]$.
The network dynamics turns out to be described by a discrete-time linear dynamical system with matrices
$
A = \mathbbm{1}_{6}\otimes A_{d} - \gamma  \mathcal{L}\otimes \mathbbm{1}_{4}
$
and
$
C = \mathbbm{1}_{6}\otimes [0,0,1,0],
$
where $A_{d} = \exp(A_{c}T_{s})$, $\mathcal{L}$ is the Laplacian matrix of the network, and $T_s = 0.1$ [s] is the sampling interval. For the sake of simplicity, we have chosen the same value $\gamma=0.02$ for the coupling constants between all the connected sites, which ensures the synchronization of the system states. Note that synchronization is reached if $\gamma < 0.31685$. The threshold values of the six binary sensors are taken, respectively, equal to $[0.5,0.2,-0.5,-0.8,-0.2,0.3]'$. In all  simulations, the initial state of each $2$-mass $2$-spring system is a uniformly distributed random variable centred around the vector $\overline{x}_{0} = [0.618,0,1,0]'$ with variations of $\pm 5$ for each component, while the measurement noise is a white sequence uniformly distributed in the interval $[-0.05,0.05]$. Moreover, the validity of Proposition~1.2 for the network is ensured by choosing $\varepsilon = 10^{-5}$ with $\overline{P} = \mathbbm{1}_{24}$. The duration of each simulation experiment is fixed to $35$ [s], and the corresponding RMSE of the proposed MHE filters is averaged over $100$ Monte Carlo trials. In Fig.~\ref{fig:RMSE_network} the RMSEs, normalized by the Euclidean norm of the true system state, of the LSMHE and PWMHE algorithms are plotted. It can be seen that, also in this case, the PWMHE filter exhibits better performance in the transient, and that the convergence of its estimation error is slower by a factor of approximately $4$ with respect to the case of the single oscillator.

\section{Probabilistic approach}

In this section, we pursue a probabilistic approach to state estimation with binary sensors by following \cite{GherardiniMAP,GherardiniFIELDmap}. In this respect, some work has recently addressed parameter identification or state estimation with binary measurements by following approaches based on the recursive propagation of conditional probability density functions~\cite{Ristic,Ribeiro1,Ribeiro2,Aslam,Capponi}. Relying on the aforementioned noise-assisted paradigm, we introduce a novel probabilistic approach to recursive state estimation based on binary measurements. These algorithms are based on a \textit{moving-horizon} (MH) approximation of \textit{Maximum A-posteriori Probability} (MAP) estimation algorithms, and they can be considered as an extension of the works in \cite{Ristic,Wong2007}, concerning recursive parameter estimation. As it will be shown later as a novel result of this thesis, if the dynamical system is linear and the noise distributions are described by log-concave probability density functions, then the proposed MH-MAP state estimator involves the solution, at each sampling interval, of a convex optimization problem, practically feasible for real-time implementation. Moreover, by exploiting a probabilistic approach, the presence of measurement noise can be helpful to enhance the amount of information coming from the sensors, leading to the effective definition of a \textit{noise-assisted paradigm} for state estimation. As it will shown later, for quantum mechanical systems the introduction of MH-MAP estimators could open the ways towards noise-assisted quantum estimation schemes.

Let us consider the nonlinear dynamical system of (\ref{1}). As before, the measurements are provided by a set of binary sensors according to the following equation, which is almost identical to (\ref{2}):
\begin{equation}
\begin{array}{rclcl}
y_{t}^{i} & = & h^{i}(z_{t}^{i}) & = &  \left\{
\begin{array}{ll} 1, & \mbox{if }  z_{t}^{i} \geq \tau^{i} \\
                            0,   & \mbox{if }  z_{t}^{i}  < \tau^{i}
\end{array}
\right.,
\end{array}
\end{equation}
where $z_{t}^{i} = g^{i}(x_{t}) + v_{t}^{i}$, with $i = 1,\ldots,p$. Moreover, it is assumed that the statistical behavior of the system is characterized by
\begin{equation}\label{23}
x_{0}\sim\mathcal{N}(\overline{x}_{0},P^{-1}),\hspace{2mm}w_{t}\sim\mathcal{N}(0,G^{-1}),\hspace{2mm}v_{t}\sim\mathcal{N}(0,R)
\end{equation}
where $\mathcal{N}(\mu,\nu)$ denote a \textit{normal distribution} with mean $\mu$ and variance $\nu$. In Eq.~(\ref{23}), $R\equiv{\rm diag}(r_{1},\ldots,r_{p})$; $\mathbb{E}[w_{j}w_{k}']=0$ and $\mathbb{E}[v_{j}v_{k}']=0$ if $j\neq k$; and $\mathbb{E}[w_{j}v_{k}']=0$, $\mathbb{E}[w_{j}x_{0}']=0$, $\mathbb{E}[v_{j}x_{0}']=0$ for any $j,k$. As before, each sensor $i$ produces a threshold measurements $y^{i}_{t}\in\{0,1\}$ depending on whether the noisy system output $z^{i}_{t}$ is below or above the threshold $\tau^{i}$.

\subsection{Maximum a-posteriori state estimation}

The probabilistic approach to state estimation with binary sensors is recast in the Bayesian framework, which exploits the \textit{Maximum A posteriori Probability (MAP)} estimation theory. In this way, we evaluate for each sensor the probability that the corresponding measurements assume one of the two binary values, in relation to the dynamical evolution of the system state we are monitoring. As a result, also a binary sensor is always characterized by an informative content, i.e. each binary measurement $y^{i}_t$ intrinsically provides information about the state $x_t$. Such information is encoded in the likelihood functions $p(y^{i}_{t}|x_{t})$ related to the $i-$th threshold sensor. The binary measurements $y^{i}_{t}$ are Bernoulli random variables, so that, for any binary sensor $i$ and any time instant $t$, the likelihood function $p(y^{i}_{t}|x_{t})$ is given by
\begin{equation}\label{24}
p(y^{i}_{t}|x_{t})~=~p(y^{i}_{t}=1|x_{t})^{y^{i}_{t}} ~p(y^{i}_{t}=0|x_{t})^{1-y^{i}_{t}},
\end{equation}
where
\begin{equation}\label{25}
p(y^{i}_{t}=1|x_{t}) = F^{i}(\tau^{i}-g^{i}(x_{t}))
\end{equation}
and
\begin{equation}\label{26}
p(y^{i}_{t}=0|x_{t})=1-p(y^{i}_{t}=1|x_{t})\equiv \Phi^{i}(\tau^{i}-g^{i}(x_{t})).
\end{equation}
The function $F^{i}(\tau^{i}-g^{i}(x_{t}))$ is the complementary cumulative distribution function (CDF) of the random variable $\tau^{i}-g^{i}(x_{t})$. Since $v^{i}_{t}\sim\mathcal{N}(0,r^{i})$, the conditional probability $p(y^{i}_{t}=1|x_{t}) = F^{i}(\tau^{i}-g^{i}(x_{t})) $ can be written in terms of a \textit{Q-function}, which describes the tail probability of a standard normal probability distribution~\cite{Wim2006}. In other words:
\begin{equation}\label{27}
F^{i}(\tau^{i}-g^{i}(x_{t}))  = \frac{1}{\sqrt{2\pi r_{i}}}\int_{\tau^{i}-g^{i}(x_{t})}^{\infty}\exp\left(-\frac{u^{2}}{2r_{i}}\right)du
= Q \left(\frac{\tau^{i}-g^{i}(x_{t})}{\sqrt{r_{i}}}\right) \, .
\end{equation}
Now, let us recall that $Y_{t}={\rm col} (y_0,\ldots,y_{t} )$ is the vector of all binary measurements collected up to time $t$ and $X_{t}\equiv {\rm col} (x_{0},\ldots,x_{t} )$ is the vector of the state trajectory. $\hat{X}_{t|t}\equiv{\rm col}( \hat{x}_{0|t},\ldots,\hat{x}_{t|t} )$, instead, collects the estimates of $X_{t}$, made at any stage $t$. Then, at each time instant $t$, given the a-posteriori probability $p(X_{t}|Y_{t})$, the estimate of the state trajectory is obtained by solving the following MAP estimation problem:
\begin{equation}\label{28}
\hat{X}_{t|t}  = \text{arg}\max_{X_{t}}p(X_{t}|Y_{t})
=\text{arg} \min_{X_{t}}  - \ln p(X_{t}|Y_{t}).
\end{equation}
From the Bayes rule
\begin{equation}\label{29}
p(X_{t}|Y_{t}) ~\propto ~p(Y_{t}|X_{t})~p(X_{t}),
\end{equation}
where $p(Y_{t}|X_{t})$ is the likelihood function of the binary measurement vector $Y_{t}$, and
\begin{equation}\label{30}
p(X_{t}) = \prod_{k=0}^{t-1}p(x_{t-k}|x_{t-k-1},\ldots,x_{0})~p(x_{0})
= \prod_{k=0}^{t-1}p(x_{t-k}|x_{t-k-1})~p(x_{0}),
\end{equation}
for which the Markov property for the dynamical system state has been taken into account. Being the initial state $x_{0}$ and the process noise $w_{t}$ normally distributed vectors, it holds that
\begin{equation}\label{31}
p(x_0) \propto \exp\left(-\frac{1}{2}\|x_{0}-\overline{x}_0 \|^{2}_{P}\right)
\end{equation}
and
\begin{equation}\label{32}
p(x_{k}|x_{k-1}) \propto \exp\left(-\frac{1}{2}\|x_{k+1}-f(x_{k},u_{k})\|^{2}_{G}\right),
\end{equation}
so that
\begin{equation}\label{33}
p(X_{t}) = \exp\left(-\frac{1}{2}\left[\|x_{0}-\overline{x}_{0}\|^{2}_{P}+\sum_{k=0}^{t}\|x_{k+1}-f(x_{k},u_{k})\|^{2}_{G}\right]\right).
\end{equation}
Now, the following assumption has to be stated:
\begin{enumerate}[\bf {A}3]
\item
Statistical independence of the threshold measurements.
\end{enumerate}
Under this assumption, the likelihood function $p(Y_{t}|X_{t})$ can be written as
\begin{eqnarray}\label{34}
p(Y_{t}|X_{t})&=&\prod_{k=0}^{t}p(y_{k}|x_{k})=\prod_{k=0}^{t}~\prod_{i=1}^{p}p(y_{k}^{i}|x_{k})\nonumber \\
&=&\prod_{k=0}^{t}~\prod_{i=1}^{p}F^{i}(\tau^{i}-g^{i}(x_{k}))^{y_{k}^{i}} ~ \Phi^{i}(\tau^{i}-g^{i}(x_{k}))^{1-y_{k}^{i}}.
\end{eqnarray}
In conclusion, the log-likelihood function, natural logarithm of the likelihood function, reads
\begin{equation}\label{35}
\ln p(Y_{t}|X_{t})=\sum_{k=0}^{t}~\sum_{i=1}^{p}\left[y_{k}^{i} \, \ln F^{i}(\tau^{i}-g^{i}(x_{k}))+(1-y_{k}^{i}) \, \ln \Phi^{i} (\tau^{i}-g^{i}(x_{k}))\right],
\end{equation}
and the cost function $- \ln p(X_{t}|Y_{t})  =-\ln p(Y_{t}|X_{t})-\ln p(X_{t})$ to be minimized in the MAP estimation problem (\ref{28}) turns out to be, up to additive constant terms, equal to
\begin{eqnarray}\label{36}
J_{t} (X_t) &=& \|x_{0}-\overline{x}_{0}\|^{2}_{P}+\sum_{k=0}^{t}\|x_{k+1}-f(x_{k},u_{k})\|^{2}_{G}\nonumber \\
&-&\sum_{k=0}^{t}~\sum_{i=1}^{p}\left[y_{k}^{i} \ln F^{i}(\tau^{i}-g^{i}(x_{k}))+(1-y_{k}^{i}) \ln \Phi^{i}(\tau^{i}-g^{i}(x_{k}))\right].\nonumber \\
&&
\end{eqnarray}
Unfortunately, a closed-form expression for the global minimum of (\ref{36}) does not exist and, hence, the optimal MAP estimate $\hat X_{t|t}$ has to be determined by resorting to some numerical optimization routine. With this respect, the main drawback is that the number of optimization variables grows linearly with time, since the vector $X_t$ has size $(t+1) \, n$. As a consequence, as $t$ grows the solution of the full information MAP state estimation problem (\ref{28}) becomes eventually unfeasible, and some approximation has to be introduced.

In this regard, we propose an approximation solution, which is based on the MHE approach to solve state estimation problems~\cite{Morari,Delgado}. If we introduce again the sliding window $\mathfrak W_t = \{t-N, t-N+1, \ldots, t\}$, then the goal of the estimation problem becomes to find an estimate of the partial state trajectory $X_{t-N:t} \equiv {\rm col} ( x_{t-N},\ldots,x_{t} )$ by using the information available in $\mathfrak W_t$. In this way, besides increasing the information content of the binary measurements as in \cite{GherardiniCDC,GherardiniAutomatica}, by adopting the MHE approach we are also able to solve state estimation problems with constrained system dynamics, such that $x_{t}\in X\subseteq\mathbb{R}^{n}$, $u_{t}\in U\subseteq\mathbb{R}^{m}$, $w_{t}\in W\subseteq\mathbb{R}^{n}$ and $v_{t}^{i}\in\mathbb{V}\subseteq\mathbb{R}^{p}$, where $X$, $U$, $W$ and $V$ are convex sets. Therefore, in place of the \textit{full information cost} $J_t (X_t)$, at each time instant $t$ the minimization of the following \textit{moving-horizon cost} is addressed:
\begin{eqnarray}\label{37}
&&J_{t}^{\rm MH} (X_{t-N:t}) = \Gamma_{t-N} (x_{t-N}) + \sum_{k=t-N}^{t}\|x_{k+1}-f(x_{k},u_{k})\|^{2}_{Q}\nonumber \\
&&- \sum_{k=t-N}^{t}~\sum_{i=1}^{p}\left[y_{k}^{i} \ln F^{i}(\tau^{i}-g^{i}(x_{k}))+(1-y_{k}^{i}) \ln \Phi^{i}(\tau^{i}-g^{i}(x_{k}))\right],\nonumber \\
&&
\end{eqnarray}
where the non-negative initial penalty function $\Gamma_{t-N} (x_{t-N}) $, known in the MHE literature as \textit{arrival cost} (see \cite{RaRaMa03,AlBaBaZavCDC10}), is introduced so as to summarize the past data $y_0,\ldots,y_{t-N-1}$ not explicitly accounted for in the objective function. The form of the arrival cost plays an important role in the behavior and performance of the overall estimation scheme. While in principle $\Gamma_{t-N} (x_{t-N})$ could be chosen so that the minimization of the moving-horizon cost (\ref{37}) yields the same estimate that would be obtained by minimizing (\ref{36}), an algebraic expression for such a true arrival cost seldom exists, even when the sensors provide continuous (non-threshold) measurements~\cite{RaRaMa03}. Hence, some approximation must be used. With this respect, a common choice~\cite{NLMHE}, also followed in this section, consists of assigning to the arrival cost a fixed structure penalizing the distance of the state $x_{t-N}$ at the beginning of the sliding window from some prediction $\overline{x}_{t-N} $ computed at the previous time instant, making the estimation scheme recursive. A natural choice is then a quadratic arrival cost of the form
\begin{equation}\label{eq:arrival}
\Gamma_{t-N} (x_{t-N}) = \|x_{t-N}-\overline{x}_{t-N}\|^{2}_{\Psi} \, ,
\end{equation}
which has been used also in the deterministic approach to state estimation with binary sensors, shown in the previous section. From the Bayesian point of view, this choice corresponds to approximating the probability density function of the state $x_{t-N}$, conditioned to all the measurements collected up to time $t-1$, with a Gaussian having mean $\overline{x}_{t-N}$ and covariance $\Psi^{-1}$. As for the choice of the weight matrix $\Psi$,
in the case of continuous measurements it has been shown that stability of the estimation error dynamics can be ensured provided that $\Psi$
is not too large, so as to avoid an overconfidence on the available estimates~\cite{NLMHE,AlBaBaZavCDC10}. Recently in \cite{GherardiniCDC,GherardiniAutomatica}, similar results have been proven to hold also in the case of binary sensors, but in the deterministic context. In practice, $\Psi$ can be seen as a design parameter which has to be tuned by pursuing a suitable trade-off between such stability considerations and the necessity of not neglecting the already available information, since in the limit for $\Psi$ going to zero the approach becomes a finite memory one.

Summing up, at any time instant $t = N,N+1,\ldots$, the following problem has to be solved. \\ \\
\textbf{Problem $E_{t}^{C}$:} Given the prediction $\overline{x}_{t-N}$, the input sequence $\{ u_{t-N}, \ldots, u_{t-1} \}$, the measurement sequences $\{ y^{i}_{t-N}, \ldots, y^{i}_t , \, i = 1, \ldots, p \}$, find the optimal estimates $\hat{x}_{t-N|t}^{\circ},\ldots,\hat{x}_{t|t}^{\circ}$ that minimize the cost function (\ref{37}) with arrival cost~(\ref{eq:arrival}). \\ \\
In order to propagate the estimation procedure from Problem $E_{t-1}^{C}$ to Problem $E_{t}^{C}$, the prediction
$\overline{x}_{t-N}$ is set equal to the value of the estimate of $x_{t-N}$ made at time instant $t-1$, i.e., $\overline{x}_{t-N} = \hat{x}_{t-N|t-1}$.
Clearly, the recursion is initialized with the a priori expected value $\overline{x}_0$ of the initial state vector. Let us observe that, in general, solving Problem $E_{t}^{C}$ entails the solution of a non-trivial optimization problem. However, when the (discrete-time) dynamical system is linear, the resulting optimization problem turns out to be convex so that standard optimization routines can be used in order to find the global minimum.
To see this, let us consider again that $f(x_t,u_t) = A x_t + B u_t$ and $ g^{i}(x_t) = C^{i} x_t$, $i=1,\ldots,p$, where $A$, $B$, $C^{i}$ are constant matrices of suitable dimensions. Then, the following result, whose proof is Appendix~\ref{chapter:appA}, holds. \\ \\
{\bf Proposition~1.3:} If assumption {\bf A3} holds, the dynamical system is linear and the noise are distributed as a Gaussian probability density function, then the CDF $\Phi^{i}(\tau^{i}-C^{i}x_{t})$ and its complementary function $F^{i}(\tau^{i}-C^{i}x_{t})$ are log-concave. Hence, the cost function (\ref{37}) with arrival cost~(\ref{eq:arrival}) is convex. \\ \\
The convexity of the cost function~(\ref{37}) is guaranteed also in the more general case in which the statistical behavior of the random variables $x_{0}$, $w_{t}$, $v_{t}$ is described by logarithmical concave distribution functions. Indeed, if a probability density function is log-concave, also its cumulative distribution function is log-concave, so that the contribution related to the threshold measurements in (\ref{37}) is effectively convex. Let us observe that the proposed MH-MAP state estimator turns out to be the optimal Bayesian filter when we want to estimate the state of a linear dynamics with a network of independent binary sensors.

\subsection{Dynamic field estimation}
\label{sec:field_estimation}

As main application of the MH-MAP estimator for macroscopic systems, we address state estimation for a spatially distributed system with a noisy measurement, which is provided by a set of binary sensors spread over the spatial domain $\Omega$ of interest. In particular, we consider the problem of reconstructing a two-dimensional \textit{diffusion} field. The diffusion process is governed by the following parabolic Partial Differential Equation (PDE):
\begin{equation}\label{38}
\dfrac{\partial c}{\partial t} - \lambda_{d}\nabla^2 c  ~=~ 0 \,\,\, \mbox{in } \Omega
\end{equation}
which models various physical phenomena, as for example the spread of a pollutant in a fluid. In this case, $c(\xi,\eta,t)$ represents the space-time dependent substance concentration, $\lambda_{d}$ denotes the constant diffusivity of the medium, and $\nabla^2 = {\partial^2 } / {\partial \xi^2} + {\partial^2 } / {\partial \eta^2}$ is the Laplace operator, $(\xi,\eta) \in \Omega$ being the 2D spatial variables. Furthermore, we assume mixed boundary conditions to the PDE (\ref{38}), i.e. a non-homogeneous Dirichlet condition $c = \psi$ on $\partial \Omega_D$, which specifies a constant-in-time value of concentration on the boundary $\partial \Omega_D$, and a homogeneous Neumann condition on $\partial \Omega_N = \partial \Omega \setminus \partial \Omega_D$, assumed impermeable to the contaminant: ${\partial c}/{\partial \upsilon} = 0$ on $\partial \Omega_N$, where $\upsilon$ is the outward pointing unit normal vector of $\partial \Omega_N$.

The objective is to estimate the values of the dynamic field of interest $c(\xi,\eta,t)$ given the measurements from a set of binary measurements in $\Omega$. The PDE system (\ref{38}) is simulated with a mesh of finite elements over $\Omega$ via the \textit{Finite Element (FE)} approximation described in \cite{TACNick,Forti_tesi}. Specifically, the domain $\Omega$ is subdivided into a suitable set of non overlapping regions, or elements, and a suitable set of basis functions $\phi_{j} (\xi,\eta)$, with $j=1,\ldots,m_\phi$, is defined on such elements. The choices of the basis functions $\phi_{j}$ and of the elements are key points of the FE method. In this specific case, we have chosen the elements of the mesh to be triangles in 2D, which define a FE mesh with vertices $({\xi}_j,\eta_j) \in \Omega, j=1,\ldots,m_\phi $. Then each basis function $\phi_{j}$ is assumed to be a piece-wise affine function, which vanishes outside the elements of the mesh in correspondence of the vertices $({\xi}_j,\eta_j)$, so that $\phi_{j}({\xi}_i,\eta_i)=\delta_{ij}$, where $\delta_{ij}$ denotes the Kronecker delta. In order to take into account also the mixed boundary conditions, the basis functions are supposed to follow a proper ordering law: $m$ points of the mesh correspond to vertices, which lie either in the interior of $\Omega$ or on $\partial \Omega_N$, while the others $m_\phi-m$ points correspond to vertices lying on the boundary $\partial\Omega_D$. Accordingly, the discretized function, modeling the substance concentration within the domain $\Omega$, is approximated as
\begin{equation}\label{EXPA}
c(\xi,\eta,t) \approx \sum_{j=1}^{m} \phi_{j}(\xi, \eta) \, c_j(t) + \sum_{j=m+1}^{m_\phi} \phi_{j}(\xi, \eta) \, \psi_j,
\end{equation}
where $c_j(t)$ are the unknown expansion coefficient of the function $c(\xi,\eta,t)$ at the time instant $t$, while $\psi_j$ is the known expansion coefficient of the boundary function $\psi(\xi,\eta)$. In this regard, let us observe that the second summation in (\ref{EXPA}) is needed in order to impose the non-homogeneous Dirichlet condition on the boundary $\partial\Omega_D$.

As stated by the FE approximation, we recast the PDE (\ref{38}) into the following integral form:
\begin{equation}\label{39}
\int_\Omega \frac{\partial c}{\partial t} \varphi \, d\xi d\eta  \, - \,
\lambda_{d} \int_\Omega ~\nabla^2 c~ \varphi \, d\xi d\eta  =0
\end{equation}
where $\varphi(\xi,\eta)$ is a generic space-dependent weight function. It is worth to point out that the function $\varphi(\xi,\eta)$ is a weight function, that is introduced as an additional degree of freedom of the method in order to ensure that \textit{on average} the solution of the PDE is effectively given by the substance concentration (\ref{EXPA}). This procedure, which spatially discretizes the diffusion field within the domain $\Omega$, relies on weighted residual methods. The interested reader to further details on the FEM theory is referred to \cite{Brenner96}. Now, by applying Green's identity, i.e.
\begin{equation}
\int_\Omega ~\nabla^2 c~ \varphi \, d\xi d\eta = \int_{\partial\Omega}\frac{\partial c}{\partial\upsilon}\varphi d\xi d\eta
- \int_{\Omega}\nabla c\nabla\varphi d\xi d\eta,
\end{equation}
one obtains:
\begin{equation}\label{eq:Green}
\int_\Omega \frac{\partial c}{\partial t} \varphi \, d\xi d\eta + \lambda_{d}
\int_\Omega \nabla^T c ~\nabla \varphi \, d\xi d\eta
- \lambda_{d}
\int_{\partial \Omega} \frac{\partial c}{\partial \upsilon} \varphi \,
d\xi d\eta
= 0 \, .
\end{equation}
Usually, the Galerkin weighted residual method~\cite{Brenner96} is then applied. It ensures that the error done by the approximation is minimal in correspondence of the nodes of the elements of the mesh, that in this case are the vertices of the triangles. According to the Galerkin method, the test function $\varphi$ is chosen equal to the basis functions $\phi_{i}(\xi, \eta)$. Hence, by exploiting the approximation given by (\ref{EXPA}), we obtain the following equation:
\begin{eqnarray}\label{eq:Galerkin}
&&\sum_{i=1}^{m}  \int_\Omega \phi_i \phi_j \, d\xi d\eta \, \dot{c}_i(t)  + \lambda_{d} \sum_{i=1}^{m}  \int_\Omega \nabla^T \phi_i ~\nabla \phi_j \, d\xi d\eta \, c_i (t)\nonumber \\
&& + \lambda_{d}  \sum_{i=m+1}^{m_\phi}   \int_\Omega \nabla^T \phi_i ~\nabla \phi_j \, d\xi d\eta \, \psi_i = 0
\end{eqnarray}
for $j = 1, \ldots, m$. Let us observe that in (\ref{eq:Galerkin}) the boundary integral in (\ref{eq:Green}) has been omitted, since it is equal to 0 due to the validity of the homogeneous Neumann condition (\ref{Nbc}) on $\partial\Omega_N$ and to the fact that, by construction, the basis functions $\phi_j$, $j = 1, \ldots, m$, are vanishing on $\partial\Omega_D$.

Now, by defining the state vector $x \equiv {\rm col} (c_1 , \ldots , c_m) $ and the vector of boundary conditions with $\gamma \equiv  {\rm col} (\psi_{m+1}, \ldots, \psi_{m_\phi})$, (\ref{eq:Galerkin}) can be written in the more compact form
\begin{equation}\label{40}
M \dot x (t) + S x(t) + S_D \gamma = 0
\end{equation}
where $S$ is the so-called stiffness matrix, $M$ the mass matrix, and $S_D$ captures the physical interconnections among the vertices affected by the boundary condition and the remaining nodes of the mesh. The expression of the matrices $S$, $M$ and $S_{D}$ can be directly derived by (\ref{eq:Galerkin}). Thus, if we apply for example the implicit Euler method, (\ref{40}) can be discretized in time, obtaining the following linear discrete-time model
\begin{equation}\label{dt-sys}
x_{t+1} = A \, x_t + B \, u + w_t
\end{equation}
where
\begin{equation*}\label{eq:def}
\begin{aligned}
A &\equiv \left[ \mathbbm{1} +\delta t ~M^{-1}S\right]^{-1} \\
B &\equiv \left[ \mathbbm{1} +\delta t ~M^{-1}S\right]^{-1}M^{-1}\delta t \\
u &\equiv - S_D ~\gamma
\end{aligned}.
\end{equation*}
Moreover, in (\ref{dt-sys}) $\delta t$ is the time integration interval of the implicit Euler method, and $w_t$ is the process disturbance taking into account the space-time discretization errors. Let us notice that the linear system (\ref{dt-sys}) has dimension $m$ equal to the number of vertices of the mesh, which do not lie on the boundary $\partial\Omega_D$, and it is assumed to be monitored by a network of $p$ threshold sensors. Each sensor, before threshold quantization is applied, directly measure the pointwise-in-time-and-space concentration of the contaminant in a point $(\xi_i,\eta_i)$ of the spatial domain $\Omega$. By exploiting (\ref{EXPA}), such a concentration can be written as a linear combination of the concentrations on the grid points in that
\begin{equation}
c(\xi_i,\eta_i,t) \approx C^{i} x_t + D^{i} \gamma,
\end{equation}
where
\begin{eqnarray}
C^{i} &\equiv& \left [ \phi_1(\xi_i,\eta_i) \; , \ldots \; , \phi_n(\xi_i,\eta_i) \right ], \\
D^{i} &\equiv& \left [ \phi_{n+1}(\xi_i,\eta_i) \; , \ldots \; , \phi_{n_\phi}(\xi_i,\eta_i) \right ].
\end{eqnarray}
Hence the resulting output function takes the form
\begin{equation}
z_{t}^{i} = C^{i} x_{t} + v_{t}^{i},\hspace{3mm}i=1,\ldots,p
\end{equation}
where the constant $D^{i} \gamma$ can be subsumed into the threshold $\tau_i$.

\begin{figure}[h!]
   \vspace{.2cm}
	\centering
	\includegraphics[scale=0.38]{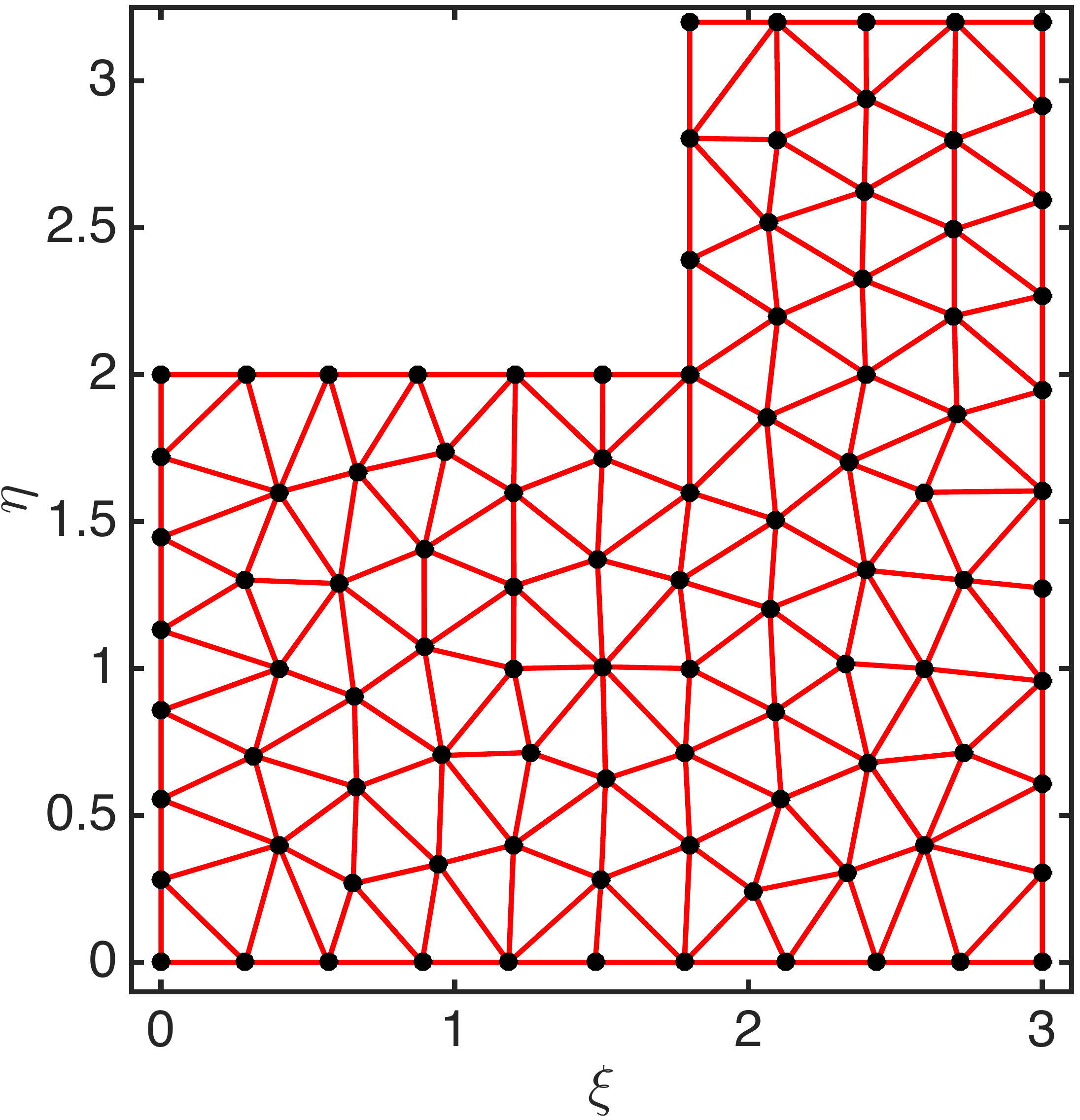}
	\caption{Mesh, used by the MH-MAP estimator, given by 152 elements (triangles), and 97 nodes (vertices).}
	\label{fig:mesh}
\end{figure}

As example, let us consider the diffusion equation (\ref{38}) with $\lambda_{d} = 0.01 ~ [m^2/s]$. It has been discretized in $1695$ triangular elements and $915$ vertices, where the field of interest is defined over a bounded 2D spatial domain $\Omega$, which covers an area of $7.44 ~ [m^2]$. Moreover, we have chosen the fixed integration step length equal to $\delta t = 1 ~ [s]$, $\gamma = 30 ~ [g/m^2]$, $x_{0} = 0_n ~ [g/m^2]$ as initial condition of the field vector, and a non-homogeneous Dirichlet boundary condition on the bottom edge and no-flux condition (Neumann boundary condition) on the remaining portions of $\partial\Omega$. As shown in Fig.~\ref{fig:mesh}, the domain $\Omega$ has a L-shape. Traditionally, L-shaped domains have been used in boundary-value problems as a basic yet challenging problem, since the non-convex corner causes a singularity in the solution of the diffusion equation. The aforementioned setting defines the \textit{ground truth} simulator of the problem, which constitutes the basis to design the corresponding MH-MAP estimator. The latter, indeed, implements a coarser mesh (in this regard, see Fig.~\ref{fig:mesh}) of ${m_\phi} = 97$ vertices ($m = 89$), and runs at a slower sample rate (i.e. $0.1 ~ [Hz]$), so that the filter is affected also by model uncertainties.
The initial condition of the estimated dynamic field is set to $\overline{x}_{0} = 5 \cdot {1}_n ~ [g/m^2]$, the moving window has size $N = 5$, and the weight matrices in (\ref{23}) are chosen as $P = 10^3 \cdot \mathbbm{1}_n$ and $Q = 10^{2} \cdot \mathbbm{1}_n$. As for the binary measurements, we first corrupted the \textit{true} concentrations (from the dynamical model of (\ref{dt-sys})) with a Gaussian noise with variance $r^i$, and, then, we applied a different threshold $\tau^i$ for each sensor $i$ of the network. Furthermore, in order to receive informative threshold measurements, the threshold $\tau^i, ~ i = 1,...,p$, are generated as uniformly distributed random numbers in the interval $[0.05,29.95]$, being $[0,30]$ the range of nominal concentration values throughout each simulation. The duration of each simulation experiment is fixed to $1200 ~ [s]$ (120 samples).
\begin{figure}[h!]
	\centering
    \includegraphics[scale=.275]{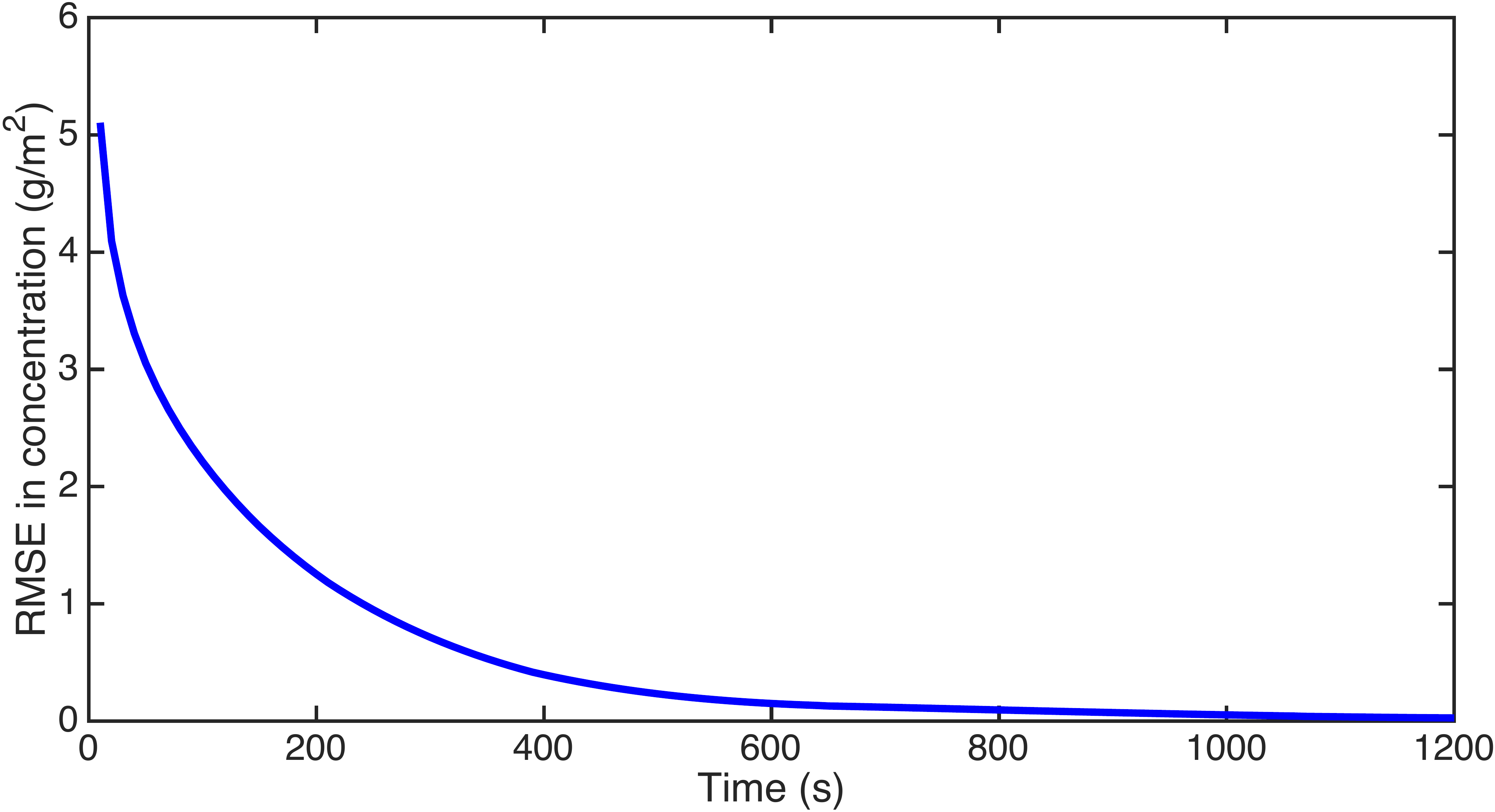}
	\caption{RMSE of the concentration estimates from the MH-MAP state estimator as a function of time, for a random network of 5 threshold sensors.}
	\label{fig:fig1_MAP}
\end{figure}

Fig.~\ref{fig:fig1_MAP} shows the performance of the proposed MH-MAP state estimators implemented in MATLAB, in terms of the RMSE of the estimated concentration field. To obtain the RMSEs plotted in Fig.~\ref{fig:fig1_MAP}, the estimation error $e_{t,j}$ at time $t$ in the $j-$th simulation run has been averaged over $304$ sampling points (evenly spread within $\Omega$) and the number of Monte Carlo realizations has been set to $\alpha=100$.
\begin{figure}[h!]
	\vspace{.2cm}
	\centering
	\includegraphics[scale=1.9]{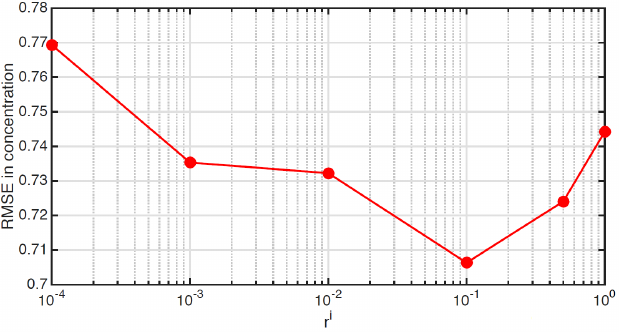}
	\caption{RMSE of the concentration estimates as a function of the measurement noise variance, for a fixed constellation of 20 threshold sensors. It is shown here that operating in a noisy environment turns out to be beneficial, for certain values of $r^i$, to the state estimation problem.}
	\label{fig:fig2_MAP}
\end{figure}
It can be observed that the proposed estimators successfully estimate the dynamic field, even with a small number of randomly deployed binary sensors. Furthermore, the effect of the measurement noise on the mean value of the RMSE can be seen in Fig.~\ref{fig:fig2_MAP}, in which it becomes apparent how for certain values of $r^i$, including an observation noise with higher variance, the quality of the overall estimates can actually be improved. Such result numerically demonstrate the validity of the above stated noise-assisted paradigm in the recursive state estimation with threshold measurements.

\subsection{Fast MH-MAP filter for large-scale problems}

In order to achieve a good approximation of the original continuous field, a large number of basis function need to be used in the expansion (\ref{EXPA}). Hence, in general the FEM-based space discretization gives rise to a large-scale system possibly characterized by thousands of state variables, equal to the number of the vertexes of the mesh not lying on the boundary $\partial\Omega$ of the domain. This means that a direct application of the MH-MAP filter involves the solution, at each time instant, of a large-scale (albeit convex) optimization problem. Although today commercial optimization software can solve general convex programs of some thousands equations, the problem becomes intractable from a computational point of view when the number of variables (that is, the number of vertexes of the FE grid) is too large. Further, even when a solution to the
large-scale optimization problem can be found the time required for finding it may not be compatible with real-time operations (recall that the MH-MAP filter requires that each optimization terminates within one sampling interval).

Here, we propose a more computationally \textit{efficient} and \textit{fast} version of the MH-MAP filter for the real-time estimation of a dynamic field that is based on the idea of decomposing the original large-scale problem into simpler subproblems by means of a two-stage estimation procedure. Such results are discussed in \cite{GherardiniFIELDmap}. The proposed method allows to efficiently solve the problem of estimating the state (ideally infinite dimensional) of a spatially-distributed dynamical system just by using sensors with minimal information content, such as a binary sensor. The improved version of the aforementioned MH-MAP filter, which can be suitable for large-scale systems, will split the estimation problem into two main steps:
\begin{enumerate}[(1)]
  \item Estimation of the local concentration correspondence of each binary sensor by means of $p$ independent MH-MAP filters. The concentration estimates provided by each local MH-MAP filter allows to recast the threshold measurements into linear pseudo-measurements.
  \item Field estimation over a mesh of finite elements defined over the (spatial) domain $\Omega$ on the basis of the linear pseudo-measurements provided by the local filters in step 1. For this purpose, any linear filtering technique suitable for large-scale systems can be used (see e.g. the finite-element Kalman filter as in \cite{TACNick}. In this step, field estimation is performed by minimizing a single quadratic MH cost function for linear systems.
\end{enumerate}
This solution turns out to be more computationally efficient as compared to a direct application of the MH-MAP filter to the dynamical system in that: (i) the number of binary sensors spread over the domain $\Omega$ is typically much smaller than the number of vertexes (i.e. $p\ll n$); (ii) as will be clarified in the following, each local MH-MAP filter in step 1 involves the solution of a convex optimization problem with a reduced number of variables.

\subsubsection{Step~1}

Let us analyze in more detail step~1 of the fast MH-MAP filter. To this end, let us denote by $\sigma_{t}^{i}$ the value of the concentration in correspondence of sensor $i$
at the $t$-th sampling instant, i.e. $\sigma_{t}^{i} = c(\xi_i,\eta_i,t)$, and let $\sigma^{i}_{k,t}$ denote the value of the $k$-th time-derivative of such a concentration, i.e.
\begin{equation}\label{eq:taylor_sigma}
\sigma^{i}_{k,t} \equiv \left . \frac{\partial^k}{\partial s^k} c(\xi_i,\eta_i,t) \right |_{s = t T_s}.
\end{equation}
Under the hypothesis of a small enough sampling time, in correspondence of each binary sensor the dynamical evolution of the propagating field can be approximated by resorting to a truncated Taylor series expansion, so that
\begin{equation}
\sigma_{t+1}^{i} \approx \sigma_{t}^{i} + \sum_{k=1}^K \frac{(T_s)^k}{k!} \sigma^{i}_{k,t}
\end{equation}
Then, the local dynamics of the concentration in correspondence of sensor $i$ can be described by a linear dynamical system with state $\chi_{t}^{i} = \left[ \sigma_{t}^{i},\sigma_{1,t}^{i} , \ldots, \sigma_{K,t}^{i}\right]'$ and state equation
\begin{equation}\label{eq:chi_step1}
\chi_{t+1}^{i} = \widetilde{A} \, \chi_{t}^{i} + w_t^{i},
\end{equation}
where the matrix $\widetilde{A}$ is obtained from (\ref{eq:taylor_sigma}) and $w_t^{i}$ is the disturbance acting on the local dynamics with zero mean and inverse covariance $\widetilde{G}$. Notice that models like (\ref{eq:chi_step1}) are widely used in the construction of filters for estimating time-varying quantities whose dynamics is unknown or too complex to model (for instance, they are typically used in tracking of moving objects~\cite{Bar-Shalom}). With this respect, a crucial assumption for the applicability of this kind of models is that the sampling interval be sufficiently smaller as compared to the time constants characterizing the variation of the quantities to be estimated. Hence, their application in the present context is justified by the fact that, in practice, (binary) concentration measurements can be taken at a high rate so that between two consecutive measurements only small variations can occur. In model (\ref{eq:chi_step1}), the simplest choice amounts to taking $K=0$ and $w^{i}_t$ as a Gaussian white noise, which corresponds to approximating the concentration as nearly constant (notice that, in this case, we have $\widetilde{A} = 1$). Instead, by taking $K=1$, we obtain a nearly-constant derivative model with state transition matrix
\begin{equation}
\widetilde{A} = \left [ \begin{array}{cc} 1 & T_s \\ 0 & 1 \end{array} \right]
\end{equation}
which is equivalent to the usual nearly constant velocity models for moving object tracking~\cite{Bar-Shalom}. Clearly, each local model (\ref{eq:chi_step1}) is related to the $i$-th binary measurement via the measurement equation
\begin{eqnarray}
z_{t}^{i} &=& \widetilde{C} \, \chi_{t}^{i} + v_{t}^{i} \nonumber \\
y_{t}^{i} &=& h^{i} \left ( z_{t}^{i} \right ) \label{meas:chi_step1}
\end{eqnarray}
where
\begin{equation}
\widetilde{C} \equiv [1 \; , 0, \cdots \; , 0] \, .
\end{equation}
Then, for each sensor $i$, at each time instant $t$ the minimization of the following MH-MAP cost function is addressed
 \begin{eqnarray}\label{eq:cost_step1}
\widetilde J_{t}^{i} (\mathfrak{X}_{t-N:t}^{i})  &=& \|\chi_{t-N}^{i} - \overline{\chi}_{t-N}^{i}\|^{2}_{\widetilde \Psi} + \sum_{k=t-N}^{t}\|\chi_{k+1}^{i} - \widetilde A \, \chi_{k}^{i} \|^{2}_{\widetilde Q}\nonumber \\
&-& \sum_{k=t-N}^{t} \left[y_{k}^{i} \ln F^{i} \left (\tau^{i}- \widetilde C \,  \chi_{k}^{i}  \right )+
(1-y_{k}^{i}) \ln \Phi^{i} \left (\tau^{i}- \widetilde C \,  \chi_{k}^{i}  \right )\right],\nonumber \\
&&
\end{eqnarray}
where $\mathfrak{X}_{t-N:t}^{i} \equiv {\rm col}\left( \chi_{k}^{i} \right)_{t-N}^{t}$ and $\overline{\chi}_{t-N}^{i}$ is the estimate of the local state at time $t-N$ computed at the previous iteration. In conclusion, at any time instant $t = N, N+1, \ldots$, for any binary sensor $i$ the following problem has to be solved. \\ \\
\textbf{Problem step~1:} Given the prediction $\overline{\chi}_{t-N}^{i}$ and the measurement sequence $\{ y^{i}_{t-N}, \ldots, y^{i}_t \}$, find the optimal estimates $\hat{\chi}_{t-N|t}^{i},\ldots,\hat{\chi}_{t|t}^{i}$ that minimize the cost function $\widetilde J_{t}^{i}(\mathfrak{X}_{t-N:t}^{i})$. \\ \\
As before, the propagation of the estimation problem from time $t-1$ to time $t$ is ensured by choosing $\overline{\chi}_{t-N}^{i} = \hat{\chi}_{t-N|t-1}^{i}$. The number of variables involved in each of such optimization problems is $(K +1)(N+1)$ and, in view of (\ref{eq:chi_step1}) and (\ref{meas:chi_step1}), the cost function $\widetilde J_{t}^{i}$ is \textit{convex} according to Proposition~1.3. Hence, basically, step 1 amounts to solving $p$ convex optimization problems of low/moderate size.

\subsubsection{Step~2}

In step 2, the concentration estimates $\hat{\sigma}_{k|t}^{i}$, $k= t-N, \ldots, t$, obtained by solving Problem of step~1 in correspondence of any binary sensor $i$, are used as linear pseudo-measurements in order to estimate the whole concentration field over the spatial domain $\Omega$. By resorting again to the FE approximation, the vector of coefficients $x_t$ can be estimated, for example, by minimizing a quadratic MH cost function of the form:
\begin{eqnarray}\label{eq:cost_function_2}
\overline J_{t}(X_{t-N:t}) &=& \|x_{t-N} - \overline{x}_{t-N}\|^{2}_{\Psi} + \sum_{k=t-N}^{t}\|x_{k+1}-Ax_{k}-Bu\|^{2}_{Q}\nonumber \\
&+& \sum_{k=t-N}^{t}\sum_{i=1}^{p}\|\hat{\sigma}_{k|t}^{i}-C^{i}x_{k} - D^{i} \gamma \|^{2}_{\Xi^{i}}
\end{eqnarray}
where the quantities $A$, $B$, $\gamma$, $C^{i}$, $D^{i}$, for $i=1,\ldots,p$, are obtained by means of the FE method as in Section \ref{sec:field_estimation}. Notice that each term weighted by the positive definite matrix $\Xi^{i}$ penalizes the distance of the concentration $C^{i} x_{k} + D^{i} \gamma$ estimated through the FE approximation from the concentration $\hat{\sigma}_{k|t}^{i}$ estimated in step~1 on the basis of the binary measurements. The prediction $\overline{x}_{t-N}$ is computed in a recursive way as previously shown. In conclusion, at any time instant $t = N, N+1, \ldots$, after the application of step~1 the following problem has to be addressed. \\ \\
\textbf{Problem step~2:} Given the prediction $\overline{x}_{t-N}$ and the optimal estimates $\{\hat{\sigma}_{t-N|t}^{i},\ldots,\hat{\sigma}_{t|t}^{i}\}$, $i = 1, \ldots, p$ obtained by solving Problem of step~1, find the optimal estimates $\hat{x}_{t-N|t},\ldots,\hat{x}_{t|t}$ that minimize the cost function $\overline J_{t}(X_{t-N:t})$.\\ \\
The above estimation problem  admits a closed-form solution since the cost function (\ref{eq:cost_function_2}) depends quadratically on the states $\{x_{t-N|t},\ldots,x_{t|t}\}$. Hence, the computational efforts needed to perform step~2 of the fast MH-MAP filter turns out to be limited, so that the overall algorithm is computationally efficient as compared to a direct application of the MH-MAP filter to the large-scale system arising from the FE discretization. Numerical investigations of the effectiveness of the proposed Fast MH-MAP estimator can be found in \cite{GherardiniFIELDmap}.

\section{Noise-assisted estimation}

The noise-assisted paradigm relies on the idea that we can extract a greater amount of information from the knowledge of the measurement noise, which unavoidably affects each binary sensor. The measurement noise, indeed, shifts (of a certain amount) the analog measurements between the system outputs and the threshold of the binary sensors, leading in this way to a sufficiently large number of additional switching instants. The increased information content given by a greater number of switching instants can be exploited for estimation purposes \textit{only} by using a probabilistic approach, whose aim is to find the optimal estimates by solving a \textit{Maximum A Posteriori (MAP)} estimation problem. Such information is contained in the likelihood function of the binary measurements, which is evaluated at each time instant of the sliding window $\mathfrak W_t$. The MAP estimation problem, then, relies on defining a stochastic cost function, which is proportional to the corresponding log-likelihood function. Let us observe, moreover, that additional information contributions from the measurements can be extracted also by virtual/fictituos switching instants, which are first obtained by propagating the state estimates at the previous step of the procedure by using the model of the system and, then, compared with the predictions given by the measurement equation within the estimation sliding window. Clearly, the moving horizon approximation is crucial to design a viable recursive procedure to state estimation with binary sensors, allowing us both to reduce the computational complexity of the problem and to derive stability results.

Let us observe that the latter considerations lose of significance if the measurement devices are linear. A linear sensor, indeed, provides the maximum amount of information when the measurement noise is absent, returning in such case the best estimates. However, when the sensors are characterized by a prominent nonlinear characteristic (as e.g. for a binary sensor), the aforestated probabilistic approach can ensure to obtain good performance in noisy environments.

\subsection*{Towards noise-assisted quantum estimation}

The probabilistic approach to state estimation with binary sensors is very promising if applied to quantum mechanics, mainly for two reasons.
\begin{itemize}
\item
We could be able to derive a generalized expression of the Heisenberg's uncertainty principle when one or more measurement devices are taken into account and an algorithm (or in general a software) is integrated to them, with the capabilities to process the information coming from the devices and provide improved accuracy (super-resolution) for quantum state estimation~\cite{ParisBook}.
\item
We could design and realize reliable estimation schemes based on the noise-assisted estimation paradigm introduced in this chapter.
\end{itemize}
Hence, to conclude the chapter we will introduce a motivational example to convince the reader about a possibly successful application of the MH-MAP estimator to quantum systems. To this end, let us start from the well-known double-slit experiment, that has demonstrated the fundamentally probabilistic nature of quantum mechanical phenomena, by showing that light and matter can exhibit features both of waves and particles~\cite{davisson1928}.

\begin{figure}[h!]
\includegraphics[scale = 2.2]{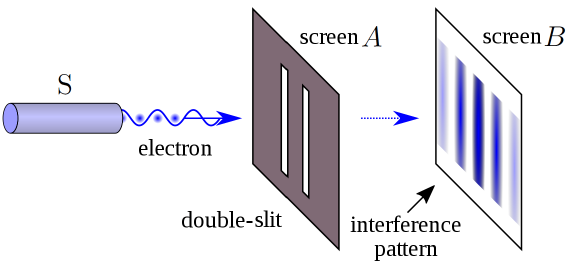}
\centering
\caption{Pictorial representation of the double-slit experiment, proving the wave-particle duality. From a laser source, one electron at a time reaches the screen $B$, by first passing through the holes $1$ and $2$ of the screen $A$. On the screen $B$ we observe an interference pattern, which is given by dark and light regions representing the constructive and destructive interference fringes of system with a wave behaviour.}
\label{fig:double_slits}
\end{figure}

\begin{figure}[h!]
\includegraphics[scale = 2.2]{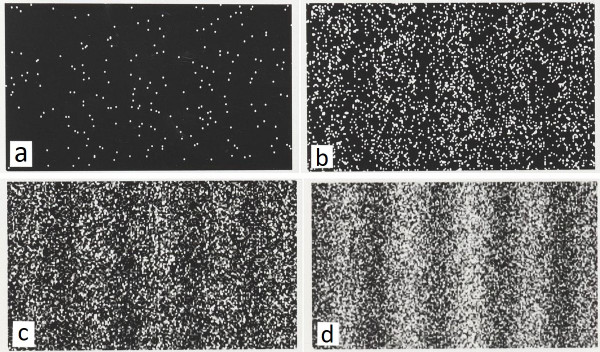}
\centering
\caption{Double-slit experiment by Tonomura et.al.~\cite{Tonomura}, which was performed by collecting $100$ (a), $3000$ (b), $20000$ (c) and $70000$ (d) electrons on screen $B$.}
\label{fig:tonomura}
\end{figure}

A pictorial representation of the double-slit experiment is shown in Fig.~\ref{fig:double_slits}, where ${\rm S}$ denotes a source of electrons, with the same energy and the same probability to impinge on the screen $A$, after coming out in all the space directions.
Such screen has two holes, $1$ and $2$, through which the electrons may pass. Moreover, behind the screen $A$ a second screen $B$ is present. In correspondence of $B$, we place a set of $p$ photo-detectors, each at various distance $d_{i}$, with $i = 1,\ldots,p$, from the center of the screen~\cite{feynman93}. The reconstruction of the interference patterns along the screen $B$ is provided by the photo-detectors, which measure the presence or the absence of an electron at the distance $d_{i}$ at the time instant $t$. In case both the holes are open, the standard intensity profile $I\propto \mathfrak{p}(d_{i})$, representing the wave interference pattern, is recovered. $I$ is the intensity of a wave, which is arriving at the screen $B$ (at distance $d_{i}$ from the center) by starting from ${\rm S}$, while $\mathfrak{p}(d_{i})$ is the probability to find an electron at such a distance. In particular, the images (a)-(d) in Fig.~\ref{fig:tonomura} from \cite{Tonomura} by Tonomura et.al.~show the interference patterns with $100$, $3000$, $20000$ and $70000$ electrons, whose mean velocity is approximately equal to $0.4c$, where $c$ is the speed of light.

A photo-detector is a binary measurement device. Indeed, by assuming that the power transferred by the laser source ${\rm S}$ is very weak, each photo-detector records a pulse representing the arrival of an electron at a different time instant $t$, so that two or more photo-detectors cannot simultaneously respond for the arrival of a particle. The latter event can occur only if the source emits two electrons within the resolving time of the detectors. However, the probability of such occurrence decreases exponentially by reducing the power of the laser source. As a result, each detector at position $d_{i}$ records the passage of a single electron, which travels from ${\rm S}$ to $d_{i}$, at different (random) times. In other words, the effective arrival of the electrons at the screen $B$ is not continuous, but corresponds to a rain of particles. As a remark, let us observe that we are implicitly assuming that each measurement device (which can be also modeled as a single qubit in quantum computing or a cavity mode in quantum electrodynamics) is not altered by the process of measurement, so that the presence of the detector can only affect the energy difference with the measured system.

Now, to introduce the noise-assisted quantum estimation paradigm, we define by $\mathfrak{p}_{1}(d_{i})$ and $\mathfrak{p}_{2}(d_{i})$ the probabilities for the electron to arrive at the screen $B$, respectively, through hole $1$ and hole $2$. Each of these probabilities can be found by measuring the change of the electron to arrive at $d_{i}$ when only the corresponding hole is open. The probability $\mathfrak{p}(d_{i})$ strictly depends on $\mathfrak{p}_{1}(d_{i})$ and $\mathfrak{p}_{2}(d_{i})$: if both holes are open, then the change of arrival at the position $d_{i}$ is \textit{not} simply given by the sum of the probabilities $\mathfrak{p}_{1}(d_{i})$ and $\mathfrak{p}_{2}(d_{i})$. It has been observed, indeed, that $\mathfrak{p}(d_{i})$ is the absolute square of a phase term $\phi(d_{i})$ (a complex number), which is the \textit{arrival amplitude} to reach the point $d_{i}$. In other words
\begin{equation}
\mathfrak{p}(d_{i}) \equiv |\phi(d_{i})|^{2},~~~\text{with}~~~\phi(d_{i}) = \phi_{1}(d_{i}) + \phi_{2}(d_{i}),
\end{equation}
where $\phi_{1}(d_{i})$ and $\phi_{2}(d_{i})$ are solutions of a wave equation, modeling the spread of an electron from ${\rm S}$ to the point $d_{i}$ by passing through the holes $1$ and $2$, respectively. As a result, $\mathfrak{p}_{1}(d_{i}) = |\phi_{1}(d_{i})|^{2}$ and $\mathfrak{p}_{2}(d_{i}) = |\phi_{2}(d_{i})|^{2}$. However, this interference pattern can no longer be observed if one of the two holes is closed, or if one observer is present behind them. In such cases, the probability $\mathfrak{p}(d_{i})$ turns out to be \textit{classical}, in the sense that it becomes equal to the sum of $\mathfrak{p}_{1}(d_{i})$ and $\mathfrak{p}_{2}(d_{i})$. Thus, the presence of an observer in correspondence of hole $1$ or $2$ radically changes the dynamics of the system. In this regard, it might be worth asking \textit{what is the best accuracy that can be achieved in detecting the presence of an observer behind one of the two holes by adopting a version of the MH-MAP state estimator, which has been properly designed for the quantum mechanical framework}. To this end, we first model the \textit{measurement equation}, by observing that the binary measurements are \textit{statistically independent} in each time interval (indeed the photo-detectors do not affect each other). Since the resolution of the photo-detector is chosen comparable with the intensity of the laser source, so that each sensor can record at most one electron in the time instants of the acquisition procedure, the outputs $y(d_{i},t)$ of the photo-detectors are impulsive signals. The latter assume value $1$ or $0$ in correspondence of the discretized time instants $t_{j} = jT_{s}$ ($T_{s}$ is the sampling time, which in this case is directly proportional to the detector resolution), respectively, when an electron is detected or not detected. If we consider again the sliding window $\mathfrak W_t = \{t-N, t-N+1, \ldots, t\}$ composed of $N$ samples, then the measurement equation will be the intensity profile $\mathfrak{p}(d_{i})$, which is evaluated within $\mathfrak W_t$:
\begin{equation}
\mathfrak{p}_{t-N:t}(d_{i}) = \frac{1}{N}\left(\sum_{j=0}^{N}y(d_{i},t-N+j)\right).
\end{equation}
Then, being each electron detected by only one photo-detector at any time instant $t$, the following constraint can be stated:
\begin{equation}
\sum_{i=1}^{p}y(d_{i},t) = 1~~\forall t,
\end{equation}
so that the probability conservation is ensured:
\begin{equation*}
\sum_{i=1}^{p}\mathfrak{p}_{t-N:t}(d_{i})=\frac{1}{N}\sum_{j=0}^{N}\sum_{i=1}^{p}y(d_{i},t-N+j)=\frac{1}{N}\sum_{j=0}^{N}1=1.
\end{equation*}
Instead, the measurement noise $v_{t}^{i}$ on the $i-$th photo-detector is modeled as a Bernoulli random variable, whose information content is encoded in the conditional probability $p(v_{t}^{i}|y_{t}^{i})$, where $y_{t}^{i}$ is the measurement outcome at time instant $t$ as if the $i-$th photo-detector were \textit{ideal}. Thus, if $y_{t}^{i}=0$, then the measurement noise $v_{t}^{i}$ can assume the values $0$ and $1$, respectively, with probabilities $p_{1}$ and $1 - p_{1}$. Conversely, if $y_{t}^{i}=1$, $v_{t}^{i}$ can be $0$ or $-1$ with probabilities $p_{2}$ and $1 - p_{2}$. More formally,
\begin{equation*}
\text{If}~~y_{t}^{i} = 1 \Rightarrow v_{t}^{i} = \begin{cases} 0, \text{with prob.} = p_{1} \\ -1, \text{with prob.} = 1 - p_{1} \end{cases}
\end{equation*}
and
\begin{equation*}
\text{if}~~y_{t}^{i} = 0 \Rightarrow v_{t}^{i} = \begin{cases} 0, \text{with prob.} = p_{2} \\ 1, \text{with prob.} = 1 - p_{2} \end{cases},
\end{equation*}
so that
\begin{equation*}
\sum_{k}p(v_{t,k}^{i}|y_{t}^{i} = y_{t,j}^{i}) = 1, \forall j,i\in\mathbb{N}~~\text{and}~~\forall t\in\mathbb{R},
\end{equation*}
where $v_{t,k}^{i}$ and $y_{t,j}^{i}$ are, respectively, the values that can be assumed by the random variables $v_{t}^{i}$ and $y_{t}^{i}$ at time $t$. Finally, also the presence of the observer will be described by a piece-wise function, entering the dynamical behaviour of the whole system. In this respect, the double-slit experiment represents the natural benchmark to test the presence and the role of an external observer in quantum experimental setups, since any attempt to determine which slit an electron has passed through destroys its interference pattern on the arrival screen (screen $B$ in Fig.~\ref{fig:double_slits}).

To summarize, by knowing the dynamical model describing the dynamics of an electron from the laser source ${\rm S}$ to the screen $B$, one could infer if an external observer is present in correspondence of the holes $1$ or $2$, i.e. if the probability profiles $\mathfrak{p}_{t-N:t}(d_{i})$ are classical or quantum. Afterwards, by taking into account the contribution of noise over the photo-detector outcomes, one could estimate not only the presence of the observer, but also the quantum state of the electron (i.e. the expectation values of its position and momentum operators) around the two holes on the screen $A$, by using the outcomes \textit{yes} or \textit{no} from the photo-detectors. We expect that, by applying a properly designed MH-MAP estimator, the estimation accuracy is largely improved.

As additional remarks, let us notice that placing a given number $p$ of photo-detectors along the screen $B$ and, then, waiting for the arrival of an electron to the screen $B$ corresponds, to all effects, to irregularly sample a portion of the quantum state of the system. Indeed, both the measurement noise and the dephasing (which randomizes the phase of the wave function of the system at any time instant $t$) contributes to make irregular the nominal sampling interval. Furthermore, it is worth noting that to define a noise-assisted estimation paradigm in the quantum mechanical context it is essential to assume that the measurements are not directly performed on the components of the system state that has to be inferred. In other words, the estimation scheme (exploiting a noise-assisted paradigm) has to be chosen among those schemes, which implement a \textit{Quantum Non-Demolition (QND) measurement}~\cite{Barchielli1991}. The latter is a special type of measurement, for which the uncertainty of the measured observable (given by the measurement process) does not increase with respect to the measured value after the subsequent evolution of the system. Also for this reason, a QND measurement is the \textit{most classical} and least disturbing type of measurement in quantum mechanics. Accordingly, also the noise-assisted estimation paradigm requires both to perform indirect sequences of measurements and to propagate (over time) the estimates of the quantum state without losing information about the dynamical behaviour of the system. In this way, the accuracy of the estimation procedure shall be improved (from a purely probability point of view), by quantitatively using the information content of the measurement outcome, and a driven optical cavity Quantum Electro-Dynamics (QED) setup could be chosen as the most suitable apparatus to effectively achieve noise-assisted quantum estimation. Such a setup, already used e.g. in \cite{Cavities2012} to probe energy transport dynamics in photosynthetic bio-molecules, is essentially given by a pump-probe scheme, in which the probe system is the cavity mode and the sample system, whose state has to be estimated, is confined inside an optical cavity. The energy injected into the system, instead, is provided by an external laser field (pump).

\section{Conclusions and contributions}

Summarizing, this chapter provides the following contributions:
\begin{itemize}
\item
Design of novel moving-horizon state estimators for discrete-time dynamical systems subject to binary (threshold) measurements using both a deterministic and a probabilistic approach.
\item
By adopting the deterministic approach, both a least-square and a piece-wise quadratic cost function to be minimized have been introduced, either including or not constraints (boundary conditions). By assuming also the presence of unknown but bounded noises affecting both the system and the measurement devices, stability results have been proved, showing that all proposed estimators, irrespectively of the cost being used and of the inclusion of constraints, guarantee an asymptotically bounded estimation error under suitable observability assumptions. Performance comparison and examples have demonstrated the effectiveness, in terms of both estimation accuracy and computational cost, of our approach, especially with respect to particle filtering. However, from a more practical point of view, it is worth noting that, depending on the system dynamics, the adoption of binary sensors with time-invariant thresholds may not be sufficient to always ensure uniform observability. Thus, it can be required that one uses a greater number of sensors or to make each threshold oscillate within the range of variability of the sensed variable with a sufficiently high frequency. In this case, the optimal value for the frequency of oscillation of the sensors threshold is strictly dependent on the value of $N$ (length of the observation window) and the eigenvalues of $A$. Indeed, one can expect that less binary sensor detections are obtained when the eigenvalues of $A$ are real (and not complex conjugate) or the input excitation is poor (given e.g. by pulse or step functions).
\item
Formulation of \textit{Moving Horizon} (MH) \textit{Maximum A posteriori Probability} (MAP) estimators to solve state estimation problems with binary sensors (probabilistic approach). It has been proved that, in case the dynamical system is linear and the threshold measurements are statistically independent, the optimization problem turns out to be convex, and, thus, solvable with computationally efficient algorithms. By recasting the estimation problem in a Bayesian framework, each binary measurement is characterized by an information content for each time instant $t$, encoded in the likelihood functions $p(y_{t}^{i}|x_{t})$, which is able to distinguish in probability if a binary switching is due to noise or to the dynamical behaviour of the system. Moreover, the simulation results have exhibited the conjectured \textit{noise-assisted} feature of the proposed estimator: starting from a null measurement noise, the estimation accuracy improves until the variance of the noise achieves an optimal value, beyond which the estimation performance deteriorates.
\item
By using the probabilistic approach and the \textit{Finite Element (FE)} approximation, field estimation of spatially distributed system with noisy binary measurements has been addressed. The PDE, modeling a given propagating field, has been discretized and simulated with a mesh of finite elements over the spatial domain $\Omega$, and a network of pointwise-in-time-and-space threshold sensors has been introduced. Then, we have proposed also an improved version of the MH-MAP estimator, which can be adopted for large-scale systems.
\item
Introduction of a noise-assisted estimation paradigm for quantum dynamical systems.
\end{itemize}

\chapter{Noise-assisted quantum transport}
\label{chap:NAT}
\fancyhead{}
\fancyhead[LEH]{\leftmark}
\fancyhead[RO]{\rightmark}

\textit{
In this chapter, we will address theoretical models for noise-assisted quantum transport, that have been confirmed and reproduced with an high degree of controllability by a scalable transport emulator based on optical fiber cavity networks, completely realized at the Consiglio Nazionale delle Ricerche (CNR) in Florence, Italy. The possibility to design a perfectly controllable experimental setup, whereby one can tune and optimize its dynamics parameters, is a challenging but very relevant task, so as to emulate the transmission of energy in light harvesting processes. Also disorder and dephasing noise can be finely tuned within the emulator until the energy transfer efficiency is maximized. In particular, we proved that the latter are effectively two control knobs allowing to change the constructive and destructive interference patterns to optimize the transport paths towards an exit site. \footnote{The part of this chapter related to noise-assisted transport has been published as ``Disorder and dephasing as control knobs for light transport in optical cavity networks'' in \textit{Scientific Reports} \textbf{6}, 37791 (2016)~\cite{GherardiniNAT}.}
}

\section*{Introduction}

Transport phenomena, i.e. the transmission of energy through interacting systems, represent a very interdisciplinary topic with applications in many fields of science, such as physics, chemistry, and biology. In particular, the study and a full understanding of such mechanisms will allow to improve in experimental and industrial setups the process of transferring classical and quantum information across complex networks, and to explain the high efficiency of the excitation transfer through a network of chromophores in photosynthetic systems~\cite{qbiobook}. Very recently theoretical and experimental studies, indeed, have shown that the remarkably high efficiency (almost $100\%$) of the excitation energy transfer in photosynthetic systems seems to be the result of an intricate \textit{interplay between quantum coherence and noise}~\cite{Lee2007SCI316,Engel2007NAT446,Mohseni2008JCP129,Plenio2008NJP10,Caruso2009JChPh131,Rebentrost2009NJP11,Collini2010NAT463,Panitchayangkoon10PNAS107,
Hildner2013SCI340}. In this regard, let us recall the concepts of quantum coherence and dephasing noise. Quantum coherence is a feature of each quantum system, and comes from their wave-like properties. As previously shown, particles such as electrons can behave as waves, which can interfere and originate peculiar patterns, given by sequences of bright and dark bands representing, respectively, constructive and destructive interference. Such wave-like behaviour, which is mathematically described by a wave function, is related to quantum coherence. Conversely, dephasing (or quantum decoherence)~\cite{Schlosshauer2005} is the mechanism which leads to the loss of coherence, and involves quantum systems that are not completely isolated, but still interacting with the environment. Then, the effects of the environment to the quantum system dynamics is to randomize the phase of its wave function.

Regarding energy transport phenomena, the presence of coherence between chromophores of the photosynthetic system leads to a very \textit{fast delocalization} of the excitation, that can hence exploit several paths to the target site, named also as sink or reaction center. However, since the destructive interference among different pathways and energy gaps between sites of the complex are obstacles to the transmission of energy, this regime is not optimal by itself. Only the additional and unavoidable presence of disorder and noise, which is usually assumed to be deleterious for the transport properties, seems to positively affect the transmission efficiency~\cite{Caruso2009JChPh131,FC2014}. This can be explained in terms of the
\begin{itemize}
  \item inhibition of destructive interference;
  \item opening of additional pathways for excitation transfer.
\end{itemize}
This mechanism is known as \textbf{Noise-Assisted Transport (NAT)}, and it has been recently observed in some physical platforms: all-optical cavity-based networks~\cite{VicianiPRL2015,Biggerstaff2015}, integrated photonic structures~\cite{FC2016}, and genetically engineered complexes~\cite{FC2016NATmat}. However, a deep analysis of the underlying contributions, such as \textit{interference}, \textit{disorder} and \textit{dephasing}, and their interplay was still missing from the experimental point of view. In this regard, the possibility to experimentally realize simple test platforms being able to mimic transport on complex networks, reproducing NAT effects, could allow \begin{itemize}
  \item to clarify the role of disorder, interference and noise contributions in the transport behavior of natural (photosynthetic) complexes;
  \item to design feasible model of different complex networks, where the role of topology can be further investigated;
  \item to engineer new artificial molecular structures where all the aforementioned control knobs (e.g. disorder and interference patterns) are optimized to achieve some desired tasks, such as the maximization of the transferred energy or its temporary storage in some part of the network.
\end{itemize}
As shown also in chapter~\ref{chapter1}, the possibility to introduce and exploit a noise-assisted paradigm relies on our understanding of the complexity features of the system, as given in this case by the \textit{network topology}. Thus, the introduction of a simple experimental setup has remarkable advantages with respect to real biological samples or expensive artificial systems, which are very difficult to manipulate both in their geometry and in the system parameters, being these aspects governed by specific bio-chemical laws.

In this chapter, we will argue the recently realized optical platform, entirely based on fiber-optic components, which has been designed to emulate the transmission of energy in light harvesting processes. In particular, we will show how an optimal combination of constructive and destructive interference with static disorder and dephasing noise can be successfully exploited as feasible control knobs to manipulate the transport behavior of coupled structures and optimize the transmission rate. It is worth noting that such a setup has provided the first experimental observation of the typical NAT predictions in the dependence of the network transmission rate as a function of the amount of noise introduced into the system~\cite{VicianiPRL2015}.

\section{Theoretical model}

In the proposed experimental setup, the coherent propagation of excitons in a $N-$site quantum network is simulated by the propagation of photons in a network of $N$ coupled optical cavities. The dynamics of an optical cavity network can be described by the following Hamiltonian:
\begin{equation}\label{eq1}
H = \sum_{i}\hbar\omega_{i}a^{\dagger}_{i}a_{i}+\sum_{(i,j)}\hbar g_{ij}\left(a^{\dagger}_{i}a_{j}
+a_{i}a^{\dagger}_{j}\right),
\end{equation}
where $a_{i}$ and $a^{\dagger}_{i}$ are the usual bosonic field operators~\cite{Reed1975}, that annihilate and create an excitation (a photon in our case) in the $i$-th site of the network, $\omega_{i}$ is the corresponding resonance frequency, and $g_{ij}$ are the coupling constants between all the connected sites. The first term in (\ref{eq1}) describes the energy structure of the system, while the second one is related to the hopping process among the network nodes. Hereafter, we will refer to the (random) energy level spacings of the network's sites as \textit{static disorder} and it will be obtained by tuning the frequencies $\omega_{i}$ of the network sites. Conversely, the presence of \textit{dephasing} noise, randomizing the photon phase during the dynamics, will be introduced in terms of dynamical disorder, i.e. time-dependent random variation of the site energies. The dephasing rate for the site $i$ is denoted with $\gamma_{i}$. To summarize:
\begin{itemize}
  \item Static disorder $\equiv$ time-independent random energy level spacings of the network's sites.
  \item Dephasing noise $\equiv$ time-dependent random variation of the site energies.
\end{itemize}
From a mathematical point of view, the effects of the presence of some dephasing noise on the quantum network dynamics is described by the so-called Lindblad super-operator $\mathcal{L}_\textrm{deph}(\rho)$~\cite{PetruccioneBook}, defined as
\begin{equation}\label{deph_operator}
\mathcal{L}_\textrm{deph}(\rho)=\sum_{i}\gamma_{i}\left(-\{a^{\dagger}_{i}a_{i},\rho\}+
2a^{\dagger}_{i}a_{i}\rho a^{\dagger}_{i}a_{i}\right) \; ,
\end{equation}
where $\rho$ denotes the \textit{density matrix} describing the state of the system. The super-operator $\mathcal{L}_\textrm{deph}$ models the interaction of a quantum system with its environment (described by an Hamiltonian term) under the validity of the Born and Markov approximations. The Born approximation relies on assuming between the system and the environment a \textit{weak coupling}, which allow us to consider the contribution of the interaction on the behaviour of the system only up to the second order of the corresponding perturbative series as a function of the coupling term. The Markov approximation, instead, is based on the hypothesis that the environment has so many degrees of freedoms to consider \textit{negligible} memory effects between the system and the environment itself. The evolution of the density matrix is given by the following differential Lindblad (Markovian) master equation:
\begin{equation}\label{master_equation_NAT}
\frac{d\rho}{dt}=-\frac{i}{\hbar}[H,\rho] + \mathcal{L}_\textrm{deph}(\rho) + \mathcal{L}_\textrm{inj}(\rho)
+ \mathcal{L}_\textrm{det}(\rho),
\end{equation}
where $[\cdot,\cdot]$ is the commutator. In this regard, let us recall that in quantum mechanics the density matrix has been introduced to describe the state of a quantum system, which does not necessarily live in a coherent superposition state corresponding to the eigenstate of a physical (Hermitian) observable. Such state is also called \textit{mixed quantum state}, and is defined as a statistical ensemble of pure states. In (\ref{master_equation_NAT}), the Lindbladian operators $\mathcal{L}_\textrm{inj}(\rho)$ and $\mathcal{L}_\textrm{det}(\rho)$ describe two distinct irreversible transfer processes, respectively, from the light source to the network (energy injection) and from the exit site to an external sink (energy detection). In particular, the injection process is modeled by a thermal bath of harmonic oscillators, whose temperature is expressed by the thermal average boson number $n_\textrm{th}$. In the Markov approximation, this process is described by the following Lindbladian term:
\begin{equation}\label{eq4}
\mathcal{L}_\textrm{inj}(\rho) = n_\textrm{th}\frac{\Gamma_{0}}{2}\left(-\{a_{0}a^{\dagger}_{0},\rho\}+
2a^{\dagger}_{0}\rho a_{0}\right) +(n_\textrm{th}+1)\frac{\Gamma_{0}}{2}\left(-\{a^{\dagger}_{0}a_{0},\rho\}+2a_{0}\rho a^{\dagger}_{0}\right),
\end{equation}
where $a^{\dagger}_{0}$ is the bosonic creation operator for the network input site (denoted as site $0$), and $\{\cdot,\cdot\}$ defines the anticommutator. The photons leaving the network, instead, are detected by the sink, that is usually denoted as the output port of the network, modeling the reaction center of a photosynthetic biological system. Each light-harvesting complex, indeed, is composed by several chromophores that turn photons into excitations and lead them to the reaction center, where the first steps of conversion into a more available form of chemical energy occurs. This part is described by another Lindbladian super-operator, that is
\begin{equation}\label{eq5}
\mathcal{L}_\textrm{det}(\rho)=\Gamma_\textrm{det}\left(2a^{\dagger}_\textrm{det}a_{k}\rho a^{\dagger}_{k}a_\textrm{det}-
\{a^{\dagger}_{k}a_\textrm{det}a^{\dagger}_\textrm{det}a_{k},\rho\}\right),
\end{equation}
where $a^{\dagger}_\textrm{det}$ refers to the effective photon creation in the detector with the subsequent absorption of excitations from the site $k$, according to operator $a_{k}$, with $\Gamma_\textrm{det}$ being the rate at which the photons reach irreversibly the detector.

Furthermore, the transferred excitation energy, reaching the sink at time $t$, is defined as
\begin{equation}\label{eq6}
E_\textrm{tr}(t) = 2\Gamma_\textrm{det}\int_{0}^{t}\text{Tr}\left[\rho(\tau)a^{\dagger}_{k}a_{k}\right]d\tau,
\end{equation}
where $\textrm{Tr}[\cdot]$ is the trace operator, $\Gamma_\textrm{det}$ denotes the rate at which the photons reach irreversibly the output detector, and $a_{k}$ refers to the effective absorption of photons from the site $k$ corresponding to the output site. To compare the theoretical results with the experimental data, where the light is continuously injected into the network (with rate $\Gamma_{0}$) and absorbed from the detector, we need to define the network transmission as the steady-state rate for the photons in the detector, i.e.
\begin{equation}\label{eq7}
\lim_{t \rightarrow \infty} 2\Gamma_\textrm{det}\text{Tr}\left[\rho(t)a^{\dagger}_{k}a_{k}\right].
\end{equation}
Let us point out that, if one repeats a single-photon experiment many times, one obtains the same statistics corresponding to an injected coherent state~\cite{Amselem2009PRL103}, but this holds just because nonlinear processes are \textit{not} present in our setup. Moreover, note that the model is able to take into account also the slight asymmetry that is present in the experimental setup by imposing different coupling rate in the two paths of the networks, i.e. $g_{01}\neq g_{02}$ and $g_{13}\neq g_{23}$. Such asymmetry is mainly caused by different loss rates in the two resonators.

\begin{figure}[h!]
\centering
\includegraphics[width=1.0\textwidth]{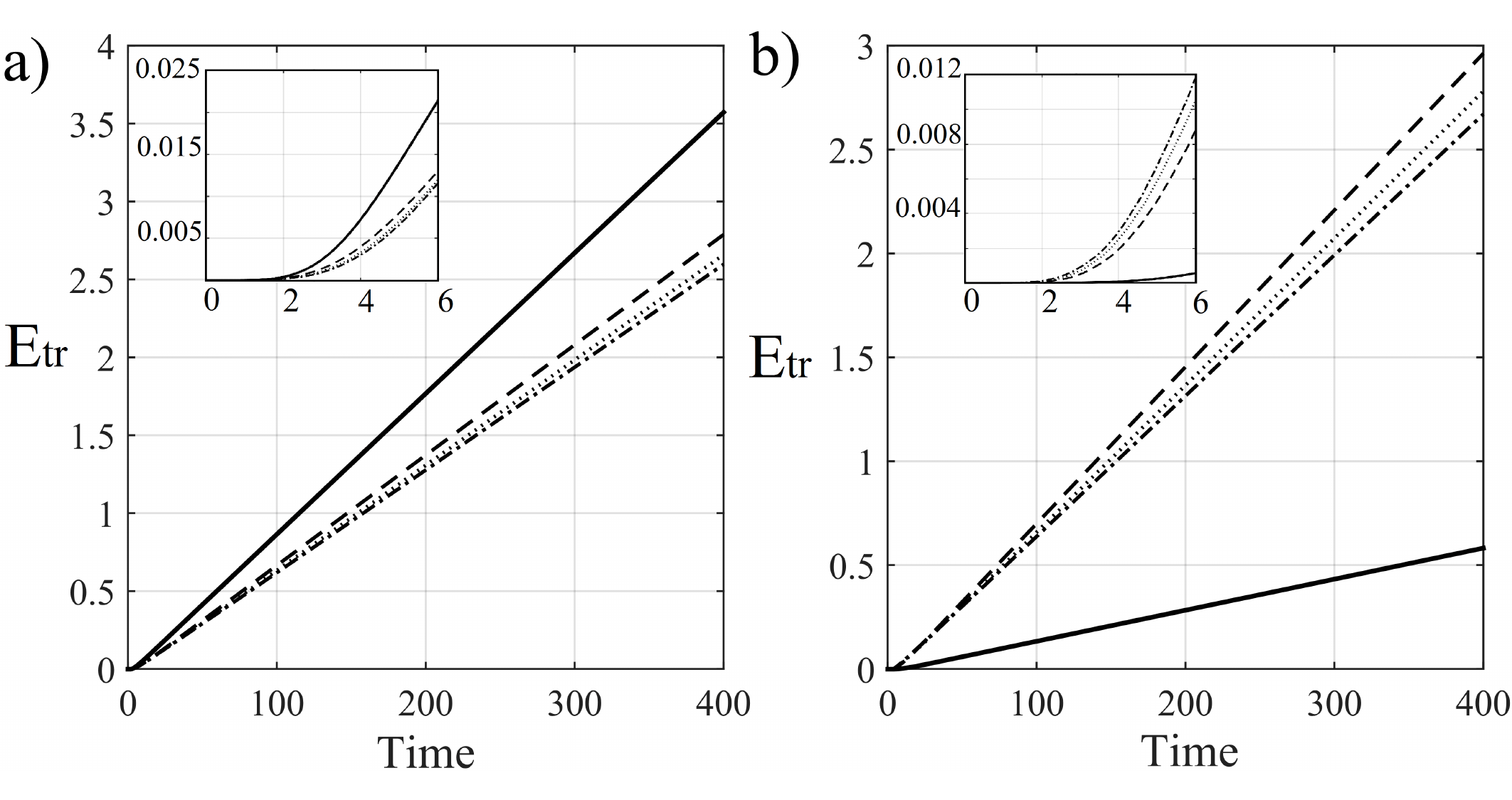}
\caption{Time evolution of the excitation transferred energy. $E_\textrm{tr}$ vs time for constructive (a) and destructive (b) interference. In both figures, the curves refer to different values of $\gamma_{2}$ and $\omega_{2}$: $(\gamma_{2},\omega_{2})=[0,0]$ and $(\gamma_{2},\omega_{2})=[0,3]$ for solid and dash-dot lines, respectively, while $(\gamma_{2},\omega_{2})=[2,0]$ and $(\gamma_{2},\omega_{2})=[2,3]$ for dashed and dotted lines. Moreover, we set $\Gamma_\textrm{det}=\Gamma_{0}=0.5$. Notice that $E_\textrm{tr}$ is measured in terms of the number of transmitted photons, while the time is in the units of the inverse coupling rates. Inset: Exponential behaviour of the transferred energy in the transient time regime.}
\label{fig:psink2}
\end{figure}

In Fig.~\ref{fig:psink2} different time behaviours of the transferred energy are shown, for different initial conditions of global constructive (\ref{fig:psink2}a) and destructive (\ref{fig:psink2}b) interference and different values of dephasing and static disorder.
The numerical results, which will be then compared with the experimental data from the setup, are obtained by implementing the master equation (\ref{master_equation_NAT}). To this aim, we have assumed that the network is initially empty, namely with no excitations inside, while a laser source continuously injects photons in the site $0$ with a rate $\Gamma_{0}$. Moreover, in order to take into account the experimental imperfections, the non-vanishing coupling constants are set in the range $[0.2, 0.5]$. Indeed, by varying such parameters, we observe a similar qualitative behaviour in agreement with the experimental observations, as we simply expect from our abstract model. The destructive interference, then, is simply obtained by introducing a phase in the hopping strength $g_{01}$, i.e. changing its sign, while the cavity resonance frequencies $\omega_{i}$ are all vanishing in the absence of static disorder that instead leads to a variation of the frequency of site $2$ within the range $[0,2]$. Similarly, the only non-zero dephasing rate $\gamma_{2}$ for the cavity $2$ (when dephasing is on) is chosen in the range $[0,1]$. Since the photon injection is continuous in time, the energy in the steady-state condition increases monotonically in time with an asymptotic linear behavior whose slope is indeed the transmission rate. An exponential behaviour is instead observed for the initial temporal regime, as shown in the inset of Fig.~\ref{fig:psink2}. Then we find that, for constructive interference both disorder and dephasing individually reduce the transferred energy (no NAT). However, in presence of some disorder inhibiting the path to constructive interference, dephasing slightly assists the transport by opening additional pathways. This can be also intuitively explained by the fact that the two cavities are not energetically on resonance (because of the disorder) but the line broadening effect, induced by dephasing, allows again the hopping between them. On the other side, destructive interference leads to very small transferred energies since the two transmission paths over the two cavities cancel each other (opposite phase). In this case, dephasing and disorder can hinder such a perfect cancellation, thus partially restoring transport, i.e. NAT behaviour.

\subsection{Computational complexity}

Here we evaluate the computational time for the single realization of the numerical dynamical evolution of networks of increasing size (i.e., number of sites/cavities $N$) with the model (\ref{deph_operator}).
\begin{figure}[h!]
 \centering
 \includegraphics[width=0.925\textwidth]{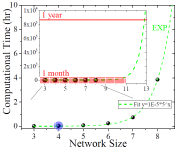}
 \caption{Computational complexity increasing the network size. Computational time (in hours) as a function of the simulated network size (i.e. $N$ number of cavities) for a single realization of the system dynamical evolution, with only one choice of the system parameters. An exponential behaviour is observed (green dashed line), hence a scheme with more than ten sites/cavities (red dashed region in the inset) becomes very hard to be simulated by a powerful workstation, corresponding on average to at least one month of simulation. Our experimental setup corresponds to the case of four cavities (blue dashed region), but can be easily extended to more cavities.}
\label{fig8}
\end{figure}
As shown in Fig.~\ref{fig8}, the computational complexity increases exponentially with $N$. In other words, already adding a few cavities to our model would take months to theoretically simulate the corresponding dynamics for a given set of parameters, and the computation becomes unfeasible if one wants to reconstruct the dynamical behaviour of the system where thousands of simulations are needed to take into account dynamical disorder, dephasing, etc. In particular, the critical number of sites above which it becomes very hard to reproduce the theoretical data is around $8$ (corresponding to around six months of simulations for one thousand realizations) -- see again Fig.~\ref{fig8}. Let us notice that, while the experimental scheme has been realized with coherent states of light that in principle allow its classical simulation time to scale polynomially with the number of optical elements, if one instead considered a full quantum regime (for example, with several single-photon walkers), the computation complexity would have indeed scaled exponentially, as observed above. On the other hand, the experimental complexity is not so affected by the network size and, at most, linearly increases in terms of both the cost of the optical components (cavities, beam-splitters, etc.) and the practical realization and observation time of the stationary behavior of the optical system. However, even if the increase of the network size is not a significant limit for the present setup employing a classical coherent source, the same would not be true when operating in a quantum regime, with one or multiple single-photon sources and coincidence detection. In such a case, the increased losses in a larger network (mainly due to the presence of several FBG resonators) could substantially reduce the efficiency of the setup and imply much larger acquisition times.

\section{Experimental setup}

\subsection{Network of fiber-optic resonators}

\begin{figure}[h!]
 \centering
 \includegraphics[width=1\textwidth]{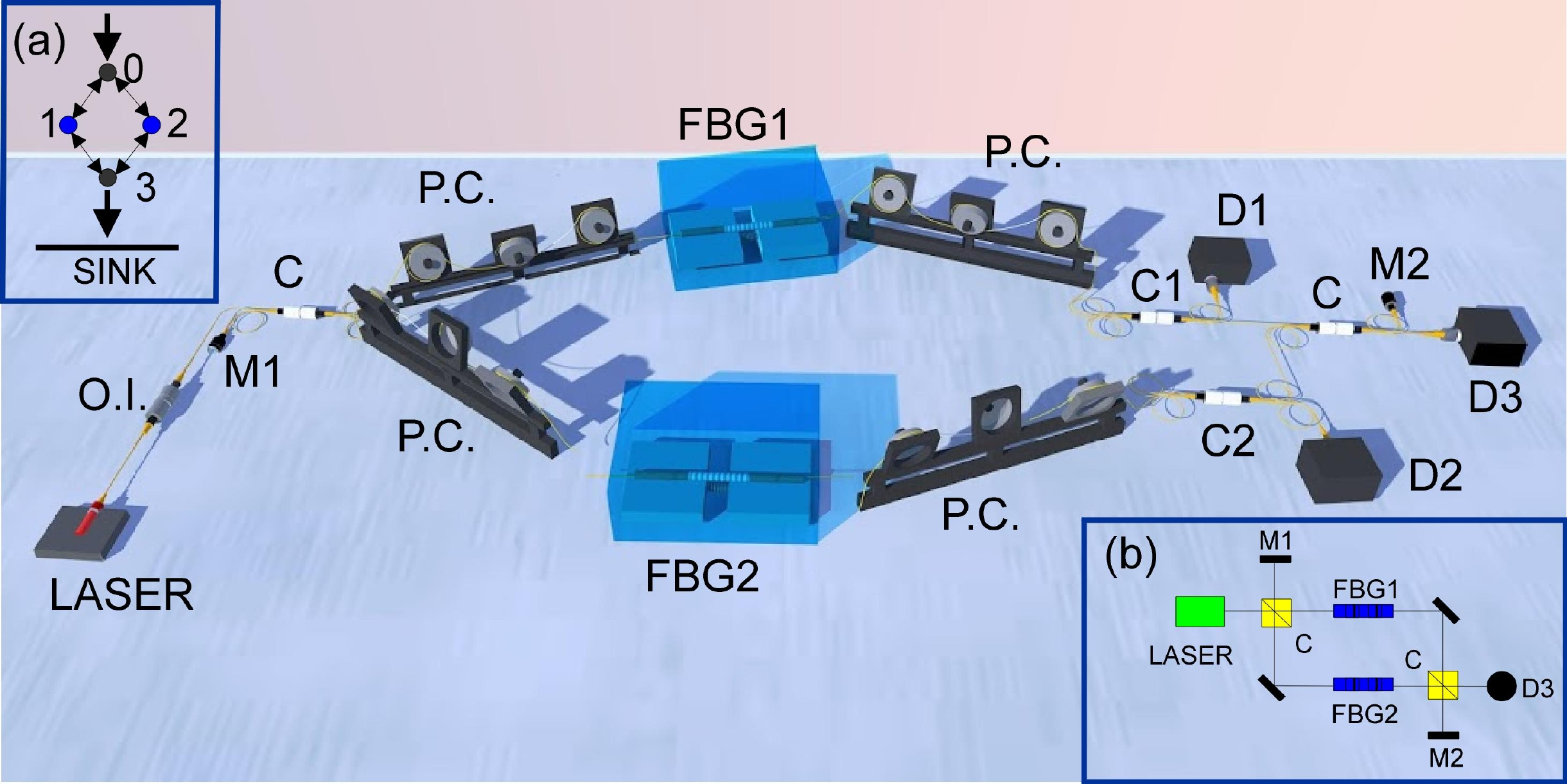}
 \caption{Experimental design of the fiber-optic cavity network. Light is injected in the network by a continuous-wave diode laser, transmitted over coupled optical cavities, and then irreversibly absorbed by a detector measuring the transmission rate. O.I., Optical Isolator; M1 and M2, mirrors; P.C., polarization controller; C, 50x50 fiber coupler; C1 and C2, 90x10 fiber couplers; FBG1 and FBG2, Fiber Bragg Grating resonators; PZT piezoelectric transducer; D1, D2 and D3, detectors. The PZT can be driven to introduce a difference between the resonance frequencies of FBG1 and FBG2, which can be either constant in time ($V_0$) or variable ($V(t)$), in order to insert disorder and/or dephasing into the network. Inset (a): scheme of the 4-site network mimicked by the optical setup. Inset (b): simplified scheme of our optical platform.}
 \label{fig1}
\end{figure}

The realized experimental fiber-optic setup, which reproduces energy transport phenomena (including the NAT effect), is given by the 4-site network shown in the inset (a) of Fig.~\ref{fig1}~\cite{VicianiPRL2015}. The detailed scheme of the experimental apparatus, instead, is shown in Fig.~\ref{fig1}.
More schematically, the scheme is reproduced by the Mach-Zehnder setup of inset (b), which presents the following two main differences with respect to a standard Mach-Zehnder interferometer:
\begin{itemize}
  \item The insertion of a FBG resonator in each path of the interferometer.
  \item The presence of two additional mirrors (M1 and M2) at the normally unused input and output port of the interferometer.
\end{itemize}
The resonators FBG1 and FBG2 represent the sites 1 and 2 with variable local excitation energy $\omega_1$ and $\omega_2$, while the role of the other two sites 0 and 3 is played by two fiber optic couplers (C), which represent two sites with fixed local excitation energy resonant with the energy of the propagating excitation $\omega_\textrm{S}$ ($\omega_0 = \omega_3 = \omega_\textrm{S}$). The presence of the two additional mirrors M1 and M2 makes it possible to couple sites 1 and 2, since the light reflected by the resonators is partially re-inserted into the network by M1, while transmitted light is partially recycled by M2.

The setup is entirely based on single-mode fiber-optic components at telecom wavelength (1550 nm). The choice of fiber components presents the following advantages:
\begin{itemize}
  \item It completely removes issues related to matching the transverse spatial mode of the fields and considerably simplifies the alignment of sources, cavities, and detectors, thus allowing one to easily adjust the network size and topology.
  \item Working at telecom wavelengths guarantees low optical losses and a low cost of the fiber components.
\end{itemize}
These two advantages are essential to achieve the \textit{scalability} of the apparatus. The FBG resonators are characterized by a straightforward alignment and easy tunability by tiny deformations of the fiber section within the Bragg mirrors. Each resonator is inserted in a home-made mounting to isolate it from environmental noise and allowing the piece of fiber containing the cavity to be stressed and relaxed in a controlled way by the contact with a piezoelectric transducer (PZT). In such a way the length of each cavity and, consequently, its resonance frequency, can be finely tuned. The laser source injects light of frequency $\omega_\textrm{S}$ into one input port of the first (50:50) fiber coupler C. Light exiting the two cavities passes through two more polarization controllers before being coupled by a second (50:50) fiber coupler C. Finally, one of the interferometer outputs is measured by detector D3.

All the parameters characterizing the network and that are described in the following sections are expressed in terms of the \textit{cavity detuning parameter} $\Delta x$, which is defined as
\begin{equation}\label{eq8}
  \Delta x \equiv \omega_2 - \omega_1.
\end{equation}
Thus, the cavity detuning parameter is equal to the difference between the resonance frequencies of the two cavities (in units of their linewidth). Finally, two additional fiber couplers (C1 and C2) are used to split a small portion (about 10\%) of the light in each interferometer arm in order to measure the transmission peaks of each single cavity before interference. The transmission signals are measured by detectors D1 and D2 while scanning the laser frequency $\omega_\textrm{S}$ over an interval including a single longitudinal mode of both cavities; from the difference between the frequency positions of the two peaks it is possible to infer $\Delta x$.

\subsection{Network parameters}

The network configuration is completely described by the following 3 characteristics:
\begin{itemize}
  \item The initial conditions related to global interference.
  \item Static disorder.
  \item Dephasing or dynamical disorder.
\end{itemize}
All these features are defined in terms of the cavity detuning parameter $\Delta x$. Let us observe that changes in the global interference of the network and in the values assumed by dephasing and static disorder will involve radical variations in the dynamical behaviour of the whole system. In this regard, the implemented network of fiber-optic resonators is one of the first experimental setups in which we can observe and reproduce the beneficial effects in controlling some parameters, that are defined by stochastic processes. Moreover, since the aforementioned model well fits the experimental data from the network transmission, future investigations about the definition of optimized routines of stochastic variables for control tasks are desirable. \\ \\
\textbf{Initial Conditions for Interference.-}
The interferometric apparatus involves no active stabilization. Consequently, the system is intrinsically unstable and the network response will be time dependent (on the time scale of the order of hundreds of ms). This intrinsic instability can be used to establish different initial conditions of global interference for our network. The system throughput when the two cavities are resonant ($\Delta x=0$) and without any kind of noise will vary between a minimum and a maximum value in correspondence of global destructive or constructive interference. The output signal measured by detector D3 in this case will thus set the initial conditions of global interference of the network.\\ \\
\textbf{Disorder or Static Disorder.-}
A network is said to be disordered if the local excitation energies of different sites are unequal ($\omega_j \neq \omega_k$, with $j \neq k $), but constant in time. In the experimental setup, the static disorder of the network is quantified with the cavity detuning parameter $\Delta x$.
\begin{figure}[h!]
 \centering
 \includegraphics[width=1.02\textwidth]{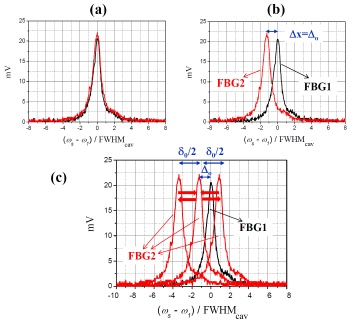}
 \caption{Definition of Disorder and Dephasing.  Transmission peaks of FBG1 (black line) and FBG2 (red line), measured respectively by detector D1 and detector D2 when the laser frequency $\omega_{S}$ is scanned over an interval including a single longitudinal mode of both cavities. a) Ordered system without dephasing: $\Delta x = 0$ and constant. b) System with static disorder $\Delta_0$: $\Delta x = \Delta_0$ and constant (no dephasing). c) System with disorder $\Delta_0$ and dephasing $\delta_0$: $\Delta x$ variable between $\Delta_0 - \delta_0/2$ and $\Delta_0 + \delta_0/2$.}
 \label{fig1bis}
\end{figure}
The case of an ordered system is illustrated in Fig.~\ref{fig1bis}a, where $\Delta x = 0$ and constant. The case of a disordered system with a static disorder $\Delta x = \Delta_0 \neq 0$ is shown in Fig.~\ref{fig1bis}b. We assume that a system presents a medium level of disorder if $\Delta x \sim 1$, i.e. $(\omega_2 - \omega_1) \sim$ $\textrm{FWHM}_\textrm{cav}$, which is defined as the full-width at half-maximum linewidth of the cavity mode, while we have a high level of disorder if $\Delta x > 1$, i.e. $(\omega_2 - \omega_1) >$ $\textrm{FWHM}_\textrm{cav}$. Thus, to summarize:
\begin{itemize}
  \item Ordered system $\longrightarrow$ $\Delta x = 0$ and constant.
  \item Medium level of disorder $\longrightarrow$ $\Delta x \sim 1$ $\longrightarrow$ $(\omega_2 - \omega_1) \sim$ $\textrm{FWHM}_\textrm{cav}$.
  \item High level of disorder $\longrightarrow$ $\Delta x > 1$ $\longrightarrow$ $(\omega_2 - \omega_1) >$ $\textrm{FWHM}_\textrm{cav}$.
\end{itemize}
\noindent
\textbf{Dephasing or Dynamical Disorder.-}
Dephasing or dynamical disorder introduces a random phase perturbation in one or more sites of the network, thus resulting in temporal fluctuations of the corresponding resonance frequencies $\omega_j$'s around their stationary values. We can introduce it into our network by slightly changing the value of $\omega_2$ during measurement by means of the piezoelectric transducer. In such a way, $\Delta x$ is not time constant but can vary within an interval of $\pm \delta x / 2$ around $\Delta x$. The amount of dephasing can be quantified by the amplitude $\delta x$ of this interval. The case of a network with disorder $\Delta_0$ and a dephasing $\delta_0$ is illustrated in Fig.~\ref{fig1bis}c, where $\Delta x$ is variable in the interval $[\Delta_0 - \delta_0 / 2,\Delta_0 + \delta_0 / 2]$. We assume that a system presents a medium level of dephasing if  $\delta x \sim 1$, and a high level of dephasing if $\delta x > 1$.\\ \\
\textbf{Network Transmission.-} Finally, the transmission of the network is defined as the output signal measured by detector D3 in correspondence of the transmission peak of cavity 1 ($\omega = \omega_1$). Technical details about the acquisition procedure can be found in \cite{GherardiniNAT}. Then, the value of the transmission is normalized to the measured value of the output signal in conditions of constructive interference, without disorder and without dephasing, i.e. for $\Delta x = \delta x=0$.

\section{Experimental results}

\begin{figure}[h!]
 \centering
 \includegraphics[width=1.05\textwidth]{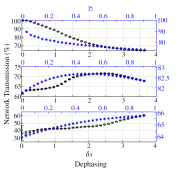}
 \caption{Role of dephasing for constructive interference. Experimental network transmission (black circles) vs dephasing $\delta x$, and numerical evaluation of the network transmission (blue stars) vs $\gamma_{2}$, for an initial condition of constructive interference and for different values of disorder: no disorder ($\Delta x =0$, and $\omega_{2} = 0$, top figure), medium disorder ($\Delta x =0.7$, and $\omega_{2} = 1$, medium figure), and large disorder ($\Delta x =2$, and $\omega_{2} = 2$, bottom figure).}
 \label{constr_deph}
\end{figure}

The network transmission has been investigated for different initial conditions of global constructive or destructive interference, as a function of both disorder and dephasing on the experimental $4$-site network of fiber-optic resonators in Fig.~\ref{fig1}. In particular, the noise effects on the amount of transferred energy will be shown from two points of view:
\begin{itemize}
  \item We analyze the network transmission as a function of dephasing for different values of the disorder.
  \item Then, the network transmission is analyzed as a function of disorder for different values of the dephasing.
\end{itemize}
The experimental data are compared with the theoretical results obtained by the theoretical model, in order to demonstrate in a full range of cases how different regimes of interference, static disorder and dephasing noise are effective control knobs to optimize energy transport processes in complex networks. Moreover, it is worth noting that in these results we are able also to show when NAT behaviours can be observed in the presence of different operating conditions. The agreement is reached in the common behaviours of the experimental data and theoretical results, so that all the different transport behaviours are well captured. Let us remind that our experimental fiber-optic setup has been designed as a simple, scalable and low-cost platform with not perfectly identical FBG resonators and significant losses within the network, even if relatively small. Furthermore, the effective Hamiltonian of the optical platform is hard to be quantified, for the energy and coupling values, and the experimental scheme has been realized with coherent states of light not in a fully quantum regime. Despite all these limitations, we achieved a sufficiently high level of control by tuning the global interference, static and dynamical disorder. As already expressed, in the model the static disorder is added by tuning the cavity frequency of site $2$, $\omega_{2}$, while dephasing is given by $\gamma_{2}$. For both the experimental data and numerical results, the network transmission is normalized to the value of the output signal in the condition of constructive interference without disorder nor dephasing.\\ \\
\textbf{Constructive Interference.-}
We start by investigating the behavior of the network transmission for initial conditions of global constructive interference. In Fig.~\ref{constr_deph}, the experimental and theoretical network transmission are shown as a function of dephasing for three different configurations of disorder.
Without disorder ($\Delta x =0$), the only effect of dephasing is to reduce the transferred energy, i.e. no NAT is observed because the different pathways do already constructively interfere. However, if the system's energy landscape presents some disorder ($\Delta x > 0.4$), NAT effects can be detected and, in particular conditions of disorder, one finds the typical bell-shaped NAT behavior with an optimal value of dephasing that maximizes the network transmission. In other terms, \textit{dephasing enhances the transport efficiency if the additional presence of disorder inhibits the otherwise fast constructive-interference path}. A similar behavior is found for the theoretical model for parameters compatible with the experimental ones, though it should be noted a discrepancy for small values of dephasing. The latter is due to the difficult experimental feasibility of perfect constructive interference, which is also hard to properly quantify in the model. Similar reasonings hold for destructive interference.
\begin{figure}[h!]
 \centering
 \includegraphics[width=1.0\textwidth]{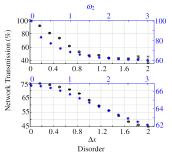}
 \caption{Role of disorder for constructive interference. Experimental network transmission (black circles) vs disorder $\Delta x$, and numerical evaluation of the network transmission (blue stars) vs $\omega_{2}$, for an initial condition of constructive interference and for different values of dephasing: no dephasing ($\delta x = 0$, and $\gamma_{2} = 0$, top figure), and large dephasing ($\delta x = 1.83$, and $\gamma_{2} = 1.8$, bottom figure).}
 \label{constr_stat}
\end{figure}\\
In Fig.~\ref{constr_stat}, the network transmission is shown as a function of disorder for two different dephasing configurations. Here, as expected, we find that disorder has a generally negative impact on the transport performance since it always leads to the suppression of the initial constructive interference. However, \textit{while a little bit of disorder quickly deteriorates transport in the case without dephasing, the presence of some dephasing noise that broadens the resonances has the effect of making the system more robust against static disorder}, with an evidently smoother decay in both the theoretical and experimental cases. \\ \\
\textbf{Destructive Interference.-} Repeating the analysis above \textit{for the case of initial destructive interference, it turns out that both dephasing and disorder independently assist transport}, i.e. NAT behavior, since they reduce the amount of interference that prevents the transmission of energy -- see Fig.~\ref{destr_deph}.
\begin{figure}[h!]
 \centering
 \includegraphics[width=1.05\textwidth]{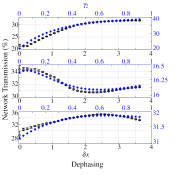}
 \caption{Role of dephasing for destructive interference. Experimental network transmission (black circles) vs dephasing $\delta x$, and numerical evaluation of the network transmission (blue stars) vs $\gamma_{2}$, for an initial condition of destructive interference and for different values of disorder: no disorder ($\Delta x =0$, and $\omega_{2} = 0$, top figure), medium disorder ($\Delta x =0.7$, and $\omega_{2} = 1$, medium figure), and large disorder ($\Delta x =2$, and $\omega_{2} = 2$, bottom figure).}
 \label{destr_deph}
\end{figure}
\begin{figure}[h!]
 \centering
 \includegraphics[width=1.05\textwidth]{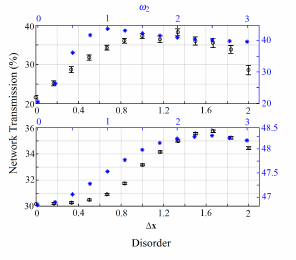}
 \caption{Role of disorder for destructive interference. Experimental network transmission (black circles) vs disorder $\Delta x$, and numerical evaluation of the network transmission (blue stars) vs $\omega_{2}$, for an initial condition of destructive interference and for different values of dephasing: no dephasing ($\delta x = 0$, and $\gamma_{2} = 0$, top figure), and large dephasing ($\delta x = 1.83$, and $\gamma_{2} = 1.8$, bottom figure).}
 \label{destr_stat}
\end{figure}
When the network is in a regime of high-disorder, again the typical bell-like NAT shape is recovered. However, if a lower amount of disorder is included, the additional contribution of dephasing does not further improve the network transmission that instead shows a minimum value for a dephasing value of $\delta x \approx 2.2$. Finally, in Fig.~\ref{destr_stat} we show the role of disorder in enhancing the transmission rate with its peak moving to higher value of disorder for increasing dephasing values.

\section{Observing and reproducing NAT}

Noise is an unavoidable feature of any system, be it physical or cyber. As it is usually known, the presence of noise usually leads to the deterioration of performance in fundamental and well-defined processes such as information processing, sensing and transport. However, noise-assisted transport phenomena occur in several physical systems, where noise can open additional transport pathways and suppress the ineffective slow ones. In the last years, indeed, this scheme has been applied to better understand energy transport in photosynthetic light-harvesting complexes, where dephasing noise remarkably enhances the transmission of an electronic excitation from the antenna complex to the reaction center in which such energy is further processed~\cite{Caruso2009JChPh131}. The basic underlying mechanisms of such a behavior are mainly due to line-broadening effects and the suppression of destructive interference, so that the interplay of quantum coherence and noise is responsible for the observed very high transport efficiency. However, while several theoretical studies have been performed, it is very challenging to actually test these ideas either in the real photosynthetic pigment-protein complexes or in artificial ones, since their structure and dynamical properties cannot be controlled or even indirectly measured~\cite{Cavities2012,Strong2015} with the required resolution. Moreover, these samples are usually quite expensive or difficult to synthesize. For these reasons, it is very convenient to reproduce such transport phenomena in a controlled system where one can tune the parameters and measure the corresponding dynamical behaviors, while also playing with the underlying network geometry. This will allow to better understand the underlying physical phenomena and to start engineering new molecular/nano-structures for more efficient and feasible technological applications.

In this regard, the experimental setup, that has been previously described, is a simple and scalable test optical platform with optimized accessible noise features for energy transport. Although the experiment could be fully described by classical optics, the role of the quantum coherence in energy transport processes, as observed for photosynthetic systems, can be mimicked by the propagation of photons in correspondence of different global interference conditions. In such a interferometric apparatus, the effective controllability of parameters is reached by exploiting the intrinsic instability of the setup, where no phase locking is used for the stabilization of the system. Hence, this setup not only allows to observe the NAT peak in the network transmission as a function of the amount of noise in the system dynamics, but, more importantly, can be exploited to monitor and control different noise sources for quantum transport dynamics with specific types of global interference. Overall, we observed that:
\begin{itemize}
  \item When constructive interference provides a very fast path to the exit site, dephasing has a detrimental effect and reduces the amount of transferred energy.
  \item If some disorder is present in the system energy levels, thus blocking the constructive interference path, dephasing represents a recovery tool to achieve again higher transport efficiency, thanks to NAT effects.
  \item In the presence of destructive interference, both dephasing and disorder are able to speed up the energy transport, whereby dephasing often provides a faster NAT mechanism.
\end{itemize}
In conclusion, the role of noise in increasing the transferred energy can be explained by considering how the pathways of energy transfer are modified: destroying the inefficient ones (given by the inhibition of destructive interference patterns in the network) or giving access to more efficient network hubs. It is worth noting that, by increasing the network size, the results presented in this chapter could be used also to observe even more complex transport behaviors, that can be very hardly simulated on a computer. Indeed, as shown also in this context, the network topology or connectivity plays a crucial role that deserves to be further studied in future investigations.

\section{Conclusions and contributions}

Summarizing, this chapter provides the following contributions:
\begin{itemize}
\item
We have numerically and experimentally validated a class of theoretical models for noise-assisted quantum transport.
\item
A quite simple, scalable and controllable experimental setup of coupled cavities (only based on single-mode fiber optic components) have been introduced as transport emulator, in which the system noise parameters can be properly tuned to maximize the transfer efficiency. These optical setups, indeed, turned out to be capable to mimic the transport dynamics as in natural photosynthetic organisms, so that it could be a very promising platform to artificially design optimal nanoscale structures for novel, more efficient, clean energy technologies. In this regard, the use of single-mode telecom optical fiber components is an essential ingredient, as it dramatically reduces costs and losses, besides canceling all issues related to the alignment and spatial mode-matching procedures that should be faced when adding other cavities to the system or changing its topology.
\item
We experimentally investigated the optimal interplay between constructive and destructive interference, static disorder and dephasing to optimize the transport paths for excitons towards an exit site. In other words, disorder and dephasing can be exploited as control knobs to manipulate the transport performance of complex networks, as for example the optimization of the final transmission rate or the capability to temporarily store energy/information in the system. In this regard, the demonstration of noise-enabled information transfer phenomena~\cite{Caruso2010PRL105,DobekPRL2011} has shown that an information transmission system exhibits optimal features when both static and dynamical disorder are tuned to specific values, so that the information transfer is considerably enhanced.
\item
As a final remark, it is worth noting that the numerical simulations have been performed by modeling each site of the network as a two-level quantum system, and, thus, all the presented results can be applied to real quantum networks, where the tensor product of sites grows exponentially with their number. In this setup, the quantum transport is favoured by the presence of noise sources on dynamical parameters. However, such property is valid on average by taking into account networks with different number of sites and topologies. In this way, the presence of noise cannot be seen as a control pulse in the system-theoretical sense. We believe that further investigations are necessary to analytically derive the conditions for the controllability of the quantum network Hamiltonian under the presence of an external stochastic driving. Indeed, noise-assisted transport features impose system symmetries, which hamper the controllability of the system and build preferential pathways for excitons towards exit sites.
\end{itemize}

\chapter{Large deviations and stochastic quantum Zeno phenomena}
\label{chap:LD}
\fancyhead{}
\fancyhead[LEH]{\leftmark}
\fancyhead[RO]{\rightmark}

\textit{
In this chapter, we will apply the LD theory~\cite{Touchette1} to open quantum systems. In particular, we will base our procedure on the modeling of the local couplings between a quantum system and the environment as projection events given by the action of one (or more) measurement operators, along the lines of the formalism of quantum jump trajectories~\cite{PlenioRMP}. The latter, resulting from the dissipative influence of the environment, are intrinsically stochastic processes, since any interaction occurs at irregular time intervals without any a-priori predictability. The stochasticity of such a measurement sequence can also be introduced by experimental noise or a randomly fluctuating classical field coupled to the system. We will analytically show that, in the limit of a large number $m$ of randomly distributed measurements, the distribution of the probability for the system to remain in the initial state assumes a large-deviation form, namely, a profile decaying exponentially in $m$. This result has allowed to obtain analytical expressions also for the most probable and the average value of such probability. The former represents what an experimentalist will measure in a single typical implementation of the measurement sequence, while the latter is given by averaging the experimental outcomes over a large (ideally infinite) number of experimental runs. Hence, by tuning the probability distribution of the time intervals between consecutive measurements, one can effectively realize a specific value for the most probable survival provability, thereby allowing to engineer novel optimal control protocols for the manipulation, e.g., of the atomic population related to a specific quantum state. Furthermore, the extension of these theoretical results to the application of multi-dimensional projection operators has made it possible to control also the amount of quantum coherence within an arbitrary Hilbert subspace. Thus, in conclusion, by characterizing the statistical space of configurations concerning the random variables, that enter into the system dynamics and describe the effects of the system-environment interactions, we are able to formulate a noise-assisted quantum control paradigm, which may require to be supported by quantum noise sensing techniques.
}
\footnote{
The results shown in this chapter have been published as ``Stochastic Quantum Zeno by Large Deviation Theory'', in \textit{New Journal of Physics}, \textbf{18(1)}, 013048 (2016)~\cite{Gherardini2016NJP}; ``Fisher information from stochastic quantum measurements'', in \textit{Physical Review A} \textbf{94}, 042322 (2016)~\cite{GherardiniFisher}; ``Stochastic quantum Zeno-based detection of noise correlations'', in \textit{Scientific Reports} \textbf{6}, 38650 (2016)~\cite{GherardiniSciRep}; ``Ergodicity in randomly perturbed quantum systems'', in \textit{Quantum Science and Technology} \textbf{2(1)}, 015007 (2017)~\cite{GherardiniQST};  ``Quantum Zeno dynamics through stochastic protocols'', in \textit{Annalen der Physik} \textbf{529(9)}, 1600206 (2017)~\cite{GherardiniAnnalen}.
}

\section*{Introduction}

The theory of \textit{Large Deviations (LD)} studies the exponential decay of probabilities concerning observables of stochastic dynamical systems~\cite{Varadhan:1984,Ellis2006,Dembo1,Touchette1}. In particular,by means of LD theory we can derive the scaling of such probabilities when the deviation of their results from the expected value is relevant. For the sake of clarity, let us consider for example the following question: Which is the probability that the single realization of the stochastic process $\frac{1}{n}\sum_{i = 1}^{n}X_{i}$ is greater than $3/4$, given that $X_{1},\ldots,X_{n}$ are Bernoulli random variable with probability $1/3$ to take the value $1$? Clearly, the occurrence of such event is a large deviation with respect to the expected result of the process (that will be around $1/3$), and its probability is exponentially small. LD theory has been largely used to identify and gather information on the occurrence of extreme or rare events, that arise from the realization of tail fluctuations (i.e. fluctuations with a low but relevant frequency of occurrence) in the system dynamics. Moreover, in LD theory the relationship between information theory and statistics is very close (see e.g. \cite{Cover2006}), and its application to dynamical systems for monitoring and control purposes is desirable. In the last years an increasing interest has led to several studies of large deviations in both classical and quantum systems. In the latter case, the LD formalism has been discussed in the context of quantum gases~\cite{Gallavotti1}, quantum spin systems~\cite{Netovcny1}, and quantum information theory~\cite{Ahlswede1} among others. Furthermore, an interesting recent application pursued in \cite{Garrahan1,Garrahan4} has invoked the LD theory to develop a thermodynamic formalism to study quantum jump trajectories \cite{PlenioRMP} of open quantum systems~\cite{PetruccioneBook}.

In this chapter we will show how to use LD theory to derive the exponential decay of the probability distribution of the probability that an arbitrary quantum system, subject to repeated sequence of quantum measurements, is confined within a given portion of the corresponding Hilbert space.

\section{LD theory and open quantum systems}

Here, the results in \cite{Gherardini2016NJP} are discussed, and the interplay between the sequence of projective measurements and stochastic contributions from an external environment on a quantum system dynamics is analyzed by using LD theory.

In the extreme case of a frequent enough series of measurements projecting the system back to the initial state, its dynamical evolution gets completely frozen. As a consequence, the probability that the system remains in the initial state approaches unity in the limit of an infinite number of measurements. This effect is known as the \textit{quantum Zeno effect (QZE)}, that was first discussed in a seminal paper by Sudarshan and Misra in 1977~\cite{Misra1}. The QZE can be understood intuitively as resulting from the collapse of the wave function corresponding to the initial state of the system due to the process of measurement. Then, it was later explored experimentally in systems of ions~\cite{Itano1990}, polarized photons~\cite{Kwiat:1995}, cold atoms~\cite{Fischer:2001}, and dilute Bose-Einstein condensed gases~\cite{Streed:2006}. In particular, in \cite{Itano1990} it was observed the inhibition of induced transitions in an RF transition between two $^{9}Be^{+}$ ground-state hyperfine levels, while in \cite{Kwiat:1995}, interaction-free measurements have been experimentally proved by using single photons in a Michelson interferometer, being the presence of an absorbing object in one of the arms of an interferometer able to modify the interference of an incident photon, which is used as a probe. Thus, the photon and the object do not need to interact one with the other, and the presence of the object is revealed by a sequence of repeated measurements, which inhibit the coherent evolution of the photon. Finally, in \cite{Fischer:2001} a system of cold sodium atoms trapped in a far-detuned standing wave of light is studied, and it has been observed that, depending on the frequency of the measurements, the decay features of the atoms are suppressed (Zeno effect) or enhanced (anti-Zeno effect) with respect to the unperturbed case. Moreover, in noisy quantum systems both the Zeno and anti-Zeno effects have been shown in \cite{kofman2000,kofman2001}, and, then, proposed for thermodynamical control of quantum systems~\cite{Kurizki2008} and quantum computation~\cite{Lidar2012}. In recent times, Zeno phenomena have assumed particular relevance in applications owing to the possibility of quantum control, whereby specific quantum states (including entangled ones) may be \textit{protected from decoherence} by means of projective measurements~\cite{Maniscalco2008,Kim2012}.

In its original formulation, QZE was defined over a sequence of repeated measurements at constant times, while only recently \cite{Shushin1} considered the case of randomly spaced in time measurements, which takes the name of \textit{Stochastic Quantum Zeno Effect (SQZE)}. According to SQZE, the survival probability that the system remains in the projected state becomes itself a random variable, that takes on different values corresponding to different realizations of the measurement sequence. In this regard, one would expect that the expectation value of the survival probability, obtained by averaging the measurement sequence over a large (ideally infinite) number of realizations, leads to the result obtained for an evenly spaced sequence under some constraints (e.g. the mean time interval between consecutive measurements is finite). However, some interesting questions, of both theoretical and experimental relevance, naturally emerges:
\begin{itemize}
  \item Is it possible to have realizations of the measurement sequence that give values of the survival probability significantly \textbf{deviated} from the mean?
  \item How typical/atypical are those realizations?
  \item Are there ways to quantify the probability measures of such realizations?
\end{itemize}
These questions assume particular importance in devising experimental protocols that on demand may slow down or speed up efficiently the transitions of a quantum system between its possible states.

In this chapter, by exploiting tools from probability theory, we propose a framework that allows an effective addressing of the questions posed above. In particular, we adapt the well-established theory of LD to quantify the dependence of the survival probability on the realization of the measurement sequence, in the case of independent and identically distributed (i.i.d.) time intervals between consecutive measurements. In doing this, our goal is twofold:
\begin{itemize}
  \item Adapt and apply the LD theory to discuss the QZE by transferring tools and ideas from classical probability theory to the arena of quantum Zeno phenomena.
  \item Analytically predict the corresponding survival probability and exploit it for a new type of control based on the stochastic features of the applied measurements.
\end{itemize}

\subsection{Sequences of repeated quantum measurements}

Here, we will argue how to model sequences of repeated quantum measurements. In this respect, let us consider a quantum mechanical system described by a finite-dimensional Hilbert space $\mathcal{H}$, which may be taken to be a direct sum of $r$ orthogonal subspaces $\mathcal{H}^{(k)}$, i.e.
$$
\mathcal{H}=\bigoplus_{k=1}^{r}\mathcal{H}^{(k)}.
$$
To each subspace is assigned a projection operator $\Pi^{(k)}$, such that
$$
\Pi^{(k)}\mathcal{H} = \mathcal{H}^{(k)}.
$$
Then, we assume that the initial state of the quantum system is described by a density matrix $\rho_{0}$, which undergoes a unitary dynamics to evolve in time $t$ to
$\exp(-iHt)\rho_{0}\exp(iHt)$, where $H$ denotes the system Hamiltonian and the reduced Planck's constant $\hbar$ has been set to unity. Observe that usually $H$ commutes with the projectors $\Pi^{(k)}$.

In this model, starting with a $\rho_{0}$ that belongs to one of the subspaces, say subspace $\tilde{r}\in 1,2\ldots,r$, so that $\rho_{0}=\Pi^{(\tilde{r})}\rho_{0}\Pi^{(\tilde{r})}$ and ${\rm Tr}[\rho_{0}\Pi^{(\tilde{r})}] = 1$, we subject the system to an arbitrary but fixed number $m$ of consecutive measurements separated by time intervals $\tau_j:~\tau_j >0$, with $j=1,\ldots,m$. During each interval $\tau_j$, the system follows a unitary evolution described by the Hamiltonian $H$, while the measurement corresponds to applying the projection operator $\Pi^{(\tilde{r})}$. \textit{We take the $\tau_j$'s to be independent and identically distributed (i.i.d.) random variables sampled from a given distribution $p(\tau)$}, with the normalization $\int p(\tau)d\tau = 1$. Moreover, we assume that $p(\tau)$ has a finite mean, which is denoted by $\overline{\tau}$. For the sake of simplicity, in the following we will represent $\Pi^{(\tilde{r})}$ and $\mathcal{H}^{(\tilde{r})}$ by $\Pi$ and $\mathcal{H}_{\Pi}$, respectively. The (unnormalized) density matrix at the end of evolution for a total time
\begin{equation}\label{T-defn}
\mathcal{T}\equiv \sum_{j=1}^m\tau_j,
\end{equation}
corresponding to a given realization of the measurement sequence $\{\tau_j\}\equiv\{\tau_j;~j=1,2,\ldots,m\}$, is
given by
\begin{eqnarray}\label{eq:Wm}
\textbf{W}_m(\{\tau_j\})&\equiv&\left(\Pi~\mathcal{U}_m\right)\ldots\left(\Pi~\mathcal{U}_1\right)\rho_{0}
\left(\Pi~\mathcal{U}_1\right)^\dagger\ldots\left(\Pi~\mathcal{U}_m\right)^\dagger\nonumber \\
&=&\textbf{R}_{m}(\{\tau_j\})\rho_{0}\textbf{R}_{m}^{\dagger}(\{\tau_j\}),
\end{eqnarray}
where we have defined
\begin{equation}
\textbf{R}_m(\{\tau_j\})\equiv\prod_{j=1}^{m}\Pi~\mathcal{U}_{j}\Pi,
\end{equation}
and $\mathcal{U}_{j}\equiv\exp\left(-iH\tau_{j}\right)$. Clearly, also $\mathcal{T}$ is a random variable that depends on the realization of the sequence $\{\tau_j\}$. Let us observe that, to obtain (\ref{eq:Wm}), we have used the relations $\Pi^{\dagger}=\Pi$, $\rho_{0} = \Pi\rho_{0}\Pi$ and $\Pi^{2}=\Pi$, which are obtained by modeling the quantum measurement with a projection operator.

The \textit{survival probability}, namely, \textit{the probability that the system belongs to the subspace} $\mathcal{H}_{\Pi}$ \textit{at the end of the evolution}, is given by
\begin{equation}\label{surv_prob}
\mathcal{P}(\{\tau_j\})\equiv\rm{Tr}\left[\textbf{W}_{m}(\{\tau_j\})\right]=\rm{Tr}\left[\textbf{R}_{m}(\{\tau_j\})\rho_{0}\textbf{R}_{m}^{\dagger}(\{\tau_j\})\right],
\end{equation}
while the final (normalized) density matrix is
\begin{equation}
\rho(\{\tau_j\})=\frac{\textbf{R}_{m}(\{\tau_j\})\rho_{0}\textbf{R}_{m}^{\dagger}(\{\tau_j\})}{\mathcal{P}(\{\tau_j\})}.
\end{equation}
Note that the survival probability $\mathcal{P}(\{\tau_j\})$ depends on the system Hamiltonian $H$, the initial density matrix $\rho_0$ and also on the probability distribution $p(\tau)$.

\subsection{Survival probability statistics}

Now, we will provide a novel method from LD theory to derive the \textit{distribution of the survival probability $\mathcal{P}(\{\tau_j\})$} with respect to different realizations of the sequence $\{\tau_j\}$. To this end, let us suppose that the system is initially in a pure state $|\psi_{0}\rangle$
belonging to $\mathcal{H}_{\Pi}$, so that $\rho_{0} = |\psi_{0}\rangle\langle\psi_{0}|$, and that the projection operator is given by $\Pi\equiv|\psi_{0}\rangle\langle\psi_{0}|$. In this way, starting with a pure state, the system evolves according to the following repetitive sequence of events: unitary evolution for a random interval, followed by a measurement that projects the evolved state into the initial state. The survival probability $\mathcal{P}(\{\tau_j\})$ is, then, evaluated by using (\ref{surv_prob}) to get
\begin{equation}\label{surv_prob_product}
\mathcal{P}(\{\tau_j\})=\prod_{j=1}^{m}q(\tau_j),
\end{equation}
where the probability $q(\tau_j)$ is defined as
\begin{equation}\label{probability_q}
q(\tau_j)\equiv\left|\langle\psi_{0}|\mathcal{U}_j|\psi_{0}\rangle\right|^{2},
\end{equation}
which takes on different values depending on the random numbers $\tau_j$. Note that, being a probability, possible values of $q(\tau)$ lie in the range $0 < q(\tau) \le 1$. Moreover, the distribution of $q(\tau_j)$ is obtained as
\begin{equation}
\textrm{Prob}\left(q(\tau_j)\right) = p(\tau_j)\left|\frac{d\tau_j}{dq(\tau_j)}\right|,
\end{equation}
where (\ref{probability_q}) gives
\begin{equation}
\left|\frac{dq(\tau_{j})}{d\tau_j}\right|=2\left|\langle\psi_{0}|H\mathcal{U}_{j}|\psi_{0}\rangle\right|.
\end{equation}
Then, from (\ref{surv_prob_product}) one derives the distribution of $\mathcal{P}$ as
\begin{equation}
\textrm{Prob}\left(\mathcal{P}\right)=\left[\prod_{j=1}^{m}\int d\tau_j~ p(\tau_j)\right]\delta\left(\prod_{j=1}^{m}q(\tau_{j})-
\mathcal{P}\right),
\end{equation}
where $\delta(\cdot)$ is the Dirac-delta distribution. In particular, one may be interested in the \textit{average value} of the survival probability, where the average corresponds to repeating a large number of times the protocol of $m$ consecutive measurements interspersed with unitary dynamics for random intervals $\tau_j$. One gets:
\begin{equation}\label{average_surv_prob}
\langle
\mathcal{P}\rangle=\prod_{j=1}^{m}\int d\tau_j~ p(\tau_j)q(\tau_{j}).
\end{equation}
In this regard, let us observe that here and in the following we will use angular brackets to denote averaging with respect to different realizations of the stochastic sequence under analysis. Additionally, let us note that writing $q(\tau)$ as
\begin{equation}\label{qmu-expansion0}
q(\tau) = 1 - \mu(\tau);~0 \le \mu(\tau) < 1,
\end{equation}
we have
\begin{equation}\label{qmu-expansion1}
\mu(\tau) = \left|\sum_{k=1}^\infty \frac{(-i\tau)^k}{k!}\langle H^k\rangle\right|^2,
\end{equation}
with
\begin{equation}
\langle H^{k}\rangle \equiv \langle \psi_0|H^k|\psi_0\rangle;~k=0,1,2,\ldots
\end{equation}
In particular, considering $\tau \ll 1$, one has, to leading order in $\tau^2$, the result
\begin{equation}\label{qmu-expansion2}
\mu(\tau) = \frac{\tau^2}{\tau_{Z}^2},
\end{equation}
where $\tau_{Z}$ is the so-called Zeno-time \cite{PascazioJPA,SmerziPRL2012} and is defined as
\begin{equation}
\begin{cases}
\tau_{Z}^{-2} \equiv \Delta^{2}H, \\
\Delta^{2}H \equiv \langle H^2\rangle - \langle H\rangle^2.
\end{cases}
\end{equation}

Let us now employ the LD formalism to derive the statistics of the survival probability $\mathcal{P}(\{\tau_j\})$ in the limit of $m \to \infty$. In this limit, (\ref{T-defn}) gives
\begin{equation}
\langle \mathcal{T}\rangle = m\overline{\tau},
\end{equation}
where we have used the fact that the $\tau_j$'s are i.i.d. random variables and $\overline{\tau}$ is a finite number. Moreover, let us consider $p(\tau)$ to be a $d$-dimensional Bernoulli distribution, namely $\tau$ takes on $d$ possible discrete values $\tau^{(1)},\tau^{(2)},\ldots,\tau^{(d)}$ with corresponding probabilities $p^{(1)},p^{(2)},\ldots,p^{(d)}$, such that $\sum_{k=1}^{d} p^{(k)}=1$. The average value of the survival probability is, then, obtained by using (\ref{average_surv_prob}) as
\begin{equation}\label{P-avg}
\langle\mathcal{P}\rangle=\exp\Big(m\ln\sum_{k=1}^d p^{(k)}q(\tau^{(k)})\Big).
\end{equation}
In order to introduce the LD formalism for the survival probability, consider the log-survival-probability
\begin{equation}\label{log_surv_prob}
\mathcal{L}(\{\tau_j\})\equiv\ln\left(\mathcal{P}(\{\tau_j\})\right)=\sum_{k=1}^{d}n_{k}\ln q(\tau^{(k)}),
\end{equation}
where
$n_{k}$ is the number of times $\tau^{(k)}$ occurs in the sequence $\{\tau_j\}$. Noting that $\mathcal{L}(\{\tau_j\})$ is a sum of i.i.d. random variables, its probability distribution is given by
\begin{eqnarray}\label{prob_L}
\textrm{Prob}(\mathcal{L})&=& \sum_{\{n_k\}:~\sum_{k}n_{k}=m}\frac{m!}{n_{1}!n_{2}!\ldots n_{d}!}(p^{(1)})^{n_1}\ldots (p^{(d)})^{n_d}
\delta\left(\sum_{k=1}^{d}n_{k}\ln q(\tau^{(k)})-\mathcal{L}\right)\nonumber \\
&=&\frac{m!}{\tilde{n}_{1}!\tilde{n}_{2}!\ldots
\tilde{n}_{d}!}\prod_{k=1}^{d}(p^{(k)})^{\tilde{n}_k},
\end{eqnarray}
where, as indicated, the summation in the first equality is over all possible values of $n_1,n_2,\ldots,n_d$ subject to the constrain $\sum_{k=1}^d
n_\alpha=m$. In the second equality, instead, $\tilde{n}_k$'s are such that
\begin{equation}\label{constraints}
\begin{cases}
\displaystyle{\sum_{k=1}^{d}\tilde{n}_{k}=m}, \\
\displaystyle{\sum_{k=1}^{d}\tilde{n}_{k}\ln q(\tau^{(k)}) = \mathcal{L}}.
\end{cases}
\end{equation}

Starting from (\ref{prob_L}) and considering the limit $m \to \infty$, the following LD form for the probability distribution $\textrm{Prob}\left(\mathcal{L}/m\right)$ can be derived (in this regard, see Appendix~\ref{chapter:appB}) as
\begin{equation}\label{prob_log_surv}
\textrm{Prob}\left(\mathcal{L}/m\right)\approx\exp\Big(-mI\left(\mathcal{L}/m\right)\Big),
\end{equation}
where the function $I(\xi)$, also called rate function~\cite{Touchette1}, is given by
\begin{equation}\label{rate_function}
I\left(\xi\right)=\sum_{k=1}^{d}f(\tau^{(k)})\ln\left(\frac{f(\tau^{(k)})}{p^{(k)}}\right),
\end{equation}
where
\begin{equation}
\begin{cases}
\displaystyle{f(\tau^{(k)})=\frac{\ln q(\tau^{(d)})-\xi}{(d-1)\Big[\ln q(\tau^{(d)})-\ln q(\tau^{(k)})\Big]};~k=1,\ldots,(d-1)}, \\
\displaystyle{f(\tau^{(d)})=1 - \sum_{k = 1}^{d-1}f(\tau^{(k)})}.
\end{cases}
\end{equation}
The approximate symbol $\approx$ in (\ref{prob_log_surv}) stands for the fact that there are subdominant $m$-dependent factors on the r.h.s. of the equation. An alternative form to (\ref{prob_log_surv}), that involves an exact equality and can be considered as the equation defining the
function $I(\xi)$, is
\begin{equation}
\lim_{m\to \infty}-\frac{1}{m}\textrm{Prob}\left(\mathcal{L}/m\right) = I\left(\mathcal{L}/m\right).
\end{equation}

The rate function $I\left(\xi\right)$ in (\ref{rate_function}) is the relative entropy or the \textit{Kullback-Leibler distance} between the set of
probabilities $\{f(\tau^{(k)})\}$ and the set $\{p^{(k)}\}$. It has the property to be positive and convex, with a single non-trivial
minimum~\cite{Cover2006}. Equation~(\ref{prob_log_surv}) implies that \textit{the value at which the function} $I(\mathcal{L}/m)$ \textit{is minimized corresponds to the most probable value} $\mathcal{L}^{\star}$ \textit{of} $\mathcal{L}$ \textit{as} $m \to \infty$. Using
$$
\left.\frac{\partial I(\mathcal{L}/m)}{\partial \ln q(\tau^{(k)})}\right|_{\mathcal{L} = \mathcal{L}^\star}=0;~
k = 1,\ldots,d
$$
we get (see Appendix~\ref{chapter:appB})
\begin{equation}\label{L_star}
\mathcal{L}^\star = m\sum_{k=1}^{d} p^{(k)}\ln q(\tau^{(k)}).
\end{equation}

As for the distribution of the survival probability, one may obtain a LD form for it in the following way:
\begin{eqnarray}\label{surv-prob-distr}
\textrm{Prob}(\mathcal{P}) &=&\int d\mathcal{L}~\textrm{Prob}(\mathcal{L})\delta(\mathcal{L}-\mathcal{\ln P})\nonumber \\
&=&\int d(\mathcal{L}/m)~\textrm{Prob}(\mathcal{L}/m)\delta(\mathcal{L}/m-\mathcal{\ln P})\nonumber \\
 &\approx&\int d(\mathcal{L}/m)\exp\left(-mI(\mathcal{L}/m)\right)\delta(\mathcal{L}/m-\mathcal{\ln P}))\nonumber \\
 &\approx&\exp\Big(-m ~{\rm
 min}_{\mathcal{L}:\mathcal{L}=m\ln\mathcal{P}}I(\mathcal{L}/m)\Big),
\end{eqnarray}
where in the third step we have considered large $m$ and have used (\ref{prob_log_surv}), while in the last step we have used the saddle point method to evaluate the integral. We, thus, obtain
\begin{equation}
\lim_{m\to \infty}-\frac{\ln(\textrm{Prob}(\mathcal{P}))}{m} = J(\mathcal{P}),
\end{equation}
with
\begin{equation}
J(\mathcal{P})\equiv{\rm min}_{\mathcal{L}:\mathcal{L} = m\ln\mathcal{P}}I(\mathcal{L}/m).
\end{equation}
\textit{The value at which $J(\mathcal{P})$ takes on its minimum value gives the most probable value of the survival probability} in the limit $m \to \infty$, which may also be obtained by utilizing the relationship between $\mathcal{L}$ and $\mathcal{P}$; one gets
\begin{equation}\label{P_star}
\mathcal{P}^{\star} = \exp\Big(m\sum_{k = 1}^{d} p^{(k)}\ln q(\tau^{(k)})\Big),
\end{equation}
which may be compared with the average value in (\ref{P-avg}). In other words, while the average value $\langle \mathcal{P} \rangle$ is determined by the logarithm of the averaged $q(\tau^{(k)})$, the most probable value $\mathcal{P}^{\star}$ is given by the average performed on the logarithm of $q(\tau^{(k)})$. The latter is the so-called \textit{log-average} or the \textit{geometric average} of the quantity $q(\tau^{(k)})$ with respect to the $\tau$-distribution.

A straightforward generalization of (\ref{P_star}) for a generic \textit{continuous} $\tau$-distribution is
\begin{equation}\label{PG-defn}
\mathcal{P}^\star = \exp\Big(m\int d\tau p(\tau)\ln q(\tau)\Big),
\end{equation}
while that for the average reads as
\begin{equation}\label{P-avg1}
\langle \mathcal{P} \rangle = \exp\Big(m\ln\int d\tau p(\tau)q(\tau)\Big).
\end{equation}
Using the so-called Jensen's inequality, namely, $\langle \exp(\xi)\rangle \ge \exp(\langle \xi\rangle)$, it immediately follows that
\begin{equation}
\langle \mathcal{P} \rangle \ge \mathcal{P}^\star,
\end{equation}
with the equality holding only when no randomness in $\tau$ (that is, only a single value of $\tau$ exists) is considered. The difference between $\mathcal{P}^\star$ and
$\langle \mathcal{P}\rangle$ can be estimated in the following way in an experiment. If we perform a large number $m$ of projective measurements on our quantum system, then:
\begin{itemize}
\item
The value of the survival probability to remain in the initial state that is measured in a single experiment will very likely be close to $\mathcal{P}^\star$, with deviations that decrease fast with increasing $m$.
\item
On the other hand, averaging the survival probability over a large (ideally infinite) number of experimental runs will yield $\langle \mathcal{P}\rangle$.
\end{itemize}

All the derivations above were based on the assumption of a fixed number $m$ of measurements, so that the total time interval $\mathcal{T}$ is a quantity fluctuating between different realizations of the measurement sequence. To obtain the LD formalism, we eventually let $m$ approach infinity, which in turn leads to an infinite $\langle\mathcal{T} \rangle$. We now consider the situation where we keep the total time $\mathcal{T}$ fixed, and let $m$ fluctuate between realizations of the measurement sequence. In this case, in contrast to (\ref{prob_L}), we have the following joint probability distribution:
\begin{small}
\begin{eqnarray}\label{joint_prob_L_T}
\textrm{Prob}(\mathcal{L},\mathcal{T})&=&\sum_{m}\sum_{n_{k}:~\sum_{k}n_{k}=m}
\frac{m!}{n_{1}!\ldots n_{d}!}\prod_{k=1}^{d}\left(p^{(k)}\right)^{n_{k}}\delta\left(\sum_{k=1}^{d}n_{k}\ln
q(\tau^{(k)})-\mathcal{L}\right)\nonumber \\
&\times&\delta\left(\sum_{k=1}^{d}n_{k}\tau^{(k)}-\mathcal{T}\right).
\end{eqnarray}
\end{small}
We thus have to find the set of $n_{k}$'s, which we now refer to as $\tilde{n}_{k}$'s, such that the following conditions are
satisfied:
\begin{equation}\label{const_2}
\begin{cases}
\displaystyle{\sum_{k=1}^{d}\tilde{n}_{k}=m}, \\
\displaystyle{\sum_{k=1}^{d}\tilde{n}_{k}\ln q(\tau^{(k)}) = \mathcal{L}}, \\
\displaystyle{\sum_{k = 1}^{d}\tilde{n}_{k}\tau^{(k)} = \mathcal{T}}.
\end{cases}
\end{equation}
The above equations have a unique solution only for $d=2$, that is, when one has a Bernoulli distribution. In this case, the solutions satisfy
\begin{equation}
\frac{\mathcal{T} - m\tau^{(2)}}{\tau^{(2)}-\tau^{(1)}} = \frac{\mathcal{L} - m\ln q(\tau^{(2)})}{\ln q(\tau^{(2)})-\ln q(\tau^{(1)})},
\end{equation}
which may be solved for $m$, for given values of $\mathcal{L}$ and $\mathcal{T}$, and then used in (\ref{joint_prob_L_T}) to determine $\textrm{Prob}(\mathcal{L},\mathcal{T})$. In the limit $m\rightarrow\infty$, provided the mean $\overline{\tau}$ of $p(\tau)$ exists, (\ref{T-defn}) together with the
law of large numbers
%\footnote{The law of large numbers states that the sum of a large number $N$ of i.i.d. random variables, when scaled by $N$, tends to the mean of
%the underlying identical distribution with probability one as $N$ approaches infinity~\cite{Papoulis1984}.}
gives:
\begin{equation}\label{T-defn1}
\mathcal{T} = m\overline{\tau}.
\end{equation}
In this case, for every $d$, one obtains an LD form for $\textrm{Prob}(\mathcal{L},\mathcal{T})$ (see Appendix~\ref{chapter:appB} for the derivation):
\begin{equation}\label{LD-PLT}
\textrm{Prob}(\mathcal{L},\mathcal{T})\approx\exp\Big(-m\mathcal{I}\left(\frac{\mathcal{L}}{m},\frac{\mathcal{T}}{m}\right)\Big),
\end{equation}
where
\begin{eqnarray}
&&\mathcal{I}\left(\xi,y\right)=\sum_{k=1}^{d}g(\tau^{(\alpha)})\ln\left(\frac{g(\tau^{(k)})}{p^{(k)}}\right)\label{rate_function_2},
\\
&&g(\tau^{(k)})=\frac{\tau^{(d)}(m\ln q\left(\tau^{(d)}\right)-\xi)}
{\left(\ln
q\left(\tau^{(d)}\right)-\xi\right)(\tau^{(d)}-\tau^{(k)})+y(d-1)\ln\left(q\left(\tau^{(d)}\right)/q\left(\tau^{(k)}\right)\right)}\label{g_tau_k};\nonumber
\\
&&~~~~~~~~~~~~~~~~~~~~~~~~~~~~~~~~~~~~~~~~~~~~~~~~~~~k=1,\ldots,(d-1), \\
&&g(\tau^{(d)})=1-\sum_{k=1}^{d-1}g(\tau^{(k)})\label{g_tau_d}.
\end{eqnarray}
The rate function (\ref{rate_function_2}) is related to the rate function (\ref{rate_function}) as follows:
\begin{equation}
I(\mathcal{L}/m)={\rm min}_{\mathcal{T}/m: \mathcal{T}=\langle
\mathcal{T}\rangle}\mathcal{I}\left(\frac{\mathcal{L}}{m},\frac{\mathcal{T}}{m}\right),
\end{equation}
and, similarly to (\ref{surv-prob-distr}), one has
\begin{equation}
\textrm{Prob}(\mathcal{P},\mathcal{T})\approx\exp\left(-m\mathcal{J}(\mathcal{P},\mathcal{T}/m)\right),
\end{equation}
where
\begin{equation}
\mathcal{J}(\mathcal{P},\mathcal{T}/m)\equiv{\rm min}_{\mathcal{L}:\mathcal{L} = m\ln\mathcal{P}}\mathcal{I}(\mathcal{L}/m,\mathcal{T}/m).
\end{equation}
Finally, as in (\ref{PG-defn}), the most probable value of the survival probability for a continuous $\tau$-distribution is given by
\begin{equation}
\mathcal{P}^\star(\mathcal{T})=\exp\Big(m\int d\tau p(\tau)\ln g(\tau)\Big).
\end{equation}

\subsection{Quantum Zeno limit}

Stochastic quantum Zeno effect has been previously introduced by applying on an arbitrary quantum system a sequence of projective measurements, which are randomly spaced in time. Therefore, to recover the exact quantum Zeno limit, we assume that the $m$ projective measurements are at times equally separated by an amount $\overline{\tau}$, so that one has
$p(\tau) = \delta(\tau - \overline{\tau})$, and $\langle \mathcal{T} \rangle = \mathcal{T} = m\overline{\tau}$. As a consequence, we obtain
\begin{equation}
\mathcal{P}^\star = \langle \mathcal{P} \rangle = \mathcal{P}(\overline{\tau}) \equiv \exp\Big(m\ln q(\overline{\tau})\Big) = \exp\left(\frac{\mathcal{T}}{\overline{\tau}}~\ln
q(\overline{\tau})\right).
\end{equation}
QZE, then, is recovered in the limit $\overline{\tau}\to 0$ with finite $\mathcal{T}$. Indeed, by using (\ref{qmu-expansion0}) and (\ref{qmu-expansion2}), one obtains
\begin{equation}
\mathcal{P}(\overline{\tau})\approx\exp\Big(-\mathcal{T}~\overline{\tau}\Delta^{2}H\Big) \approx 1,
\end{equation}
provided that $\Delta^{2}H$ is finite, as it is the case for a finite-dimensional Hilbert space.

Let us, now, discuss QZE for a general $p(\tau)$. Note that in this case it is natural in experiments to keep the number of measurements $m$ fixed at a large value, with the total time $\mathcal{T}$ fluctuating between different sequences of measurements $\{\tau_j\}$. From (\ref{PG-defn}) and (\ref{P-avg1}), with the use of (\ref{qmu-expansion0}), and the Taylor expansion of $\log(1+\xi)$ for $\xi < 1$, we get
\begin{equation}
\mathcal{P}^\star = \exp\Big(-m\sum_{n=1}^\infty \frac{\langle\mu^{n}\rangle}{n}\Big) \approx \exp\Big(-m\langle\mu\rangle\Big)\label{QZE-1}
\end{equation}
and
\begin{equation}
\langle \mathcal{P} \rangle = \exp\Big(-m \sum_{n=1}^\infty \frac{\langle\mu\rangle^n}{n}\Big) \approx \exp\Big(-m\langle\mu\rangle\Big),\label{QZE-2}
\end{equation}
where
\begin{equation}
\langle \mu^{k} \rangle \equiv \int d\tau p(\tau)\mu^{k}(\tau);~k=1,2,3,\ldots
\end{equation}
From (\ref{QZE-1}) and (\ref{QZE-2}), it follows that in the limit of very frequent measurements so that $m \to \infty$, provided that $\langle \mu \rangle \approx 0$, one recovers the QZE condition, i.e.
\begin{equation}\label{Zeno_condition}
\mathcal{P}^\star = \langle\mathcal{P}\rangle \approx 1.
\end{equation}
Thus, the condition to obtain QZE in the case of stochastic measurements is
\begin{equation}\label{Zeno_condition_2}
\langle \mu \rangle = \frac{\int d\tau p(\tau)\tau^{2}}{\tau_Z^{2}}\approx 0 \; ,
\end{equation}
which, considering that $\tau_Z$ is finite, reduces to the requirement
\begin{equation}
\int d\tau p(\tau)\tau^{2} \approx 0.
\end{equation}
For instance, for a quite general probability distribution $p(\tau)$ with \textit{power-law tails}, namely,
$$
p(\tau) \sim \left(\frac{\tau_{0}}{\tau}\right)^{1+\alpha},
$$
with $\alpha>0$ and $\tau_{0}$ being a given time scale, QZE is achieved  for $\tau_{0} \ll 1$ and $\alpha > 2$, corresponding to a finite second moment of $p(\tau)$.

\subsection{Illustrative example - Zeno-protected entangled states}

\begin{figure}[h!]
\centering
\includegraphics[width=135mm]{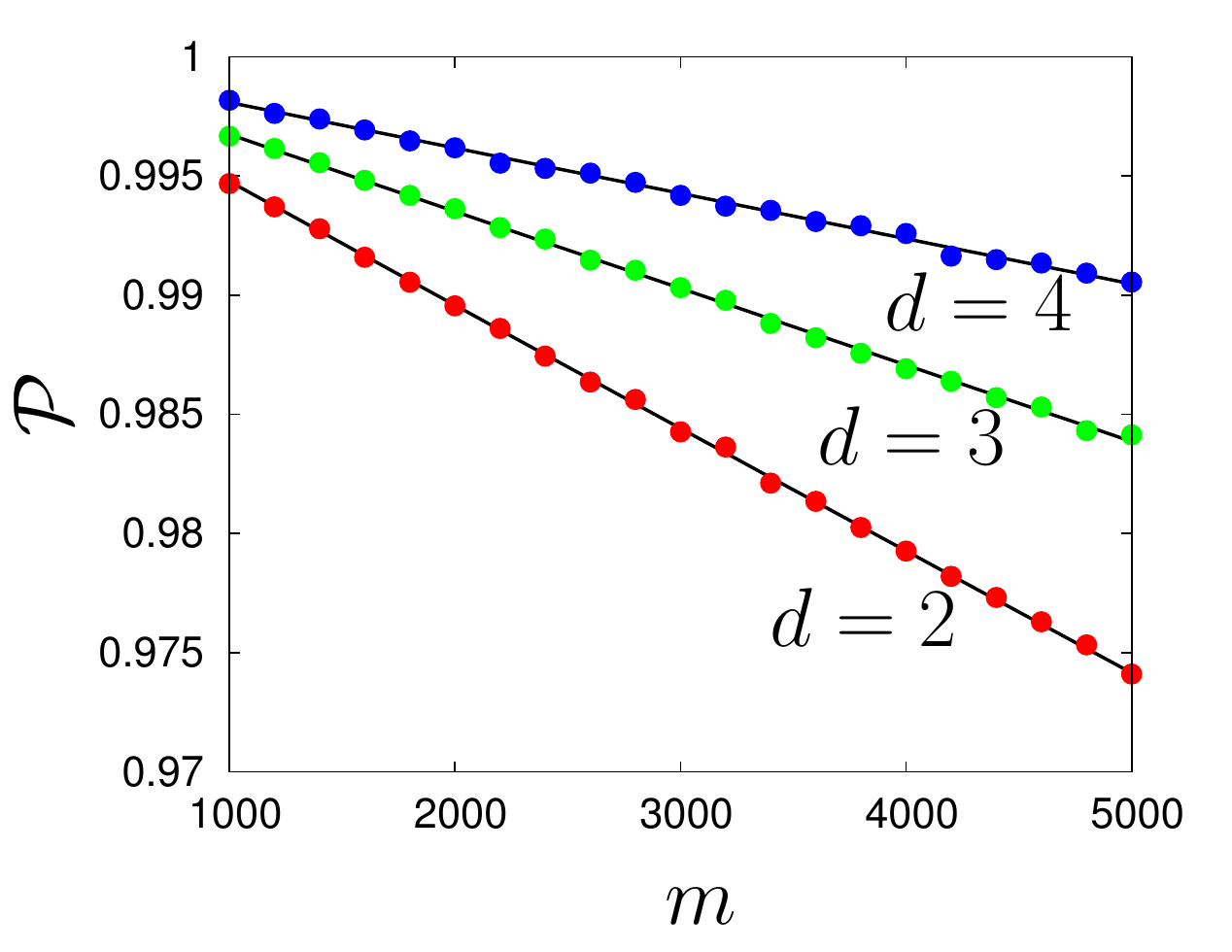}
\caption{Survival probability $\mathcal{P}$ as a function of the number of measurements $m$ for a $d$-dimensional Bernoulli distribution for the $\tau_j$'s, with $d=2,3,4$. Specifically, we have chosen for $d=4$ the values $p^{(1)}=0.3, ~p^{(2)}=0.2, ~p^{(3)}=0.05, ~p^{(4)}=0.45$, and
$\tau^{(1)}=\tau_{0},~\tau^{(2)}=3\tau_{0},~\tau^{(3)}=2\tau_{0},~\tau^{(4)}=0.5\tau_{0}$, with $\tau_{0}=1$ ns. For $d=3$, the values are $p^{(1)}=0.3, ~p^{(2)}=0.2, ~p^{(3)}=0.5$, and $\tau^{(1)}=\tau_{0},~\tau^{(2)}=3\tau_{0},~\tau^{(3)}=2\tau_{0}$, while for $d=2$, we have taken $p^{(1)}=0.3, ~p^{(2)}=0.7$, and $\tau^{(1)}=\tau_{0},~\tau^{(2)}=3\tau_{0}$. Here, the points denote the values obtained by evaluating (\ref{surv_prob}) numerically for a typical realization of the measurement sequence $\{\tau_j\}$, while the lines denote the asymptotic most probable values obtained by using (\ref{P_star}).}
\label{dnomial-1}
\end{figure}

\begin{figure}[h!]
\centering
\includegraphics[width=136mm]{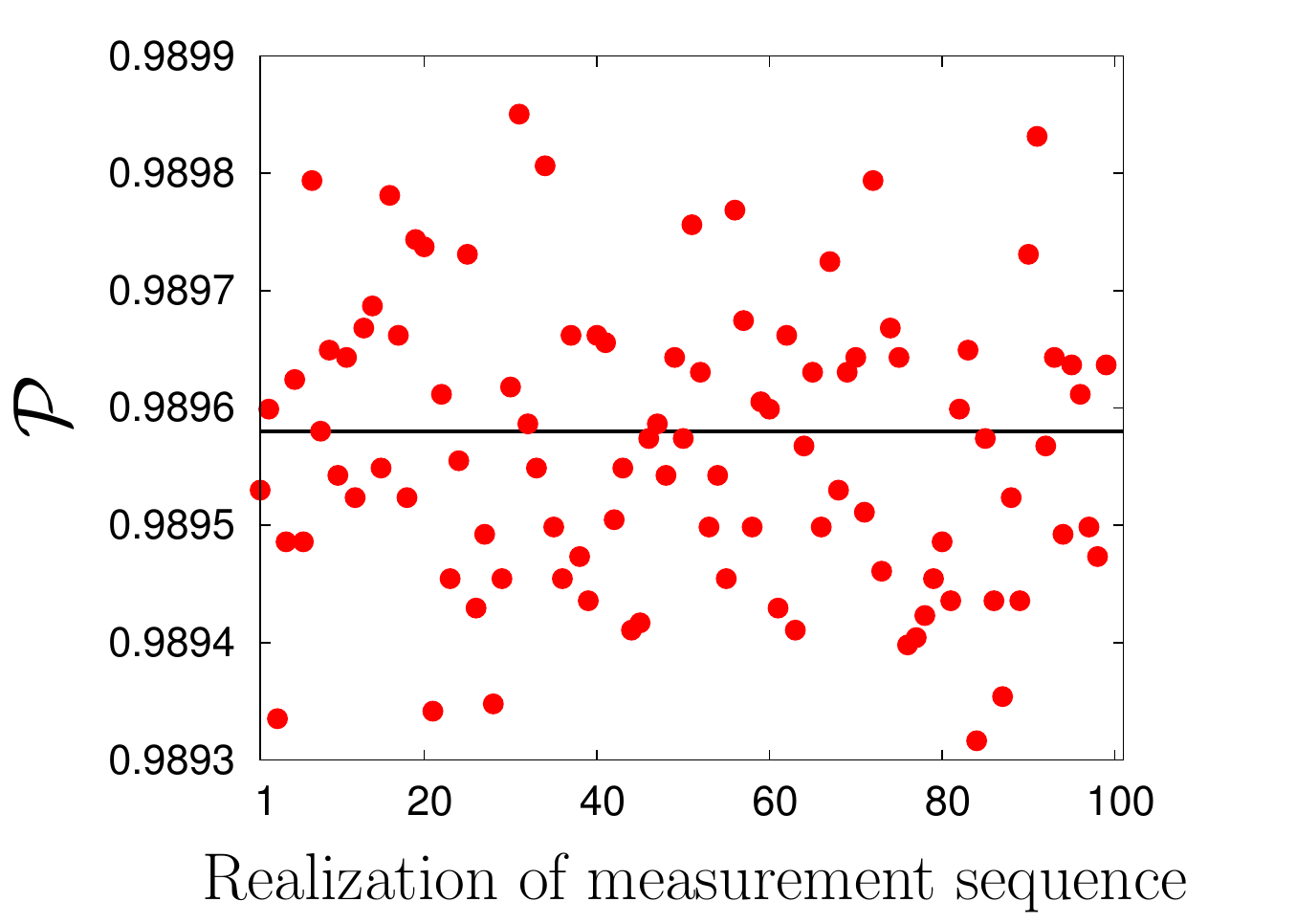}
\caption{Comparison between the survival probability $\mathcal{P}$ obtained by evaluating (\ref{surv_prob}) numerically for $100$ typical
realizations of the measurement sequence (points) and the most probable value $\mathcal{P}^{\star}$ (line) obtained by using (\ref{P_star}), for the case $d=2$ in Fig.~\ref{dnomial-1} and for the number of measurements $m=2000$.}
\label{dnomial-2}
\end{figure}

Here, in order to test our analytical results, we numerically simulate the dynamical evolution of a generic $n$-level quantum system governed by the following Hamiltonian:
\begin{equation}\label{Hamiltonian-LD}
H = \sum_{j=1}^{n}\omega_{j}|j\rangle\langle j|+\sum_{j=1}^{n-1}\Omega\left(|j\rangle\langle j+1| + |j+1\rangle\langle j|\right).
\end{equation}
Here, $|j\rangle \equiv |0 \dots 1 \dots 0 \rangle$, with $1$ in the $j-$th place and $0$ otherwise, denotes the state for the $j$-th level with $\omega_{j}$ the corresponding energy, while $\Omega$ is the coupling rate between nearest-neighbor levels. For simplicity, we take $n=3$, $\Omega=2\pi f$, with $f=100$ kHz, and $\omega_{j}=2\pi f_{j}$, with $f_{1}=30$ kHz, $f_{2}=20$ kHz and $f_{3}=10$ kHz. We choose the initial state $|\psi_0\rangle$ to be the following entangled (with respect to the bipartition $1|23$) pure state
\begin{equation}\label{initial-state-LD}
|\psi_{0}\rangle \equiv \frac{1}{\sqrt{2}}(|100\rangle+|001\rangle).
\end{equation}
Under these conditions, we obtain the survival probability $\mathcal{P}$ as a function of the number of measurements $m$ for a $d$-dimensional Bernoulli distribution for the $\tau_j$'s, with $d=2,3,4$ -- see Fig.~\ref{dnomial-1}.
We find a perfect agreement between the numerical evaluation of (\ref{surv_prob}) for a typical realization of the measurement sequence $\{\tau_j\}$ and the asymptotic most probable values obtained by using (\ref{P_star}). Moreover, a comparison between these two quantities for $d=2$, $m=2000$, and $100$ typical realizations of the measurement sequence is shown in Fig.~\ref{dnomial-2}.

\begin{figure}[h!]
\centering
\includegraphics[width=135mm]{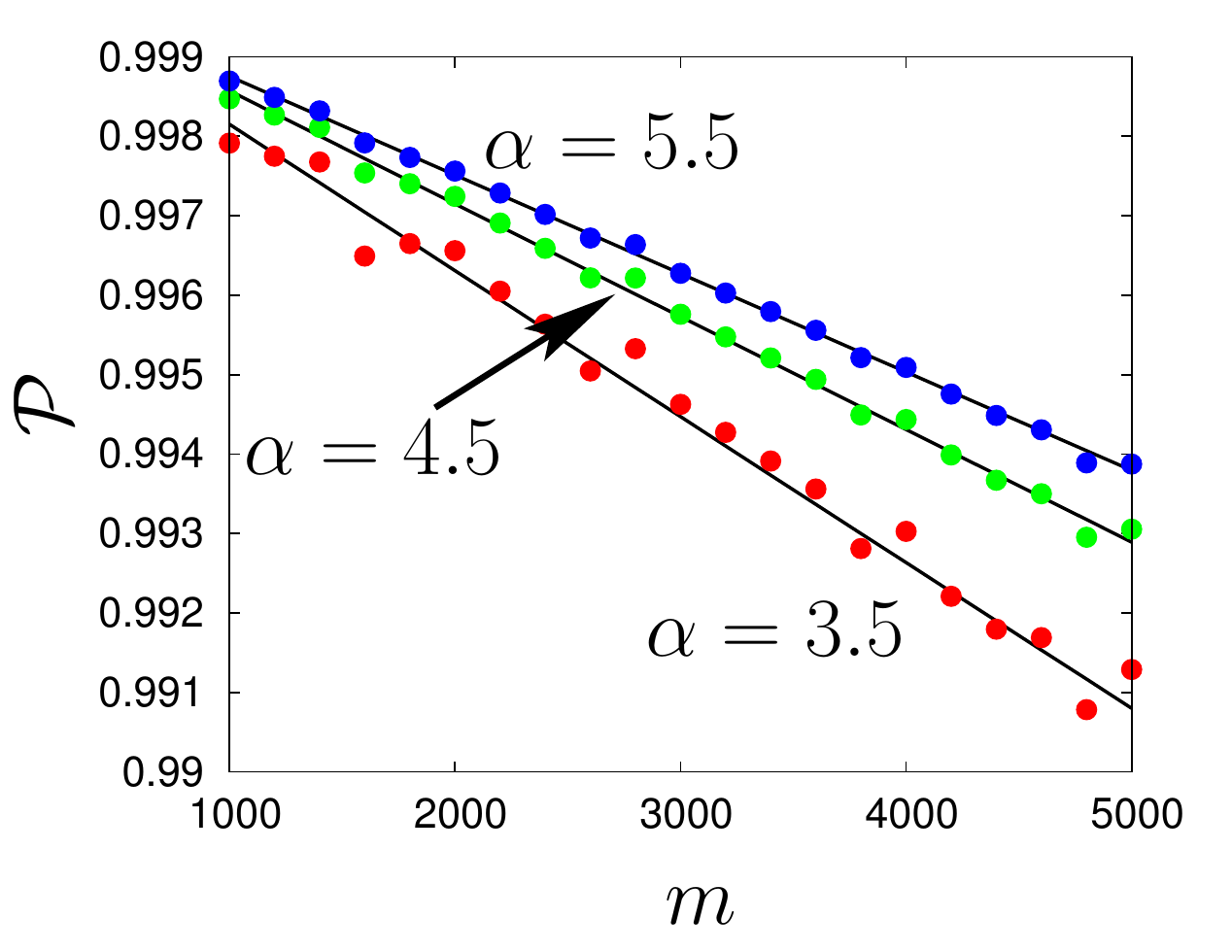}
\caption{Survival probability $\mathcal{P}$ as a function of the number of measurements $m$ for the distribution $p(\tau)= \alpha\frac{\tau_{0}^{\alpha}}{\tau^{1+\alpha}}$, with $\alpha>0$ and $\tau \in [\tau_{0},\infty]$. Here, $\tau_{0}$ is a time scale set to $1$ ns. Besides, we choose values of $\alpha$ such that $p(\tau)$ has a finite mean and a finite second moment, i.e. $\alpha > 2$. As in the preceding figures, the points denote the values obtained by evaluating numerically (\ref{surv_prob}) for a typical realization of the measurement sequence $\{\tau_j\}$, while the lines denote the asymptotic most probable values obtained by using (\ref{PG-defn}).}
\label{levy}
\end{figure}

Furthermore, to test our analytical predictions for a continuous $\tau$-distribution, we have considered the following distribution for
the $\tau_j$'s, namely,
$$
p(\tau)= \alpha\frac{\tau_{0}^{\alpha}}{\tau^{1+\alpha}},
$$
with $\alpha>0$ and $\tau \in [\tau_0,\infty)$. The corresponding survival probability shown in Fig.~\ref{levy} further confirms our analytical predictions.
Note that the decrease of fluctuations around the most probable value with increasing $\alpha$ is consistent with the concomitant smaller fluctuations of $\tau$ around the average $\overline{\tau}$. In all the cases discussed here, we observe excellent agreement with our estimate of the most probable value based on the LD theory. Moreover, our analytical predictions are numerically confirmed also for any coherent superposition state $|\psi_{0}\rangle \equiv a_1|100\rangle+ a_2|001\rangle$ with $|a_1|^2+|a_2|^2 = 1$.

Finally, we want to address the following question. \textit{Does the presence of disorder in the sequence of measurement time intervals enhance the survival probability?} To address it, we consider a $d$-dimensional Bernoulli $p(\tau)$ with $d=2$, and a given fixed value of the average $\overline{\tau}=p^{(1)}\tau^{(1)}+p^{(2)}\tau^{(2)}$. Then, in the first scenario, we apply $m$ projective measurements at times equally spaced by the amount $\overline{\tau}$, while in the second we sample this time interval from $p(\tau)$. As previously shown, the absence of randomness on the values of $\tau$ trivially leads to $\mathcal{P}^\star=\langle \mathcal{P} \rangle$. In the second scenario, the most probable value $\mathcal{P}^\star$ is given by (\ref{P_star}), from which
\begin{equation}
\mathcal{P}^{\star}=\exp\Big(m\Big[p^{(1)}\ln q(\tau^{(1)})+(1-p^{(1)})\ln q(\tau^{(2)})\Big]\Big),
\end{equation}
with $\tau^{(2)}=\frac{\overline{\tau}-p^{(1)}\tau^{(1)}}{p^{(2)}}$. Thus, \textit{the question arises as to whether for given fixed} $\overline{\tau}$ \textit{and} $\tau^{(1)}$ \textit{a random sequence of measurement yields a larger value of the survival probability than the one obtained by performing equally spaced measurements}.
\begin{figure}[h!]
\centering
\includegraphics[width=135mm]{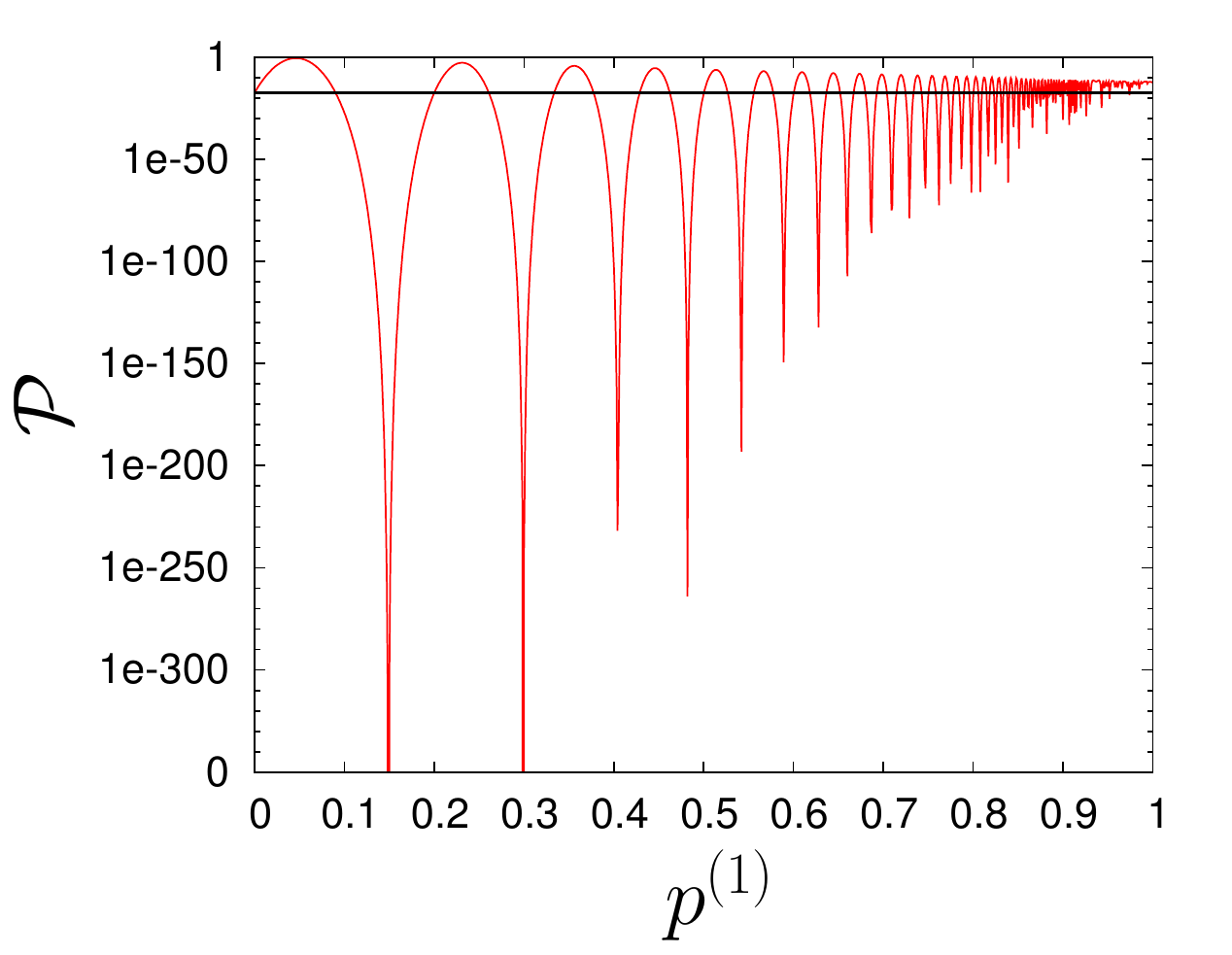}
\caption{Most probable value $\mathcal{P}^{\star}$ (red lines) for a $d$-dimensional Bernoulli distribution $p(\tau)$ with $d=2$, given fixed values of the average $\overline{\tau}=p^{(1)}\tau^{(1)}+p^{(2)}\tau^{(2)}$ and $\tau^{(1)}$, $m=100$. The black line denotes the value $\mathcal{P}^{\star}$ in the case of projective measurements equally spaced in time, with $m=100$. We have considered $\overline{\tau}=2.4\tau_{0}$, $\tau^{(1)}=\tau_{0}$, and $\tau_{0}=10~\mu$s.}
\label{ratio-fig}
\end{figure}
For the Hamiltonian (\ref{Hamiltonian-LD}) and the initial state (\ref{initial-state-LD}), we show in Fig.~\ref{ratio-fig} the behavior of $\mathcal{P}^\star$ as a function of $p^{(1)}$ at fixed values of $\overline{\tau}=2.4\tau_{0}$ and $\tau^{(1)}=\tau_{0}$, with $\tau_{0} = 10~\mu$s. A comparison with $\mathcal{P}(\overline{\tau})$ shows that while in the Zeno limit, such a disorder is deleterious, there are instances where random measurements are beneficial in enhancing the survival probability. Moreover, as shown in Fig.~6 in Ref.~\cite{Gherardini2016NJP}, an effective survival probability enhancement is reached in every dynamical evolution regime (except that in the Zeno limit) also in the behaviour of the ratio $\mathcal{P}^{\star}/\mathcal{P}(\overline{\tau})$ as a function of $\mathcal{P}(\tau^{(1)})$ at fixed values of $p^{(1)}=0.99$, $m=100$ and $\overline{\tau}=2.4\tau^{(1)}$, with $\tau^{(1)}\in[1,250]~ns$. Interestingly enough, these regimes might be particularly relevant from the experimental side when the ideal Zeno condition is only partially achieved.

\section{Experimental realization with BECs}

In this section, an experimental demonstration of the theoretical results in \cite{Gherardini2016NJP}, which have been previously introduced, is shown~\cite{GherardiniQST}. The experimental platform is given by a \textit{Bose-Einstein condensate (BEC)} of Rubidium ($^{87}$Rb) atoms, prepared on an atom chip~\cite{SchaferZeno}. Atom chips give the possibility to realize a trapping field for cold-atom clouds with current-carrying wires, whose induced magnetic field is compensated with a constant magnetic field (the bias) perpendicularly to the wire. The presence of a bias field, indeed, generates a zero of the total magnetic field on the axis parallel to the wire direction at a given (fixed) weight. Around such direction the field can be approximated with a quadrupole, which thus constitutes a linear guide for the atoms~\cite{PetrovicCHIP}. Further details on the setup will be discussed in this section, together with the experimental results.

By applying the LD theory to a sequence of randomly-distributed projective measurements, we have proved the equivalence between the analogous of the time and ensemble averages of the configurations assumed by the stochasticity in the time interval between measurements, when the system approaches the quantum Zeno regime. The observation of such equivalence corresponds to prove the \textit{ergodicity} of the interaction modes between the system and the environment. This result could pave the way towards the development of new feasible schemes to control quantum systems by tunable and usually deleterious stochastic noise. Before showing in detail the experimental results, let us recall that the mathematical definition of ergodicity was initially introduced by von Neumann~\cite{Neumann,Neumann2}: his ergodic theorem ensures that only rarely an observable of the system deviates considerably from its average value. In accordance with von Neumann's theory, Peres defined ergodicity in quantum mechanics as the equality between the time and ensemble averages of an arbitrary quantum operator~\cite{Peres1}. Recently, in \cite{NeilNatPhys} ergodic dynamics have been proved in a small quantum system consisting of three superconducting qubits, which was realized as a general framework for investigating non-equilibrium thermodynamics.

\begin{figure}[h!]
\centering
\includegraphics[width=0.70\textwidth,angle=0]{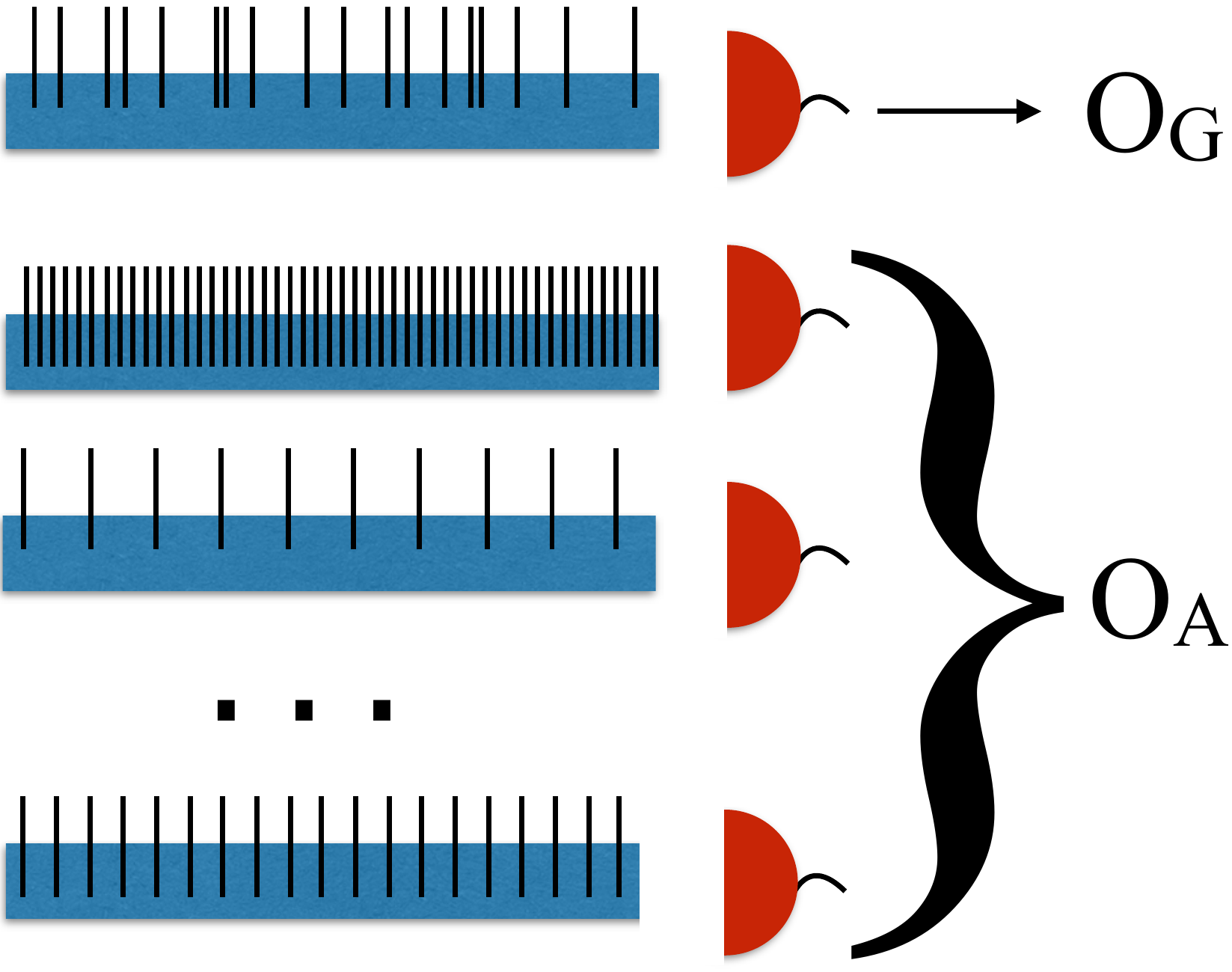}
\caption{Pictorial representation of the two measurement schemes corresponding, respectively, to here called geometric $O_{G}$ and arithmetic $O_{A}$ averages of a generic observable $O$ of an open quantum system. The black lines represent the interaction with the external environment, that is mimicked by a sequence of consecutive projective measurements, where the total time of a given realization of the measurements sequence is a random variable depending on how stochasticity is realized.}
\label{Fig1_QST}
\end{figure}

Now, given an observable $O$, we introduce two schemes to take into account the presence of stochastic noise in terms of an external environment by applying consecutive quantum projective measurements to the system, as shown in Fig.~\ref{Fig1_QST}.
The first scheme is based on the measurement of $O$ after a single dynamical evolution of the system, that interacts with the environment at stochastically distributed times (geometric average $O_G$). The second scheme, instead, consists of averaging the final observable outcomes over different dynamical realizations of the system each periodically interacting with the environment (arithmetic average $O_A$). Then, different realizations of the measurement sequence correspond to different time intervals in the
system-environment interaction, but extracted from the same probability distribution as in the first scheme. Moreover, the total number of times in which the time interval is sampled by the corresponding probability distribution establishes the relative weight to compute the arithmetic average $O_A$.

More formally, let us consider a sequence of repeated quantum measurements, which are separated by random time intervals sampled from a given probability density function $p(\tau)$. The most probable value $\mathcal{P}^{\star}(m)$ of the survival probability (probability that the system belongs to the measurement subspace after $m$ measurements) is given by (\ref{PG-defn}), while the \textit{average value} $\langle\mathcal{P}(m)\rangle$ by (\ref{P-avg1}). It can be observed that $\mathcal{P}^{\star}(m)$ is identically equal to the \textit{geometric average} $\overline{\mathcal{P}}_{g}$ of the survival probability, weighted by $p(\tau)$. Indeed
\begin{equation}\label{def_P_g}
\overline{\mathcal{P}}_g \equiv \prod_{\{\tau\}}q(\tau)^{mp(\tau)} = \exp\left(m\int_{\tau}d\tau p(\tau)\ln(q(\tau))\right)
= \mathcal{P}^{\star}(m),
\end{equation}
where the index of multiplication assumes all possible values of $\{\tau\}$, i.e. the ones in the support of $p(\tau)$. In the limit of a large number of measurements $M$, the geometric average $\overline{\mathcal{P}}_{g}(m)$ is the value to which the \textit{time average}
\begin{equation}
\hat{\mathcal{P}}_{M} (m) \equiv \frac{1}{M}\sum_{k=1}^{M}\mathcal{P}(\{\tau_{j}\}_{j=1}^{k})^{\frac{m}{k}}
\end{equation}
of the survival probability $\mathcal{P}(\{\tau_{j}\})$ converges, so that
\begin{equation}
\hat{\mathcal{P}}(m)\equiv\lim_{M\rightarrow\infty}\frac{1}{M}\sum_{k=1}^{M}\mathcal{P}(\{\tau_{j}\}_{j=1}^{k})^{\frac{m}{k}}=\overline{\mathcal{P}}_{g}(m).
\end{equation}
As a matter of fact, the value of the survival probability after $m$ measurements can be \textit{estimated} by using the corresponding value $\mathcal{P}(\{\tau_j\}_{j=1}^{k})$ after $k$ measurements by using the relation
\begin{equation}
\mathcal{P}(\{\tau_j\}_{j=1}^{m})\approx\mathcal{P}(\{\tau_j\}_{j=1}^{k})^{\frac{m}{k}}.
\end{equation}
This value, then, if averaged for $k = 1,\ldots,M$, converges to the geometric average $\overline{\mathcal{P}}_{g}(m)$ in the limit of large $M$. Finally, one can consider the \textit{ordered case of periodic projective measurements}, i.e. $\tau_{j}=\tau$, but with $\tau$ being selected according to $p(\tau)$. This leads to the definition of the \textit{arithmetic average}, i.e.
\begin{equation}
\overline{\mathcal{P}}_{a} \equiv \int_\tau d\tau p(\tau)q(\tau)^{m} = \exp\left(\ln\int_{\tau}d\tau p(\tau)q(\tau)^{m}\right).
\end{equation}
Using the Jensen's inequality and considering that $\langle \xi\rangle^{m}\leq\langle \xi^{m}\rangle$ for any $\xi\in [0,1]$ and $m\in\mathbb{N}$, it follows that
\begin{equation}\label{inequality_P}
\overline{\mathcal{P}}_{g}\leq\langle\mathcal{P}\rangle\leq\overline{\mathcal{P}}_{a},
\end{equation}
i.e.
\begin{equation}
\exp\left(m\int_{\tau}d\tau p(\tau)\ln(q(\tau))\right)\leq\left(\int_{\tau}d\tau p(\tau)q(\tau)\right)^{m}\leq
\exp\left(\ln\int_{\tau}d\tau p(\tau)q(\tau)^{m}\right).
\end{equation}

As main result, it is has been proved in \cite{GherardiniQST} that \textit{in the \textit{Zeno regime} it is sufficient to examine the series of constant $\tau$ in order to determine} $\overline{\mathcal{P}}_{g}$, $\overline{\mathcal{P}}_{a}$ \textit{and} $\langle\mathcal{P}\rangle$. The \textit{Zeno regime} is defined by the relation
\begin{equation}
m\int d\tau p(\tau)\ln q(\tau) = m\langle\ln q(\tau)\rangle\ll 1,
\end{equation}
in perfect agreement with (\ref{Zeno_condition}). Within this limit all the three averages are \textit{equal}; indeed
\begin{equation}\label{eq:equal-averages}
\overline{\mathcal{P}}_{a}\approx\left\langle 1 + m\ln q\right\rangle = 1 + m\langle\ln q\rangle\approx\overline{\mathcal{P}}_{g},
\end{equation}
and, as a consequence of the relation (\ref{inequality_P}), the equality holds also for the ensemble average $\langle\mathcal{P}\rangle$. Since in the geometric average the noise is averaged over time while the other two averages are computed over different realizations of the measurement sequence, the validity of this equality proves the ergodicity of the system-environment interaction modes. More specifically, let us define the \textit{normalized discrepancy} $\mathcal{D}$ between $\overline{\mathcal{P}}_{g}$ and $\overline{\mathcal{P}}_{a}$ as
\begin{equation}
\mathcal{D}\equiv\frac{\overline{\mathcal{P}}_{a}-\overline{\mathcal{P}}_{g}}{\overline{\mathcal{P}}_{a}} = 1-e^{-\Delta q(\tau,m)}\approx \Delta q(\tau,m),
\end{equation}
where
\begin{equation}
\Delta q(\tau,m)\equiv\ln\langle q(\tau)^{m}\rangle - \langle\ln q(\tau)^{m}\rangle.
\end{equation}
$\Delta q(\tau,m)$ is equal to zero only within the Zeno regime, while outside the equality (\ref{eq:equal-averages}) (under second-order Zeno approximation) breaks down. As a matter of fact, the leading term in $\mathcal{D}$ is of fourth order in $\tau$, i.e.
\begin{equation}\label{discrepancy}
\Delta q(\tau,m)\approx\frac{m^{2}}{2}(\Delta^{2}H)^{2}\left(\langle\tau^{4}\rangle - \langle\tau^{2}\rangle^{2}\right),
\end{equation}
which is determined by the second and the fourth moment of $p(\tau)$ ($\langle\tau^{2}\rangle\equiv\int_{\tau}d\tau p(\tau)\tau^{2}$ and $\langle\tau^{4}\rangle\equiv\int_{\tau}d\tau p(\tau)\tau^{4}$, respectively) and by the variance of the energy $\Delta^2 H$ (see Appendix~\ref{chapter:appB} for further details).

For the sake of clarity, let us consider a bimodal distribution for $p(\tau)$, with values $\tau^{(1)}$ and $\tau^{(2)}$ and probability $p_{1}$ and $p_{2} = 1 - p_{1}$. The survival probability in the Stirling approximation for $m$ sufficiently large is distributed as
\begin{equation}\label{eq:prob-distribution1}
\mathrm{Prob}(\mathcal{P}) \approx \frac{1}{\sqrt{2\pi m p_1 p_2}}\exp\left(-\frac{(k(\mathcal{P})-m p_1)^2}{2m p_1 p_2}\right),
\end{equation}
where
\begin{equation}\label{eq:prob-distribution2}
k(\mathcal{P}) = \frac{\ln \mathcal{P}-m\ln q(\tau^{(2)})}{\ln q(\tau^{(1)})-\ln q(\tau^{(2)})},
\end{equation}
is the frequency of the event $\tau^{(1)}$. Also the derivation of (\ref{eq:prob-distribution1}) can be found in Appendix~\ref{chapter:appB}.
\begin{figure}[h!]
\centering
\includegraphics[width=0.9\textwidth,angle=0]{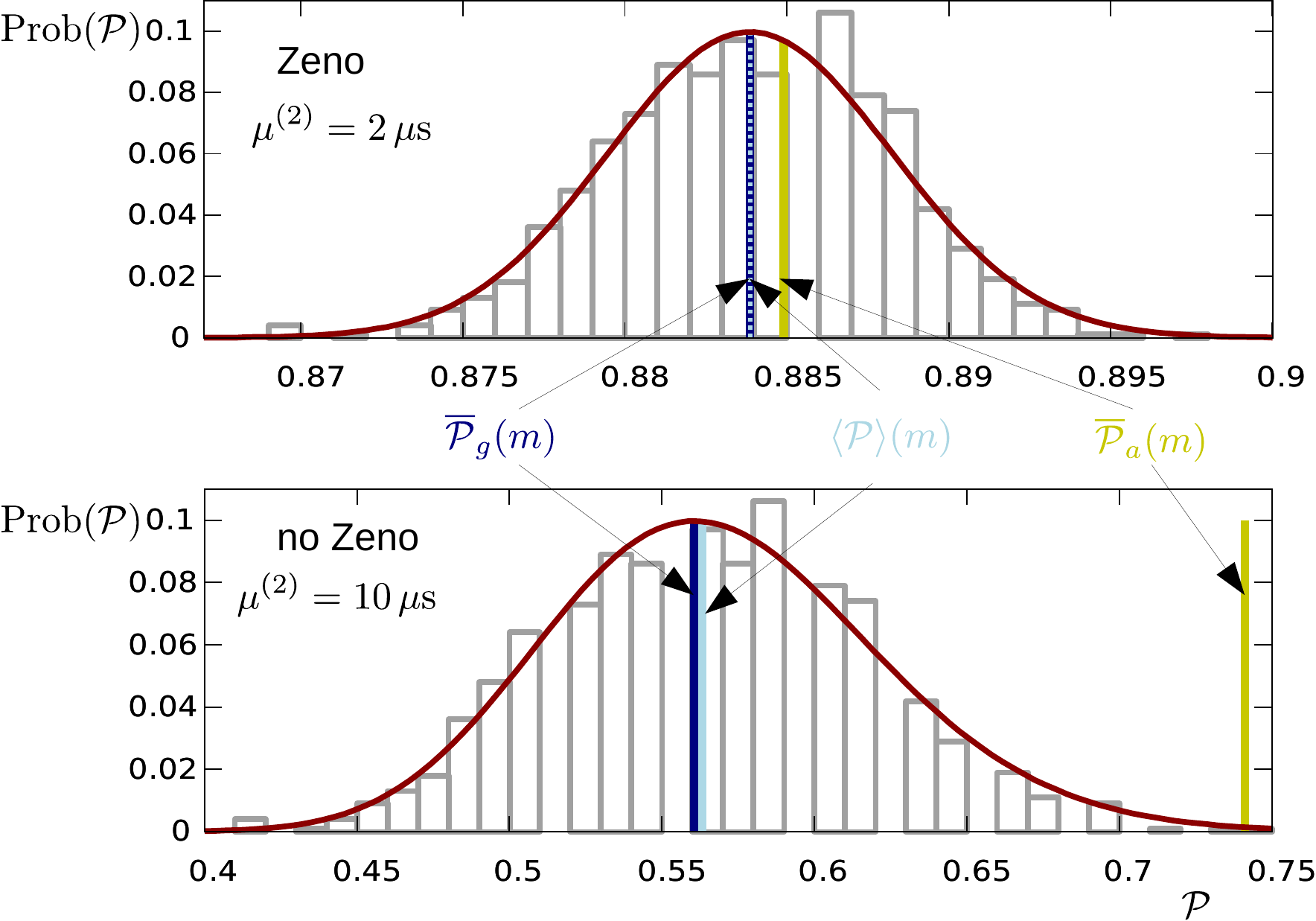}
\caption{Probability distribution $\mathrm{Prob}(\mathcal{P})$ of the survival probability. The grey boxes refers to the relative frequencies of $\mathcal{P}$, obtained by evaluating numerically $1000$ random realizations of the survival probability after $m=100$ measurements. They are compared to the expected distribution (\ref{eq:prob-distribution1}) shown in dark red. The top panel shows the results in the Zeno regime, the lower panel for violated Zeno condition. As it can be seen, the values of the geometric average $\overline{\mathcal{P}}_g$ and of the expectation value $\langle\mathcal{P}\rangle$ are very similar. In the Zeno limit, also $\overline{\mathcal{P}}_a$ is very close to these two values, unlike the lower panel where the Zeno condition does not hold, hence demonstrating the \textit{ergodic hypothesis} for a randomly perturbed quantum system only in the Zeno regime.}
\label{Fig:distribution_QST}
\end{figure}
Fig.~\ref{Fig:distribution_QST} shows the distribution $\mathrm{Prob}(\mathcal{P})$ of the survival probability for this bimodal distribution. The grey boxes refer to the relative frequencies of $\mathcal{P}$ with $\tau^{(1)} = 1$ $\mu$s after $m=100$ measurements for $1000$ random realizations, as compared to the expected distribution (\ref{eq:prob-distribution1}) in dark red. The top panel displays the results for $\tau^{(2)} = 2$ $\mu$s (satisfying both the Zeno condition and the ergodic hypothesis), the lower panel for $\tau^{(2)} = 10$ $\mu$s (not satisfying both the Zeno condition and the ergodic hypothesis). We select two values of $\tau^{(2)}$ (i.e., $2$ $\mu$s and $10$ $\mu$s) in order to show two different scenarios closely related to the experimental observations (in this regard, see also Fig.~\ref{Fig:experimentalresults}). $\tau^{(2)} = 2$ $\mu$s, indeed, is close to the minimal time interval that is experimentally feasible and leads to Zeno dynamics in a regime where the geometric average $\overline{\mathcal{P}}_{g}$ can also be different from $1$ (depending on the choice of $p_{2}$). On the other side, $\tau^{(2)} = 10$ $\mu$s guarantees that the Zeno condition is violated with $\overline{\mathcal{P}}_{g}$ being however significantly larger than zero. In both scenarios $\overline{\mathcal{P}}_{g}$ is the maximal value assumed by $\mathrm{Prob}(\mathcal{P})$ and the expectation value $\langle\mathcal{P}\rangle$ is very close to it. In the Zeno limit also $\overline{\mathcal{P}}_a$ is very close to these two values, while in the lower panel, where the Zeno condition is violated, it assumes a different value, confirming the analytical results. The other parameters are $\Delta H = 2\pi\cdot 2.5$ kHz, $p_{1}=0.8$ and $p_{2}=0.2$. Qualitatively similar behaviours have been observed for other parameter values.

Finally, in Table~\ref{Table_1} we show the difference between the values of $\overline{\mathcal{P}}_{a}$ and $\overline{\mathcal{P}}_{g}$ for a bimodal distribution when varying the probability $p_{1}$ but with $\tau^{(2)} = 10$ $\mu$s. Outside the Zeno regime, the arithmetic average $\overline{\mathcal{P}}_{a}$ is always different from the geometric average $\overline{\mathcal{P}}_{g}$. Such discrepancy disappears when the stochasticity in the time interval between the measurements vanishes, i.e. for $p_1 = 0$ (complete leakage) and for $p_1 = 1$ (standard Zeno regime). As it will be shown in the following, all the theoretical predictions are well corroborated by the experimental data.
\begin{table}[h!]
\centering
\begin{tabular}{|c|c|c|}
\hline
$p_1$          & $\overline{\mathcal{P}}_{a}$    &   $\overline{\mathcal{P}}_{g}$       \\
\hline\hline
0.01            & 0.0905     & 0.0842         \\
0.05            & 0.1234     & 0.0927         \\
0.2             & 0.2470     & 0.1329         \\
0.5             & 0.4941     & 0.2729         \\
0.8             & 0.7412     & 0.5606         \\
0.95            & 0.8648     & 0.8035         \\
0.99            & 0.8977     & 0.8845         \\
\hline
\end{tabular}
\caption{Arithmetic and geometric averages $\overline{\mathcal{P}}_{a}$ and $\overline{\mathcal{P}}_{g}$ as a function of the probability $p_1$ for a bimodal distribution $p(\tau)$, expressed with four decimal digits. In the simulations we have chosen $\tau^{(1)} = 1$ $\mu$s, $\tau^{(2)} = 10$ $\mu$s, $\Delta H = 2\pi\cdot 2.5$ kHz, and $m = 100$.}
\label{Table_1}
\end{table}

\subsection*{Experimental setup and methods}

The aforementioned theoretical results have been tested with a Bose-Einstein condensate of $^{87}$Rb produced in a magnetic micro-trap realized with an atom chip. The trap has a longitudinal frequency of $46~{\rm Hz}$ and a radial trapping frequency of $950~{\rm Hz}$. The BEC has typically $8\cdot10^4$ atoms, a critical temperature of $0.5~\mu{\rm K}$ and is at $300~\mu{\rm m}$ from the chip surface. The magnetic fields for the micro-trap are provided by a Z-shaped wire on the atom chip and an external pair of Helmholtz coils, while the RF fields for the manipulation of the Zeeman states are produced by two further conductors also integrated on the atom chip.

\begin{figure}[h!]
\centering
\includegraphics[width=0.7\textwidth,angle=0]{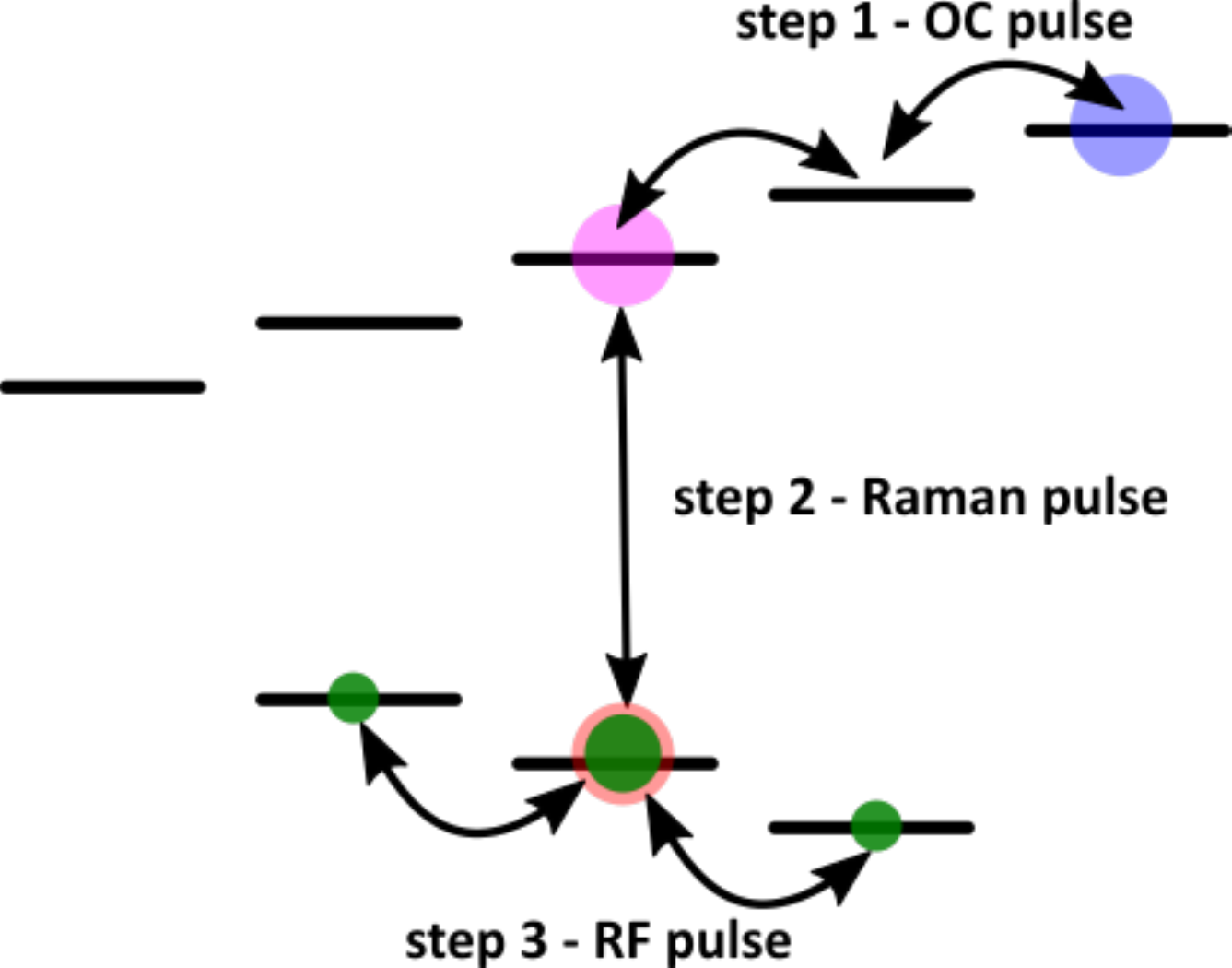}
\caption{State preparation sequence for the experiment on \textit{stochastic quantum Zeno effect}. After the condensation in the pure state $|F=2,m_F=+2\rangle$, in the first step the atoms are transferred to the state $|F=2, m_F=0\rangle$ with fidelity $\sim 95\%$. In the second step, by two Raman lasers the atoms in this sub-level are transferred to the lower state $|F=1, m_F=0\rangle$, which is the initial state $\rho_0$ for our experiment. In the third and last step, a fixed amount of population is transferred into the side sub-levels $|F=1,m_F=\pm 1\rangle$. These atoms will be used as a benchmark to compute the survival probability after the experiment.}
\label{fig_5_QST}
\end{figure}

Let us recall that the ground state of $^{87}$Rb is a hyperfine doublet separated by $6.834\rm\,GHz$ with total spin $F=2$ and $F=1$, respectively. To prepare the atoms for the experiment, the condensate is released from the magnetic trap and allowed to expand freely for $0.7 \rm\,ms$, while a constant magnetic field bias of $6.179 \rm\,G$ is applied in a fixed direction. This procedure ensures that the atom remains oriented in state $|F=2, m_F=+2\rangle$ and strongly suppresses the effect of atom-atom interactions by reducing the atomic density. The preparation consists of three steps (see Fig.~\ref{fig_5_QST}):
\begin{itemize}
  \item In the first step all the atoms are brought into the $|F=2,m_F=0\rangle$ state with high fidelity ($\sim95\%$). This is obtained applying a $50\rm\, \mu s$ long frequency modulated RF pulse designed with an Optimal Control (OC) strategy~\cite{LovecchioControl}.
  \item After the RF pulse we transfer the whole $|F=2, m_F=0\rangle$ population into the $|F=1, m_F=0\rangle$ sub-level by shining in bichromatic (Raman) laser light. This is the initial state $\rho_0$ for our experiment. Note that, with our choice of laser polarizations and thanks to the presence of the homogeneous bias field shifting away from resonance other magnetic sub-levels, the bichromatic light does not alter the population of the other magnetic sub-levels.
  \item The preparation is completed by applying another RF pulse to place some atomic population in the $|F=1,m_F=\pm 1\rangle$ states for normalization of the imaging procedure. Atoms in these last states will be not affected during the actual experiment, so they can be used as a control sample population.
\end{itemize}

In order to check each step of the preparation procedure, we record the number of atoms in each of the 8 $m_F$-states by applying a Stern-Gerlach method. In this regard, an inhomogeneous magnetic field is applied along the quantization axis for $10~{\rm ms}$. This causes the different $m_F$-sub-levels to spatially separate. After a total time of $23~{\rm ms}$ of expansion, a monochromatic light in resonance with the $|F=2\rangle\rightarrow|F'=3\rangle$ transition is used for $200 \rm\,\mu s$, so to push away all atoms in the $F=2$ sub-levels and recording the shadow cast by these atoms onto a CCD camera. We let the remaining atoms expand for further $1 \rm\,ms$ and, then, apply a bichromatic pulse containing light resonant to the $|F=2\rangle\rightarrow|F'=3\rangle$ and $|F=1\rangle\rightarrow|F'=2\rangle$ transitions, effectively casting onto the CCD the shadow of the atoms in the $F=1$ sub-levels. Another two CCD images to acquire the background complete the imaging procedure.

\begin{figure}[h!]
\centering
\includegraphics[width=0.9\textwidth,angle=0]{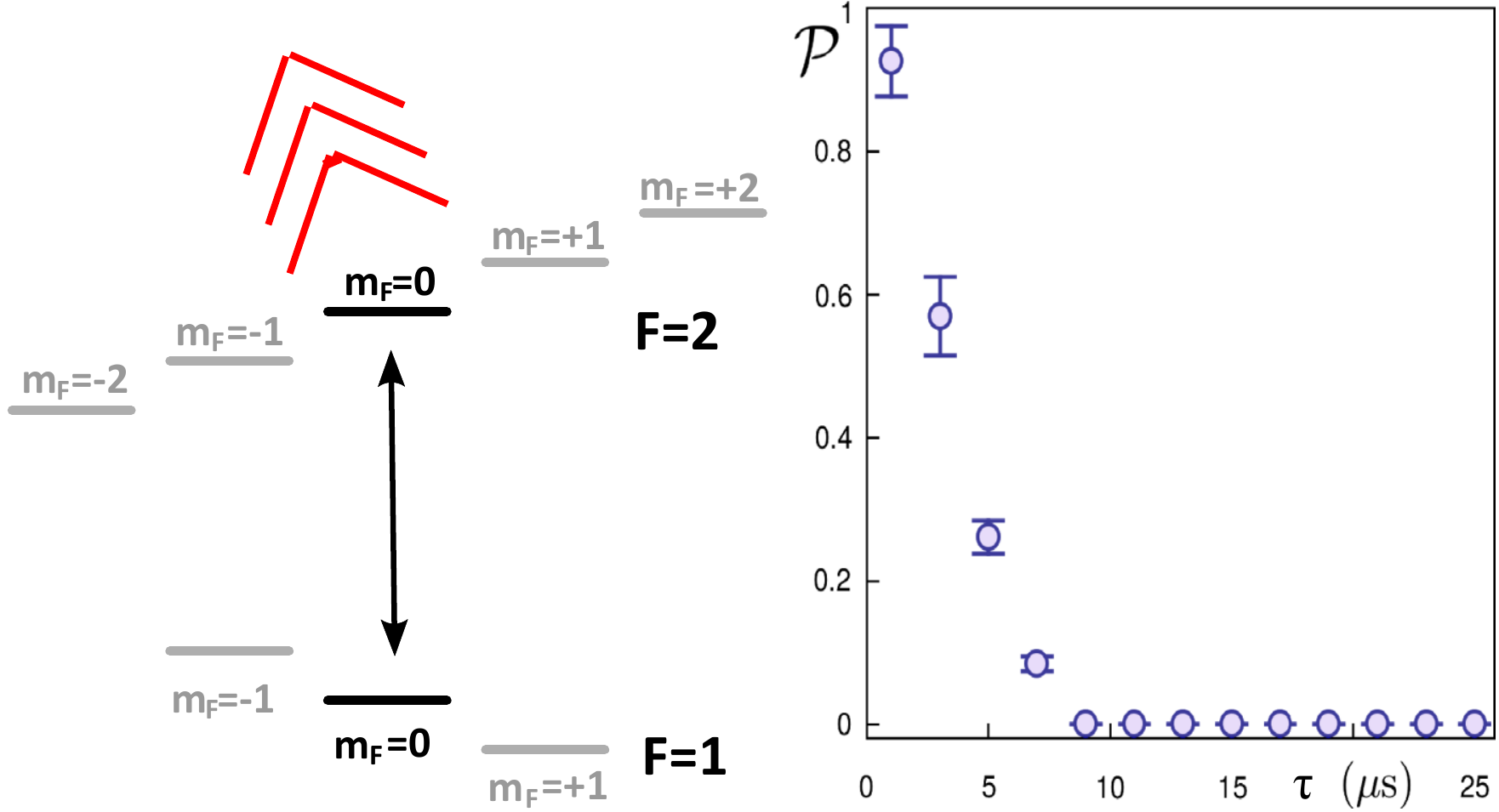}
\caption{Confinement induced by pulsed quantum Zeno effect. The ground state structure of the $^{87}$Rb in presence of a magnetic field consists of two hyperfine levels ($F=1$ and $F=2$), with no internal degeneracy. A laser induced Raman transition couples the sub-levels $|F=1,m_F=0\rangle$ and $|F=2, m_F=0\rangle$, while a laser on resonance with the transition $|F=2\rangle\rightarrow|F'=3\rangle$ (red arrows in the picture) depletes the population of the former. If the laser is strong enough, this equates to a projective measurement. On the right we show the typical exponential decay of the survival probability of the atoms in the $|F=1, m_F=0\rangle$ sub-level while the Raman coupling is on, and simultaneously the laser resonant to the $|F=2\rangle\rightarrow|F'=3\rangle$ transition is pulsed 100 times. The survival probability is plotted as a function of the interval $\tau$ between two pulses.}
\label{fig_3pr_QST}
\end{figure}

The experiments are performed by coupling the $|F=1,m_F=0\rangle$ and $|F=2, m_F=0\rangle$ with a Raman transition driven at a Rabi frequency of $5 \rm\, kHz$ by a bichromatic laser beam, as shown in Fig.~\ref{fig_3pr_QST}.
Since we are working with ground state atoms, with our choice of laser polarizations and thanks to the presence of the homogeneous bias field (shifting away from resonance other magnetic sub-levels) and selection rules for Raman transitions, we have effectively isolated a closed 2-level system. The projective measurements $\Pi = |\psi_0\rangle\langle\psi_0|$, then, are realized by shining the atoms with a $1\rm\,\mu s$ pulse of light resonant with the $|F=2\rangle\rightarrow|F'=3\rangle$ component of the Rubidium $D2$ line. Note that from the excited state $|F'=3\rangle$ atoms will immediately decay outside the condensate and will not be seen by our imaging system. Under constant coupling by the Raman beams, we apply $100$ projective measurements $\Pi$ after variable intervals of free evolution $\tau_{j}$. At the end of the sequence we measure the population remaining in state $|F=1,m_F=0\rangle$ and normalize it by comparison with the population in states $|F=1,m_F=\pm1\rangle$. This allows to measure, in a single shot, the survival probability $\mathcal{P}$ of the atoms in the initial state. Each experimental sequence is repeated $7$ times to obtain averages and standard deviations.

\subsection{Ergodicity of the system-environment interaction modes}

\begin{figure}[h!]
\centering
\includegraphics[width=0.9\textwidth]{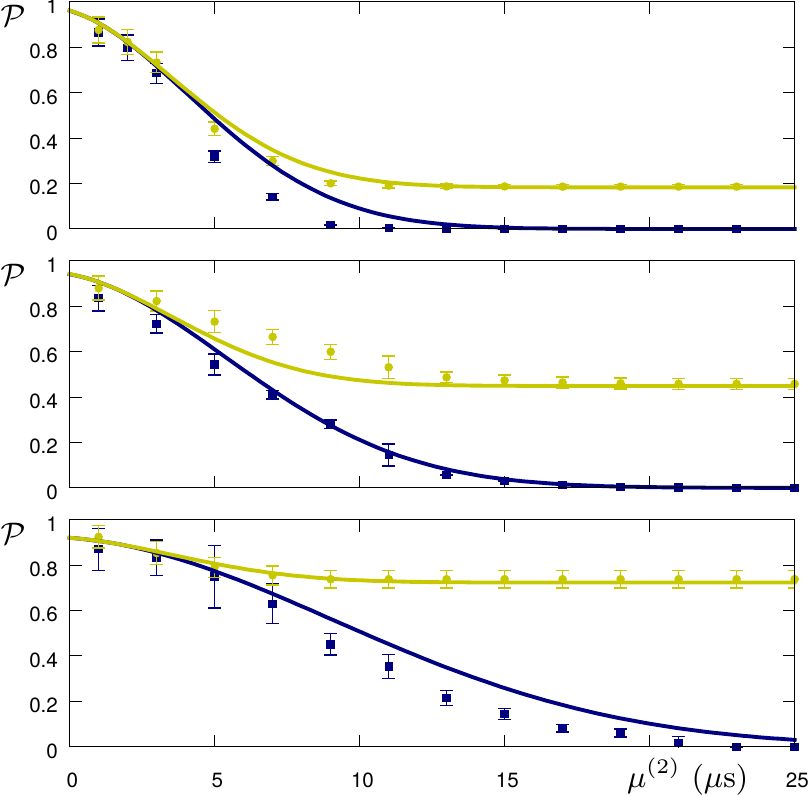}
\caption{Scaling of $\overline{\mathcal{P}}_{g}$ (blue) and $\overline{\mathcal{P}}_{a}$ (yellow) with the interval length $\tau^{(2)}$ of a bimodal distribution for $p(\tau)$. In all three cases $\tau^{(1)} = 2$  $\mu$s and $\Delta H = 2.5$ kHz. The probabilities of the bimodal distribution $(p_{1},p_{2})$ are, respectively, $(0.2,0.8)$ (upper panel), $(0.5, 0.5)$ (middle panel), and $(0.8, 0.2)$ (lower panel). The solid lines are the theoretical curves, while the single points are experimental values where the error bars indicate the standard deviation of the experimental error. The Zeno regime corresponds to vanishing $\tau_2$.}
\label{Fig:experimentalresults}
\end{figure}

To realize the theoretical predictions we experimentally measure the geometric and arithmetic averages of the survival probability $\mathcal{P}$ by assuming $p(\tau)$ as a bimodal distribution, where we take $\tau^{(1)}=2\rm\,\mu s$ to be fixed and $\tau^{(2)}$ variable between $2\rm\,\mu s$ and $25\rm\,\mu s$. Overall, the experiment can be synthesized in two sets of data acquisitions:
\begin{itemize}
  \item In a first set of experiments we measure the arithmetic average $\overline{\mathcal{P}}_a$ by fixing the intervals of free evolution $\tau_j$ to be all the same and equal to $\tau\in\{\tau^{(1)},\tau^{(2)}\}$ and we determine $\mathcal{P}(\tau)$, i.e the probability for an atom to remain in the initial state as a function of $\tau$. As shown in Fig.~\ref{fig_3pr_QST}, $\mathcal{P}(\tau)$ displays the characteristic \textit{exponential decay}, which becomes negligible, in our case, after $9\rm\,\mu s$. After measuring $\mathcal{P}(\tau^{(1)})$ and $\mathcal{P}(\tau^{(2)})$, we then calculate the arithmetic average of the two data with statistical weights $p_{1}$ and $p_{2}$, respectively. In this way we obtain $\overline{\mathcal{P}}_{a}(\tau^{(2)})$ which represents the statistical mean averaged over the two possible system configurations as a function of the variable time $\tau^{(2)}$. In Fig.~\ref{Fig:experimentalresults} we report as yellow dots the results of three choices $(0.2, 0.8)$, $(0.5, 0.5)$, and $(0.8, 0.2)$ for the statistical weights $(p_1,p_2)$.
  \item In order to determine the geometric average $\overline{\mathcal{P}}_g$ of a single realization, we perform a second set of experiments. In each experimental sequence we now choose the intervals of free evolution $\tau_j$ from the bimodal probability density function given by $\tau^{(1)}$ and $\tau^{(2)}$ with probabilities $(p_{1},p_{2})$. The results of these experiments give the geometrical average $\overline{\mathcal{P}}_{g}(\tau^{(2)})$ of the survival probability as a function of the parameter $\tau^{(2)}$. We choose again the probabilities $(0.2, 0.8)$, $(0.5, 0.5)$, and $(0.8, 0.2)$  and the experimental results are shown as blue squares in Fig.~\ref{Fig:experimentalresults}.
\end{itemize}
As it can be observed in the figure, the agreement of theoretical predictions and experiments is generally very good, although some deviations go beyond the error bars and are systematic. Indeed, in the model the measurement has been assumed to be instantaneous while in the experiment it is a dissipative process of a duration of about $1\,\mathrm{\mu s}$. Furthermore, we can see in Fig.~\ref{Fig:experimentalresults} that for small values of $\tau^{(2)}$, i.e. in the Zeno regime, the two averages $\overline{\mathcal{P}}_g$ and $\overline{\mathcal{P}}_a$ practically coincide, and this has been predicted by approximating the discrepancy between the two quantities with $\Delta q(\tau,m) \approx \frac{m^2}{2}(\Delta^2 H)^2(\langle\tau^4\rangle-\langle\tau^2\rangle^{2})$, which is of fourth order in $\tau^{(1)},\tau^{(2)}$. Finally, it is worth noting that Fig.~\ref{Fig:distribution_QST} corresponds to two cases of the lower panel of Fig.~\ref{Fig:experimentalresults}.

\section{Fisher information from stochastic quantum measurements}

In the previous section, we have shown how the interaction between a quantum system and the external noisy environment can be modeled with a sequence of stochastic measurements, i.e. measurements separated by random time intervals. Here, we analytically study the \textit{distinguishability}~\cite{Wootters1981,Braunstein1994} of two different sequences of stochastic measurement in terms of the \textit{Fisher Information (FI)} measure~\cite{Cover2006}, as given in \cite{GherardiniFisher}. Indeed, if we want to characterize the dynamics and the statistics of a randomly perturbed quantum system by measuring its state after a given evolution time, it becomes important to investigate how much two arbitrary states, obtained by propagating different stochastic contributions, can be distinguished by the measurement process. In this regard, a key role is played again by the quantum Zeno effect, whereby \textit{the largest interval such that two quantum states remain indistinguishable under an arbitrary evolution is given by the Zeno time}. As proved in~\cite{SmerziPRL2012}, the Zeno time can be written in terms of the Fisher information computed as a function of the conditional probability that the state of the system (after a free evolution) is projected into the Zeno subspace. In this context, a FI measure has been recently introduced to investigate the realizability of quantum Zeno phenomea, when non-Markovian noise is also included~\cite{Zhang2015}, but, as in \cite{SmerziPRL2012}, the small parameter of the theory is the constant time interval between two consecutive measurements. Conversely, within the formalism of stochastic quantum measurements, we will introduce a \textit{Fisher information operator}, for which the dynamical small parameters are defined by the statistical moments of the stochastic noise acting on the quantum system.

\subsection{Fisher information operator}

\begin{figure}[h!]
\centering
\includegraphics[width=0.9\textwidth]{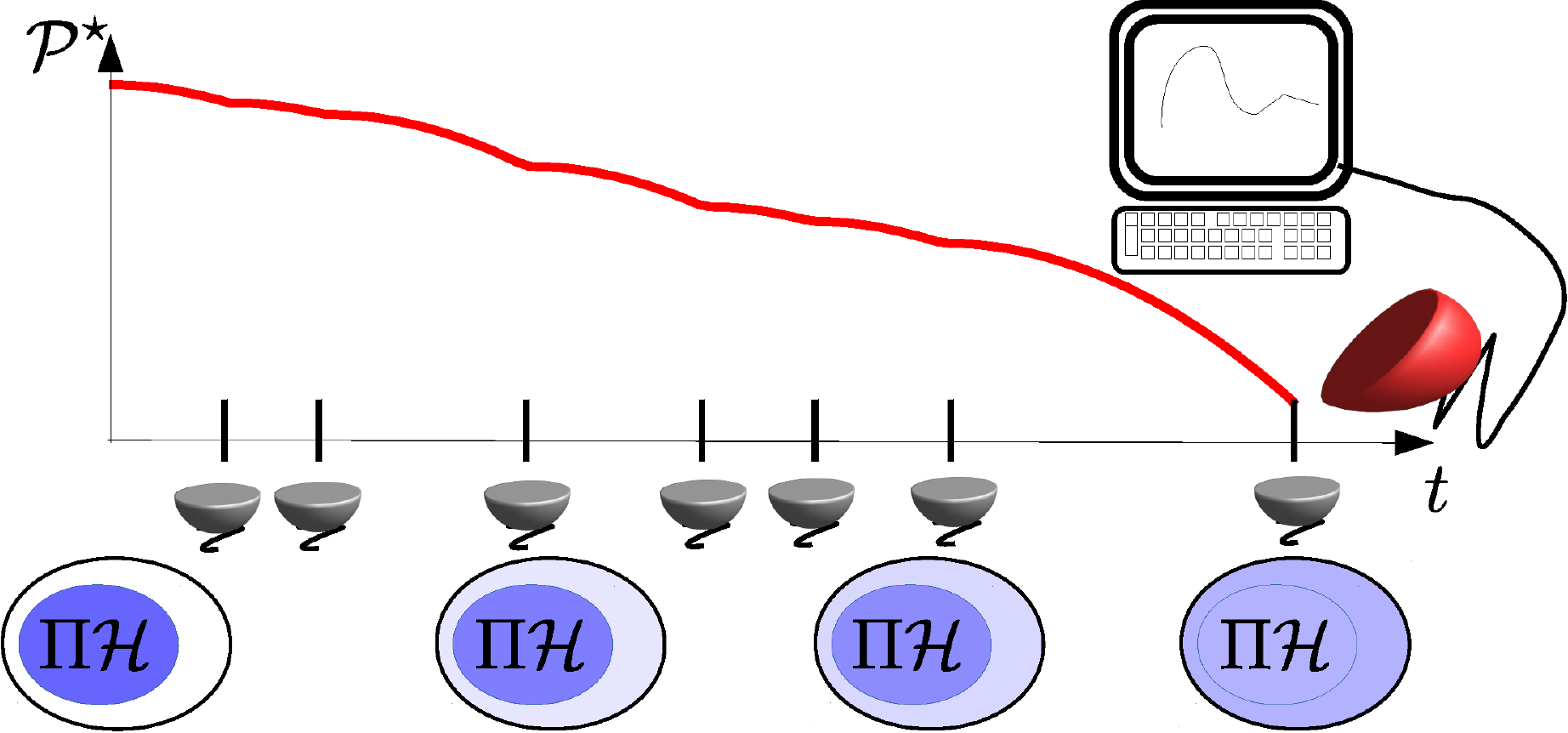}
\caption{Decay of the survival probability $\mathcal{P}^\star$ for a quantum system to remain in an Hilbert subspace when subjected to a stochastic sequence of measurements. As the time goes on, the population slowly leaks out of the subspace ($\Pi\mathcal{H}$, where $\Pi$ is the measurement projector and $\mathcal{H}$ is the full Hilbert space) as illustrated by the blue shades in the lower panel. After each measurement $\mathcal{P}^\star$ evolves quadratically in time. Only the final survival probability is registered by a (red) detector.}
\label{fig1_Fisher}
\end{figure}

As shown before, the survival probability $\mathcal{P}$ of a quantum system subject to a sequence of $m$ random measurements is a random variable, which converges to the corresponding most probable value $\mathcal{P}^{\star}$ in the limit of a large number of measurements. In particular, Fig.~\ref{fig1_Fisher} shows how the survival probability decays with ongoing time, slowed down by the intermediate measurements. At final time, after $m$ projective measurements, we make a final measurement and we register its outcome -- survival or not. For large enough $m$, the repetition of the experiment will allow us to determine the most probable value $\mathcal{P}^\star$, to which the survival probability converges for every single realization of the $\tau_j$'s.
By introducing a perturbation $\delta p(\tau)$ of the probability density function $p(\tau)$, we are interested in investigating the \textit{sensitivity of the survival probability most probable value} $\mathcal{P}^{\star}$ with respect to such perturbation, which induces a change of $\mathcal{P}^{\star}$ by the quantity
\begin{equation}
\delta\mathcal{P}^{\star} = m\mathcal{P}^\star\int_{\tau}d\tau\delta p(\tau)\ln q(\tau).
\end{equation}
In other words, $\delta\mathcal{P}^{\star}$ quantifies how sensitive is $\mathcal{P}^{\star}$ to a change of $p(\tau)$, and corresponds to the following functional derivative:
\begin{equation}\label{functional_derivative}
   \frac{\delta\mathcal{P}^\star}{\delta p}(\cdot)=m\mathcal{P}^\star\int_{\tau}d\tau(\cdot)\ln q(\tau).
\end{equation}
It is worth noting that, formally, this functional derivative is an element of the dual space with respect to that of the probability density functions $p(\tau)$, and thus a linear mapping from the admissible changes $\delta p(\tau)$ to the real number $\delta \mathcal{P}^{\star}$. We can express this fact by introducing the ket notation $\langle\cdot|$, such that the functional derivative (\ref{functional_derivative}) is given by
\begin{equation}
\Big\langle\frac{\delta\mathcal{P}^\star}{\delta p}\Big| = m\mathcal{P}^\star\langle\ln q |.
\end{equation}
Observe that for two arbitrary functions $f$ and $g$ the application of a bra to a ket leads to the scalar product
\begin{equation}
\langle f|g\rangle = \int_{\tau}d\tau f(\tau)g(\tau),
\end{equation}
where the bra $\langle f|$ is an element of the dual space and defines a linear mapping of the ket $|g\rangle$ onto the space of real numbers (through the scalar product operation).

If the projective measurements are frequent enough, the system evolution is effectively limited to the subspace given by the measurement projector $\Pi$, such that in the limit of infinite measurement frequency, the survival probability given by its most probable value $\mathcal{P}^\star$ converges to one. By using the bra-ket notation, the small deviation from this ideal scenario can be approximated by the following relation:
\begin{equation}\label{eq:Zeno_FI}
\mathcal{P}^\star\approx 1 + \Big\langle\frac{\delta\mathcal{P}^\star}{\delta p}\Big| p\Big\rangle,
\end{equation}
whereby the \textit{quality} of the Zeno confinement is determined by the \textit{sensitivity} of the survival probability $\mathcal{P}^\star$ with respect to a perturbation $\delta p(\tau)$. Such sensitivity is closely linked to the corresponding Fisher information, which quantifies the information on $p(\tau)$ that can be extracted by a statistical measurement of $\mathcal{P}^{\star}$. When dealing with a single estimation parameter $\theta$ and possible measurement results $\eta$, the Fisher information is defined as
\begin{equation}
F(\theta) \equiv \int_{\eta}\frac{1}{p(\eta|\theta)}\left(\frac{\partial p(\eta|\theta)}{\partial\theta}\right)^{2}d\eta,
\end{equation}
where $p(\eta|\theta)$ is the conditional probability to observe the result $\eta$ given a known value of the parameter $\theta$. In the case of a \textit{binary} event, i.e. $\eta\in\{\text{yes},\text{no}\}$, the integral reduces to a sum over the two events, and since $p(\text{no}|\theta) = 1 - p(\text{yes}|\theta)$ such that $\displaystyle{\left[\frac{\partial p(\text{yes}|\theta)}{\partial\theta}\right]^{2} = \left[\frac{\partial p(\text{no}|\theta)}{\partial\theta}\right]^{2}}$, the FI simplifies to
\begin{equation}\label{FI_theta}
F(\theta) = \frac{1}{p(\text{yes}|\theta)(1-p(\text{yes}|\theta))}\left(\frac{\partial p(\text{yes}|\theta)}{\partial\theta}\right)^{2}.
\end{equation}
The Fisher information (\ref{FI_theta}) quantifies the information that we obtain on $\theta$ when an event yes or no occurs. Thus, let us now consider the case where we perform $m$ projective measurements on the quantum system and we keep only the result of the last measurement. As shown in Fig.~\ref{fig1_Fisher}, we measure survival or not, hence one of two possible events with respective probabilities $\mathcal{P}^\star$ and $1-\mathcal{P}^\star$. Given two different probability density functions $p(\tau)$ characterized by their statistical moments, one can ask how they can be distinguished by a proper measurement. Since the probability depends on the function $p(\tau)$ (instead of a single parameter $\theta$), we approach this problem by generalizing the Fisher Information Matrix (FIM)
\begin{equation}
F_{ij}(\underline{\theta})\equiv\frac{1}{p(\text{yes}|\underline{\theta})(1-p(\text{yes}|\underline{\theta}))}\left(\frac{\partial p(\text{yes}|\underline{\theta})}
{\partial\theta_i}\right)\left(\frac{\partial p(\text{yes}|\underline{\theta})}{\partial\theta_j}\right),
\end{equation}
depending on the vector $\underline{\theta}\equiv(\theta_1,\theta_2,\dots)'$, to a \textit{Fisher Information Operator (FIO)}, which involves the functional derivatives of $\mathcal{P}^\star$. We get
\begin{equation}
F(p)\equiv\left|\frac{\delta\mathcal{P}^\star}{\delta p}\Big\rangle \Big\langle\frac{\delta\mathcal{P}^\star}{\delta p}\right|
=m^2\frac{\mathcal{P}^\star}{1-\mathcal{P}^\star}|\ln q\rangle\langle \ln q|\,.
\end{equation}
The Fisher information operator has the following three properties:
\begin{enumerate}[(1)]
  \item Since also $\mathcal{P}^\star$ depends on $m$, in the Zeno limit the FIO linearly scales with $m$:
\begin{equation}
F(p) \approx \frac{m}{|\langle\ln q|p\rangle|}|\ln q\rangle\langle \ln q|\,.
\end{equation}
  \item The FIO is a rank one operator, since binary measurement outcomes determine just $\mathcal{P}^\star$ and not its distribution. As a consequence, it is characterized by the single eigenvector $|v\rangle=|\ln q\rangle$ corresponding to the non-zero eigenvalue
\begin{equation}
\Lambda_v=m^2\frac{\mathcal{P}^\star}{1-\mathcal{P}^\star}\Vert \ln q\Vert^{2},
\end{equation}
with $\Vert \cdot \Vert$ being the $L_2$-norm, which is defined as
\begin{equation}
\Vert \ln q\Vert^{2} = \int_{\tau}d\tau|\ln q(\tau)|^{2}.
\end{equation}
  \item The FIO can be transformed into a FIM, if it is expressed in a certain basis, and the corresponding FIM in the generic basis $\{|f_i\rangle\}$ is given by the relation
\begin{equation}
 F_{ij} = m^{2}\frac{\mathcal{P}^\star}{1-\mathcal{P}^\star}\langle f_i|\ln q\rangle\langle \ln q|f_j\rangle.
\end{equation}
In particular, we might be interested in expressing the FIO in terms of the statistical moments
\begin{equation}
 \langle\tau^{k}\rangle \equiv \int_{\tau}d\tau p(\tau)\tau^{k}
\end{equation}
of the probability density function $p(\tau)$. As shown in Appendix~\ref{chapter:appB}, the corresponding FIM reads as
\begin{equation}\label{eigenvalue_F}
\widetilde{F}_{ij} = m^{2}\frac{\mathcal{P}^{\star}}{(1-\mathcal{P}^{\star})}\frac{\beta_{i}\beta_{j}}{i!j!},
\end{equation}
where
\begin{equation}
\beta_{k} \equiv \left.\frac{\partial^{k}\ln(q(\tau))}{\partial\tau^{k}}\right|_{\tau = 0}.
\end{equation}
\end{enumerate}
The third property of the Fisher information operator implies that, in principle, we can distinguish two probability density functions that differ by a single statistical moment or a linear combination of them. As main result, \textit{the highest sensitivity of such a distinguishability problem is found for a difference in the statistical moments of} $p(\tau)$ \textit{along the (single) eigenvector} $\underline{v}$ (corresponding to the non-zero eigenvalue $\lambda_{v}$) of the FIM (\ref{eigenvalue_F}). The non-zero eigenvalue is given by
\begin{equation}\label{eq:eigenvalue}
\lambda_{v} = m^{2}\frac{\mathcal{P}^{\star}}{1-\mathcal{P}^{\star}}\sum_{k}\left(\frac{\beta_k}{k!}\right)^2.
\end{equation}
%Since the basis functions $\{|f_{k}\rangle\}$ were not normalized, $\lambda_{v}$ is different from the eigenvalue $\Lambda_{v}$ of the FIO.
Moreover, the $k-$th element of the (non-normalized) eigenvector $\underline{v}$ is equal to $\underline{v}_{k} = \beta_{k}/k!$. Therefore, the most probable value $\mathcal{P}^\star$ can be expressed also as a function of $\Lambda_{v}$ ($\lambda_v$) and $|v\rangle$ ($\underline{v}$), such that
\begin{equation}
\mathcal{P}^\star = \frac{\Lambda_{v}}{\Lambda_{v} + m^{2}\|v\|^{2}} = \frac{\lambda_{v}}{\lambda_{v} + m^{2}\|\underline{v}\|^{2}}
\end{equation}
or equivalently
\begin{equation}
\mathcal{P}^\star = \exp\left(m\langle\underline{v}|\underline{\mu}\rangle\right),
\end{equation}
where the functions
\begin{equation}
\|\underline{v}\| \equiv \sqrt{\sum_{k}(v_{k})^{2}}
\end{equation}
and
\begin{equation}
\langle\underline{v}|\underline{\mu}\rangle = \langle\underline{v}|{\rm col}(\langle\tau^{k}\rangle)_{k}\rangle \equiv \sum_{k}v_{k}\langle\tau^{k}\rangle
\end{equation}
are, respectively, the Euclidian norm of $\underline{v}$ and the scalar product between $\underline{v}$ and $\underline{\mu}$, which collects the statistical moments of $p(\tau)$. As final remark, it is worth noting that the eigenvector of the FIM (\ref{eigenvalue_F}) depends only on the system properties ($\{\beta_{k}\}$), while the corresponding eigenvalue depends on both the system ($\{\beta_{k}\}$) and the probability density function $p(\tau)$ through the quantities $\{\langle\tau^{k}\rangle\}$.

\section{Stochastic quantum Zeno dynamics}

\begin{figure}[t]%h!
\centering
 \includegraphics[width=0.925\textwidth]{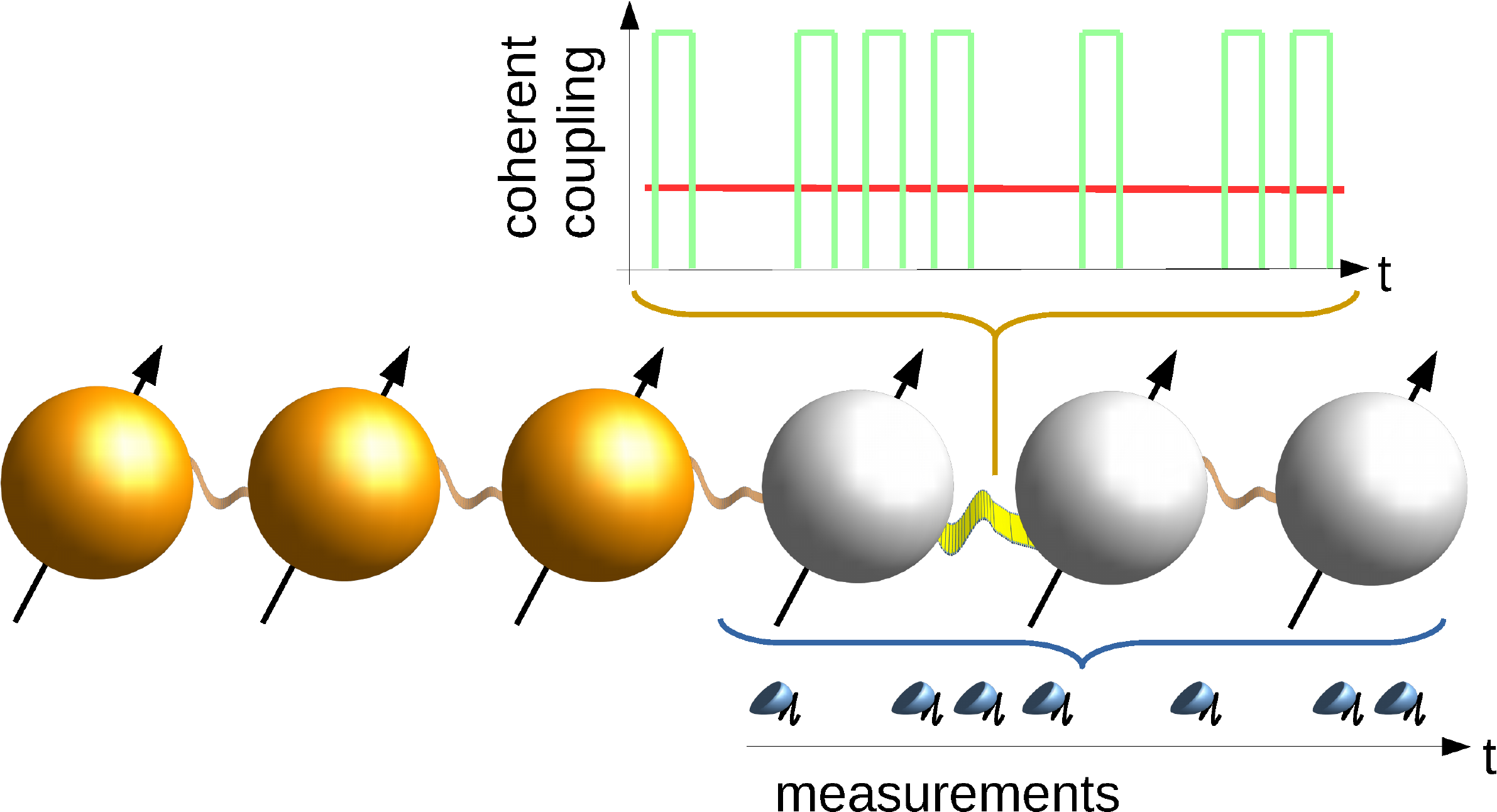}
 \caption{Pictorial representation of the observation protocols for stochastic quantum Zeno dynamics. A subsystem (orange, left) exhibits quantum Zeno dynamics when decoupled from the rest of the system by frequent measurements (blue ``detectors'', randomly spaced on the time axis) of the (population) leakage from the subspace or alternatively by a strong coherent coupling effectively locking the dynamics of the border site. The blue curly bracket indicates that the occupation in the grey part of the chain is measured to determine the leakage out of the (orange) subsystem, while the yellow thicker link indicates where the coherent coupling acts. The coherent coupling in its temporal behaviour can be continuous (red) or pulsed (green), as shown in the graph in the upper panel of the figure (coupling strength vs. time). The measurements as well as the coupling pulses can be spaced randomly, thus making the leakage stochastic.}
 \label{fig:overview_Annalen}
\end{figure}
The generalization of the QZE is given by the so-called \textit{Quantum Zeno Dynamics (QZD)}, which is achieved by applying sequences of projective measurements onto a multi-dimensional Hilbert subspace~\cite{PascazioPRL2002}. In this case, the system evolves away from its initial state, but remains confined in the subspace defined by the measurement operator~\cite{PascazioJPA}. At the very heart of QZD there is the quantum mechanical concept of the \textit{measurement back-action}, which is the ability to drive a given quantum state along specific paths by measuring the system: if the measurements are frequent enough, then the system is continuously projected back to its initial state, and the back-action confines its dynamics within the measurement subspace. The QZD has been confirmed first in an experiment with a Rubidium Bose-Einstein condensate in a five-level Hilbert space~\cite{SchaferZeno} and later in a multi-level Rydberg state structure~\cite{SignolesZeno}. In particular, \cite{SchaferZeno} realizes confinement of the atom dynamics in a subspace of a 5-level hyperfine manifold through four different coherent and dissipative protocols. Instead, \cite{SignolesZeno} examines a 51-dimensional angular momentum space, where the observation protocol allows to adjust the size of the accessible subspace and the confinement can be used to produce ``Schr\"{o}dinger cat'' states.

In this section, we investigate how the stochasticity in the time intervals between a series of projective measurements modifies the probability of a quantum system to be \textit{confined in an arbitrary Hilbert subspace}, by generalizing the LD formalism for SQZE~\cite{Gherardini2016NJP} to \textit{Stochastic Quantum Zeno Dynamics (SQZD)}. These results are discussed also in \cite{GherardiniAnnalen}. Moreover, since both theoretically~\cite{PascazioJPA} and experimentally~\cite{SchaferZeno} it has been demonstrated that QZD evolutions can be equivalently achieved not only by frequent projective measurements, but also by strong continuous coupling or fast coherent pulses, we will analyze also the accessibility to quantum Zeno dynamics if stochastic coherent or dissipative protocols are taken into account (see Fig. \ref{fig:overview_Annalen}). The aim of using protocols, which rely on quantum Zeno dynamics, is to constrain the quantum system dynamics to remain within a given Hilbert subspace, also called \textit{Zeno subspace}. The perfect (ideal) implementation of such a protocol forbids the system to go beyond the Zeno subspace, so that the system dynamics is described exclusively by the projected Hamiltonian $\Pi H\Pi$ (Zeno Hamiltonian). In this case, the dynamical evolution of the system is determined by the propagator
\begin{equation}\label{propagator_ZD}
\mathcal{U}^{(\Pi)}(t)\equiv\hat{T}\exp\left(-i\int_0^t \Pi H(\xi)\Pi d\xi\right)
\end{equation}
so that $\rho^{(\Pi)}(t)=\mathcal{U}^{(\Pi)}(t)\rho_0\big( \mathcal{U}^{(\Pi)}(t)\big)^\dagger$, where $\hat{T}$ denotes the time ordering operator, while $\rho^{(\Pi)}(t)$ is the density matrix describing the state of the system within the Zeno subspace.

\subsection{Zeno protocols}

\subsubsection*{Stochastic projective measurements protocol}

To realize quantum Zeno dynamics, the \textit{standard} observation protocols are given by applying a sequence of repeated projective measurements separated by \textit{constant} small intervals, in which the system freely evolves with unitary dynamics. In this way, the quantum state is projected onto the multidimensional subspace $\mathcal{H}_{\Pi}\equiv \Pi\mathcal{H}$ by the measurement operator $\Pi$, which usually does not commute with the system Hamiltonian $H$~\cite{PascazioPRL2002,PascazioJPA}. However, if we consider a stochastic distribution $p(\tau)$ of the time intervals between the measurements, the QZD can be described also in the case of temporal noise within the protocol. Moreover, the presence of some stochasticity introduces the possibility to engineer the dynamics by varying the underlying probability density function $p(\tau)$. Indeed, by controlling the functional behaviour of $p(\tau)$, we can influence the \textit{strength of confinement} of the system and, in principle, vary its time behaviour by means of a sophisticated interplay with the system internal dynamics. This could allow to explore the whole Hilbert space of a quantum system, by dynamically engineering the measurement operator and, thus, slowly moving the population from one portion of the Hilbert space to another. To this end, let us consider again a sequence of $m$ projective measurements separated by random time intervals $\tau_{j}$, $j=1,\dots,m$, which are assumed to be independent and identically distributed random variables. Accordingly, by generalising the results of \cite{Gherardini2016NJP}, the survival probability $\mathcal{P}_{m}(\{\mu_{j}\})\equiv\text{Prob}\left(\rho_{m}\in\mathcal{H}_{\Pi}\right)$ that the system belongs to the Zeno subspace $\mathcal{H}_{\Pi}$ after $m$ projective measurements (at the total time $\mathcal{T}$) is equal to
\begin{equation}
\mathcal{P}_{m}(\{\tau_{j}\}) = \prod_{j=1}^{m}q_{j}(\tau_{j}),
\end{equation}
where
\begin{equation}
q_{j}(\tau_{j}) \equiv \text{Tr}[\Pi~\mathcal{U}_{j}\Pi\rho_{j-1}\Pi~\mathcal{U}_{j}^{\dagger}\ \Pi]
\end{equation}
is the probability to find the system in the Zeno subspace at the $j-$th measurement. As shown also in \cite{SmerziPRL2012}, for small $\tau_j$ the single survival probability $q_j(\tau_j)$ can be expanded as
\begin{equation}
q_j(\tau_j) = 1 - \Delta^{2}_{\rho_{j-1}}H_{\Pi} \tau_{j}^{2},
\end{equation}
where $\Delta^2_{\rho_{j-1}}H_{\Pi}$ is the variance of the Hamiltonian
\begin{equation}
H_{\Pi} \equiv H - \Pi H\Pi
\end{equation}
with respect to the state $\rho_{j-1}$.

In the case the measurement subspace is unidimensional (as given in the previous sections) or more generally when $\mathcal{T}$ is small compared to the system dynamics within the Zeno subspace, the survival probability $q_j(\tau_{j})$ reduces to
\begin{equation}
q(\tau_{j}) = \text{Tr} [\Pi~\mathcal{U}_{j}\Pi\rho_{0}\Pi~\mathcal{U}_{j}^\dagger\Pi] =
1 - \Delta^2_{\rho_{0}}H_{\Pi}\tau^{2}_{j},
\end{equation}
where the variance is now calculated with respect to the initial state. With this simplification, the most probable value $\mathcal{P}^\star$ of the survival probability $\mathcal{P}_{m}(\{\tau_{j}\})$ is
\begin{equation}\label{eq:P_geom}
\mathcal{P}^\star = \prod_{\{\tau\}} q(\tau)^{mp(\tau)} = \exp\left(m\int_{\tau}d\tau p(\tau)\ln(q(\tau))\right),
\end{equation}
as given by (\ref{PG-defn}) and (\ref{def_P_g}). Now, the following theorem can be stated: \\ \\
\textbf{Theorem~3.1}: \textit{Given a stochastic sequence of $m$ projective measurements separated by random time intervals $\{\tau_{j}\}$, the most probable value $\mathcal{P}^\star$ of the survival probability $\mathcal{P}$ can be expressed as
\begin{equation}\label{eq:P-strictZeno3}
\mathcal{P}^\star \approx 1 - m\Delta^{2}_{\rho_{0}}H_{\Pi}(1+\kappa)\overline{\tau}^{2},
\end{equation}
under the \textbf{strong Zeno limit}
\begin{equation}\label{strong_ZL}
m\Delta^2 H_{\rho_0}(1+\kappa)\overline{\tau}^2\ll 1,
\end{equation}
with $\kappa \equiv \Delta^{2}\tau/\overline{\tau}^2$, where $\overline{\tau}$ and $\Delta^{2}\tau$ are, respectively, the expectation value and the variance of the probability density function $p(\tau)$. Conversely, if the \textbf{weak Zeno limit}
\begin{equation}\label{eq:skewness_condition}
\langle\tau^3\rangle \equiv \int_{\tau}d\tau p(\tau)\tau^{3}\ll\frac{1}{mC},
\end{equation}
is valid, where $C$ is a positive constant so that
\begin{equation}
\left|\frac{1}{6}\frac{\partial^{3}\ln(q(\tau))}{\partial\tau^{3}}\big|_{\tau = \xi\in[0,\tau]}\right|\leq C,
\end{equation}
then $\mathcal{P}^\star$ can be approximated as
\begin{equation}\label{eq:P_zeno2}
\mathcal{P}^\star\approx\exp\left(-m\Delta^2_{\rho_{0}}H_{\Pi} (1+\kappa)\overline{\tau}^{2}\right).
\end{equation}}
The proof of Theorem~3.1 can be found in Appendix~\ref{chapter:appB}. \\ \\
Theorem~3.1 defines two approximated expressions for the survival probability's most probable value $\mathcal{P}^\star$, which quantifies the confinement of the quantum system dynamics within the Zeno subspace. The first is obtained under the so-called \textit{strong Zeno limit}~(\ref{strong_ZL}), which requires a tight condition for the square of the expectation values of $p(\tau)$, and ensures an ideal Zeno confinement also when a stochastic sequence of measurements is applied to the quantum system. As a matter of fact, if we set $\kappa = 0$ (i.e. we consider a sequence of equally-distributed measurements), we directly recover the survival probability for standard quantum Zeno dynamics~\cite{SmerziPRL2012}. Conversely, the \textit{strong Zeno limit}~(\ref{eq:skewness_condition}) provides an expression for $\mathcal{P}^\star$ when the confinement is \textit{good but not perfect}, allowing to model system dynamics outside the Zeno subspace due to large deviations of $p(\tau)$ with respect to the average behaviour of the system. Indeed, (\ref{eq:P_zeno2}) does not depend on the variance of the probability distribution $p(\tau)$, but on its degree of skewness.

In the more general case that the measurement subspace has dimension greater than one and the dynamics within the subspace plays a role, the previous simplification $q_j(\tau_j) = q(\tau_j)$ is no longer valid, so that we cannot substitute $\rho_{j-1}$ with $\rho_0$ within the equation $q_j(\tau_j) = 1-  \Delta^2_{\rho_{j-1}}H_{\Pi}\tau_j^2$. However, a different approximation can be made:
\begin{itemize}
  \item First, approximate the state of the system with $\rho^{(\Pi)}(t)$, which denotes the dynamics for perfect Zeno confinement.
  \item Secondly, assume that the system Hamiltonian (in general time-dependent) is constant in the small time interval between two measurements.
\end{itemize}
As a consequence, the survival probability $q_{j}(\tau_{j})$ for small enough $\tau_j$ can be expanded as
\begin{equation}
q_{j}(\tau_{j}) \approx \widetilde{q}(\tau_{j},c_{j}) = 1 - c_{j}^2\tau_{j}^2,
\end{equation}
where
\begin{equation}
c_{j} \equiv \Delta_{\rho^{(\Pi)}_{j-1}}H_{\Pi}(t_{j-1}),
\end{equation}
and $\displaystyle{\Delta^{2}_{\rho^{(\Pi)}_{j-1}}H_{\Pi}(t_{j-1})}$ is the variance of $H_{\Pi}(t_{j-1})$ with respect to the density matrix $\rho_{j-1}^{(\Pi)}$. Moreover, for the coefficients $c_{j}$ we introduce the artificial probability density function $\widetilde{p}(c)$, that properly takes into account the average influence of the system dynamics on the \textit{leakage} (out of the Zeno subspace) by requiring that
\begin{equation}
\int_{c}\widetilde{p}(c)c^{2}dc = \frac{1}{\mathcal{T}}\int_{0}^{\mathcal{T}}\Delta^{2}_{\rho^{(\Pi)}(t)}H_\Pi(t)dt.
\end{equation}
In this way, the most probable value of the survival probability for quantum Zeno dynamics can be written as
\begin{equation}\label{eq:SQZD}
\mathcal{P}^{\star} = \prod_{\{c\}}\prod_{\{\tau\}}\left(\prod_{j=1}^{m}\widetilde{q}(\tau_{j},c_{j})\right)^{p(\tau)\widetilde{p}(c)} = \exp\left(m\int_{\tau,c}d\tau dc p(\tau)\widetilde{p}(c)\ln(\widetilde{q}(\tau,c))\right).
\end{equation}
Finally, under the hypothesis that the quantum system is in the weak Zeno limit and that the Hamiltonian varies only slowly in time (compared to the time scale of the measurement intervals), we can state that $\Delta^{2}_{\rho^{(\Pi)}(t)}H_\Pi(t)$ changes slowly with respect to the measurement frequency. Hence, by making the approximation
\begin{equation}
\ln(\tilde q(\tau,c)) \approx 1 - \tilde q(\tau,c),
\end{equation}
the integral in (\ref{eq:SQZD}) can be easily solved, so as to obtain
\begin{equation}\label{eq:SQZD2}
\mathcal{P}^\star \approx \exp\left( -\frac{m\overline{\tau}^{2}(1+\kappa)}{\mathcal{T}}\int_{0}^{\mathcal{T}}\Delta^{2}_{\rho^{(\Pi)}(t)}H_\Pi(t)dt \right).
\end{equation}
As main result, (\ref{eq:SQZD2}) is the generalization of (\ref{eq:P_zeno2}) for \textit{stochastic quantum Zeno dynamics} and time-dependent Hamiltonian.

\subsubsection*{Coherent protocols}

In the previous section, we have considered how to realize stochastic Zeno dynamics by means of instantaneous projective measurements. However, projective measurements are difficult to be experimentally implemented, since the duration of a single measurement might be comparable to or even larger than the time scale of the system dynamics.

Accordingly, quantum Zeno dynamics can be alternatively achieved via coherent couplings~\cite{PascazioPRL2002,FacchiPRA2004,FacchiPRA2005,PascazioJPA}: \textit{continuous coupling} (c.c.) or \textit{pulsed coupling} (p.c.). To this end, we add to the system Hamiltonian $H$ the additional coupling Hamiltonian $gH_{c}$, that acts on the complement $\mathcal{H}_{\mathbbm{1}-\Pi}$ of the Zeno subspace. For the continuous coupling protocol, the coupling strength $g$ is constant over time and in the limit of strong coupling strength $g$ different regions of the system Hilbert space can be dynamically disjointed. Conversely, for the pulsed coupling protocol, the coupling is switched \textit{on} and \textit{off} repeatedly to perform fast unitary kicks (with high coupling strength $g$), which are followed by time intervals of switched-off coupling.

\begin{figure}[h!]
\centering
\includegraphics[width=1\linewidth]{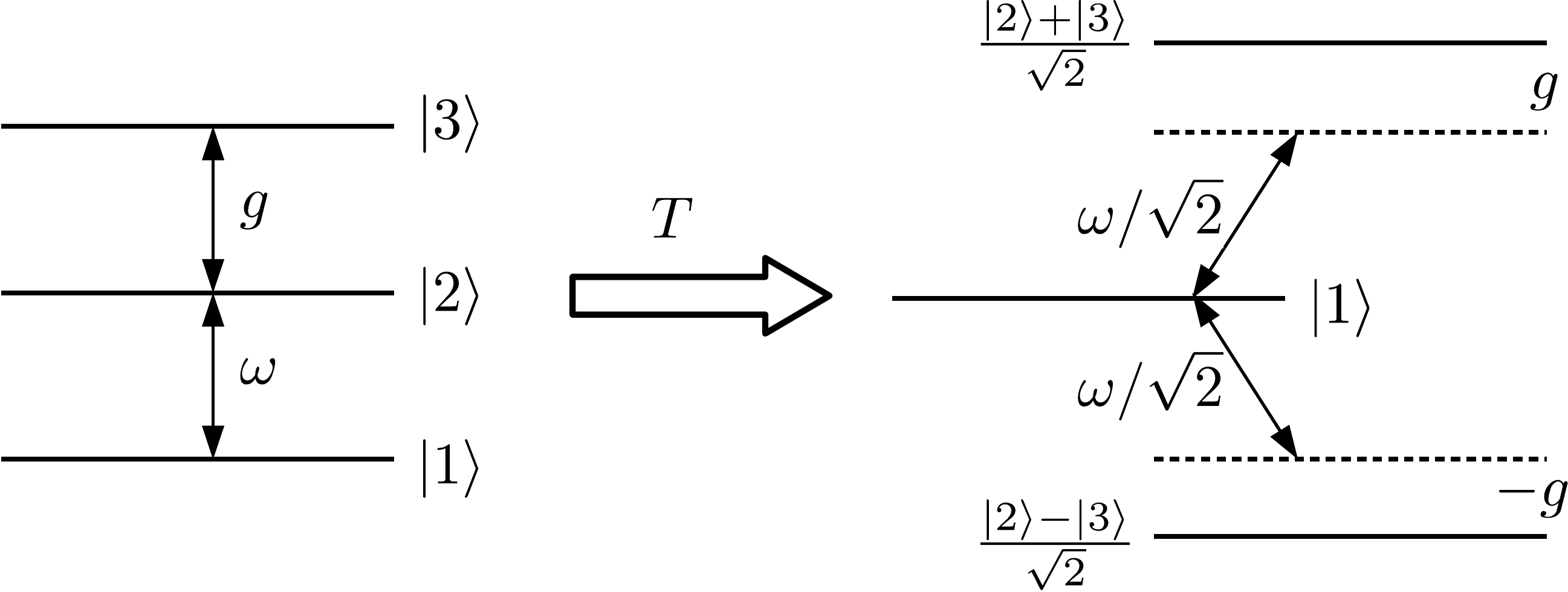}
\caption{Coherent coupling protocol in the paradigmatic three level system. Initially the system is prepared in state $|1\rangle$ (left). This state is coupled to the state $|2\rangle$ by a Rabi frequency of $\omega$. This level $|2\rangle$, in turn, is coupled to a third level $|3\rangle$ with constant coupling strength $g$. Under the basis change $G$ we get the picture on the right hand side. Level $|1\rangle$ is now coupled to the two new basis states $(|2\rangle+|3\rangle)/\sqrt{2}$ and $(|2\rangle-|3\rangle)/\sqrt{2}$, where both couplings are detuned. The detuning has an absolute value of $|g|$ and, thus, by increasing the coupling between the levels $|2\rangle$ and $|3\rangle$, we effectively lock the population in level $|1\rangle$.}
\label{fig:3level}
\end{figure}

\textit{The time intervals between two unitary kicks can allow for the same stochasticity as the time-disordered measurements}. These unitary kicks or instantaneous rotations, indeed, are given by the propagator
\begin{equation}
\mathcal{U}^{(p.c.)} = \exp\left(-i H_{c}s\right),
\end{equation}
where the time $s$ denotes the rotation angle. This rotation angle is given by the pulse area (i.e. the coupling strength integrated over the duration of the pulse) of a coupling pulse in a finite time realization. As in the case of quantum bang-bang controls for dynamical decoupling tasks~\cite{Viola1999}, we assume that the pulse area is \textit{finite} and that arbitrarily strong coupling kicks lead to practically \textit{instantaneous} rotations. Similarly to the time-disordered sequence of projective measurements, also the Zeno protocol based on pulsed coupling is intrinsically stochastic if the pulses are separated by random time intervals $\tau_{j}$ sampled from $p(\tau)$. Accordingly, in order to compare the results from the two coherent coupling schemes, we require that \textit{on average the pulse area of the two coherent coupling protocols is the same}. Then, the survival probability is evaluated by computing
\begin{equation}
\mathcal{P} = \text{Tr}(\Pi\rho^{(c.c.)})~~~\text{or}~~~\mathcal{P} = \text{Tr}(\Pi\rho^{(p.c.)}),
\end{equation}
where $\rho^{(c.c.)}$ and $\rho^{(p.c.)}$ are the normalised density matrices of the system at the end, respectively, of the continuous and pulsed coupling Zeno protocols. It is worth noting that a closed expression for the survival probability as a function of the coupling strength $g$ cannot be trivially calculated; however, we can derive the scaling of $\mathcal{P}$ with respect to $g$. For this purpose, let us consider, without loss of generality, the continuous coupling method, and, then, the total system Hamiltonian $H_{{\rm tot}}$ (with the additional term coupling term $gH_{c}$), which can be decomposed as
\begin{equation}
H_{{\rm tot}} = \Pi H\Pi\otimes\mathbbm{1} + \mathbbm{1}\otimes \left[g H_{c} + (\mathbbm{1}-\Pi)H(\mathbbm{1}-\Pi)\right] + H_{{\rm int}}.
\end{equation}
In this regard, we have assumed that $H_{c}$ acts only outside the Zeno subspace, and $H_{{\rm int}}$ is the interaction Hamiltonian term between the subspace and its complement. By transforming the total Hamiltonian in a basis where $H_{c}$ is diagonal, the coupling between the Zeno subspace and its complement is effectively a \textit{driving}, that is off-resonant by a term proportional to $g$. As a consequence, the confinement error $1-\mathcal{P}$ within the Zeno subspace scales as $||H_{int}||^2/g^2$. This becomes clearer if we consider the paradigmatic three level system (see also Fig.~\ref{fig:3level}), given by the Hamiltonian
$$
H_{{\rm tot}} = \omega(|1\rangle\langle 2| + |2\rangle\langle 1|) + g(|2\rangle\langle 3| + |3\rangle\langle 2|).
$$
\\
The coupling rate (with strength $g$) to the upper level $|3\rangle$ plays the role of the measurement, and the Zeno subspace is assumed to be the state $|1\rangle$. The coupling Hamiltonian $H_{c}$, thus, is given by the term $g(|2\rangle\langle 3| + |3\rangle\langle 2|)$, as it is shown on the left hand side of Fig.~\ref{fig:3level}. Then, let us introduce a linear transformation $G$, which diagonalizes $H_{c}$ and makes the coupling diagonal. In the canonical matrix representation, $G$ can be chosen equal to
\begin{equation}
G = \begin{pmatrix}
    1 & 0 & 0 \\
    0 & \frac{1}{\sqrt{2}} & \frac{1}{\sqrt{2}} \\
    0 & \frac{1}{\sqrt{2}} & -\frac{1}{\sqrt{2}}
    \end{pmatrix},
\end{equation}
so that the transformed Hamiltonian is
\begin{equation}
G^{\dagger}H G= \begin{pmatrix}
                 0 & \frac{\omega}{\sqrt{2}} & \frac{\omega}{\sqrt{2}} \\
                 \frac{\omega}{\sqrt{2}} & g & 0 \\
                 \frac{\omega}{\sqrt{2}} & 0 & -g
                \end{pmatrix}.
\end{equation}
The system and the Hamiltonian after the transformation $G$ are sketched on the right hand side of Fig.~\ref{fig:3level}. We can observe that, if the initial state of the system is taken in the Zeno subspace $\mathcal{H}_{\Pi}$, then the coupling makes extremely difficult the transfer of the system dynamics outside $\mathcal{H}_{\Pi}$, since the transition to the rest of the Hilbert space (here, driven by the Rabi frequency $\omega$) is moved out of resonance by a factor $g$. As a consequence, the effective driving is reduced to $\omega^2/g^2$. When $g \rightarrow \infty$, we obtain an ideal confinement of the quantum system in the measurement subspace. This can be easily seen by solving the model, and computing the corresponding survival probability
\begin{equation}
\mathcal{P}(t)=\left[1-\frac{2\omega^2}{\omega^2 + g^2}\sin^2\left(\frac{\sqrt{\omega^2 + g^2}\, t}{2}\right)\right]^2
\end{equation}
in the Zeno subspace. In conclusion, the confinement error scales with one over the square of the coupling strength, as it can later observed in the inset of Fig.~\ref{fig:lambda-methods}.

\subsection{Illustrative example - Quantum spin chains}

The dynamics within the Zeno subspace can be characterized also by collective behaviours originating from inter-particle interactions. In this regard, let us consider a chain of $N$ qubits, whose dynamics is described by the following Hamiltonian:
\begin{equation}
H_N = \gamma_{1}\sum_{i=1}^{N}\sigma_z^{i} + \frac{\gamma_{2}}{2}\sum_{i=1}^{N-1} \left(\sigma_x^i\sigma_x^{i+1} + \sigma_y^i\sigma_y^{i+1} \right)\,,
\end{equation}
where $\sigma_z^i$ is the Pauli z-matrix acting on the $i$-th site, and $\sigma_{x/y}^i\sigma_{x/y}^{i+1}$ are the interaction terms, which couple spins $i$ and $i+1$ through the tensor product of the respective Pauli matrices~\cite{Baxter1}. Moreover, $\gamma_{1}$ is an external magnetic field, while $\gamma_{2}$ denotes the coupling strength of the interaction. Here, we are interested in a dynamical regime, whereby the measurement projectors restrict the dynamics to excitations of the first $\nu$ spins, which thus define a $2^{\nu}$-dimensional measurement subspace. If we measure the excitations outside this subspace, both the Hamiltonian evolution and the negative measurement outcomes (which give the absence of population in the rest of the chain) preserve the number of excitations. In particular in the following, by neglecting states with more than one excited spin, we will limit the dynamics of the spin chain to the single excitation sector, and only to pure states of the form
\begin{equation}\label{eq:psi_example}
|\psi(t)\rangle = \sum_{i=1}^{N}\phi_{i}(t)|1_i\rangle.
\end{equation}
In (\ref{eq:psi_example}) $|1_i\rangle=|0..010..0\rangle$ denotes the state with one excitation at site $i$, while the coefficients of the initial state of the chain will be chosen so that $\phi_{k}(0)=0$ for $k>\nu$. Under these assumptions, the probability to find the system in the measurement subspace after the $j$-th measurement is equal to
\begin{equation}
 q_j(\tau_j)= 1 - \gamma_{2}^{2}\tau_j^2 |\phi_{\nu}(t_{j-1})|^2\,,
\end{equation}
where
$$
\gamma_{2}^2|\phi_{\nu}(t_{j-1})|^2 = \Delta _{|\psi_{j-1}\rangle}^2 H_{\Pi},
$$
and, as before, the variance $\Delta _{|\psi_{j-1}\rangle}^2 H_{\Pi}$ is computed with respect to the state $|\psi_{j-1}\rangle$. In other words, the probability $q_j(\tau_{j})$ can be directly computed just by observing the modulus of the state $|1_{\nu+1}\rangle$ at time $t_j$, corresponding to the leakage outside the measurement subspace.

In the case the initial condition of the dynamics is given by an eigenstate of the Zeno-Hamiltonian $\Pi H_N\Pi \equiv H_{\nu}$ (which is the spin chain Hamiltonian with $\nu$ spins) and we re-normalise the system state after every measurement, then the coefficient $|\phi_{\nu}(t)|$ is approximately constant and equal to $\phi_{\nu}$, such that
$$
q_j(\tau_j) = 1 - \gamma_{2}^2 \tau_j^2 |\phi_{\nu}(t_{j-1})|^2 = 1 - \gamma_{2}^2\tau_j^2 \phi_{\nu}^2 = q(\tau_j),
$$
In this way, the quantum mechanical probability of finding the system in the subspace upon measurements depends just on the length of the interval $\tau_{j}$, and from (\ref{eq:P_zeno2}) we have
\begin{equation}\label{eq:P-const-exc}
\mathcal{P}^\star = \exp\left(-\gamma_{2}^{2}\phi_{\nu}^{2}m \overline{\tau}^2(1+\kappa)\right).
\end{equation}

However, in a more general case the time dependence of $|\phi_{\nu}(t)|^2$ has to be taken into account, and $\mathcal{P}^{\star}$ can be computed either numerically (by simulating the sequence of repeated measurement on the $N$ spin chain) or analytically (by using the approximation given by (\ref{eq:SQZD2}) for stochastic quantum Zeno dynamics). In the latter case, we have
\begin{equation}\label{eq:P-average-exc}
\mathcal{P}^\star \approx \exp\left( -\frac{m \gamma_{2}^{2}\overline{\tau}^{2}(1+\kappa)}{\mathcal{T}}\int_0^{\mathcal{T}}|\phi_{\nu}(t)|^2 dt \right).
\end{equation}

\begin{figure}[h!]
\centering
\includegraphics[width=0.9\linewidth]{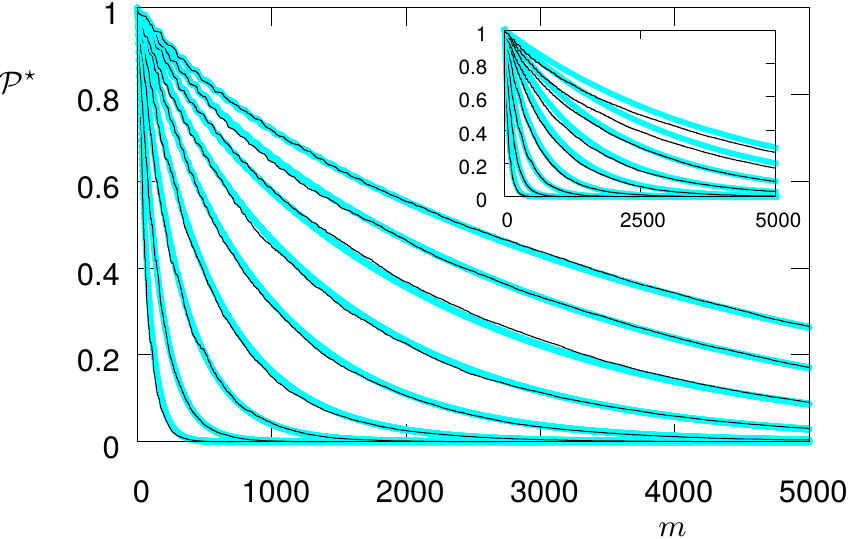}
\caption{The quantum spin chain is initially prepared in the W-state (\ref{w_state}). In the figure, we show one realisation of $\mathcal{P}$ (for each $\nu = 1,\dots,9$, from bottom to top) as a function of the number of measurements $m$ (black lines) compared to $\mathcal{P}^\star$, calculated by using (\ref{eq:P-average-exc}) (cyan lines). Inset -- The same realizations (black lines) compared to $\mathcal{P}^\star$, calculated by (\ref{eq:P-const-exc}) (cyan lines). The probability density function is bimodal with $p_1 = p_2 = 0.5$, $\tau^{(1)}=1\,\mathrm{\mu s}$, and $\tau^{(2)}=5\,\mathrm{\mu s}$.}
\label{fig:W-P-exc}
\end{figure}

In the following, numerical results for a chain of $N = 12$ spins are presented. All the results are evaluated for a bimodal distribution of the measurement intervals. First, we examine the behaviour of the survival probability when the system is subjected to a stochastic protocol of projective measurements and two different initial states are considered. In particular, the initial state of the system is prepared, respectively, as an \textit{entangled W-state} (i.e. a delocalized excitation) and then as a state where the excitation is localized in the \textit{left-most spin} of the chain. For each set of parameters we consider a single realization of random time intervals $\tau_j$ and we calculate the survival probability as $\mathcal{P}=\prod_{j}q_j(\tau_j)$, where $q_j(\tau_j)$ is the probability (numerically calculated) to find the population in the subspace after the $j-$th measurement. For the coherent Zeno protocols, instead, the survival probability is given by $\mathcal{P}=\mathcal{P}(t_j)$, that is the population of the system within the measurement subspace at time $t_j=\sum_j \tau_j$. \\ \\
\textbf{W-state}: Let us prepare the quantum system in the entangled state
\begin{equation}\label{w_state}
|\psi_{\nu}(t)\rangle = \frac{1}{\sqrt{\nu}}\sum_{i=1}^{\nu}|1_i\rangle.
\end{equation}
In Fig.~\ref{fig:W-P-exc} we show the survival probability (i.e. $\mathcal{P}=\prod_{j}q_j(\tau_j)$, black lines) obtained by numerical simulations of a sequence of random measurements for $\nu = 1,\dots, 9$ (bottom to top), compared to (\ref{eq:P-average-exc}) (cyan lines).
An excellent agreement is observed: the numerical values and the theoretical approximation practically coincide, confirming thus the validity of the approximation. Although the initial state (\ref{w_state}) is not an eigenstate of $H_{\nu}$, the dynamics of the system approximately converges to such a state, as observed in the numerical simulations. Hence, we can compare the survival probability $\mathcal{P}$ (black lines), obtained by the numerical simulation, to $\mathcal{P}^\star$ computed from (\ref{eq:P-const-exc}) (cyan lines), where $|\phi_{\nu}(t)|^2$ is assumed to be constant. In this regard, the inset of Fig.~\ref{fig:W-P-exc} shows the comparison between this analytical approximation and the numerical values. The agreement is better for small $\nu$, where the discrepancy between the initial state and the eigenstates of the subspace Hamiltonian $H_{\nu}$ is small (in particular, for $\nu = 1,2$ the initial state is an eigenstate of $H_{\nu}$). \\ \\
\textbf{Left-most qubit excited}: By starting from $|1_1\rangle$, the excitation travels towards the edge of the subspace, where it is reflected. Hence, apart from the spreading, the excitation oscillates between the edge of the chain and the edge of the subspace, with a velocity $\varsigma$ given by the Lieb-Robinson bound~\cite{LR1972}. The velocity $\varsigma$ can be determined by evaluating (for $\nu = 2,\dots,10$) the time when the excitation first peaks at the \textit{edge qubit} $\nu$, which is the one qubit belonging to the subspace that directly interacts with the rest of the chain. We numerically obtain $\varsigma\approx0.06\,$sites$/$ms, in good agreement with the theoretical bound given by the norm of the interaction operator~\cite{Kliesch}, i.e.
$$
\varsigma \leq e\left\Vert\frac{\gamma_{2}}{2}(\sigma_{x}^{\nu}\sigma_{x}^{\nu + 1} + \sigma_{y}^{\nu}\sigma_{y}^{\nu + 1})\right\Vert
\approx 0.085\,\text{sites}/\text{ms}.
$$
\begin{figure}[h!]
\centering
\includegraphics[width=0.9\linewidth]{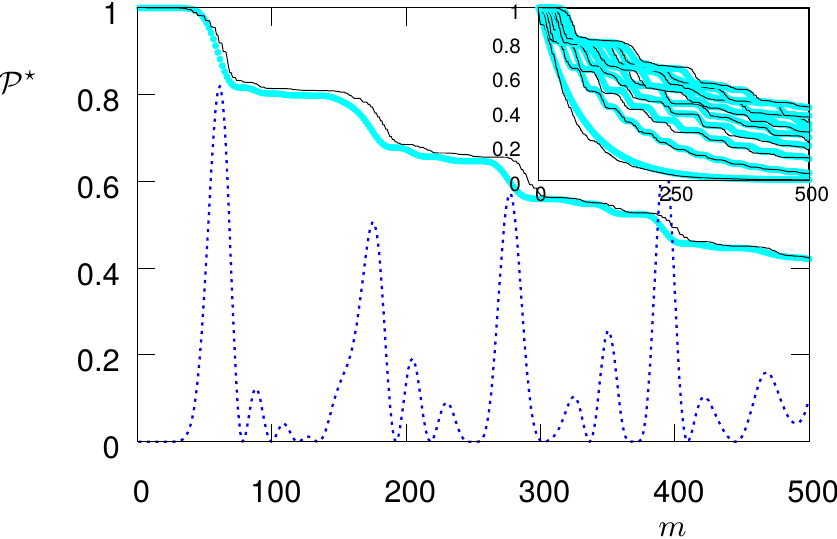}
\caption{The quantum spin chain is prepared in the state $|1_1\rangle$. We plot the numerical value of $\mathcal{P}^\star$ (black line) as a function of the number of measurements $m$ compared to (\ref{eq:P-average-exc}) (cyan lines) for $\nu=9$. The blue dashed line is $|\phi_{9}(t)|^2$ as obtained by a simulation with $H_{9}$, while in the inset of the figure we show $\mathcal{P}^{\star}$ for $\nu = 1,\dots,9$ (from bottom to top). The probability density function is bimodal with $p_1=p_2=0.5$, $\tau^{(1)}=1\,\mathrm{\mu s}$, and $\tau^{(2)}=5\,\mathrm{\mu s}$.}
\label{fig:fe-P-average-exc}
\end{figure}\\
Fig.~\ref{fig:fe-P-average-exc} shows the survival probability $\mathcal{P}=\prod_{j}q_j(\tau_j)$ (black lines), obtained by numerical simulations and compared to (\ref{eq:P-average-exc}) (cyan lines) for $\nu = 9$ (in the inset the most probable value $\mathcal{P}^{\star}$ is shown for $\nu = 1,\dots,9$, bottom to top). The plateaus correspond to zero or very little excitation of the edge qubit (with $|\phi_{\nu}|$ very small), while the steps correspond to a considerable excitation located at the edge qubit. This excitation (i.e. $|\phi_{\nu}|^2$ for $\nu = 9$) of the edge qubit, plotted as a blue dashed line, oscillates between the edge of the chain and the edge of the subspace and the peaks indicate the time instances where the excitation is practically located at the edge qubit. The remnant plateaus for $\nu = 1$ occur only in the numerical simulation and are absent in the model, since they do not come from an oscillation of the excitation in the 1-qubit subspace, but from repetitive measurements after the smaller time interval $\tau^{(1)}$. Thus, it is an effect that is averaged out in the model. For $\nu>1$, instead, the plateaus are originated also by the dynamics within the subspace and, thus, are present both in the single realizations (numerics) and in the averaged model (theory). \\ \\
\textbf{Coherent Couplings}: Without applying a sequence if quantum measurements on the system, we can include the coupling of the system with a coherent driving by means of the following additional (coupling) Hamiltonian:
\begin{equation}
H_c(\nu) = \left(\sigma_x^{\nu + 1}\sigma_x^{\nu + 2} + \sigma_y^{\nu + 1}\sigma_y^{\nu + 2} \right).
\end{equation}
The coupling is chosen so that $\displaystyle{g = \frac{\pi}{2\overline{\tau}}}$ in the case of continuous coupling, and $\displaystyle{s = \frac{\pi}{2}}$ in the case of pulsed coupling. Thus, on average in both cases the pulse area of the coupling is the same, and for the pulsed coupling the projective measurement is substituted by an excitation flip between the qubits $\nu + 1$ and $\nu + 2$. The performance of the Zeno protocols are evaluated by introducing the \textit{Uhlmann fidelity}~\cite{Uhlmann, Jozsa}, defined as
\begin{equation}\label{eq:fidelity}
\mathcal{F}^{({\rm protocol})} = \text{Tr}\sqrt{\sqrt{\rho_m^{(\Pi)}} \rho_m^{({\rm protocol})} \sqrt{\rho_m^{(\Pi)}}}.
\end{equation}
$\mathcal{F}^{({\rm protocol})}$ compares the evolved density matrices to the density matrix $\rho_{m}^{(\Pi)}\equiv\rho^{(\Pi)}(t = \mathcal{T})$, which is obtained by exact subspace evolutions. The superscript ({\rm protocol}), instead, refers to the examined Zeno protocols given by projective measurements $(p.m.)$, continuous coupling $(c.c.)$ or pulsed coupling $(p.c.)$.

\begin{figure}[h!]
\centering
 \includegraphics[width=0.9\linewidth]{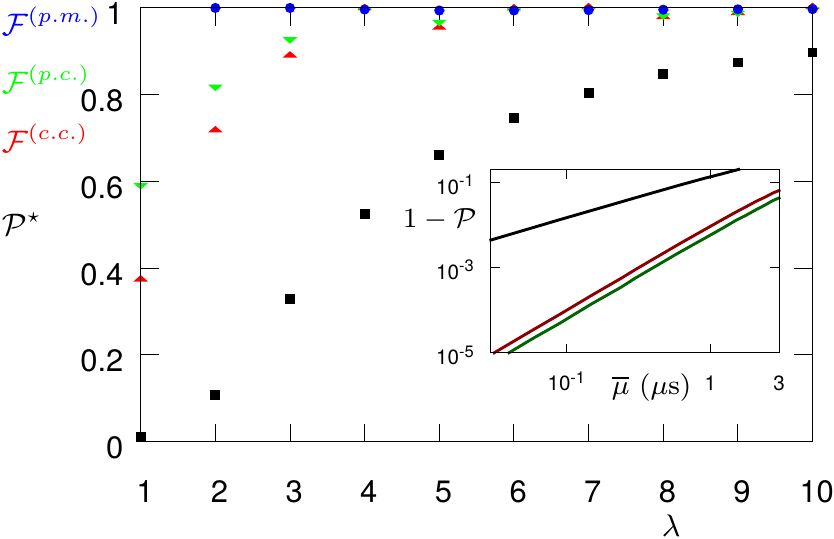}
 \caption{Performance of the Zeno protocols as a function of the subspace size $\nu$. The red upper triangles, green lower triangles and blue circles show the fidelities, respectively, for continuous coupling, pulsed coupling and projective measurements. Instead, the black squares show the survival probability. The simulations where carried out for the initial W-state and a bimodal probability density function with $p_1 = p_2 = 0.5$, $\tau^{(1)}=3\,\mathrm{\mu s}$ and $\tau^{(2)}=5\,\mathrm{\mu s}$. The inset shows how the system behaves for $\nu = 5$ when $m\overline{\tau}$ is constant, and the interaction (given by the number of measurements $m$ or the coherent coupling strength $g$) is varied: As we approach the Zeno limit the confinement error $1-\mathcal{P}$ vanishes for all the three Zeno protocols (from top to bottom: p.m. (black), c.c. (dark red), p.c. (dark green)), and the scaling with respect to $\overline{\tau}$ is \textit{linear} for the protocol based on projective measurements and \textit{quadratic} for the coherent coupling methods.}
 \label{fig:lambda-methods}
\end{figure}

Fig.~\ref{fig:lambda-methods} shows the fidelity $\mathcal{F}$ of the respective dynamics as a function of the number of qubits $\nu$ within the subspace.
While projective measurements $(p.m.)$ yield the highest fidelity, \textit{all three Zeno protocols show a similar scaling behaviour with respect to} $m$ \textit{and} $\nu$. It should be noted though that, due to the \textit{probabilistic} nature of the projective measurements given by the survival probability $\mathcal{P}^{\star}$, the coherent methods show the better \textit{deterministic} performance with a slight advantage for pulsed coupling $(p.c.)$ over coherent coupling $(c.c.)$. For increasing $\nu$, we approach higher values of fidelity and survival probability, since the edge qubit is on average less populated, so that we have less leakage. The inset of Fig.~\ref{fig:lambda-methods} shows the leakage $1-\mathcal{P}$ for the three protocols, projective measurements (black), pulsed coupling (dark green) and continuous coupling (dark red), when approaching the Zeno limit, by setting $m\overline{\tau}$ to be a constant value and decreasing $\overline{\tau}$ while at the same time $m$ is increasing. The results are shown for $\nu = 5$, $p_1=1$, and $\tau^{(1)}=3\,\mathrm{\mu s}$. While the projective measurements approach shows a \textit{linear} scaling with $\overline{\tau}\propto 1/m$, the coherent coupling protocols exhibit a \textit{quadratic} scaling (in this regard, see the inset of Fig.~\ref{fig:lambda-methods}). The linear scaling in the first case is a direct consequence of (\ref{eq:P-strictZeno3}), while the quadratic scaling in the latter case corresponds to the prediction of the off-resonant driving model.
\begin{figure}[h!]
\centering
 \includegraphics[width=0.9\linewidth]{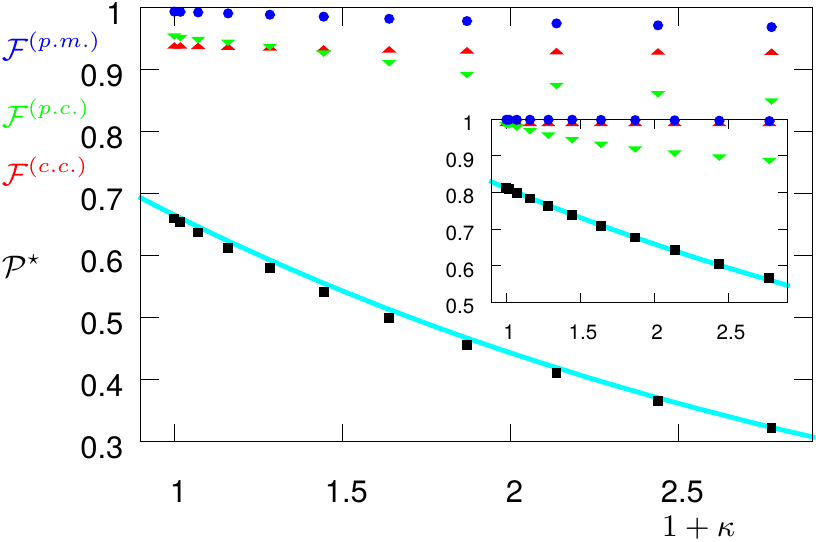}
 \caption{Performance of the three protocols as a function of the time disorder $1 + \kappa$. The red upper triangles, green lower triangles and blue circles show the fidelities, respectively, for continuous coupling, pulsed coupling and projective measurements. The black squares show the survival probability. The system was initially prepared in the W-state. The cyan curves are the theoretical values obtained by (\ref{eq:P-average-exc}), where $|\phi_{\nu}(t)|$ has been taken from the time evolution with $H_{\nu}$. The probability density function is bimodal with $p_1=0.8$, $p_2=0.2$, $\overline{\tau}=3\,\mathrm{\mu s}$, $\tau^{(1)}\in[1,3]\,\mathrm{\mu s}$ and $\tau^{(2)}\in[3,11]\,\mathrm{\mu s}$, corresponding to $\kappa\in[0,1.778]$. Inset: Performance of the three protocols when only the left-most spin was initially excited.}
 \label{fig:kappa-methods}
\end{figure}\\
Finally, in Fig.~\ref{fig:kappa-methods} the performance of the Zeno protocols as a function of the time disorder $1 + \kappa$ are shown. As it can be observed, we find a decrease in the fidelity $\mathcal{F}$ both for the protocol based on projective measurement (p.m.), and for the coherent pulsed coupling (p.c.), while, trivially, no change occurs for continuous coupling $(c.c.)$. At the same time the survival probability for the projective measurement protocol decreases to about half its ordered value ($\kappa=0$) over the plotted range of disorder. Fig.~\ref{fig:kappa-methods} shows the behaviour of these quantities for the case when the system is initially prepared in the W-state, while in the inset it is shown that the behaviour is very similar when the system is initially prepared with an excitation in the left-most spin.

\section{Stochastic sequences of correlated quantum measurements}

Usually the environment is unknown and very hard to be characterized. In particular, it can be distinguished according to whether the system to which it is coupled can generate \textit{Markovian} or \textit{non-Markovian} dynamics~\cite{Rivas}. In this regard, time correlations in the noisy environment can potentially generate non-Markovian dynamics within the quantum system, depending on the structure and energy scale of the system Hamiltonian. In particular, also classical environments exhibiting non-Gaussian fluctuations (i.e. characterized by non-Gaussian probability density functions) can lead to non-Markovian quantum dynamics, as shown in Ref.~\cite{Benedetti2014,Benedetti2016}.

In this section, we will consider a quantum system subject to a sequence of projective measurements, where each measurement (defined by the projector $\Pi$) occurs after a fixed time interval $\tau$ and the system driving is given by a random classical field. More specifically, we will study a quantum system that is coupled to a bath that effectively acts on the system via a time fluctuating classical field $\Omega(t)$, according to the following Hamiltonian:
\begin{equation}
H_{{\rm tot}}(t) = H_{0} + \Omega(t)H_{{\rm noise}} = H_0 + [\langle\Omega\rangle + \omega(t)]H_{{\rm noise}},
\end{equation}
where $H_{0}$ is the Hamiltonian of the unperturbed system, while $H_{noise}$ describes the coupling of the environment with the system. Moreover, we assume that $\Omega(t)$ takes real values with mean $\langle\Omega\rangle$, whereby $\omega(t)$ is the fluctuating part of the field with vanishing mean value. The system dynamics for a given realization of the random field $\Omega(t)$, then, is described by the \textit{stochastic Schr{\"o}dinger equation}
\begin{equation}
\dot{\rho}(t)=-i[H_0 + \langle\Omega\rangle H_{{\rm noise}},\rho(t)] - i\omega(t)[H_{{\rm noise}},\rho(t)],
\end{equation}
that, if averaged over the statistics of the field $\Omega(t)$ as shown in Appendix~\ref{chapter:appB}, takes the form of the following \textit{master equation}:
\begin{equation}
\langle\dot{\rho}(t)\rangle = -i[H_0 + \langle\Omega\rangle H_{{\rm noise}},\rho(t)] - \int_{0}^{t} \langle\omega(t)\omega(t')\rangle [H_{{\rm noise}},[H_{{\rm noise}},\rho(t')]] dt',
\end{equation}
where $\langle\omega(t)\omega(t')\rangle$ denotes the \textit{second-order time correlation function} or \textit{memory kernel} of the random field $\omega(t)$, and $[\cdot,\cdot]$ is the commutator. If the classical field is a white noise, the second-order time correlation function turns out to be a Dirac-delta distribution, i.e. $\langle\omega(t)\omega(t')\rangle\propto\delta(t-t')$, and the \textit{standard Lindblad-Kossakowski master equation}~\cite{PetruccioneBook} is obtained. Otherwise, a different memory kernel can lead to non-Markovian dynamics depending on the structure and time scale of the Hamiltonian, as for example demonstrated for random telegraph noise (RTN) and $1/f$-noise~\cite{Benedetti2014,Benedetti2016}. We denote the single measurement quantum survival probability (i.e. the probability for the system to remain confined within the measurement subspace) as $q(\Omega)$, that depends on the value of $\Omega$ during the time interval $\tau$ and thus is a random variable. Accordingly, the survival probability for the whole time duration is given by
\begin{equation}\label{surv_prob_SciRep}
\mathcal{P}_{k}(m) = \prod_{j=1}^{m}q(\Omega_{j,k}) \; ,
\end{equation}
where $k = 1,\dots N$ labels the realization of a trajectory, $j$ represents the time order of the $m$ measurements, and $\Omega_{j,k}(t)$ is the corresponding fluctuating field. Moreover, in (\ref{surv_prob_SciRep}) the single measurement quantum survival probability $q(\Omega_{j,k})$ is defined as
\begin{equation}
q(\Omega_{j,k}) = \textrm{Tr}\left[\rho_{k}(j\tau)\Pi\right],
\end{equation}
where $\rho_{k}(j\tau)$ is the $k-$th realization of the system density matrix at time $t_{j} = j\tau$.

Also in this case, where the stochasticity is given by the random classical field $\Omega(t)$, the survival probability becomes a random variable described by the stochastic quantum Zeno dynamics formalism. In this regard, in the following we propose a way to \textit{probe} the presence of noise correlations in the environment, by analyzing the time and ensemble average of the system survival probability. In particular, we will demonstrate how such environmental time correlations can determine whether the two averages do coincide or not~\cite{GherardiniSciRep}. It is worth noting that also this method relies on the very recent idea of the so-called \textit{quantum probes}, whereby their fragile properties, as coherence and entanglement, are strongly affected by the environment features and can be used for detection purposes. Examples of such physical systems, which are used to probe environments like biological molecules or surfaces of solid bodies or amorphous materials, are quantum dots, atom chips and nitrogen vacancy centers in diamond~\cite{Taylor2008,Balasubramanian2008,Maze2008,McGuinness2015,Hollenberg2009,Hofferberth2008,Gierling2011,Rossi2015}. Finally, especially in this framework, the introduction of noise quantum filtering techniques can be required to achieve the following two main goals:
\begin{itemize}
  \item To improve the effectiveness of the predictions given by applying the LD theory to open quantum systems.
  \item To design robust quantum devices for information processing and take advantage at most of the presence of an external environment.
\end{itemize}
As a matter of fact, robust control of a quantum system is crucial to perform quantum information processing, which has to be protected from decoherence or noise contributions originating from the environment. The decay of the coherence of an open quantum system depends in a peculiar way both on the spectrum of the bath and the driving terms of the system. In this regard, as shown in \cite{Degen2017,kofman2001,Paz-Silva2014,Norris2016}, the application of different control functions lies at the core of the so-called \textit{filter function approach} to spectrally resolve quantum sensing, that however can undergo the problem of \textit{spectral leakages}. Most protocols, indeed, investigate the noise fluctuations only in a finite frequency band, while the interaction of the probe with the environment has contributions also \textit{outside} this band, leading thus to a decreasing of the measurement precision. In solving this issue, we proposed in \cite{MuellerSensing} a \textit{fast} and \textit{robust} estimation strategy (based on filter function orthogonalization, optimal control filters and multi-qubit entanglement) for the characterization of the spectral properties of classical and quantum dephasing environments within the whole frequency band. The robustness of such sensing procedure is quantified in terms of a directional Fisher information operator~\cite{GherardiniFisher}, and then optimal control theory is employed to construct filter functions that maximize the sensitivity of the filter with respect to the noise spectrum. The two methods (i.e. the optimal multi-probe method and the Zeno-based one) not only are complementary, but, being designed on two different quantum system behaviours, could be in principle used to validate the results coming from both of them.

\subsection{Time and ensemble averages vs. noise correlations}

\begin{figure}[h!]
\centering
 \includegraphics[width=1.05\linewidth]{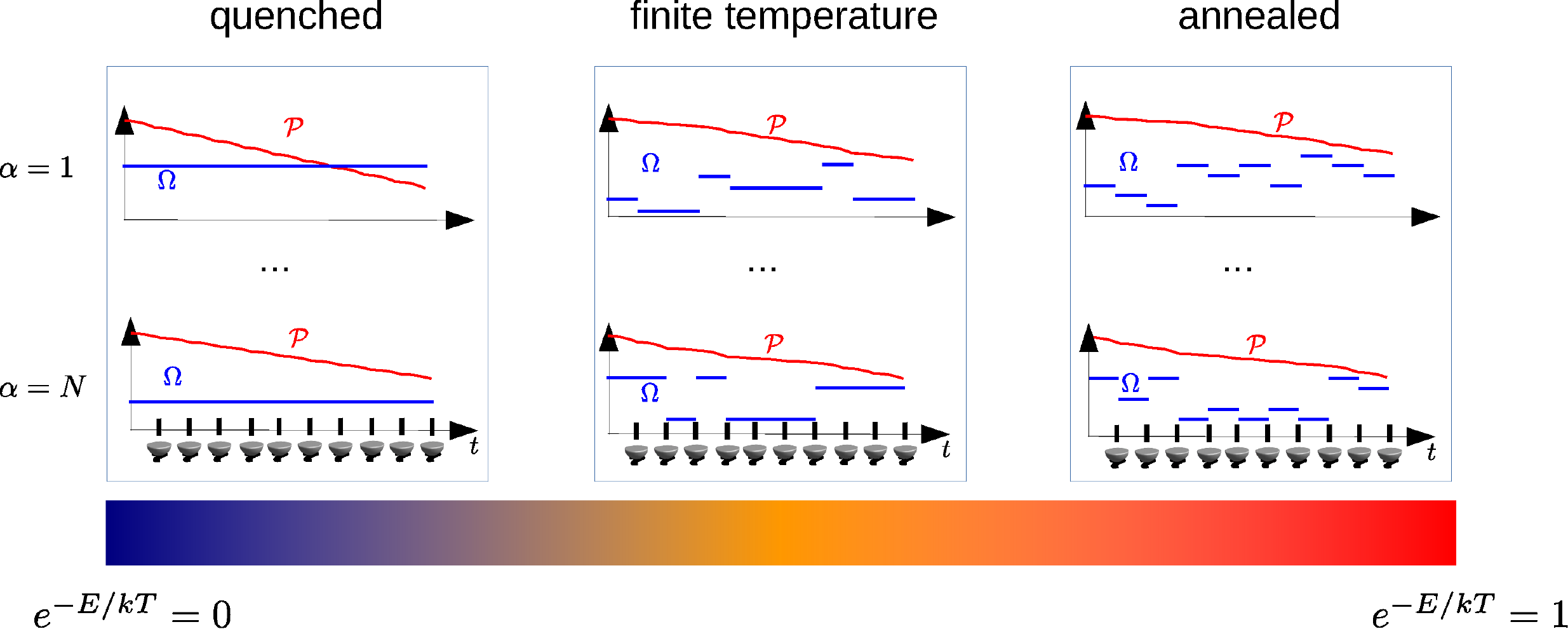}
 \caption{Schematic view of the field fluctuations and their influence on the survival probability during the measurement sequence. The driving field $\Omega$ fluctuates in time and with increasing temperature the time correlations vanish going from quenched disorder to annealed disorder. The survival probability $\mathcal{P}$ decreases in time at a rate depending on the fluctuating value of the field. For annealed disorder the effect of the field fluctuations over a couple of time intervals is averaged out and for each realization $\mathcal{P}$ converges to the same value. If we decrease the temperature, the time correlation of the fluctuation grows and this convergence slows down. In the limit of $T=0$ the fluctuations degenerate to a random offset value that determines the behavior of $\mathcal{P}$ that is now different for each realization.}
 \label{fig:sketch_SciRep}
\end{figure}

To characterize the survival probability $\mathcal{P}_{k}(m)$, two natural quantities arise: The time-average and the ensemble average. In this case, the \textit{time average} is defined as
\begin{equation}\label{eq:time_average1}
\hat{\mathcal{P}}_{k}(m)\equiv\lim_{M\rightarrow \infty}\frac{1}{M}\sum_{j=1}^{M}\mathcal{P}_{k}(j)^{\frac{m}{j}}.
\end{equation}
As before, by using the measured value of the survival probability after the $j-$th measurement, one can estimate the corresponding value after $m$ measurements as
$$
\mathcal{P}_{k}(m) \approx \mathcal{P}_{k}(j)^{\frac{m}{j}}.
$$
This value, then, is averaged for $j= 1,\dots,M$, and the limit of a large number of measurements $M$ is performed. Note that this limit will depend on the realization $k$ of the fluctuating field, and, in particular, on the strength of the noise correlation. The \textit{ensemble average}, instead, is defined as
\begin{equation}
\langle\mathcal{P}(m)\rangle\equiv\lim_{N\rightarrow \infty}\frac{1}{N}\sum_{k=1}^{N}\mathcal{P}_{k}(m),
\end{equation}
where the average of $\mathcal{P}_{k}(m)$ is performed over a large number of realizations $N$. In the limit of infinite realizations, the average does not depend on the single realization but on their probability distribution. Now, let us make the following \textit{assumption}: For each realization $k$ of the stochastic process the fluctuating field between two measurements assumes a \textit{constant} value, i.e. $\Omega_{j,k}(t)\rightarrow \Omega_{j,k}$, which is sampled from the probability density function $p(\Omega)$. Fig.~\ref{fig:sketch_SciRep} shows (in the right upper panel) how the fluctuating field $\Omega$ causes the survival probability $\mathcal{P}$ to decrease at a fluctuating rate.
Observe that within each time interval between two measurements the decrease of $\mathcal{P}$ is \textit{quadratic} in the time interval and the field strength. While the field fluctuations are random, after a few measurements the influence of these fluctuations on $\mathcal{P}$ is averaged out and the decay of $\mathcal{P}$ behaves similarly for each realization. When the field fluctuations are correlated, however, the decay of the survival probability depends much stronger on the realization because the probability distribution for $\Omega_{j+1,k}$ depends on the value of $\Omega_{j,k}$, and potentially also on the previous history. \textit{This means that the convergence of the time average can be much slower with respect to the uncorrelated case, since a random deviation will influence not only a single time interval but a range of them, according to the relaxation time} $\tau_{c}$ \textit{of the noise correlations}. For this reason, the results, that will be shown later, about the behaviours of the time and ensemble averages as a function of the noise correlation will depend just on the statistics of $q(\Omega)$ and not on the actual dependence of $q$ on $\Omega$, so that $\Omega$ will be treated as a parameter describing the statistics of $q(\Omega)$ via the probability density function $p(\Omega)$. In accordance with the aforestated assumption, we sample $\Omega_{j+1,k}$ from $p(\Omega)$ with probability $\mathfrak{p}$, and $\Omega_{j+1,k}=\Omega_{j,k}$ otherwise, where the \textit{update probability} $\mathfrak{p}$ can be associated to a temperature $T$ according to the relation $\mathfrak{p} = \mathrm{e}^{-E/kT}$. In Fig.~\ref{fig:sketch_SciRep}, the temperature grows from left to right yielding different types of disorder. For $T=0$, one has $\mathfrak{p}=0$, i.e. the value of the field $\Omega$ is chosen only once randomly and then remains always the same. Hence, the relaxation time $\tau_{c}$ is infinite and the time average does always converge to the same value. It is worth noting that this scenario simulates the interaction of the system with an environment that exhibits \textit{quenched disorder}. Depending on the value of $\Omega$ in correspondence of the $k-$th realization, the decay of the survival probability $\mathcal{P}_{k}(m)$ can be faster or slower, while for infinite temperature we have $\mathfrak{p}= 1$, representing an \textit{annealed disorder} environment. Between these two extreme regimes, i.e. for finite temperature, we have
$\mathfrak{p}\in[0,1]$, hence a mixture of both behaviours. As explained in~\cite{Edwards1975}, quenched disorder means a scenario with a static noise that depends on the initial random configuration of the environment, whereas annealed disorder means that the environment changes its configuration randomly in time.

Let us write now the expressions for the time and ensemble averages when also environmental time correlations are taken into account. In particular, for the time average $\hat{\mathcal{P}}_{k}(m)$ we introduce the expected frequencies $m\, n_{\Omega}$ that the event $\Omega$ occurs in one realization of the stochastic sequence of measurements. Then, the time average is given by
\begin{equation}
\hat{\mathcal{P}}_{k}(m) = \lim_{M\rightarrow \infty}\frac{1}{M}\sum_{j=1}^{M}\prod_{\{\Omega\}}(q(\Omega)^{j n_{\Omega}})^{\frac{m}{j}} = \prod_{\{\Omega\}} q(\Omega)^{m\,n_{\Omega}},
\end{equation}
where the product is over all possible values of $\Omega$ and $n_{\Omega}$. For independent (thus \textit{uncorrelated}) and identically distributed (i.i.d.) random variables $\Omega_{j,k}$ the expected frequencies correspond directly to the underlying probability density function $p(\Omega)$. Instead, for \textit{correlated} $\Omega_{j,k}$ the convergence of the time average might not be unique or not even exist. The latter consideration is very relevant, since it is linked to the \textit{Markov property and recurrence} of a stochastic process~\cite{Lamperti1960}, as explained in more detail below by introducing the theoretical expressions for the time average in different correlated dynamical regimes. In this regard, let us recall that a Markovian stochastic process does not imply Markovian quantum system dynamics, since a Markovian fluctuating field can generate non-Markovianity through its time-correlations. The ensemble average, instead, is the expectation value of the survival probability $\mathcal{P}$, i.e.
\begin{equation}\label{eq:ensemble-average-annealed}
\langle\mathcal{P}(m)\rangle \equiv \int_{\mathcal{P}}d\mathcal{P}{\rm Prob}(\mathcal{P})\mathcal{P}
= \int_{\Omega_1}d\Omega_1\dots\int_{\Omega_m}d\Omega_m\prod_{j=1}^{m} p_j(\Omega_j|\Omega_1,\dots \Omega_{j-1})q(\Omega_j),
\end{equation}
where ${\rm Prob}(\mathcal{P})$ is the probability distribution of the survival probability $\mathcal{P}_{k}(m)$ (which is by itself a random variable depending on the field fluctuations) and $p_j(\Omega_j|\Omega_1,\dots\Omega_{j-1})$ is the conditional probability of the event $\Omega_j$ given the process history. In the case of i.i.d. random variables $\Omega_j$, (\ref{eq:ensemble-average-annealed}) becomes
\begin{equation}
\langle\mathcal{P}(m)\rangle
= \int_{\Omega_1}d\Omega_1\dots\int_{\Omega_m}d\Omega_m\prod_{j=1}^m p(\Omega_j)q(\Omega_j) = \left(\int p(\Omega)q(\Omega)\right)^m.
\end{equation}

\begin{figure}[h!]
 \centering
 \includegraphics[width=0.9\textwidth]{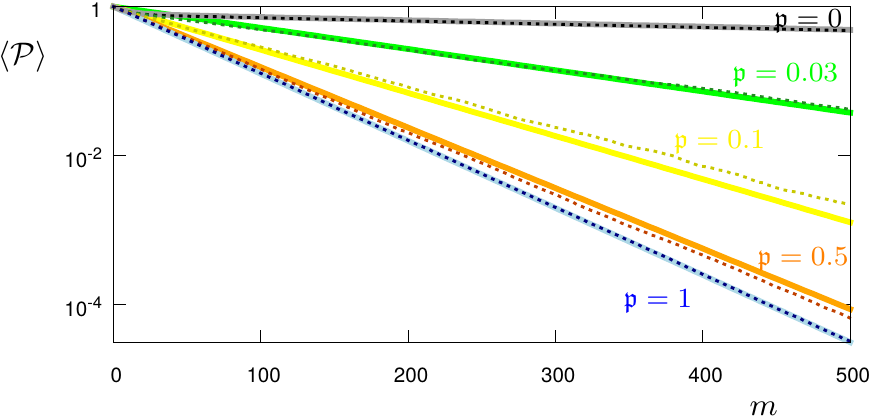}
 \caption{Ensemble Averages for $\mathfrak{p}=0,0.5,0.1,0.03,1$ (black, green, yellow, red, blue). The dashed lines correspond to the values calculated from 1000 realizations of the stochastic process, while the solid lines correspond to the respective theory curves.}
 \label{fig:ensemble-average}
\end{figure}

Finally, we compute the time and ensemble averages as a function of $\mathfrak{p}$ in three different regimes: (i) Annealed Disorder ($\mathfrak{p} = 1$), (ii) a finite temperature case with $\mathfrak{p}\in[0,1]$ and a number $m$ of measurements such that at least $5-10$ jumps occur, and (iii) quenched disorder ($\mathfrak{p} = 0$). In the case of annealed disorder ({\rm an}), i.e. uncorrelated noise, the two averages follow straightforwardly from the definitions, namely
\begin{equation}
\hat{\mathcal{P}}_{k}(m)_{{\rm an}} = e^{m\langle\ln q(\Omega)\rangle}
\end{equation}
for the time average and
\begin{equation}
\langle\mathcal{P}(m)\rangle_{{\rm an}} = e^{m\ln\langle q(\Omega)\rangle}
\end{equation}
for the ensemble average. Conversely, in the case of quenched disorder ({\rm qu}), each realization has constant $q(\Omega)$ and, thus, survival probability $q(\Omega)^m$. Accordingly, the ensemble average is the arithmetic average of these outcomes:
\begin{equation}
\langle\mathcal{P}(m)\rangle_{{\rm qu}} = e^{\ln\langle q(\Omega)^m\rangle} = \langle q(\Omega)^m\rangle.
\end{equation}
Instead, the time average for quenched disorder \textit{does not take a single value} but splits into several branches, i.e.
\begin{equation}
\hat{\mathcal{P}}_{k}(m)_{{\rm qu}} \in \{q(\Omega)^m\, |\,\Omega\in \mathrm{supp} (p(\Omega))\},
\end{equation}
since the underlying stochastic process is not recurrent, in the sense that given the value of $\Omega$ in the first interval, all the other values of the support of $p(\Omega)$, $\mathrm{supp} (p(\Omega))$, cannot be reached anymore within the same realization of the process. Finally, for the finite temperature ({\rm fT}) regime the problem is more difficult, but not for the time average, which is the same of the annealed disorder case:
\begin{equation}
\hat{\mathcal{P}}_{k}(m)_{{\rm fT}} = \hat{\mathcal{P}}_{k}(m)_{{\rm an}} = e^{m\langle\ln q(\Omega)\rangle}.
\end{equation}
The reason is that, despite of the time correlations, the time average is equal to the weighted geometric average of the quantity $q(\Omega)^m$ with respect to $p(\Omega)$, being computed over all possible configurations of $\{\Omega_{j}\}$ independently from the history of the process. Indeed, only in the quenched disorder case the $\sigma-$algebra of the random variable $\Omega$ is drastically decreased, and also $\mathfrak{p}$ is independent of the current value of the field. Thus, for a sufficiently long time the frequency of occurrence for the single measurement quantum survival probability $q(\Omega)$ converges to the expected values $n_{\Omega} = p(\Omega)$. Conversely, in order to derive the ensemble average we have to take into account the correlations and examine (i) the occurrence of the sequences of constant $\Omega(t)$'s over several time intervals and (ii) the updates of their values according to $\mathfrak{p}$. If the length of such a sequence is labelled by $l$, then $l$ is distributed by the Poisson distribution
\begin{equation}
r(l,\lambda_{P}) \equiv \frac{\lambda_{P}^{l}}{l!} e^{-\lambda_{P}},
\end{equation}
where $\lambda_{P} \equiv 1/\mathfrak{p}$. Thus, the expectation value of the survival probability $\mathcal{P}_{{\rm CF}}$ for this sequence of constant field values $\Omega$'s is given by
\begin{eqnarray}
\langle\mathcal{P}_{{\rm CF}}(l,\Omega,\mathfrak{p})\rangle_{l,\Omega} &\equiv& \sum_{l=0}^{\infty}r(l,\lambda_{P})\langle\mathcal{P}_{{\rm CF}}(l,\Omega,\mathfrak{p})\rangle_{\Omega} = \sum_{l=0}^{\infty}r(l,\lambda_{P})\int_{\Omega}d\Omega p(\Omega)q(\Omega)^{l}\nonumber \\
&=& \int_{\Omega}d\Omega p(\Omega)e^{-\frac{1}{\mathfrak{p}}}\left(\sum_{l=0}^{\infty}\frac{(\lambda_{P}q(\Omega))^l}{l!}\right) =
\int p(\Omega)\mathrm{e}^{\frac{q(\Omega)-1}{\mathfrak{p}}} d\Omega.\nonumber \\
&&
\end{eqnarray}
Moreover, also the update frequency of the constant $\Omega$'s is Poisson distributed, with expectation value $\mathfrak{p}m$. Hence, the ensemble average of the system survival probability in case of time-correlated random fields is equal to
\begin{eqnarray}
\langle\mathcal{P}(m)\rangle_{{\rm fT}} &=& \langle\mathcal{P}_{{\rm CF}}(l,\Omega,\mathfrak{p})\rangle_{l,\Omega,\mathfrak{p}}
= e^{-\mathfrak{p}m}\sum_{n = 0}^\infty \frac{(\mathfrak{p}m)^n}{n!} \langle\mathcal{P}_{{\rm CF}}(l,\Omega,\mathfrak{p})\rangle_{l,\Omega}\nonumber \\
&=&e^{\mathfrak{p}m(\langle\mathcal{P}_{{\rm CF}}(l,\Omega,\mathfrak{p})\rangle_{l,\Omega} - 1)}.
\end{eqnarray}
To summarize, $\langle\mathcal{P}(m)\rangle_{{\rm fT}}$ has been derived by means of the following two steps:
\begin{itemize}
\item
First, we have first computed the expectation value of the system survival probability with a repeated sequence of projective measurements, characterized by constant values of $\Omega(t)$ over the time intervals of the sequence. For such derivation, we have assumed that the length $l$ of the sequence is a Poisson distributed random variable, whose mean value has been calculated with respect to $l$ and $\Omega$.
\item
Secondly, also the update frequency of the $\Omega$'s has been modeled as a Poisson random variable, so that the ensemble average of the system survival probability turns out to be equal to the expectation value of $\langle\mathcal{P}_{{\rm CF}}(l,\Omega,\mathfrak{p})\rangle_{l,\Omega}$ with respect to $\mathfrak{p}$.
\end{itemize}
Fig.~\ref{fig:ensemble-average} shows the above calculated ensemble averages together with numerical values from the realization of $N = 1000$ stochastic processes for different values of $\mathfrak{p}$.
In all cases, for $p(\Omega)$ we have used a bimodal distribution with $p_1=0.8$, $p_2=1-p_1=0.2$ and corresponding single measurement quantum survival probabilities $q_1=0.999$, $q_2=0.9$. By decreasing (increasing) $q_1$ and $q_2$, the decay becomes faster (slower). The same happens if we increase (decrease) $p_2$, that is the probability associated with $q_2<q_1$. Note that the probabilities $p_1$, $p_2$, $q_1$, $q_2$ and the update frequency $\mathfrak{p}$ fully define the time and ensemble average of $\mathcal{P}$, so that we do not have to specify the Hamiltonian of the system.

\subsection{Detection of noise correlations}

In this section, we will show how to probe time correlations of a noisy environment coupled to a quantum system used as probe.

\subsubsection*{Accumulated standard deviation}

Let us evaluate the variance of the probability distribution ${\rm Prob}(\mathcal{P})$, which is defined as
\begin{equation}
\Delta^2\mathcal{P}(m)\equiv\langle\mathcal{P}(m)^2\rangle - \langle\mathcal{P}(m)\rangle^2,
\end{equation}
where $\Delta\mathcal{P}$ is the corresponding standard deviation. Thus, to derive the variance $\Delta^2\mathcal{P}(m)$, we still need to calculate the second moment of the probability distribution ${\rm Prob}(\mathcal{P})$. In the case of infinite temperature or annealed disorder, it is given by
\begin{equation}
\langle\mathcal{P}^2(m)\rangle_{{\rm an}}
= \int_{\Omega_1}d\Omega_1\dots\int_{\Omega_m}d\Omega_m\prod_{j=1}^m p(\Omega_j)q(\Omega_j)^2
= e^{m\ln\left(\int_{\Omega}d\Omega p(\Omega)q(\Omega)^2\right)}=e^{m\ln \langle q(\Omega)^2\rangle},
\end{equation}
so that the normalized variance is equal to
\begin{eqnarray}
\frac{\Delta^{2}\mathcal{P}(m)_{{\rm an}}}{\langle\mathcal{P}(m)\rangle_{{\rm an}}^2} &=&
\frac{\langle\mathcal{P}(m)^2\rangle_{{\rm an}} - \langle\mathcal{P}(m)\rangle_{{\rm an}}^2}{\langle\mathcal{P}(m)\rangle_{{\rm an}}^2}
=e^{m\left(\ln\langle q(\Omega)^2\rangle - \ln\langle q(\Omega)\rangle^2 \right)}-1\nonumber \\
&\approx& m\left(\ln\langle q(\Omega)^2\rangle - \ln\langle q(\Omega)\rangle^2\right),
\end{eqnarray}
and the normalized standard deviation reads as
\begin{equation}\label{SD_annealed}
\frac{\Delta\mathcal{P}(m)}{\langle\mathcal{P}(m)\rangle}
\approx\sqrt{m}\sqrt{\ln \langle q(\Omega)^2\rangle - \ln\langle q(\Omega)\rangle^2}
\approx \sqrt{m}\Delta^2 H\tau^2\sqrt{\langle\Omega^4\rangle - \langle\Omega^2\rangle^2}.
\end{equation}
Note that the r.h.s. of (\ref{SD_annealed}) is given by a second order expansion in the interval length $\tau$.

For finite temperature, instead, let us consider again the statistical ensemble composed by the sequences of projective measurements with constant $\Omega$'s, whose second statistical moment is
\begin{equation}
\langle\mathcal{P}_{{\rm CF}}(l,\Omega,\mathfrak{p})^2\rangle_{l,\Omega}
=\sum_{l=0}^{\infty}r_(l,\lambda_{P})\int_{\Omega}p(\Omega)q(\Omega)^2l d\Omega
=\int_{\Omega}p(\Omega)e^{\frac{q(\Omega)^2 - 1}{\mathfrak{p}}}d\Omega.
\end{equation}
Then, being also $\mathfrak{p}$ a Poisson random variable, the second statistical moment of the system survival probability turns out to be
\begin{eqnarray}
\langle\mathcal{P}(m)^2\rangle_{{\rm fT}} &=& \langle\mathcal{P}_{{\rm CF}}(l,\Omega,\mathfrak{p})\rangle_{l,\Omega,\mathfrak{p}}
= e^{-\mathfrak{p}m}\sum_{n=0}^{\infty}\frac{(\mathfrak{p}m)^n}{n!} \langle\mathcal{P}_{{\rm CF}}(l,\Omega,\mathfrak{p})^2\rangle_{l,\Omega}^n\nonumber \\
&=& e^{\mathfrak{p}m(\langle\mathcal{P}_{{\rm CF}}(l,\Omega,\mathfrak{p})^2\rangle_{l,\Omega} - 1)}\,,
\end{eqnarray}
and the normalized variance reads as
\begin{equation}
\frac{\Delta^2 \mathcal{P}(m)_{{\rm fT}}}{\langle \mathcal{P}(m)\rangle_{{\rm fT}}^2}
= e^{\mathfrak{p}m\left(\langle\mathcal{P}_{{\rm CF}}(l,\Omega,\mathfrak{p})^2\rangle_{l,\Omega} - 2\langle\mathcal{P}_{{\rm CF}}(l,\Omega,\mathfrak{p})\rangle_{l,\Omega} + 1\right)}-1
\end{equation}
i.e.
\begin{equation}
\frac{\Delta^2 \mathcal{P}(m)_{{\rm fT}}}{\langle \mathcal{P}(m)\rangle_{{\rm fT}}^2}
\approx \mathfrak{p}m\left(\langle\mathcal{P}_{{\rm CF}}(l,\Omega,\mathfrak{p})^2\rangle_{l,\Omega} - 2\langle\mathcal{P}_{{\rm CF}}(l,\Omega,\mathfrak{p})\rangle_{l,\Omega} + 1\right),
\end{equation}
leading to the following normalized standard deviation:
\begin{equation}\label{SD_fT}
\frac{\Delta\mathcal{P}(m)_{{\rm fT}}}{\langle \mathcal{P}(m)\rangle_{{\rm fT}}}
\approx\sqrt{m}\sqrt{1+\frac{1}{\mathfrak{p}}}\,\Delta^2 H\tau^{2}\sqrt{\langle\Omega^4\rangle}\,.
\end{equation}

Finally, for the quenched disorder case one has
\begin{equation}
\langle\mathcal{P}^2(m)\rangle_{{\rm qu}}
=\int_{\Omega}d\Omega p(\Omega)q(\Omega)^{2m}
=e^{\ln\langle q(\Omega)^{2m}\rangle}\,,
\end{equation}
where the normalized variance is given by
\begin{equation}
\frac{\Delta^2 \mathcal{P}(m)_{{\rm qu}}}{\langle\mathcal{P}(m)\rangle_{{\rm qu}}^2}
=e^{\ln\langle q(\Omega)^{2m}\rangle - \ln\langle q(\Omega)^m\rangle^2 } - 1
\approx \ln\langle q(\Omega)^{2m}\rangle - \ln\langle q(\Omega)^m\rangle^2.
\end{equation}
As a consequence, the normalized standard deviation is
\begin{equation}\label{SD_quenched}
\frac{\Delta\mathcal{P}(m)_{{\rm qu}}}{\langle\mathcal{P}(m)\rangle_{{\rm qu}}}
\approx  \sqrt{\ln \langle q(\Omega)^{2m}\rangle - \ln\langle q(\Omega)^m\rangle^2 }
\approx m\Delta^2 H\tau^{2}\sqrt{\langle\Omega^4\rangle - \langle\Omega^2\rangle^2}\,,
\end{equation}
where the latter expression is given again by a second order expansion in the interval length $\tau$.

\begin{figure}[h!]
\centering
  \includegraphics[width=0.9\textwidth]{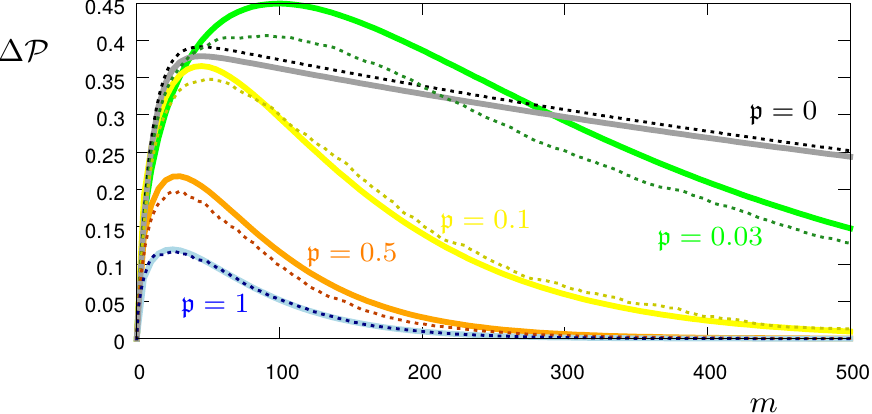}
 \caption{Standard deviation for $\mathfrak{p}=0,0.5,0.1,0.03,1$ (black, green, yellow, red, blue). The dashed lines correspond to the value calculated from 1000 realizations of the stochastic process, while the solid lines correspond to the respective theory curve.}
 \label{fig:stdev}
\end{figure}

\begin{figure}[h!]
\centering
  \includegraphics[width=0.9\textwidth]{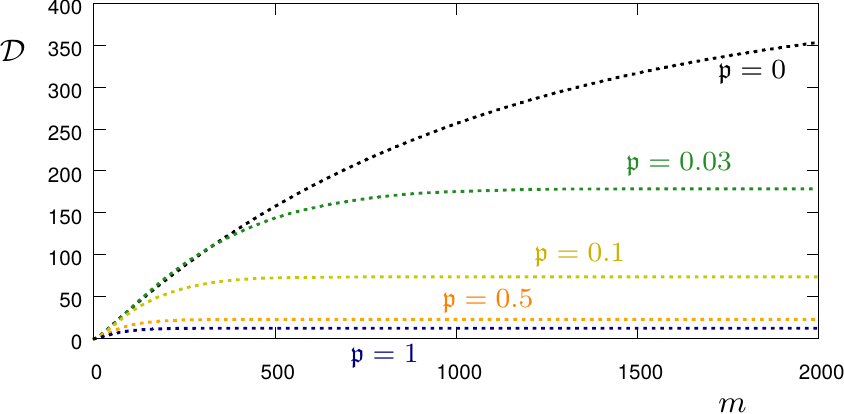}
 \caption{Accumulated standard deviation $\mathcal{D}(m) =\sum_{j=1}^m \Delta\mathcal{P}(j)$ for $\mathfrak{p}=0,0.5,0.1,0.03,1$ (black, green, yellow, red, blue). The dashed lines correspond to the values calculated from 1000 realizations of the stochastic process. For a relatively high number of measurements $m > 300$ there is a clear \textbf{monotonicity} of $\mathcal{D}$ as a function of the degree of the noise time-correlations.}
 \label{fig:int-stdev}
\end{figure}

Fig.~\ref{fig:stdev} shows the standard deviations $\Delta\mathcal{P}$ (without normalization) together with the values from the realization of 1000 stochastic processes for the chosen value of $\mathfrak{p}$, i.e. $\mathfrak{p}=0,0.03,0.1,0.5,1$.
We find that the larger is the time-correlation (the smaller $\mathfrak{p}$), the larger is the standard deviation $\Delta\mathcal{P}$ of the survival probability $\mathcal{P}$, i.e. the more the outcome depends on the single realization. To average out the \textit{non-monotonic} behaviour of $\Delta\mathcal{P}$, we introduce the \textit{accumulated standard deviation}
\begin{equation}
\mathcal{D}(m) \equiv \sum_{j=1}^m \Delta\mathcal{P}(j),
\end{equation}
given by summing up the standard deviation values for an increasing number $j = 1,\dots,m$ of measurements. The result is shown in Fig.~\ref{fig:int-stdev}.
For relatively large values of $m$ ($ > 300 $) $\mathcal{D}(m)$ \textit{monotonically} increases with the amount of time-correlations, which is directly proportional to the quantity $1-\mathfrak{p}$. Hence, we propose $\mathcal{D}(m)$ as the natural figure of merit to infer the strength of such a noise time-correlation.

\subsubsection*{Ergodicity breaking of interaction modes}

\begin{figure}[h!]
\centering
\includegraphics[width=1\textwidth]{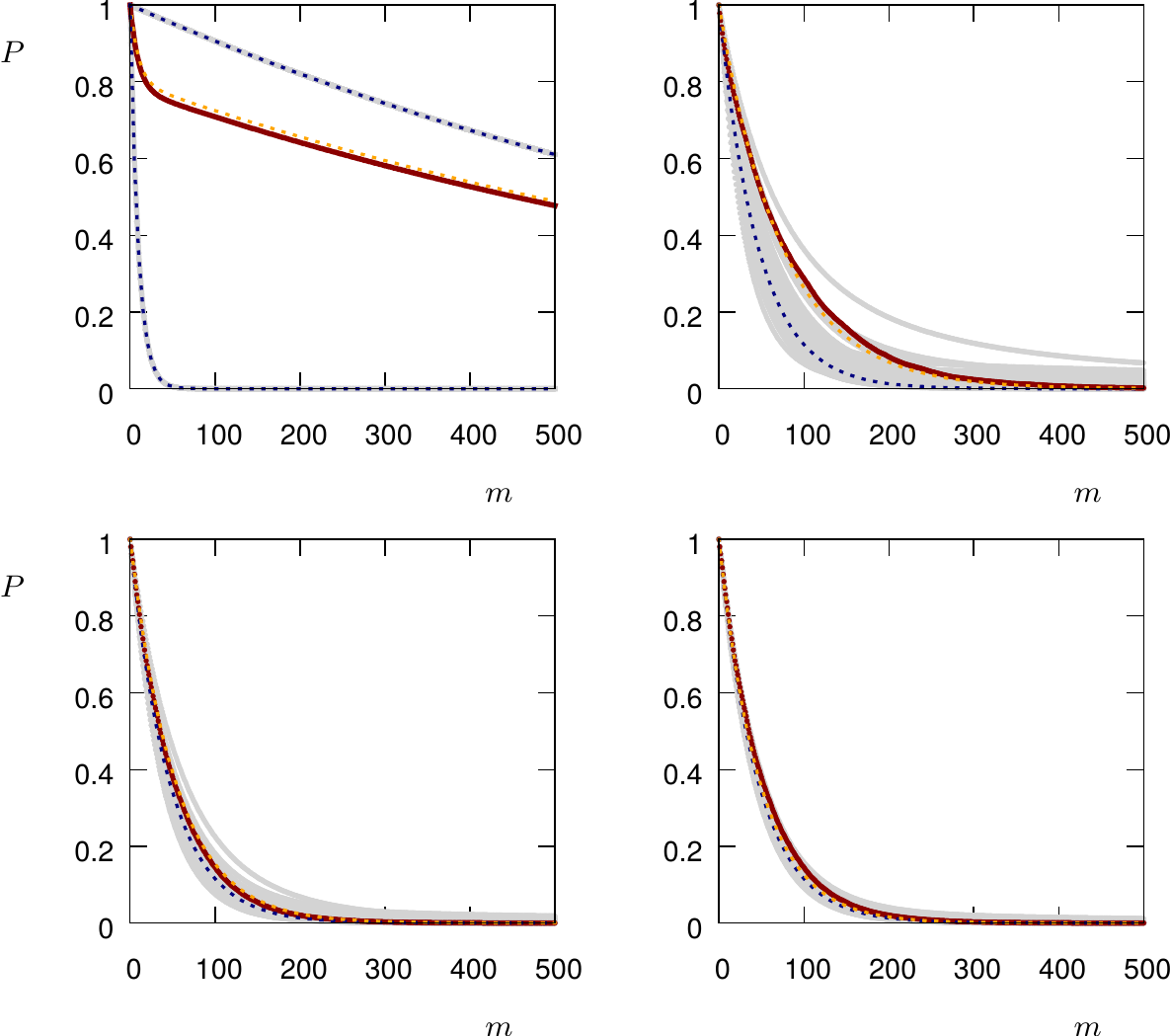}
\caption{In the numerical simulations we have considered $50$ realizations of the time average with $M = 2000$ (grey solid lines), along with the ensemble average calculated from $1000$ realizations of the stochastic process (red solid lines). These are compared to the theoretical curves for the time average (dark blue dashed) and ensemble average (orange dashed). Top left: quenched; top right: $\mathfrak{p}=0.1$; bottom left: $\mathfrak{p}=0.5$; bottom right: annealed.}
\label{fig:phasetransition}
\end{figure}

As shown before, the time and ensemble averages of the system survival probability $\mathcal{P}$ strictly depend on the update frequency $\mathfrak{p}$. Only for large values of $m$ and $N$ (i.e. many measurements and many realizations), the frequency of each event $q(\Omega)$ is $mNp(\Omega)$, independently of $\mathfrak{p}$. If we compare the expressions for such averages as a function of the noise time correlation (i.e. for different temperatures $\mathfrak{p}$), we find that the ensemble average will grow until it takes the maximum in the quenched disorder limit, which is given by the arithmetic average of the quantity $q(\Omega)^m$. In other words, one get
\begin{equation}\label{eq:averages-hierachy}
\hat{\mathcal{P}}_{k}(m)_{{\rm an}} \leq \langle\mathcal{P}(m)\rangle_{{\rm an}} \leq \langle\mathcal{P}(m)\rangle_{{\rm fT}}\leq \langle\mathcal{P}(m)\rangle_{{\rm qu}},
\end{equation}
so that the following conclusions can be stated:
\begin{itemize}
\item
For the case of annealed disorder the time and ensemble averages practically coincide: we refer to this equality as an ergodic property of the system environment interaction, as shown in \cite{GherardiniQST}.
\item
However, the more the $q(\Omega_{j,k})$ are correlated, the more the ensemble average moves away from the time average and the ergodicity is broken. This can be seen in Fig.~\ref{fig:phasetransition} where time and ensemble averages are simulated for a bimodal distribution $p(\Omega)$ for quenched and annealed disorder, and for two values of finite temperature. Also for this simulation, as well as for Figs.~\ref{fig:ensemble-average},~\ref{fig:stdev} and~\ref{fig:int-stdev}, we have used a bimodal distribution with $p_1=0.8$, $p_2=0.2$ and corresponding single measurement quantum survival probabilities $q_1=0.999$, $q_2=0.9$. As given by (\ref{SD_annealed}),~(\ref{SD_fT}) and~(\ref{SD_quenched}), the non-ergodic behaviour depends essentially on the second and fourth moment of $p(\Omega)$. In other terms, this effect will decrease if we choose $p_1\approx p_2$ or $q_1\approx q_2$. The same happens if we change the bimodal distribution into a multimodal or continuous distribution.
\end{itemize}
From an application point of view, \textit{this allows to detect correlations in a fluctuating field by measuring and comparing to each other the time and ensemble averages of the survival probability. Furthermore, by changing the time interval} $\tau$ \textit{between two measurements, we can explore the occurrence time scale of these correlations}.

In order to test our method for a real quantum system, let us now consider the following two-level Hamiltonian
$$
H_{\rm tot} = \Delta \ \sigma_z + \Omega(t)\sigma_x\,,
$$
where $\sigma_x,\sigma_z$ are Pauli matrices, $\Omega(t)$ is the (fluctuating) driving of the system (e.g. an unstable classical light field), and $\Delta$ is a detuning term. We set $\Delta=2\pi\times 5\,$MHz and $\Omega\in 2\pi\times\{1,5\}$ MHz as a fluctuating RTN field with equal probability for both values. We initially prepare the system in the ground state $|0\rangle$ and perform projective measurements in this state spaced by intervals of constant length $\tau = 100$ ns. Such scheme may be implemented on many different experimental platforms and, very recently, has been realized to prove the stochastic quantum Zeno effect with a Bose-Einstein condensate on an atom-chip~\cite{GherardiniQST}.
\begin{figure}[h!]
\centering
\includegraphics[width=1\textwidth]{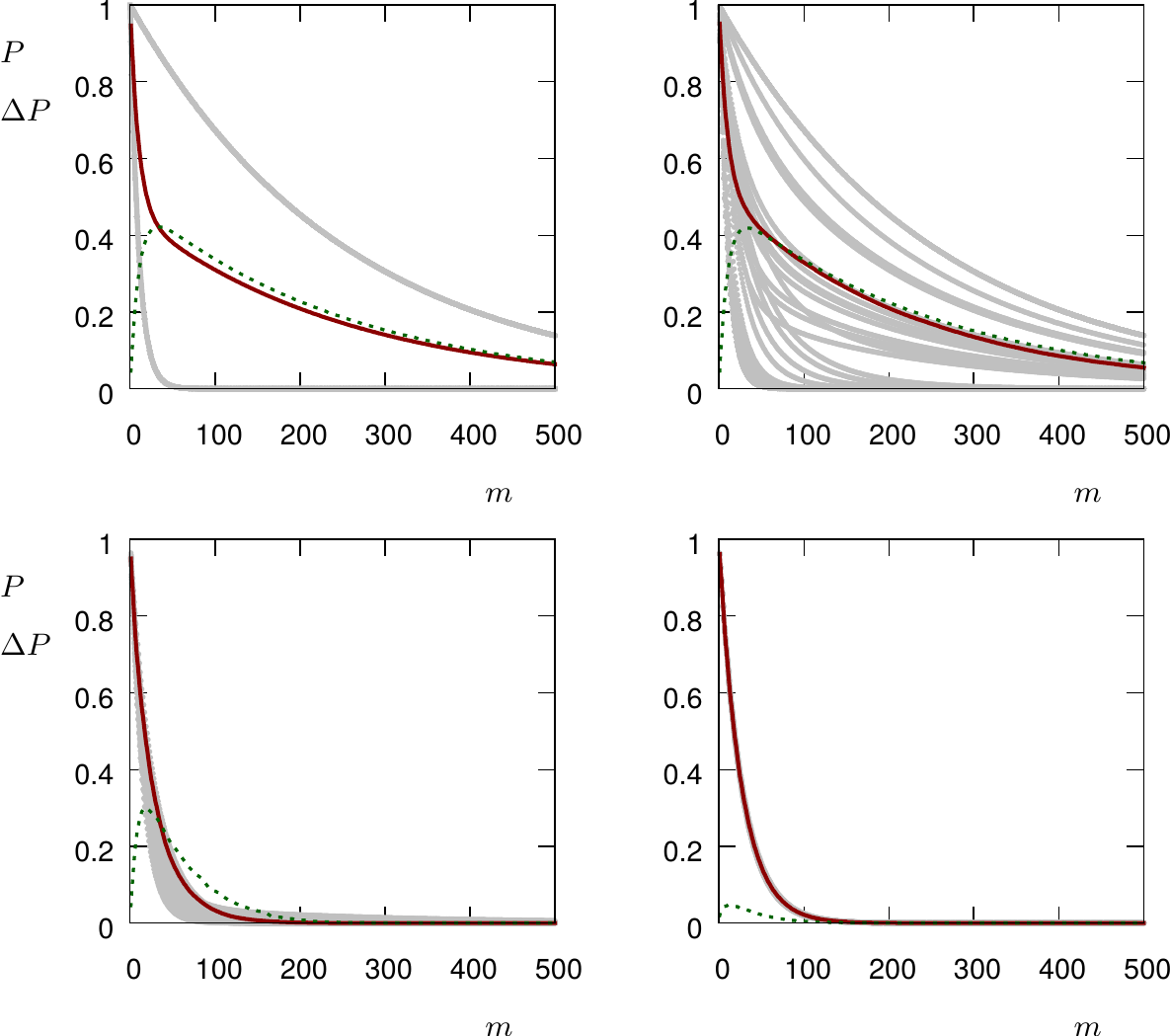}
\caption{Time and ensemble averages of the system survival probability, along with the standard deviation $\Delta\mathcal{P}$ for the two-level Hamiltonian with a fluctuating RTN (classical) field. For the numerical simulations, we have considered $50$ realizations of the time average (grey solid lines) with $M = 2000$, along with the ensemble average calculated from $1000$ realizations of the stochastic process (red solid line). The dark green dashed lines show the standard deviation $\Delta\mathcal{P}$ of the single realizations. The time scale $\tau_{c}$ of the correlation decreases from left to right and from top to bottom, ranging from perfectly correlated (quenched) disorder to uncorrelated (annealed) noise.}
\label{fig:RTN}
\end{figure}
Note that the second order time correlation function for the RTN is exponential in time, so that
\begin{equation}
\langle\omega(t)\omega(t')\rangle \propto e^{-\frac{2(t-t')}{\tau_{c}}},
\end{equation}
where the relaxation time $\tau_{c}$ is equal to the average time between two field switches. In this regard, Fig.~\ref{fig:RTN} shows the time and ensemble averages, together with the corresponding standard deviation, for an average time between the fluctuating field switches equal, respectively, to $10$, $10^3$, $10^5$, $10^7$ $ns$. It can be clearly seen how a relaxation time $\tau_{c}$ longer than the time interval $\tau$ generates a growing standard deviation $\Delta\mathcal{P}$, which can then be exploited as a \textit{witness of the strength of noise correlations}.

\section{Conclusions and contributions}

Summarizing, this chapter provides the following contributions:
\begin{itemize}
\item
We have analyzed \textit{stochastic quantum Zeno phenomena} by means of the LD theory. In particular, for an arbitrary quantum system with unitary dynamical evolution subject to a sequence of random-distributed measurements projecting it into a fixed initial state, we have analytically shown that (in the limit of a large number $m$ of measurements) the distribution of the (survival) probability to remain in the initial state assumes a \textit{large-deviation form}, namely, a profile decaying exponentially in $m$ with a positive multiplying factor. Such a profile is given by the so-called rate function, which is a function only of the survival probability. Our analytical results, then, have been tested in the case of Zeno-protected entangled states. We have shown that the presence of disorder in the sequence of time intervals between consecutive measurements is deleterious in reaching the Zeno limit. Nevertheless, the disorder does enhance the survival probability when the latter is \textit{not} exactly one, which, interestingly enough, corresponds to the typical experimental situation. Furthermore, it is worth noting that, since the decoherence may correspond to a continuous monitoring from the environment (repetitive random measurements), our formalism allows one to predict the occupation probability of an arbitrary quantum state by the knowledge of the probability distribution of the system-environment interaction times.
\item
The application of the LD theory to open quantum systems allowed to obtain \textit{analytical expressions for the most probable and the average value of the survival probability}. While the most probable value represents what an experimentalist will measure in a single typical implementation of the measurement sequence, the average value
corresponds instead to an averaging over a large (ideally infinite) number of experimental runs. Hence, by tuning the probability distribution of the time intervals between consecutive measurements, one can achieve a specific value for the most probable survival provability, thereby allowing to engineer novel optimal control protocols.
\item
We analytically and experimentally demonstrated the occurrence in the Zeno regime of an \textit{ergodic property for the interaction modes between a quantum system and the external environment}, by measuring the system at random times. In particular, by using the large deviation theory we have proved that the most probable value of the probability for the system to remain in a given quantum state is equal to the corresponding arithmetic average, computed over a statistical ensemble of ordered sequences of measurements, when the system approaches the quantum Zeno regime. These results have been experimentally tested using a Bose-Einstein condensate of Rubidium atoms, which are trapped on an atom chip. They are expected to represent further steps towards controlled manipulations of quantum systems via dissipative interactions~\cite{Schindler2013}, whereby one can control the noisy environment or part of it to perform desired challenging tasks.
\item
By exploiting again LD theory, we have analytically derived under which conditions one can distinguish two different noise probability density functions $p(\tau)$ of a stochastic sequence of quantum measurements, by evaluating the corresponding survival probability at the end of the sequence. In particular, we have introduced a Fisher information operator, which is expressed in terms of the statistical moments of the corresponding noise probability density function. This has allowed us to quantify how sensitive is the survival probability's most probable value to an arbitrary perturbation $\delta p(\tau)$ of $p(\tau)$, and to distinguish the difference between the perturbed and unperturbed noise distribution. Such a tool might become a promising method to analyze the temporal behaviour of an unknown environment when coupled to the measured quantum system.
\item
The accessibility to Zeno dynamics for a quantum system in random interaction with the environment has been analyzed. In this regard, when noise contributions in quantum Zeno protocols are taken into account, the accessibility to system dynamics becomes more difficult, so that the \textit{confinement within the measurement subspace is effectively realized only if the stochasticity is compensated by a stronger observation}, that e.g. can be realized by a sequence of measurements occurring at random times but on average more frequently. To achieve this result, we have extended the large deviation theory approach, used to introduce the stochastic quantum Zeno effect, to the description of survival probabilities from QZD by considering also the dynamics within the measurement subspace. The new approach has allowed us to find a less restrictive condition for the confinement of quantum Zeno dynamics (the weak Zeno limit), for which the observations are fast enough to ensure that the dynamics in the subspace follow closely the dynamics of a perfectly truncated system, while the survival probability decays with increasing time.
\item
Besides the stochastic quantum Zeno protocol based on projective measurements, we have shown that SQZD can be equivalently achieved with high fidelity also by applying fast \textit{random unitary kicks} or \textit{strong continuous couplings}, that have the advantage to be fully deterministic and easy to implement. Since only by modelling with enough accuracy the nature of such interactions with the environment we can effectively control a quantum dynamics in a well-defined Hilbert space portion, we believe that the results about SQZD will provide a new tool in quantum information processing and quantum computation not only for controlling the amount of quantum coherence by means of Zeno-protection protocols, but also to design engineered quantum paths within the system Hilbert space. To all effects, when this framework, originated from the application of the large deviation theory to open quantum systems, will be combined with optimization methods to derive control pulses, it could be denoted as \textit{noise-assisted quantum control} paradigm.
\item
Stochastic sequences of correlated quantum measurements have been analyzed. In particular, we have quantified stochastic quantum Zeno phenomena in time-correlated environments and we have shown how the \textit{ergodicity breaking} of the system-environment interaction modes depends on the time scale of the noise correlations. Indeed, the deviation between the time and ensemble averages of the system survival probability monotonically grows for increasing values of $\mathfrak{p}$, which quantifies the strength of such correlations. In doing this, we have introduced a novel method to probe time correlations in random classical fields coupled to the quantum probing system. The advantage of this method is that it does not rely on quantum state or process tomography but on a simple Zeno-based measurement scheme. In this way, by realizing different initial states and measurement operators, one has also the possibility to probe the effect of the environment on different subspace of the system. We believe that this approach will further contribute to the development of new schemes for quantum sensing technologies, where nanodevices may be exploited to image external structures or biological molecules via the surface field they generate. As final remark, it is worth noting that the introduction of (quantum) noise sensing schemes is crucial also to improve the reliability of the predictions provided by the LD theory about the fluctuation profile of specified system observables.
\item
As final remark, note that the results we have shown in this chapter rely on the hypothesis that the introduction of fluctuating semi-classical fields can model a wide class of noise sources, external to the dynamics of the system. Accordingly, the underlying noise-assisted properties follow from our ability to model the fluctuations within the system dynamics, i.e. to correctly predict the occurrence probabilities of the values taken by each system dynamical variable. Such assumption becomes realistic if, before manipulating the system, noise sensing techniques (see e.g. Ref. \cite{Paz-Silva2014,Norris2016,MuellerSensing}) are used to evaluate (also approximately) the shape and the intensity of the noise sources affecting the system. Thus, to make \textit{stable}, or \textit{robust}, the noise-assisted properties for a given system, the adoption of noise sensing techniques, together with LD predictions, appears to be the most efficient solution. Otherwise, the presence of unmodeled noise is expected to invalidate the positive effects of noise-assisted phenomena.
\end{itemize}

\chapter{Quantum thermodynamics}\label{chap:QTherno}

\begin{quote}
\textit{In this chapter, we will address the characterization and reconstruction of general thermodynamical quantities, such as work, internal energy and entropy for a quantum system in interaction with an external environment, not necessarily thermal. Indeed, in the quantum regime the dynamics of nanoscale systems is highly stochastic, in the sense that thermal and/or quantum fluctuations become of the same order of magnitude as the averages of the physical quantities, which define for example the Hamiltonian of the quantum system. Therefore, the analysis of the energetic and informative content of such fluctuations in terms of non-equilibrium statistical mechanics is crucial to understand the role and the effects of the stochasticity given by random system-environment interactions. On one side, our results allow to quantify the energy that is absorbed by a quantum system due to the presence of stochastic fluctuations, and, on the other side, to infer the environment structure by characterizing the thermodynamic irreversibility of a given quantum process. Moreover, we aim also to clarify the relation between the concepts of entropy and disorder in the sense of stochasticity, by starting from a quantum mechanical microscopic derivation of the entropy production until to derive a macroscopic definition given by the second law of thermodynamics.}
\footnote{The results shown in this chapter have been published as ``Reconstruction of the stochastic quantum entropy production to probe irreversibility and correlations'', in \textit{Eprint arXiv:1706.02193}, 2017 (submitted to the International Journal \textit{Quantum Science and Technology - IOPscience})~\cite{GherardiniEntropy}; ``Non-equilibrium quantum-heat statistics under stochastic projective measurements'', in \textit{Eprint arXiv:1805.00773}, 2018 (submitted to the International Journal \textit{Physical Review E})~\cite{GherardiniWORK}.}
\footnote{Part of this work was conducted while the author was a visiting Ph.D. student at SISSA, Scuola Internazionale Superiore di Studi Avanzati, in January and July 2016, and in January, March and July 2017, Trieste (Italy).}
\end{quote}

\section*{Introduction}

The stochastic behaviour of a dynamical system is linked to the presence of non-equilibrium dynamics due to the interaction of the system with an external environment. Such interactions are not necessarily known by an external observer, and, in general, are well modelled by random (fluctuating) couplings. Also the environment, indeed, is a dynamical system, and only rarely it is uniquely determined by some macroscopic variables, as e.g. the temperature, as it happens when the environment is a thermal bath. As a consequence, we can deduce that, if we only observe the evolution of a system (for example by tracing out the environment), then each transformation performed on it is generally \textit{irreversible}, since in principle such a transformation cannot be reversed by taking back both the system and the environment in their initial conditions, without using a greater amount of energy with respect to that used to realized it. Only in few cases a system transformation can be defined \textit{reversible}, i.e. when it is realized by infinitesimal and quasi-static variations, that preserve the system in an equilibrium state in each time instant. As a matter of fact, in classical mechanics the solutions of the equations of motion are unique and the motion along the trajectories in system phase space can be, in principle, always inverted to retrieve all the states previously occupied by the system~\cite{Seifert2012Review}. However, the time inversion in experiments with a macroscopic number of particles cannot be practically performed, due to some information losses and the evidence that for a system is very unlikely to occupy the same state at a later time within the dynamics. Similarly, in quantum mechanics the dynamics of the system wave function and more generally of the density matrix cannot always be reversed in time, and it ensues the corresponding need to characterize and quantify, where possible, irreversible quantum processes~\cite{Esposito2009,Campisi2011}. The typical instance is given by the thermalization of an open system, where the dissipative processes taking place due to the interaction of the system with its environment degrade the quantum nature of the system and the coherence of the quantum states~\cite{Gogolin2016}.

Accordingly, the following questions naturally emerge: How far a dynamical system can be led from an equilibrium regime by means of external interactions? What is the corresponding energy dispersion (loss of information) while performing a non-equilibrium transformation? How much energy is required to maintain the system in a non-equilibrium regime, induced by interactions with the environment? In this regard, in 1865 Rudolf Clausius first introduced the concept of \textit{entropy} production, which quantifies the unavailability of a system to produce useful work. More formally, entropy (which is not directly measurable) is a state function of a system in thermodynamic equilibrium, and is proportional to the number of microscopic configurations assumed by the system, while it is approaching to a state as specified by one or more macroscopic variables. In its formulation of thermodynamics, Clausius proved that for a thermodynamic system the entropy production is always larger than the heat exchange by the system with its surroundings. Such a statement, known as \textit{Clausius inequality}, is valid for irreversible and reversible processes, as well as for isolated and open systems. Moreover, the concept of \textit{entropy} is crucial not only in thermodynamics, where it allows to characterize irreversibility of a quantum process (for both classical and quantum systems), but also in information theory, where it is used to quantify the amount of lost information within a communication channel~\cite{Cover2006}.

In particular, in the present chapter we will address the following topics:
\begin{itemize}
  \item We will investigate the statistics of the quantum-heat absorbed by a quantum system subject to a sequence of projective measurements applied at random times, in order to characterize from an energy point of view the effects produced by the presence of some fluctuating fields due to the random interaction between the system and the environment.
  \item By starting from the derivation of the quantum fluctuation theorem for open (decoherent) systems, we will introduce an efficient protocol (relying on a two-time quantum measurements scheme) to reconstruct the entropy production of a quantum given process. In this way, we will be able to (i) understand how much the energetic configuration of the system is altered by the interaction with an arbitrary environment, and (ii) characterize the structure and the features of the external environment.
\end{itemize}

\section{Quantum-heat}

In this section, the results in \cite{GherardiniWORK} about non-equilibrium quantum-heat statistics under stochastic projective measurements are discussed. In the last decades, a growing interest in the thermodynamic properties of quantum dynamical systems has emerged \cite{Esposito2009,Campisi2011}. One of the main goals of such research activity is to devise and implement more efficient engines by exploiting quantum resources~\cite{Scully03Science299,KimPRL2011,AbahPRL2012,RossnagelPRL2014,Kosloff14ARPC65,Uzdin15PRX5,Campisi15NJP17}. In particular, it has focussed the attention on the exploration of the role of non-thermal states~\cite{Solinas2015} and the capability of characterizing the statistics of the energy, which is exchanged by a quantum system in interaction with an external environment and/or measurement apparata~\cite{Campisi2010PRL,Campisi2011PRE,Yi2013,WatanabePRE2014,Talkner16PRE93,Yuanjian16PRE94}.

Previously in the thesis, we have introduced models relying on sequences of stochastic quantum measurements~\cite{Gherardini2016NJP,Mueller2017ADP} with the aim to model the random interaction between the environment and the system within the framework of open quantum systems~\cite{Petruccione2003}. Indeed, randomness may appear in a measurement process, not only in the outcome of the measurement, but also in the time of its occurrence. Quantum measurements, at variance with classical measurements, are invasive and are accordingly accompanied by stochastic energy exchanges between the measurement apparatus and the measured system. In this work we shall adopt the convention to call such energy exchanges \emph{quantum-heat}~\cite{Elouard2016}, and denote it by the symbol $Q_q$, to distinguish it from the heat proper $Q$ (i.e., the energy exchanged with a thermal bath) and the work $W$ (i.e., the energy exchanged with a work source).

In this paper, we study the statistics of the energy exchanged between a quantum system and a measurement apparatus, under the assumption that the interaction can be modelled by a sequence of projective measurements occurring instantly and at random times. Our system does not interact with a thermal bath nor with a work source, hence the energy exchanges is all quantum-heat. Specifically, the following main results will be shown:
\begin{itemize}
\item
A direct consequence of previous studies~\cite{Campisi2010PRL,Campisi2011PRE} is that when the projective measurements occur at predetermined times, the Jarzynski equality of quantum-heat is obeyed. Here we observe that the same is true when there is randomness in the waiting time distribution between consecutive measurements. This can be understood based on the fact that the dynamics that dictate the evolution of the quantum system density matrix are unital~\cite{Albash2013,RasteginJSM2013,Kafri2012,Campisi17NJP19}. We investigate both the case when the randomness is distributed as a quenched disorder and as annealed disorder~\cite{Mezard1987}, for which we present the expression of the characteristic function.
\item
Our general analysis is illustrated for a repeatedly measured two-level system. We focus on the impact of randomness of waiting times on the average quantum-heat absorbed by the system. As compared with the case of no-randomness, the two-level system exchanges more quantum heat in the presence of randomness, when the average time between consecutive measurements is sufficiently small compared to its inverse resonance frequency. More quantum-heat is absorbed by the two level system when randomness is distributed as quenched noise as compared to annealed noise.
\item
Finally, we find that even an infinitesimal amount of randomness is sufficient to induce a non-null quantum-heat transfer when many measurements on the system Hamiltonian are performed.
\end{itemize}
These results have allowed us also to verify a phenomenon of \textit{noise-induced quantum-heat transfer} as the result of the presence of external (semi-classical) stochasticity. As further remark, it is worth pointing out how this formalism might be easily exploited even when some parameters of the Hamiltonian are fluctuating variables.

\subsection{\label{sec:model}Protocol of stochastic projective measurements}

We consider a quantum mechanical system $\mathcal{S}$ described by a finite dimensional Hilbert space $\mathcal{H}$. We assume that the system is initially at $t = 0^{-}$ in an arbitrary quantum state given a density matrix $\rho_0$. The system Hamiltonian $H$ is time-independent and reads:
\begin{equation}
H = \sum_{n}E_{n}|E_{n}\rangle\langle E_{n}|,
\label{eq:H}
\end{equation}
where $E_{n}$ and $|E_{n}\rangle$ are its eigenvalues and eigenstates, respectively. The eigenstates of $H$ are non-degenerate.

At time $t=0$ a first projective energy measurement occurs projecting the system in the state $\rho_{n} = \ket{E_{n}}\bra{E_{n}}$, with probability $p_{n} = \langle E_{n}|\rho_{0}|E_{n}\rangle$. Accordingly, the corresponding energy of the system at $t = 0^{+}$ is $E_{n}$. Afterwards, the system $\mathcal{S}$ is repeatedly subject to an arbitrary but fixed number $m$ of consecutive projective measurements of a generic observable $\mathcal O$
\begin{equation}\label{observable_o}
\mathcal{O} \equiv \sum_{k}o_{k}\Pi_{k},
\end{equation}
Here $o_{k}$'s are the possible outcomes of the observable $\mathcal{O}$, while the set $\{\Pi_{k}\}$ are the projectors belonging to the measured eigenvalues. The projectors are Hermitian and idempotent unidimensional operator satisfying the relations $\Pi_{k}\Pi_{l} = \delta_{kr}\Pi_{r}$ and $\sum_{k}\Pi_{k} = \mathbb{I}$. According to postulates of quantum measurement~\cite{Sakurai1994}, the state of the quantum system after a projective measurement is given by one of the projectors $\Pi_{k}$. We denote by $\tau_i$ the waiting time between the $(i-1)^\text{th}$ measurement and the $i^\text{th}$ of the observable $\mathcal O$. Between those measurements the system undergoes the unitary dynamics generated by its Hamiltonian (\ref{eq:H}), that is $\mathcal{U}(\tau_i)= e^{-iH\tau_i}$, where the reduced Planck's constant $\hbar$ has been set to unity. The waiting times $\tau_i$ are random variables and so is the total time $\mathcal{T} = \sum_{j = 1}^{M}\tau_{j}$, when the last, i.e. the $M^\text{th}$, measurement of $\mathcal O$ occurs. This is immediately followed by a second measurement of energy that projects the system on the state $\rho_{l} = \ket{E_{l}}\bra{E_{l}}$. The quantum-heat $Q_q$ absorbed by the system is accordingly:
\begin{equation}
Q_q = E_l - E_n
\end{equation}
In the following we shall adopt the notation $\vec{\tau}=(\tau_{1},\ldots,\tau_{m})$ for the sequence of waiting time distributions, and $\vec{k} = (k_{1},\ldots,k_{m})$ for the sequence of observed outcomes of the measurement of $\mathcal O$ in a realisation of the measurement protocol. Given the sequences $\vec{k},\vec{\tau}$, density matrix $\rho_n$ is mapped at time $\mathcal T$ into
\begin{equation}\label{rho_tau_meno}
\widetilde{\rho}_{n,\vec{k},\vec{\tau}} = \frac{\mathcal{V}(\vec{k},\vec{\tau})\rho_{n}\mathcal{V}^{\dagger}(\vec{k},\vec{\tau})}
{\mathcal{P}(\vec{k},\vec{\tau})},
\end{equation}
where  $\mathcal{V}(\vec{k},\vec{\tau})$ is the super-operator
\begin{equation}\label{eq:ubar}
\mathcal{V}(\vec{k},\vec{\tau}) \equiv \Pi_{k_{m}}\mathcal{U}(\tau_m) \cdots \Pi_{k_{1}}\mathcal{U}(\tau_1)
\end{equation}
and $\mathcal{P}(\vec{k},\vec{\tau}) \equiv {\rm Tr}\left[ \mathcal{V}(\vec{k},\vec{\tau})\rho_{n}\mathcal{V}^{\dagger}(\vec{k},\vec{\tau}) \right]$.

\subsection{Quantum-heat statistics}

$Q_{q}$ is a random variable due to the randomness inherent to measurements outcomes $\vec{k}$, stochastic fluctuations in the sequence of waiting times $\vec{\tau}$, as well as from the initial statistical mixture $\rho_0$.
Its statistics reads
\begin{equation}\label{eq:pheat}
P(Q_{q}) = \sum_{n,l}\delta(Q_{q}-E_{l} + E_{n})p_{l|n}~p_{n},
\end{equation}
where $p_{l|n}$ is the transition probability to obtain the final energy $E_{l}$ conditioned to have measured $E_{n}$ in correspondence of the first energy measurement. Denoting as $p_{l|n}(\vec{k},\vec{\tau})$ the probability to make a transition from $n$ to $l$, conditioned on the waiting time and outcomes sequences $\vec{\tau} ,\vec{k}$, the overall transition probability $p_{l|n}$ reads
\begin{equation}\label{conditional_prob_an}
p_{l|n} = \int\sum_{\vec{k}}d^{m}\vec{\tau}p(\vec{\tau})p_{l|n}(\vec{k},\vec{\tau}),
\end{equation}
where $p(\vec{\tau})$ is the joint distribution for the sequence of waiting times $\vec{\tau}$. The conditioned transition probability $p_{l|n}(\vec{k},\vec{\tau})$ is expressed in terms of the evolution super-operator $\mathcal{V}(\vec{k},\vec{\tau})$, i.e.
\begin{equation}
p_{l|n}(\vec{k},\vec{\tau}) = {\rm Tr}\left[\Pi_{l}\mathcal{V}(\vec{k},\vec{\tau})\Pi_{n}\mathcal{V}^{\dagger}(\vec{k},\vec{\tau})\Pi_{l}\right].
\end{equation}

The quantum-heat statistics is completely determined by the quantum-heat characteristic function
\begin{equation}\label{eq:G(u)}
G(u) \equiv \int P(Q_{q})e^{iuQ_{q}}dQ_{q},
\end{equation}
where $u\in\mathbb{C}$ is a complex number. Such characteristic function could be directly measured by means of Ramsey interferometry of single qubits~\cite{Dorner2013,MazzolaPRL2013,CampisiNJP2014}, or by means of methods from estimation theory~\cite{GherardiniEntropy}. Accordingly, plugging (\ref{conditional_prob_an}) into (\ref{eq:pheat}) the quantum-heat statistics becomes
\begin{small}
\begin{equation}\label{eq:pheat_2}
P(Q_{q}) = \int d^{m}\vec{\tau}  p(\vec{\tau})\sum_{n,\vec{k},l}{\rm Tr}\left[\Pi_{l}\mathcal{V}(\vec{k},\vec{\tau})\Pi_{n}\mathcal{V}^{\dagger}(\vec{k},\vec{\tau})\Pi_{l}\right]p_{n}
\end{equation}
\end{small}
Furthermore, substituting (\ref{eq:pheat_2}) in the definition (\ref{eq:G(u)}) and using ${\rm Tr}\left[\Pi_{l}\mathcal{V}\Pi_{n}\mathcal{V}^{\dagger}\Pi_{l}\right]=\bra{E_{l}}\mathcal{V}\ket{E_{n}}\bra{E_{n}}\mathcal{V}^{\dagger}\ket{E_{l}}$ we obtain
\begin{equation}
G(u) = \int d^{m}\vec{\tau}  p(\vec{\tau})\sum_{n,\vec{k},l}
\bra{E_{l}}\mathcal{V}\ket{E_{n}}\bra{E_{n}}\rho_{0}\ket{E_{n}}\cdot\bra{E_{n}}e^{-iuH}\mathcal{V}^{\dagger}e^{iuH}\ket{E_{l}}.
\end{equation}
Finally, being $e^{iuE_{l}}\ket{E_{l}} = e^{iuH}\ket{E_{l}}$ and $\bra{E_{n}}e^{-iuE_{n}} = \bra{E_{n}}e^{-iuH}$, we obtain
\begin{equation}\label{G_u_finale}
G(u) = \overline{\left\langle {\rm Tr}\left[e^{iuH}\mathcal{V}(\vec{k},\vec{\tau})e^{-iuH}\rho_{0}\mathcal{V}^{\dagger}(\vec{k},\vec{\tau})\right] \right\rangle}
\end{equation}
where the angular brackets mean quantum-mechanical expectation $\langle \cdot \rangle = {\rm Tr} (\cdot) \rho_0$, and the overline stands for the average over noise realisations $\overline{(\cdot)}=\int d^m\vec{\tau}p(\vec{\tau}) (\cdot)$.

In the special case when there is no randomness in the waiting times, i.e. if $p(\vec{\tau}) = \delta^m(\vec{\tau} - \vec{\tau}_0)$, where $\vec{\tau}_0 \equiv (\tau_0, \tau_0, \dots, \tau_0)$ and $\delta^m(\vec{x})$ denotes the $m$-dimensional Dirac delta, the characteristic function $G(u)$ reduces to
\begin{equation}\label{eq:G(u)_regular}
G(u) = \sum_{\vec{k}}
{\rm Tr}\left[e^{iuH}\mathcal{V}(\vec{k},\vec{\tau}_0)e^{-iuH}\rho_{0}\mathcal{V}^{\dagger}(\vec{k},\vec{\tau}_0)\right],
\end{equation}
in agreement with the expression in Ref.~\cite{Yi2013}.

The statistical moments of the quantum-heat are obtained, by following the general rule, from the derivatives of the quantum-heat generating function, according to the formula
\begin{equation}\label{work_moments}
\overline{\langle Q^{n}_{q}\rangle }=
\left.(-i)^{n}\partial^{n}_{u}G(u)\right|_{u=0},
\end{equation}
where $\partial^{n}_{u}$ denotes the $n-$th partial derivative with respect to $u$. Explicit expressions for $G(u)$ and $\overline{\langle Q^{n}_{q}\rangle}$ will be derived in the following section for the paradigmatic case of a two-level quantum system.

As a side remark we observe that, since the characterization of the measurement operators is encoded in the super-operator $\mathcal{V}(\vec{k},\vec{\tau})$, Eq.~(\ref{G_u_finale}) is valid also when a protocol of POVMs (excluding the first and the last measurements, performed on the energy basis) is applied to the quantum system. In such a case, the measurement projectors $\Pi_k$ are replaced by a set of Kraus operators $\{\mathcal{B}_{l}\}$, such that $\sum_{l}\mathcal{B}_{l}^{\dagger}\mathcal{B}_{l} = \mathbb{I}$.

\subsection{Fluctuation Relation}

It is a known fact that, when a quantum system is subject to a time dependent forcing protocol and as well to a predetermined number of quantum projective measurements occurring at predetermined times $\vec{\tau}$, the following holds (Jarzynski equality):
\begin{equation}
\langle e^{-\beta_{T}(\tilde{E}_l-E_n)}\rangle= e^{-\beta_{T}\Delta F},
\end{equation}
where $\tilde{E}_l$ are the final eigenvalues of the time-dependent system Hamiltonian $H(t)$, $\Delta F \equiv - \beta_{T}^{-1}\ln {\rm Tr}[e^{-\beta_{T} H(\mathcal T)}]/ {\rm Tr}[e^{-\beta_{T} H(0)}]$ denotes the free-energy difference, and the initial state of the system has the Gibbs form $\rho_0=e^{-\beta_{T} H(0)} / {\rm Tr}[e^{-\beta_{T} H(0)}]$ \cite{Campisi2011PRE}. $Z \equiv {\rm Tr}[e^{-\beta_{T} H(0)}]$ is also called partition function. If turning off the time-dependent forcing, as in the present investigation, this implies that with \textit{fixed} waiting times $\vec{\tau}$ one has:
\begin{equation}\label{eq:av-exp(-Qq)}
\langle e^{-\beta_{T} Q_q}\rangle = 1,
\end{equation}
because, without driving, all the energy change in the quantum system can be ascribed to quantum-heat and, being the Hamiltonian time-independent, in that case $\Delta F = 0$. For the sake of clarity, we recall that the notation $\langle e^{-\beta_{T} Q_q}\rangle$ denotes a purely quantum-mechanical expectation with fixed waiting time sequence $\vec{\tau}$.

However, as main result, we can easily prove that this continues to hold also if the times between consecutive measurements are random. Indeed, using (\ref{G_u_finale}), we obtain
\begin{align}
&\overline{\langle e^{-\beta_{T} Q_q} \rangle} = G(i\beta_{T}) \nonumber \\
& = \int d^{m}\vec{\tau}  p(\vec{\tau})\sum_{\vec{k}}
{\rm Tr}\left[e^{-\beta_{T} H}\mathcal{V}(\vec{k},\vec{\tau})e^{\beta_{T} H}\frac{e^{-\beta_{T} H}}{Z}\mathcal{V}^{\dagger}(\vec{k},\vec{\tau})\right] \nonumber \\
&={\rm Tr}\left[\frac{e^{-\beta_{T} H}}{Z} \int d^{m}\vec{\tau}  p(\vec{\tau}) \sum_{\vec{k}}\mathcal{V}(\vec{k},\vec{\tau})\mathcal{V}^{\dagger}(\vec{k},\vec{\tau})\right] =
\frac{{\rm Tr}\left[e^{-\beta_{T} H}\right]}{Z} = 1,
\label{G_u_finale_appendice}
\end{align}
where we have used the property
\begin{align}
\int d^{m}\vec{\tau}p(\vec{\tau}) \sum_{\vec{k}}\mathcal{V}(\vec{k},\vec{\tau})\mathcal{V}^{\dagger}(\vec{k},\vec{\tau})=\mathbb{I},
\end{align}
which follows from the normalisation $\int d^{m}\vec{\tau}  p(\vec{\tau})=1$, idempotence of projectors $\Pi_k\Pi_k=\Pi_k$, ciclyicity of the trace operation, and the unitarity of the quantum evolutions between consecutive measurements. Its mathematical significance is that the quantum channel that describes the unconditioned evolution from $t=0$ to $t=\mathcal T$
\begin{align}
\rho \mapsto \int d^{m}\vec{\tau}  p(\vec{\tau}) \sum_{\vec{k}}\mathcal{V}(\vec{k},\vec{\tau})\, \rho \, \mathcal{V}^{\dagger}(\vec{k},\vec{\tau})
\end{align}
is \emph{unital}, i.e. it has the identity $\mathbb{I}$ as a fixed point. It is this mathematical property that ensures the validity of the fluctuation relation (\ref{G_u_finale}) \cite{Albash2013,RasteginJSM2013,Kafri2012,Campisi17NJP19}.

The fluctuation relation (\ref{G_u_finale}) can also be understood by noticing that, from Eq. (\ref{eq:av-exp(-Qq)}), it is $\langle e^{-\beta_{T} Q_q}\rangle = 1$, in which the average is restricted to the sole realisations where the sequence $\vec{\tau}$ occurs. The double average remains therefore equal to one:
$\overline{\langle e^{-\beta_{T} Q_q} \rangle} = \int d^m{\vec \tau} p(\vec{\tau})\langle e^{-\beta_{T} Q_q}\rangle=1$. Accordingly, we have shown, from one side, that the fluctuation relation is \textit{robust} against the presence of randomness in the waiting times $\vec{\tau}$, and, on the other side, that such stochasticity shall not be a-posterior revealed by a measure of $\overline{\langle e^{-\beta_{T} Q_q} \rangle}$ with $\rho_0$ Gibbs thermal state, whatever are the values assumed by $\vec{\tau}$ and $p(\vec{\tau})$.

Moreover, from an experimental point of view, $\overline{\langle e^{-\beta_{T} Q_{q}}\rangle}$ can be obtained by repeating for a sufficiently large number $N$ of times the foregoing protocol of projective measurements, so that
\begin{equation}
\overline{\langle e^{-\beta_{T} Q_{q}}\rangle} = \frac{1}{N}\sum_{j=1}^{N}e^{-\beta_{T} Q_{q}^{(j)}},
\end{equation}
where $Q_{q}^{(j)}$ is the value of quantum-heat, which is measured after the $j-$th repetition of the experiment.

\subsection{Noise-induced quantum heat transfer}

Below, we will analyze in detail $\overline{\langle e^{-\beta_{T} Q_{q}}\rangle}$ and the mean quantum-heat $\overline{\langle Q_{q}\rangle}$, when a \textit{stochastic} sequence of projective quantum measurements is performed on a two-level-system. Let $E_{+}$ and $E_{-}$ denote its two energy eigenvalues. We assume the initial density matrix is diagonal in the energy eigenbasis:
\begin{equation}\label{eq:rho_0_2_level_system}
\rho_{0} = c_{1}\ket{E_{+}}\bra{E_{+}} + c_{2}\ket{E_{-}}\bra{E_{-}},
\end{equation}
with $c_1,c_2 \in[0,1]$ and $c_2=1 - c_1$. We denote the eigenstates of the measured observable $\mathcal O$ as $\{|\alpha_{j}\rangle\}$, $j = 1,2$, so that is $\Pi_j = |\alpha_{j}\rangle \langle \alpha_j|$. They can be generally expressed as a linear combination of the  energy eigenstates, i.e.
\begin{equation}
\begin{split}
&\ket{\alpha_1}=a\ket{E_{+}}-b\ket{E_{-}}\\
&\ket{\alpha_2}=b\ket{E_{+}}+a\ket{E_{-}}
\label{eq:abasis}
\end{split}
\end{equation}
where $a,b\in\mathbb{C}$, $|a|^2+|b|^2=1$ and $a^{\ast}b = ab^{\ast}$.

\subsubsection{Fixed waiting times sequence}

We begin by considering the standard case where the waiting time $\overline{\tau}$ between two consecutive measurements is constant. In this case
$p(\vec{\tau}) = \prod_{i=1}^m \delta(\tau_i-\bar\tau)$, where $\delta(\cdot)$ denotes the Dirac delta. By computing the characteristic function (\ref{eq:G(u)_regular}) in $u = i\beta_{T}$ for the two-level system, we obtain
\begin{eqnarray}\label{G_i}
G(i\beta_{T})&=&\begin{pmatrix}
|a|^{2}e^{-\beta_{T} E}+|b|^{2}e^{\beta_{T} E}\\
|a|^{2}e^{\beta_{T} E} + |b|^{2}e^{-\beta_{T} E}
\end{pmatrix}'\cdot\begin{pmatrix}
1-\overline{\nu} & \overline{\nu} \\ \overline{\nu} & 1-\overline{\nu}
\end{pmatrix}^{m-1}\nonumber \\
&\cdot&\begin{pmatrix}
|a|^{2}c_{1}e^{\beta_{T} E}+|b|^{2}c_{2}e^{-\beta_{T} E}\\
|a|^{2}c_{2}e^{-\beta_{T} E} + |b|^{2}c_{1}e^{\beta_{T} E}
\end{pmatrix},
\end{eqnarray}
where the transition probability $\overline{\nu} = \nu(\overline{\tau})$ is expressed in terms of the function
\begin{equation}
\nu(t) \equiv |\bra{\alpha_{2}}\mathcal{U}(t)\ket{\alpha_{1}}|^2=|\bra{\alpha_{1}}\mathcal{U}(t)\ket{\alpha_{2}}|^2 = 2|a|^{2}|b|^{2}\sin^{2}(2 t E),
\end{equation}
The explicit calculation is reported in the Appendix.

In Fig. \ref{fig:QJI} we report the quantity $G(i\beta_{T})= \langle e^{-i\beta_{T} Q_q} \rangle$ as a function of $c_1$ for various values of $a$, which have been chosen to be real. We first observe that $G(i\beta_{T})$ is a linear function of $c_1$. This is confirmed by the numerical simulations of $\langle e^{-i\beta_{T} Q_q} \rangle $ from the underlying protocol, which is in agreement with the analytical formula (\ref{G_i}), except for some finite size errors.
\begin{figure}[h!]
\centering
\includegraphics[scale = 0.62]{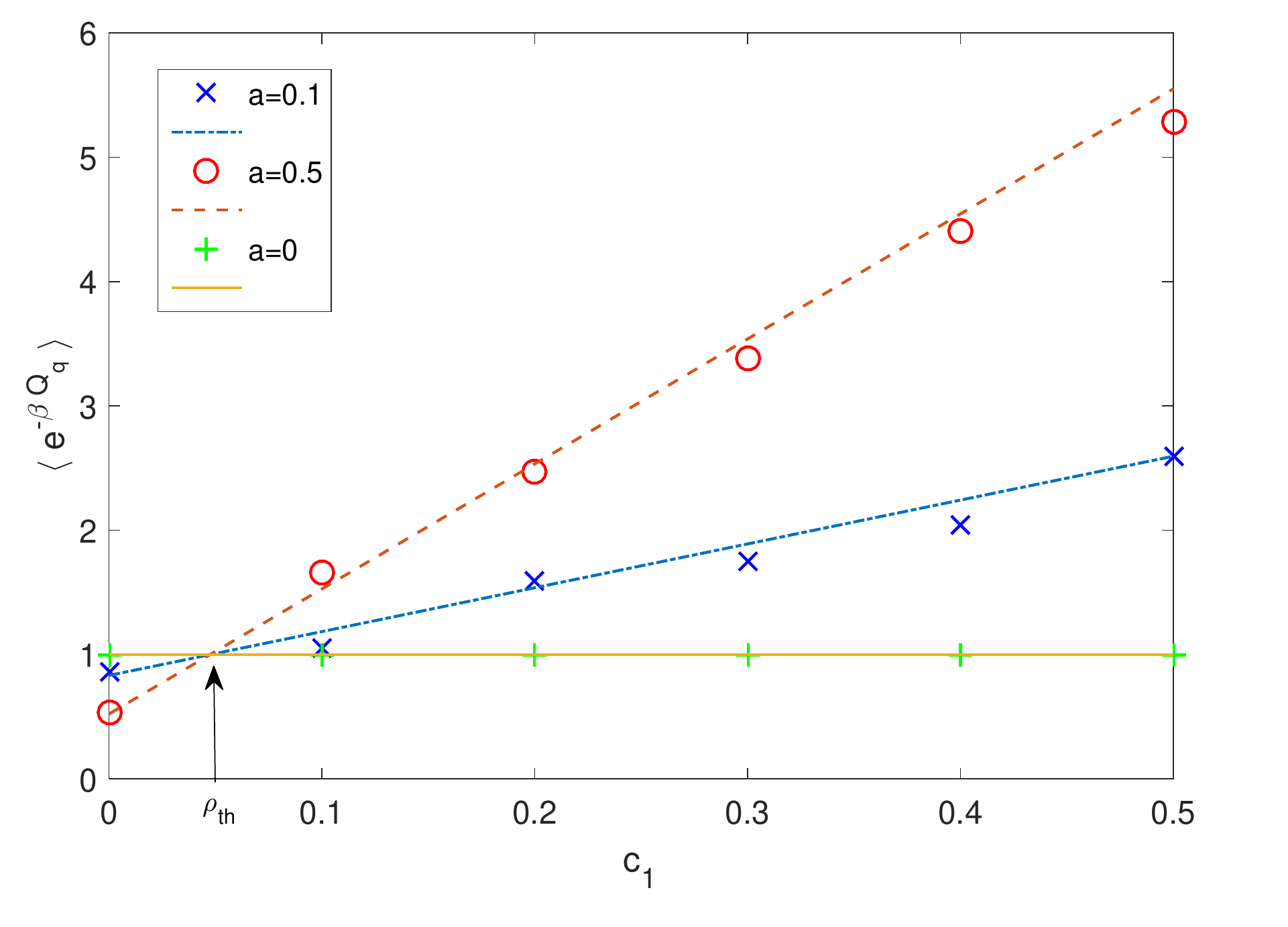}
\caption{Analytic form of $G(i\beta_{T})$ (solid yellow, dotted blue and dashed red lines) as a function of $c_{1}$, which parameterizes the initial density matrix of the system, for three real values of $a$ (respectively, $a = 0, 0.1, 0.5$). The analytical predictions are compared with the numerical simulations (green crosses, blue x-marks and red circles). The simulations have been performed by applying protocols of $m = 5$ projective measurements, averaged over 1000 realizations in order to numerically derive the mean of the exponential of work, with $E_{\pm} = \pm 1$. The point, in which all the analytical lines are crossing, corresponds to the initial thermal state $\rho_0=e^{-\beta_{T} H}/Z$ with $\beta_{T} = 1$.}
\label{fig:QJI}
\end{figure}
We further observe that, for an arbitrary value of $a$, $G(i\beta_{T})$ is identically equal to $1$ in correspondence of the value of $c_{1}$ for which $\rho_0=e^{-\beta_{T} H}/Z$, in agreement with Eq.~(\ref{eq:av-exp(-Qq)}). In Fig.~\ref{fig:QJI} such condition is realized in the point where all the analytical lines are crossing.

\subsubsection{Stochastic waiting times sequence}

\textit{Quenched disorder:} By quenched disorder it is meant that the time between consecutive measurements within a given sequence is fixed and only varies between distinct sequences.
The joint distribution $p(\vec{\tau})$ reads $p(\vec{\tau}) = p(\tau_1) \prod_{i=2}^m \delta(\tau_i-\tau_1)$. In other words, only the first waiting time of a sequence is chosen randomly from $p(\tau)$ and then that waiting time repeats within the sequence. For the sake of simplicity, we assume that $p(\tau)$ is a bimodal probability density function, with values $\tau^{(1)}$, $\tau^{(2)}$ and probabilities $p_{1}$ and $p_{2} = 1 - p_{1}$. Accordingly, from Eq. (\ref{G_u_finale}) we have that
\begin{footnotesize}
\begin{eqnarray}\label{eq:G_quenched_2_LS}
G(i\beta_{T})&=&\begin{pmatrix}
|a|^{2}e^{-\beta_{T} E}+|b|^{2}e^{\beta_{T} E}\\
|a|^{2}e^{\beta_{T} E} + |b|^{2}e^{-\beta_{T} E}
\end{pmatrix}'\cdot\left[\displaystyle{\sum_{j = 1}^{d_{\tau}}}
\begin{pmatrix}
1-\nu(\tau^{(j)}) & \nu(\tau^{(j)}) \\
\nu(\tau^{(j)}) & 1-\nu(\tau^{(j)})
\end{pmatrix}^{m-1}p_{j}\right]\nonumber \\
&\cdot&\begin{pmatrix}
|a|^{2}c_{1}e^{\beta_{T} E}+|b|^{2}c_{2}e^{-\beta_{T} E}\\
|a|^{2}c_{2}e^{-\beta_{T} E} + |b|^{2}c_{1}e^{\beta_{T} E}
\end{pmatrix}
\end{eqnarray}
\end{footnotesize}
where $d_{\tau} = 2$ is the number of values that can be assumed by the random variable $\tau$.

\textit{Annealed disorder:} By annealed disorder it is meant that the waiting times, $(\tau_{1},\ldots,\tau_{m})= \vec{\tau}$ are random variables sampled from one and the same probability distribution $p(\tau)$. Accordingly, the joint distribution of the waiting times is $p(\vec{\tau}) = \prod_{j=1}^{m}p(\tau_{j})$. Assuming $p(\tau)$ to be bimodal as above,
the characteristic function at $u=i\beta_{T}$ reads (see Appendix):
\begin{footnotesize}
\begin{eqnarray}\label{eq:G_annealed_2LS}
G(i\beta_{T})&=&\begin{pmatrix}
|a|^2e^{-\beta_{T} E}+|b|^2e^{\beta_{T} E}\\
|a|^2e^{\beta_{T} E}+|b|^2e^{-\beta_{T} E}
\end{pmatrix}'\cdot\left[\displaystyle{\sum_{j = 1}^{d_{\tau}}}\begin{pmatrix}
1-\nu(\tau^{(j)}) & \nu(\tau^{(j)}) \\ \nu(\tau^{(j)}) & 1-\nu(\tau^{(j)})
\end{pmatrix}p_{j}\right]^{m-1}\nonumber \\
&\cdot&\begin{pmatrix}
|a|^2c_{1}e^{\beta_{T} E}+|b|^2c_{2}e^{-\beta_{T} E}\\
|a|^2c_{2}e^{-\beta_{T} E}+|b|^2c_{1}e^{\beta_{T} E}
\end{pmatrix}
\end{eqnarray}
\end{footnotesize}
\begin{figure}[h!]
\centering
\includegraphics[scale = 0.675]{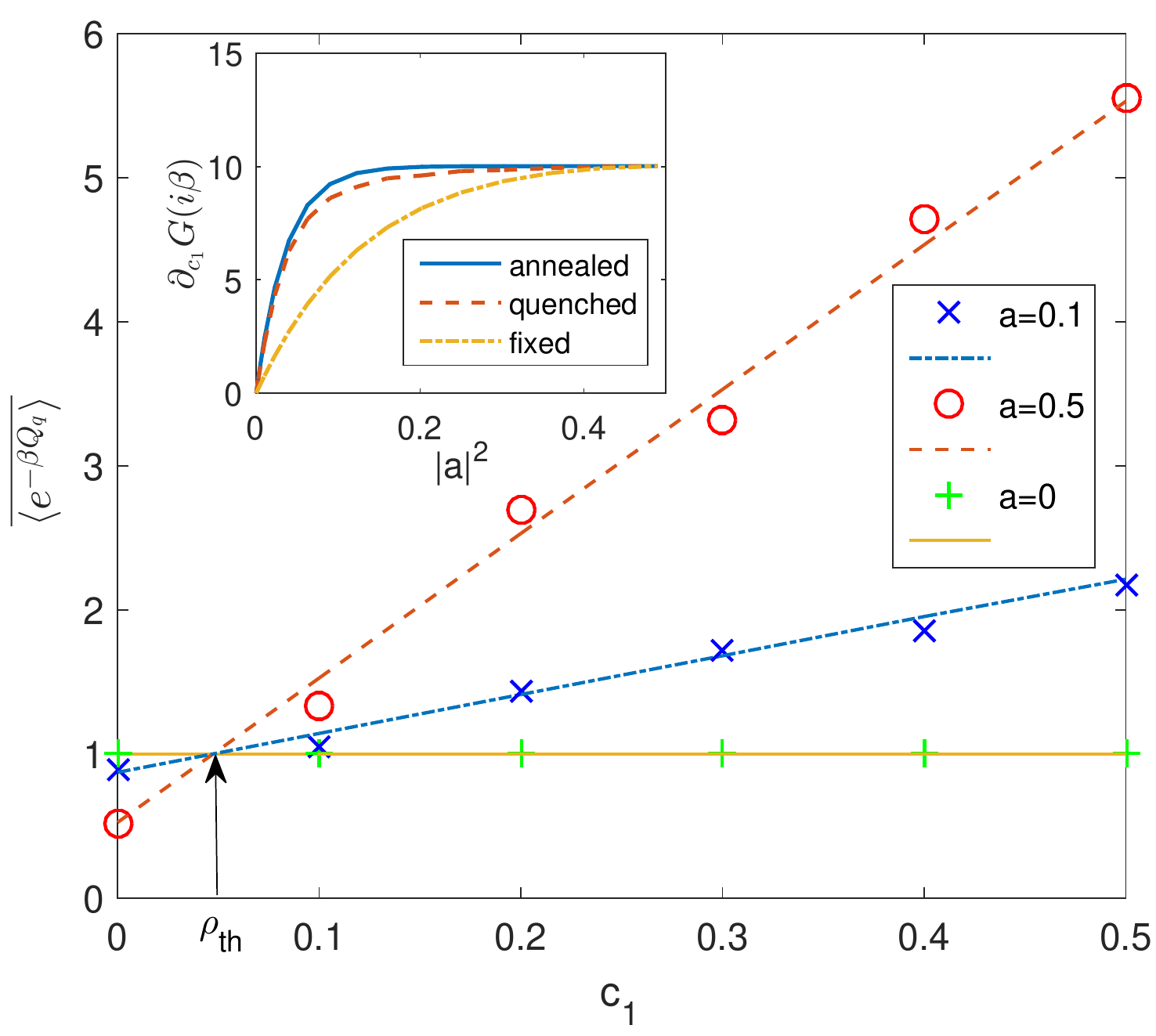}
\caption{Plot of $G(i\beta_{T})$ (solid yellow, dotted blue and dashed red lines) as a function of $c_{1}$ for three real values of $a$ ($a = 0, 0.1, 0.5$, respectively). In this case, the stochasticity in the time intervals between measurements is distributed as annealed disorder. Again the analytical predictions are compared to the numerical simulations (green crosses, blue x-marks and red circles) for the three values of $a$. Also in this case, the point in which all the lines are crossing corresponds to the thermal state. Inset: Slope of $G(i\beta_{T})$ as a function of $c_1$, i.e. $\partial_{c_1}G(i\beta_{T})$, for different values of the parameter $|a|^2$ with resolution of $|a| = 0.05$. The curves have been performed by applying protocols of $m = 5$ projective measurements, averaged over 1000 realizations, with $E_{\pm} = \pm 1$ and $\beta_{T} = 1$. Instead, for $p(\tau)$ we have chosen a bimodal probability density function, with values $\tau^{(1)}=0.01$, $\tau^{(2)}=3$ and $p_{1}=0.3$.}
\label{fig:QJIrand}
\end{figure}
In Fig.~\ref{fig:QJIrand} we plot it as a function of $c_{1}$. The presence of the disorder does not affect the linear dependence of $G(i\beta_{T})$ on  $c_1$, and it still equals $1$ in correspondence of the initial state to be thermal with temperature $1/\beta_{T}$. What the stochasticity effectively changes is the slope of $G(i\beta_{T})$ when it is plotted as a function of $c_1$. In this regard, in the inset of Fig.~\ref{fig:QJIrand} we show how the slope of $G(i\beta_{T})$ as a function of $c_1$, given the partial derivative of $G(i\beta_{T})$ with respect to $c_1$ (i.e. $\partial_{c_1}G(i\beta_{T})$), depends on $|a|^2$ for both the case of fixed and stochastic waiting times sequence with $m = 5$ measurements. The values of $\partial_{c_1}G(i\beta_{T})$ are identically equal when $|a|^2 = 0$ and $0.5$, and in the range $0 \leq |a|^2 \leq 0.5$ they are symmetric with respect to the ones in the range $0.5 \leq |a|^2 \leq 1$.

\subsubsection{Mean quantum-heat}

By substituting $u = 0$ in $\partial_{u}G(u)$ (in the appendix, we show the analytical expression of the $n-$th partial derivative of $G(u)$ for the two-level system), we find the mean value $\overline{\langle Q_{q}\rangle}$, which is a linear function in the parameter $c_1$ both in the ordered and the stochastic case. In particular,
\begin{equation}
\overline{\langle Q_{q}\rangle} = -\phi\left[2c_1 - 1\right],
\label{eq:heat_normal}
\end{equation}
where
\begin{equation}
\phi \equiv E\left[1 - \overline{\lambda(\tau)}\right].
\end{equation}
Accordingly, $\phi$ depend on the average (w.r.t. the values that can be assumed by the waiting time $\vec{\tau}$ in a given sequence of the protocol according to $p(\vec{\tau})$) of the parameter $\lambda(\tau)$, which is given by the following relation:
\begin{equation}
\lambda(\tau) = (1 - 2|a|^{2})^{2}(1 - 2\nu(\tau))^{m-1}\leq 1.
\end{equation}
Being $\phi\geq 0$, the maximum value of $\overline{\langle Q_{q}\rangle}$, i.e. $\overline{\langle Q_{q}\rangle}_{\rm max}$, occurs at $\overline{\langle Q_{q}\rangle} = \phi$ when $c_1 = 0$; while $\overline{\langle Q_{q}\rangle} = 0$ when $c_1 = 1/2$ for any value of $m$, $a$ and $p(\vec{\tau})$. Moreover, when $a=0$ or $a=1$ then $\overline{\langle Q_{q}\rangle} = 0$. This can be understood by noticing that the condition $a=0,1$ implies that the measured observable $\mathcal O$ coincides with the system Hamiltonian. In this case, the system after the initial projection onto the state $|E_{\pm}\rangle$ only acquires a phase during the free evolution while the subsequent measurements have no effect on the state. Accordingly the quantum-heat would be always null $Q_{q} = 0$ and so will be its average.

For a sequence of measurements at fixed times $\overline{\lambda(\tau)} = \lambda(\overline{\tau})$, while in the quenched and annealed disorder instance it is respectively equal to
\begin{small}
\begin{equation}
\overline{\lambda(\tau)}^{(\rm qu)} = \sum_{j = 1}^{d_{\tau}}\lambda(\tau^{(j)})p_j = (1 - 2|a|^2)^{2}\sum_{j = 1}^{d_{\tau}}[1 - 2\nu(\tau^{(j)})]^{m-1}p_j
\end{equation}
\end{small}
and
\begin{equation}
\overline{\lambda(\tau)}^{(\rm an)} = (1 - 2|a|^2)^{2}\left[\displaystyle{\sum_{j=1}^{d_{\tau}}}[1-2\nu(\tau^{(j)})]p_{j}\right]^{m-1}.
\end{equation}
Thus, we will denote the mean quantum-heat in such cases respectively as $\overline{\langle Q_{q}\rangle}^{(\rm qu)}$ and $\overline{\langle Q_{q}\rangle}^{(\rm an)}$. In general, by changing the initial density matrix $\rho_{0}$ (i.e. $c_1$), the parameter $a$ (related to the measurement bases) or the number $m$ of measurements, the mean value of the quantum-heat can assume a value within the range $[-\phi,\phi]$; and when the initial state is thermal then $\overline{\langle Q_{q}\rangle} = \beta_{T} E(1 - \overline{\lambda(\tau)})\tanh(\beta_{T} E)$, as shown also in Ref. \cite{Yi2013} for a sequence of measurements at fixed times.

Let us observe that $\overline{\langle Q_{q}\rangle} \geq 0$ if $0\leq c_1\leq 0.5$, while it is always $\leq 0$ for $0.5\leq c_1\leq 1$. These two conditions correspond to two distinct regimes: quantum-heat absorption by the two-level system and quantum-heat emission. Then, being $\overline{\langle Q_{q}\rangle}$ a linear function passing through $c_1 = 1/2$, we can study the quantum-heat transfer (heat absorption/emission) by comparing the absolute value of the maximum quantum-heat, i.e. $\left|\overline{\langle Q_{q}\rangle}_{\rm max}\right| = \phi$, for sequence of measurements at fixed and stochastic times. This implies to analyze which is the relations between $\lambda(\overline{\tau})$, $\overline{\lambda(\tau)}^{(\rm qu)}$ and $\overline{\lambda(\tau)}^{(\rm an)}$. We find that
\begin{equation}\label{inequalities_quenched}
\left|\overline{\langle Q_{q}\rangle}^{({\rm qu})}\right| \geq \left|\langle Q_{q}\rangle\right| \Longleftrightarrow (1-2\nu)^{m-1} \geq \overline{[1-2\nu(\tau)]^{m-1}}^{({\rm qu})}.
\end{equation}
and
\begin{equation}
\left|\overline{\langle Q_{q}\rangle}^{({\rm an})}\right| \geq \left|\overline{\langle Q_{q}\rangle}^{({\rm qu})}\right|,
\end{equation}
being $\overline{\lambda(\tau)}^{(\rm an)} \leq \overline{\lambda(\tau)}^{(\rm qu)}$. Eq.~(\ref{inequalities_quenched}) sets the condition allowing for the transfer on average of a greater amount of quantum-heat under the case of quenched noise as compared to the case of no noise. To better understand its physical meaning, let us consider $\tau^{(j)}\Delta E \ll 1$, $j = 1,2$. We derive that
\begin{equation}\label{inequalities_quenched_2}
\left|\overline{\langle Q_{q}\rangle}^{({\rm qu})}\right| \geq \left|\langle Q_{q}\rangle\right| \Longleftrightarrow \overline{\tau}^{2} \geq \overline{\tau^2},
\end{equation}
where $\overline{\tau^2}$ is the second statistical moment of $p(\tau)$. If the condition (\ref{inequalities_quenched_2}) is not verified, then the application of a sequence of measurements at fixed times will lead to a greater amount of transferred quantum-heat. Instead, for a given choice of $p(\tau)$ and total number of measurements $m$ more quantum-heat is absorbed/emitted by the two-level system in the case of annealed noise as compare to the quenched noise case. This agrees with the intuition that the system should heat-up more in case it is subject to higher noise, and the annealed disorder is ``more noisy'' than the quenched one. This evidences a phenomenon of \textit{noise-induced quantum-heat transfer} which we will be investigating further elsewhere.

As final remark, it is worth mentioning that in recent studies on stochastic quantum Zeno dynamics \cite{GherardiniQST,GherardiniSciRep}, it has been shown that the survival probabilities that the system remains frozen in its initial state after performing ordered and stochastic sequences of measurements behave in the opposite way: the better the Zeno confinement is, the less quantum-heat is transferred by the system.

\subsubsection{The $m\rightarrow \infty$ limit}

For $m\rightarrow\infty$ the characteristic function tends to $G_\infty(u) = (1+e^{2iuE})/2 - c_{1}\sinh(2iuE)$ for each value of $a\neq 0$ and is exactly equal to $1$ for $|a|^2=0,1$.. That is the $m\rightarrow \infty$ asymptotic characteristic function $G_\infty(u)$ presents a discontinuity at $|a|^2=0,1$. Such a discontinuity is present also in the mean quantum-heat $\overline{\langle Q_{q}\rangle}$: when $|a|^2\rightarrow 0,1$ and $m$ is finite, $\overline{\langle Q_{q}\rangle}\rightarrow 0$ for any value of $c_1$, while for $m\rightarrow\infty$ and $|a|^2 \neq 0,1$ we get $\overline{\langle Q_{q}\rangle}\rightarrow E(1 - 2c_1) = \overline{\langle Q_{q}\rangle}_{\infty}$. In this way, the $m\rightarrow \infty$ asymptotic mean quantum-heat $\overline{\langle Q_{q}\rangle}_\infty$ can be easily expressed in terms of the $m \rightarrow \infty$ asymptotic characteristic function $G_\infty(u)$, so that
\begin{equation}
G_\infty(u) = \frac{\sinh(2iuE)}{E}\overline{\langle Q_{q}\rangle}_\infty + [\cosh(2iuE) + 1].
\end{equation}

The existence of this discontinuity is a mathematical feature that is physically relevant when one performs many measurements ($m\rightarrow\infty$) of the Hamiltonian ($|a|^2\rightarrow 0,1$). Perfect measurements of the Hamiltonian ($|a|^2\rightarrow 0,1$) are accompanied by null quantum heat $\overline{\langle Q_{q}\rangle}$, however even an infinitesimal amount of noise in the measurement process will result, in the limit of many measurements, in the finite amount of quantum-heat $\overline{\langle Q_{q}\rangle}_\infty=E(1 - 2c_1)$. Note that the latter is positive (negative) if the initial state is at positive (negative) temperature $c_1 >(<) c_2$.

\section{Stochastic quantum entropy production}

In this section, we discuss the results obtained in \cite{GherardiniEntropy}, about the reconstruction of the stochastic quantum entropy production from a quantum system in interaction with the external environment. One of the major goals of the \textit{quantum thermodynamics} is the definition and characterization of irreversibility in quantum processes. This could have a significant impact on technological applications for the possibility of producing work with heat engines at high efficiency using systems where quantum fluctuations are important. In this regard, several studies have shown how to derive the quantum version of the fluctuation-dissipation theorem, both for closed and open quantum systems~\cite{Kurchan2001,Campisi2009,Campisi2010PRL,Kafri2012,RasteginJSM2013,Albash2013,Manzano2015}, and recently, in~\cite{Aberg2016} a fully quantum fluctuation theorem has been formulated, explicitly including the reservoir exchanging energy with the system and a control system driving its dynamics.

Considerable efforts have been made in measuring irreversibility, and, consequently, the stochastic entropy production in quantum thermodynamics~\cite{DeffnerPRL2011,BatalhaoPRL2015,Brunelli2016}. The ratio between the probability to observe a given quantum trajectory and its time reversal is related to the amount of heat exchanged by the quantum system with the environment. Lately it has been experimentally proved, moreover, that irreversibility in quantum non-equilibrium dynamics can be partially rectified by the presence of an \textit{intelligent observer}, identified by the well-known Maxwell's demon~\cite{GooldJPA2016}, which manages to assess additional microscopic information degrees of freedom due to a proper feed-forward strategy~\cite{CamatiPRL2016}. As previously introduced an shown in \cite{Dorner2013,MazzolaPRL2013,CampisiNJP2014,GooldPRE2014,FuscoPRX2014,Peterson2016}, the reconstruction of the fluctuation properties of general thermodynamical quantities for open quantum systems can be well-performed by adopting an interferometric setting for the measurement of the characteristic function of the work distribution.

In this section we will mainly address the following three issues:
\begin{itemize}
  \item We discuss how to relate the stochastic entropy production to the quantum fluctuation theorem, generalizing the Tasaki-Crooks theorem for open systems. This relation is obtained via the evaluation of the irreversibility of the quantum dynamics, hence highlighting the quantum counterpart of the second law of the thermodynamics at zero temperature.
  \item We propose a procedure to reconstruct the stochastic entropy production of an open quantum system by performing repeated two-time measurements, at the initial and final times of the system transformation. In particular, we will present a novel measurement scheme, that relies on quantum estimation theory~\cite{ParisBook}, able to infer the work and entropy statistics with a minimal number of measurements. The proposed algorithm requires to determine the characteristic functions of the stochastic quantum entropy distribution, so that, by adopting a parametric version of the integral quantum fluctuation theorem, we can derive the corresponding statistical moments. We will show, moreover, that the number of the required measurements scales linearly with the system size.
  \item
      By assuming that the quantum system is bipartite, we apply the reconstruction procedure both for the two subsystems and for the composite system by performing measurements, respectively, on local and global observables. The comparison between the local and the global quantity will allow us to probe the presence of correlations between the partitions of the system
\end{itemize}

\subsection{Quantum fluctuation theorem}

The fluctuations of the stochastic quantum entropy production obey the quantum fluctuation theorem, that can be derived by evaluating the forward and backward protocols for a non-equilibrium process according to a two-time quantum measurement scheme. To this end, let us consider an open quantum system that undergoes a transformation in the interval $[0,\mathcal{T}]$ consisting of measurement, dynamical evolution and second measurement. We call this forward process and then study also its time-reversal, which we call backward process:
\begin{eqnarray*}
\text{FORWARD}:~\rho_{0}\underbrace{\longmapsto}_{\{\Pi^{\textrm{in}}_{m}\}}\rho_{\textrm{in}}\underbrace{\longmapsto}_{\Phi}\rho_{\textrm{fin}}
\underbrace{\longmapsto}_{\{\Pi^{\textrm{fin}}_{k}\}}\rho_{\mathcal{T}} \\
\text{BACKWARD}:~\widetilde{\rho}_{\mathcal{T}}\underbrace{\longmapsto}_{\{\widetilde{\Pi}^{\textrm{ref}}_{k}\}}\widetilde{\rho}_{\textrm{ref}}
\underbrace{\longmapsto}_{\widetilde{\Phi}}\widetilde{\rho}_{\textrm{in}'}\underbrace{\longmapsto}_{\{\widetilde{\Pi}^{\textrm{in}}_{m}\}}\widetilde{\rho}_{0'}
\end{eqnarray*}
At time $t=0^-$ the system is prepared in a state $\rho_{0}$ and then subjected to a measurement of the observable
$$
\mathcal{O}_{\textrm{in}} = \sum_{m}a^{\textrm{in}}_{m}\Pi^{\textrm{in}}_{m},
$$
where $\Pi^{\textrm{in}}_{m}\equiv|\psi_{a_{m}}\rangle\langle\psi_{a_{m}}|$ are the projector operators given in terms of the eigenvectors $|\psi_{a_{m}}\rangle$ associated to the eigenvalues $a^{\textrm{in}}_{m}$ (the $m-$th possible outcome of the first measurement). After the first measurement (at $t = 0^{+}$), the density operator describing the ensemble average of the post-measurement states becomes
\begin{equation}
\rho_{\textrm{in}} =
\sum_{m}p(a^{\textrm{in}}_{m})|\psi_{a_{m}}\rangle\langle\psi_{a_{m}}|,
\end{equation}
where $p(a^{\textrm{in}}_{m}) = \textrm{Tr}\left[\Pi^{\textrm{in}}_{m}\rho_{0}\Pi^{\textrm{in}}_{m}\right] = \langle\psi_{a_{m}}|\rho_{0}|\psi_{a_{m}}\rangle$
is the probability to obtain the measurement outcome $a^{\textrm{in}}_{m}$. Then, the system undergoes a time evolution, which we assume described by a \textit{unital} completely positive, trace-preserving (CPTP) map $\Phi:L(\mathcal{H})\rightarrow L(\mathcal{H})$, with $L(\mathcal{H})$ denoting the sets of density operators (non-negative operators with unit trace) defined on the Hilbert space $\mathcal{H}$. Quantum maps (known also as quantum channels) represent a very effective tool to describe the effects of the noisy interaction of a quantum system with its environment~\cite{PetruccioneBook,CarusoRMP14}. A CPTP map is unital if it preserves the identity operator $\mathbbm{1}$ on $\mathcal{H}$, i.e. $\Phi(\mathbbm{1}) = \mathbbm{1}$. The assumption of a unital map covers a large family of quantum physical transformations not increasing the purity of the initial states, including, among others, unitary evolutions and decoherence processes. The time-evolved ensemble average is then denoted as
\begin{equation}
\rho_{\textrm{fin}} \equiv \Phi(\rho_{\textrm{in}}).
\end{equation}
For example, in case of unitary evolution with Hamiltonian $H(t)$, the final quantum state at $t = \mathcal{T}^{-}$ equals $\rho_{\textrm{fin}} = \Phi(\rho_{\textrm{in}}) = \mathcal{U}\rho_{\textrm{in}}\mathcal{U}^{\dagger}$, where $\mathcal{U}$ is, as before, the unitary evolution operator. After the time evolution, at time $t = \mathcal{T}^{+}$, a second measurement is performed on the quantum system according to the observable
$$
\mathcal{O}_{\textrm{fin}} = \sum_{k}a^{\textrm{fin}}_{k}\Pi^{\textrm{fin}}_{k},
$$
where $\Pi^{\textrm{fin}}_{k} \equiv |\phi_{a_{k}}\rangle\langle\phi_{a_{k}}|$, and $a^{\textrm{fin}}_{k}$ is the $k-$th outcome of the second measurement (with eigenvectors $|\phi_{a_{k}}\rangle$). Consequently, the probability to obtain the measurement outcome $a_{k}^{\textrm{fin}}$ is $p(a^{\textrm{fin}}_{k}) = \textrm{Tr}\left[\Pi^{\textrm{fin}}_{k}\Phi(\rho_{\textrm{in}})\Pi^{\textrm{fin}}_{k}\right] =
\langle\phi_{a_{k}}|\rho_{\textrm{fin}}|\phi_{a_{k}}\rangle$. Thus, the resulting density operator, describing the ensemble average of the post-measurement states after the second measurement, is
\begin{equation}
\rho_{\mathcal{T}} = \sum_{k}p(a^{\textrm{fin}}_{k})|\phi_{a_{k}}\rangle\langle\phi_{a_{k}}|.
\end{equation}
The joint probability that the events ``measure of $a^{\textrm{in}}_{m}$'' and ``measure of $a^{\textrm{fin}}_{k}$'' both occur for the forward process, denoted by $p(a^{\textrm{fin}} = a^{\textrm{fin}}_{k}, a^{\textrm{in}} = a^{\textrm{in}}_{m})$, is given by
\begin{equation}
p(a^{\textrm{fin}}_{k},a^{\textrm{in}}_{m}) = \textrm{Tr}\left[\Pi^{\textrm{fin}}_{k}\Phi(\Pi^{\textrm{in}}_{m}\rho_{0}\Pi^{\textrm{in}}_{m})\right].
\end{equation}

To study the backward process, we first have to introduce the concept of time-reversal. Time-reversal is achieved by the time-reversal operator $\Theta$ acting on $\mathcal{H}$. The latter has to be an antiunitary operator. An antiunitary operator $\Theta$ is anti-linear, i.e.
\begin{equation}
\Theta (x_1|\varphi_1\rangle + x_2|\varphi_2\rangle)= x_1^\star\Theta|\varphi_1\rangle + x_2^\star\Theta|\varphi_2\rangle
\end{equation}
for arbitrary complex coefficients $x_1$, $x_2$ and $|\varphi_1\rangle$, $|\varphi_2\rangle$ $\in$ $\mathcal{H}$, and it transforms the inner product as
$\langle \widetilde{\varphi}_1|\widetilde{\varphi}_2\rangle=\langle\varphi_2|\varphi_1\rangle$
for $|\widetilde{\varphi}_1\rangle=\Theta|\varphi_1\rangle$, and $|\widetilde{\varphi}_2\rangle=\Theta|\varphi_2\rangle$.
Antiunitary operators satisfy the relations $\Theta^{\dagger}\Theta = \Theta\Theta^{\dagger} = \mathbbm{1}$. The antiunitarity of $\Theta$ ensures the time-reversal symmetry~\cite{Sozzi2008}. We define the time-reversed density operator as $\widetilde{\rho}\equiv\Theta\rho\Theta^\dagger$, and we consider the time-reversal version of the quantum evolution operator, i.e. our unital CPTP map $\Phi$. Without loss of generality, it admits an operator-sum (or Kraus) representation:
$$
\rho_{\textrm{fin}} = \Phi(\rho_{\textrm{in}}) = \sum_{u}E_{u}\rho_{\textrm{in}}E_{u}^{\dagger}
$$
with the Kraus operators $E_{u}$ being such that $\sum_{u}E_{u}^{\dagger} E_{u} = \mathbbm{1}$ (trace-preserving)~\cite{PetruccioneBook,CarusoRMP14}. For each Kraus operator $E_{u}$ of the forward process we can define the corresponding time-reversed operator $\widetilde{E}_{u}$~\cite{CrooksPRA2008,Manzano2015}, so that the time-reversal $\widetilde{\Phi}$ for the CPTP quantum map $\Phi$ is given by
\begin{equation}\label{reversed_map}
\widetilde{\Phi}(\rho) = \sum_{u}\widetilde{E}_{u}\rho\widetilde{E}_{u}^{\dagger},
\end{equation}
where
$$
\widetilde{E}_{u} \equiv \mathcal{A}\pi^{1/2}E^{\dagger}_{u}\pi^{-1/2}\mathcal{A}^{\dagger},
$$
$\pi$ is an invertible fixed point (not necessarily unique) of the quantum map (such that $\Phi(\pi) = \pi$), and $\mathcal{A}$ is an arbitrary (unitary or anti-unitary) operator. Usually, the operator $\mathcal{A}$ is chosen equal to the time-reversal operator $\Theta$. If the density operator $\pi$
is a positive definite operator, as assumed in \cite{CrooksPRA2008,Horowitz2013}, then also the square root $\pi^{1/2}$ is positive definite and the inverse $\pi^{-1/2}$ exists and it is unique. Since our map is unital we can choose $\pi^{1/2}=\pi^{-1/2}=\mathbbm{1}$. Thus, from (\ref{reversed_map}), we can observe that also $\widetilde{\Phi}$ is a CPTP quantum map with an operator sum-representation, such that $\sum_{u}\widetilde{E}_{u}^{\dagger}\widetilde{E}_{u} = \mathbbm{1}$.
Summarizing, we have
\begin{equation*}
\widetilde{E}_{u} = \Theta E^{\dagger}_{u}\Theta^\dagger,
\end{equation*}
so that
\begin{equation*}
\widetilde{\Phi}(\rho) = \sum_{u}\widetilde{E}_{u}\rho\widetilde{E}_{u}^{\dagger}=\Theta\left(\sum_u   E^{\dagger}_{u} \widetilde{\rho} E_{u}\right) \Theta^\dagger.
\end{equation*}
We are now in a position to define the backward process. We start by preparing the system (at time $t=\mathcal{T}^{+}$) in the state $\widetilde{\rho}_{\mathcal{T}} = \Theta\rho_{\mathcal{T}}\Theta^\dagger$, and measure the observable
$$
\widetilde{\mathcal{O}}_{\textrm{ref}} \equiv \sum_{k}a^{\textrm{ref}}_{k}\widetilde{\Pi}^{\textrm{ref}}_{k},
$$
with $\widetilde{\Pi}^{\textrm{ref}}_{k} = |\widetilde{\phi}_{a_{k}}\rangle\langle\widetilde{\phi}_{a_{k}}|$ and $|\widetilde{\phi}_{a_{k}}\rangle\equiv\Theta|\phi_{a_{k}}\rangle$, that is we choose this first measurement of the backward process to be the time-reversed version of the second measurement of the forward process. If we call the post-measurement ensemble average $\widetilde{\rho}_{\textrm{ref}}$, as a consequence $\widetilde{\rho}_{\mathcal{T}}=\widetilde{\rho}_{\textrm{ref}}$, or equivalently $\rho_{\mathcal{T}}=\rho_{\textrm{ref}}$, where the latter is called \textit{reference state}. \\ \\
\textbf{Remark:} Although the quantum fluctuation theorem can be derived without imposing a specific operator for the reference state~\cite{Sagawa2014}, the latter has been chosen to be identically equal to the final density operator after the second measurement of the protocol. This choice appears to be the most natural among the possible ones to design a suitable measuring scheme of general thermodynamical quantities, consistently with the quantum fluctuation theorem. \\ \\
Accordingly, the spectral decomposition of the time-reversed reference state is given by
\begin{equation}\label{rho_ref}
\widetilde{\rho}_{\textrm{ref}} = \sum_{k}p(a^{\textrm{ref}}_{k})|\widetilde{\phi}_{a_{k}}\rangle\langle\widetilde{\phi}_{a_{k}}|,
\end{equation}
where
\begin{equation}
p(a^{\textrm{ref}}_{k}) = \textrm{Tr}[\widetilde{\Pi}^{\textrm{ref}}_{k}\widetilde{\rho}_{\mathcal{T}}\widetilde{\Pi}^{\textrm{ref}}_{k}] = \langle\widetilde{\phi}_{a_{k}}|\widetilde{\rho}_{\mathcal{T}}|\widetilde{\phi}_{a_{k}}\rangle
\end{equation}
is the probability to get the measurement outcome $a_{k}^{\textrm{ref}}$. The reference state, then, undergoes the time-reversal dynamical evolution, mapping it onto the initial state of the backward process $\widetilde{\rho}_{\textrm{in}'}=\widetilde{\Phi}(\widetilde{\rho}_{\textrm{ref}})$. At $t = 0^{+}$ the density operator $\widetilde{\rho}_{\textrm{in}'} = \widetilde{\Phi}(\widetilde{\rho}_{\textrm{ref}})$ is subject to the second projective measurement of the backward process, whose observable is given by
$$
\widetilde{\mathcal{O}}_{\textrm{in}} = \sum_{m}a^{\textrm{in}}_{m}\widetilde{\Pi}^{\textrm{in}}_{m},$$
with $\widetilde{\Pi}^{\textrm{in}}_{m} = |\widetilde{\psi}_{a_{m}}\rangle\langle\widetilde{\psi}_{a_{m}}|$, and $|\widetilde{\psi}_{a_{m}}\rangle\equiv\Theta|\psi_{a_{m}}\rangle$. As a result, the probability to obtain the outcome $a^{\textrm{in}}_{m}$ is
$p(a^{\textrm{in}}_{m}) = \textrm{Tr}[\widetilde{\Pi}^{\textrm{in}}_{m}\widetilde{\Phi}(\widetilde{\rho}_{\textrm{ref}})\widetilde{\Pi}^{\textrm{in}}_{m}] =
\langle\widetilde{\psi}_{a_{m}}|\widetilde{\rho}_{\textrm{in}'}|\widetilde{\psi}_{a_{m}}\rangle$, while the joint probability $p(a^{\textrm{in}}_{m}, a^{\textrm{ref}}_{k})$ is given by
\begin{equation}
p(a^{\textrm{in}}_{m}, a^{\textrm{ref}}_{k}) = \textrm{Tr}[\widetilde{\Pi}^{\textrm{in}}_{m}\widetilde{\Phi}(\widetilde{\Pi}^{\textrm{ref}}_{k}\widetilde{\rho}_{\mathcal{T}}
\widetilde{\Pi}^{\textrm{ref}}_{k})].
\end{equation}
The final state of the backward process is instead $\widetilde{\rho}_{0'}=\sum_{m} p(a^{\textrm{in}}_{m}) \widetilde{\Pi}^{\textrm{in}}_{m}$. Let us observe again that the main difference of the two-time measurement protocol that we have introduced here, compared to the scheme in \cite{Sagawa2014}, is to perform the $2$nd and $1$st measurement of the backward protocol, respectively, on the same basis of the $1$st and $2$nd measurement of the forward process after a time-reversal transformation.

The irreversibility of the two-time measurement scheme is, thus, analyzed by studying the stochastic quantum entropy production $\sigma$ defined as:
\begin{equation}\label{general_sigma}
\sigma(a^{\textrm{fin}}_{k},a^{\textrm{in}}_{m}) \equiv \ln\left(\frac{p(a^{\textrm{fin}}_{k}, a^{\textrm{in}}_{m})}{p(a^{\textrm{in}}_{m}, a^{\textrm{ref}}_{k})}\right)
= \ln\left(\frac{p(a^{\textrm{fin}}_{k}|a^{\textrm{in}}_{m})p(a^{\textrm{in}}_{m})}
{p(a^{\textrm{in}}_{m}|a^{\textrm{ref}}_{k})p(a^{\textrm{ref}}_{k})}\right),
\end{equation}
where $p(a^{\textrm{fin}}_{k}|a^{\textrm{in}}_{m})$ and $p(a^{\textrm{in}}_{m}|a^{\textrm{ref}}_{k})$ are the conditional probabilities of measuring, respectively, the outcomes $a^{\textrm{fin}}_{k}$ and $a^{\textrm{in}}_{m}$, conditioned on having first measured $a^{\textrm{in}}_{m}$ and $a^{\textrm{ref}}_{k}$. Its mean value
\begin{equation}\label{general_mean_sigma}
\langle\sigma\rangle = \sum_{k,m}p(a_k^{\textrm{fin}},a_m^{\textrm{in}})\ln\left(\frac{p(a_k^{\textrm{fin}},a_m^{\textrm{in}})}{p(a_k^{\textrm{in}}, a_m^{\textrm{ref}})}\right)
\end{equation}
corresponds to the classical relative entropy (or Kullback-Leibler divergence) between the joint probabilities $p(a^{\textrm{fin}}, a^{\textrm{in}})$ and $p(a^{\textrm{in}}, a^{\textrm{ref}})$, respectively, of the forward and backward processes~\cite{Cover2006,Umegaki}. The Kullback-Leibler divergence is always non-negative and as a consequence
\begin{equation}
\langle\sigma\rangle\geq 0.
\end{equation}
As a matter of fact, $\langle\sigma\rangle$ \textit{can be considered as the amount of additional information that is required to achieve the backward process, once the quantum system has reached the final state} $\rho_{\mathcal{T}}$. Moreover, $\langle\sigma\rangle = 0$ if and only if $p(a^{\textrm{fin}}_{k}, a^{\textrm{in}}_{m}) = p(a^{\textrm{in}}_{m}, a^{\textrm{ref}}_{k})$, i.e. if and only if $\sigma = 0$. To summarize, the transformation of the system state from time $t=0^{-}$ to $t=\mathcal{T}^{+}$ is then defined to be thermodynamically irreversible if $\langle\sigma\rangle > 0$. If, instead, all the fluctuations of $\sigma$ shrink around $\langle\sigma\rangle \simeq 0$ the system comes closer and closer to a reversible one. We observe that a system transformation may be thermodynamically irreversible also if the system undergoes unitary evolutions with the corresponding irreversibility contributions due to applied quantum measurements. Also the measurements back-actions, indeed, lead to energy fluctuations of the quantum system, as recently quantified in \cite{Elouard2016}. In case there is no evolution (identity map) and the two measurement operators are the same, then the transformation becomes reversible. We can now state the following theorem: \\ \\
\textbf{Theorem~4.1}: \textit{Given the two-time measurement protocol described above and an open quantum system dynamics described by a unital CPTP quantum map} $\Phi$, \textit{it can be stated that:}
\begin{equation}\label{equality_cond_prob}
p(a^{\textrm{fin}}_{k}|a^{\textrm{in}}_{m}) = p(a^{\textrm{in}}_{m}|a^{\textrm{ref}}_{k}).
\end{equation}
The proof of Theorem~4.1 can be found in Appendix~\ref{chapter:appC}. \\ \\
Throughout this article we assume that $\Phi$ is unital and this property of the map guarantees the validity of Theorem~4.1. Note, however, that \cite{Horowitz2013,Manzano2015} present a fluctuation theorem for slightly more general maps, that however violate (\ref{equality_cond_prob}).

As a consequence of Theorem~4.1 we obtain:
\begin{equation}\label{sigma}
\sigma(a^{\textrm{fin}}_{k},a^{\textrm{in}}_{m}) = \ln\left(\frac{p(a^{\textrm{in}}_{m})}{p(a^{\textrm{ref}}_{k})}\right) =\ln\left(\frac{\langle\psi_{a_{m}}|\rho_{0}|\psi_{a_{m}}\rangle}
{\langle\widetilde{\phi}_{a_{k}}|\widetilde{\rho}_{\mathcal{T}}|\widetilde{\phi}_{a_{k}}\rangle}\right).
\end{equation}
providing a general expression of the quantum fluctuation theorem for the described two-time quantum measurement scheme. Let us introduce, now, the entropy production $\widetilde{\sigma}$ for the backward processes, i.e.
$$
\widetilde{\sigma}(a^{\textrm{in}}_{m},a^{\textrm{ref}}_{k})\equiv\ln\left(\frac{p(a^{\textrm{in}}_{m}, a^{\textrm{ref}}_{k})}{p(a^{\textrm{fin}}_{k}, a^{\textrm{in}}_{m})}\right) = \ln\left(\frac{p(a^{\textrm{ref}}_{k})}{p(a^{\textrm{in}}_{m})}\right),
$$
where the second identity is valid only in case we can apply the results deriving from Theorem~4.1. Hence, if we define $\textrm{Prob}(\sigma)$ and $\textrm{Prob}(\widetilde{\sigma})$ as the probability distributions of the stochastic entropy production, respectively, for the forward and the backward processes, then it can be shown (see e.g. \cite{Sagawa2014}) that
\begin{equation}\label{qft}
\frac{\textrm{Prob}(\widetilde{\sigma} = -\Gamma)}{\textrm{Prob}(\sigma = \Gamma)} = e^{-\Gamma},
\end{equation}
where $\Gamma$ belongs to the set of values that can be assumed by the stochastic quantum entropy production $\sigma$. The identity (\ref{qft}) is usually called \textit{quantum fluctuation theorem}. By summing over $\Gamma$, we recover the \textit{integral quantum fluctuation theorem}, or \textit{quantum Jarzynski equality}, $\langle e^{-\sigma}\rangle = 1$, as previously shown e.g. in \cite{Kurchan2001,Sagawa2014}.

\subsection{Mean entropy production vs quantum relative entropy}

Following Ref.~\cite{Sagawa2014}, the essential ingredient is the non-negativity of the quantum relative entropy and its relation to the stochastic quantum entropy production. As a generalization of the Kullback-Leibler information~\cite{Umegaki}, the quantum relative entropy between two arbitrary density operators $\nu$ and $\mu$ is defined as $S(\nu\parallel\mu)\equiv\textrm{Tr}[\nu\ln\nu] - \textrm{Tr}[\nu\ln\mu]$. The Klein inequality states that the quantum relative entropy is a non-negative quantity~\cite{Vedral2002}, {\em i.e.} $S(\nu\parallel\mu)\geq 0$, where the equality holds if and only if $\nu = \mu$ - see {\em e.g.}~\cite{Sagawa2014}. In the following we will show the relation between the quantum relative entropy of the system density matrix at the final time of the transformation and the stochastic quantum entropy production for unital CPTP quantum maps. Accordingly, the following theorem can be stated: \\ \\
\textbf{Theorem~2}: \textit{Given the two-time measurement protocol described above and an open quantum system dynamics described by a unital CPTP quantum map} $\Phi$, \textit{the quantum relative entropy} $S(\rho_{\textrm{fin}}\parallel\rho_{\tau})$ \textit{fulfills the inequality}
\begin{equation}\label{eq:entropy-positivity}
0\leq S(\rho_{\textrm{fin}}\parallel\rho_{\tau})\leq\langle\sigma\rangle,
\end{equation}
\textit{where the equality} $S(\rho_{\textrm{fin}}\parallel\rho_{\tau}) = 0$ \textit{holds if and only if} $\rho_{\textrm{fin}} = \rho_{\tau}$. \textit{Then, for} $[\mathcal{O}_{\textrm{fin}},\rho_{\textrm{fin}}]=0$ \textit{one has} $\langle\sigma\rangle = S(\rho_{\tau}) - S(\rho_{\textrm{in}})$, \textit{so that}
\begin{equation}\label{eq:entropy-of-map}
0= S(\rho_{\textrm{fin}}\parallel\rho_{\tau})\leq\langle\sigma\rangle=S(\rho_{\textrm{fin}})-S(\rho_{\textrm{in}}),
\end{equation}
\textit{where $S(\cdot)$ denotes the von Neumann entropy of $(\cdot)$. Finally,} $S(\rho_{\textrm{fin}}\parallel\rho_{\tau}) = \langle\sigma\rangle$ \textit{if} $\mathcal{S}$ \textit{is a closed quantum system following a unitary evolution}. A proof of Theorem~2 is in Appendix~\ref{chapter:appC}. \\

While Eq.~(\ref{eq:entropy-positivity}) is more general and includes the irreversibility contributions of both the map $\Phi$ and the final measurement, in Eq.~(\ref{eq:entropy-of-map}) due to a special choice of the observable of the second measurement we obtain $\rho_{\textrm{fin}}=\rho_{\tau}$ and, thus, the quantum relative entropy vanishes while the stochastic quantum entropy production contains the irreversibility contribution only from the map. This contribution is given by the difference between the von Neumann entropy of the final state $S(\rho_{\textrm{fin}})$ and the initial one $S(\rho_{\textrm{in}})$\footnote{Let us assume that the initial density matrix $\rho_{\textrm{in}}$ is a Gibbs thermal state at inverse temperature $\beta_{T}$, {\em i.e.} $\rho_{\textrm{in}}\equiv e^{\beta_{T}\left[F(0)\mathbbm{1}_\mathcal{S}- H(0)\right]}$, where $F(0) \equiv -\beta_{T}^{-1}\ln\left\{\textrm{Tr}[e^{-\beta_{T} H(t = 0)}]\right\}$ and $H(0)$ are, respectively, equal to the Helmholtz free-energy and the system Hamiltonian at time $t=0$. Accordingly, the von Neumann entropy $S(\rho_{\textrm{in}})$ equals the thermodynamic entropy at $t = 0$, {\em i.e.} $S(\rho_{\textrm{in}}) = \beta_{T}(\langle H(0)\rangle - F(0))$, where $\langle H(0)\rangle \equiv \textrm{Tr}[\rho_{\textrm{in}}H(0)]$ is the average energy of the system in the canonical distribution. More generally, we can state that given an arbitrary initial density matrix $\rho_{\textrm{in}}$ the thermodynamic entropy $\beta_{T}(\langle H(0)\rangle - F(0))$ represents the upper-bound value for the von Neumann entropy $S(\rho_{\textrm{in}})$, whose maximum value is reached only in the canonical distribution. To prove this, it is sufficient to consider $S(\rho_{\textrm{in}}\parallel e^{\beta_{T}\left(F(0)\mathbbm{1}_\mathcal{S} - H(0)\right)}) = \beta_{T}\left(F(0) - \langle H(0)\rangle\right) - S(\rho_{\textrm{in}})$, from which, from the positivity of the quantum relative entropy, one has $S(\rho_{\textrm{in}})\leq\beta_{T}(\langle H(0)\rangle - F(0))$.}.

To conclude, in case the environment $\mathcal{E}$ is not thermal, so as not to induce the thermalisation of the system dynamics, or not directly accessible from the outside, i.e. partially controllable only in its own macroscopic properties, the stochastic quantum entropy production represents a very general measurable thermodynamic quantity, encoding information about the interaction between the system and the environment also in a fully quantum regime. Therefore, its reconstruction becomes really relevant, not only for the fact that we cannot longer adopt energy measurements on $\mathcal{S}$ to infer $\sigma$ and its fluctuation properties, but also because in this way we could manage to measure the mean heat flux exchanged by the partitions of $\mathcal{S}$ in case it is a multipartite quantum system, as shown in the following sections.

\subsection{Open bipartite systems}

In this section, our intent is to define and, then, reconstruct the fluctuation profile of the stochastic quantum entropy production $\sigma$ for an open multipartite system (for simplicity we will analyze in detail a bipartite system), so as to characterize the irreversibility of the system dynamics after an arbitrary transformation. At the same time, we will also study the role played by the performance of measurements both on local and global observables for the characterization of $\textrm{Prob}(\sigma)$ in a many-body context, and evaluate the efficiency of reconstruction in both cases. In particular, as shown by the numerical examples, by comparing the mean stochastic entropy productions $\langle\sigma\rangle$ obtained by local measurements on partitions of the composite system and measurements on its global observables, we are able to detect (quantum and classical) correlations between the subsystems, which have been caused by the system dynamics.

To this end, let us assume that the open quantum system $\mathcal{S}$ is composed of two distinct subsystems ($A$ and $B$), which are mutually interacting, and we denote by $A-B$ the composite system $\mathcal{S}$. However, all the presented results can be in principle generalized to an arbitrary number of subsystems. As before, the initial and final density operators of the composite system are arbitrary (not necessarily equilibrium) quantum states, and the dynamics of the composite system is described by a unital CPTP quantum map. The two-time measurement scheme on $A - B$ is implemented by performing the measurements locally on $A$ and $B$ and we assume, moreover, that the measurement processes at the beginning and at the end of the protocol are \textit{independent}. Since the local measurement on $A$ commutes with the local measurement on $B$, the two measurements can be performed simultaneously. This allows us to consider the stochastic entropy production for the composite system by considering the \textit{correlations} between the measurement outcomes of the two local observables. Alternatively, by disregarding these correlations, we can consider separately the stochastic entropy production of each subsystem.

The composite system $A - B$ is defined on the finite-dimensional Hilbert space $\mathcal{H}_{A-B}\equiv\mathcal{H}_{A}\otimes\mathcal{H}_{B}$ (with $\mathcal{H}_A$ and $\mathcal{H}_B$ the Hilbert spaces of system $A$ and $B$, respectively), and its dynamics is governed by the time-dependent Hamiltonian
$H(t) = H_{A}(t)\otimes\mathbbm{1}_{B} + \mathbbm{1}_{A}\otimes H_{B}(t) + H_{A-B}(t)$. $\mathbbm{1}_{A}$ and $\mathbbm{1}_{B}$ are the identity operators acting, respectively, on the Hilbert spaces of the systems $A$ and $B$, while $H_{A}$ is the Hamiltonian of $A$, $H_{B}$ the Hamiltonian of system $B$, and $H_{A-B}$ is the interaction term. We denote the initial density operator of the composite quantum system $A - B$ by $\rho_{0}$ (before the first measurement), which is assumed to be a product state, then the ensemble average after the first measurement (at $t = 0^{+}$) is given by the density operator $\rho_{\textrm{in}}$, which can be written as:
\begin{equation}\label{rho_in}
\rho_{\textrm{in}}=\rho_{A,\textrm{in}}\otimes\rho_{B,\textrm{in}},
\end{equation}
where
\begin{equation}
\begin{cases}
\rho_{A,\textrm{in}}=\sum_{m}p(a_{m}^{\textrm{in}})\Pi^{\textrm{in}}_{A,m} \\
\rho_{B,\textrm{in}}=\sum_{h}p(b_{h}^{\textrm{in}})\Pi^{\textrm{in}}_{B,h} \\
\end{cases}
\end{equation}
are the reduced density operators for the subsystems $A$ and $B$, respectively. The projectors
$\Pi^{\textrm{in}}_{A,m} \equiv |\psi_{a_{m}}\rangle\langle\psi_{a_{m}}|$ and $\Pi^{\textrm{in}}_{B,h}\equiv|\psi_{b_{h}}\rangle\langle\psi_{b_{h}}|$ are the projectors onto the respective eigenstates of the local measurement operators for the subsystems $A$ and $B$: the observables $\mathcal{O}^{\textrm{in}}_{A}=\sum_m a_m^{\textrm{in}}\Pi^{\textrm{in}}_{A,m}$ on system $A$ and $\mathcal{O}^{\textrm{in}}_{B}=\sum_h b_h^{\textrm{in}}\Pi^{\textrm{in}}_{B,h}$ on system $B$, with possible measurement outcomes $\{a^{\textrm{in}}_{m}\}$ and $\{b^{\textrm{in}}_{h}\}$, upon measurement of $\rho_0$. After the measurement, the composite system $A - B$ undergoes a time evolution up to the time instant $t = \mathcal{T}^{-}$, described by the unital CPTP quantum map $\Phi$, such that $\rho_{\textrm{fin}}=\Phi(\rho_{\textrm{in}})$. Then, a second measurement is performed on both systems, measuring the observables $\mathcal{O}^{\textrm{fin}}_{A}=\sum_k a_k^{\textrm{fin}}\Pi^{\textrm{fin}}_{A,k}$ on system $A$ and $\mathcal{O}^{\textrm{fin}}_{B}=\sum_l b_l^{\textrm{fin}}\Pi^{\textrm{fin}}_{B,l}$ on system $B$, where $\{a_k^{\textrm{fin}}\}$ and $\{b_l^{\textrm{fin}}\}$ are the eigenvalues of the observables, and the projector $\Pi^{\textrm{fin}}_{A,k}\equiv|\phi_{a_{k}}\rangle\langle\phi_{a_{k}}|$ and $\Pi^{\textrm{fin}}_{B,l}\equiv|\phi_{b_{l}}\rangle\langle\phi_{b_{l}}|$ are given by the eigenstates $|\phi_{a_{k}}\rangle$ and $|\phi_{b_{l}}\rangle$, respectively. After the second measurement, we have to make a distinction according to whether we want to take into account correlations between the subsystems or not.

If we disregard the correlations, the ensemble average over all the local measurement outcomes of the state of the quantum system at $t = \mathcal{T}^{+}$ is described by the following product state $\rho_{A,\mathcal{T}}\otimes\rho_{B,\mathcal{T}}$, where
\begin{equation}
\begin{cases}
\rho_{A,\mathcal{T}}=\sum_{k}p(a_{k}^{\textrm{fin}})\Pi^{\textrm{fin}}_{A,k} \\
\rho_{B,\mathcal{T}}=\sum_{l}p(b_{l}^{\textrm{fin}})\Pi^{\textrm{fin}}_{B,l} \\
\end{cases}.
\end{equation}
The probabilities $p(a_{k}^{\textrm{fin}})$ to obtain outcome $a_{k}^{\textrm{fin}}$ and
$p(b_{l}^{\textrm{fin}})$ to obtain the measurement outcome $b_{l}^{\textrm{fin}}$ are given by
\begin{equation}
\begin{cases}
p(a_{k}^{\textrm{fin}}) = \textrm{Tr}_{A}\left[\Pi^{\textrm{fin}}_{A,k}\textrm{Tr}_{B}\left[\rho_{\textrm{fin}}\right]\right] \\
p(b_{l}^{\textrm{fin}}) = \textrm{Tr}_{B}\left[\Pi^{\textrm{fin}}_{B,l}\textrm{Tr}_{A}\left[\rho_{\textrm{fin}}\right]\right] \\
\end{cases},
\end{equation}
where $\textrm{Tr}_{A}\left[\cdot\right]$ and $\textrm{Tr}_{B}\left[\cdot\right]$ denote, respectively, the operation of partial trace with respect to the quantum systems $A$ and $B$. Conversely, in order to keep track of the correlations between the simultaneously performed local measurements, we have to take into account the following global observable of the composite system $A-B$:
\begin{equation}\label{eq:ovservable_A-B}
\mathcal{O}^{\textrm{fin}}_{A - B} = \sum_{k,l}c_{kl}^{\textrm{fin}}\Pi^{\textrm{fin}}_{A - B,kl}~,
\end{equation}
where $\Pi^{\textrm{fin}}_{A - B,kl} \equiv \Pi^{\textrm{fin}}_{A,k}\otimes \Pi^{\textrm{fin}}_{B,l}$ and $\{c^{\textrm{fin}}_{kl}\}$ are the outcomes of the final measurement of the protocol. The state of the system after the second measurement at $t = \mathcal{T}^{+}$ is then described by an ensemble average over all outcomes of the joint measurements:
\begin{equation}
\rho_{\mathcal{T}} = \sum_{k,l}p(c^{\textrm{fin}}_{kl})\Pi^{\textrm{fin}}_{A - B,kl}~,
\end{equation}
where $p(c^{\textrm{fin}}_{kl}) = \textrm{Tr}\left[\Pi^{\textrm{fin}}_{A - B,kl}~\rho_{\textrm{fin}}\right]$. In both cases, consistently with the previous assumptions, we choose $\rho_{\mathcal{T}}$ as the reference state of the composite system. The measurement outcomes of the initial and final measurement for the composite system $A - B$ are, respectively, $c^{\textrm{in}}_{mh}\equiv(a_{m}^{\textrm{in}},b_{h}^{\textrm{in}})$ and $c^{\textrm{fin}}_{kl}\equiv(a_{k}^{\textrm{fin}},b_{l}^{\textrm{fin}})$. These outcomes occur with probabilities $p(c^{\textrm{in}}_{mh})$ and $p(c^{\textrm{fin}}_{kl})$, which reflect the correlation of the outcomes of the local measurements. As a result, the stochastic quantum entropy production of the composite system reads
\begin{equation}
\sigma_{A - B}(c^{\textrm{in}}_{mh},c^{\textrm{fin}}_{kl}) = \ln\left(\frac{p(c^{\textrm{in}}_{mh})}{p(c^{\textrm{fin}}_{kl})}\right).
\end{equation}
In the same way, we can define the stochastic quantum entropy production separately for each subsystem, i.e. $\sigma_A$ for subsystem $A$ and $\sigma_B$ for subsystem $B$:
\begin{equation}
 \sigma_A(a^{\textrm{in}}_{m},a^{\textrm{fin}}_{k})=\ln\left(\frac{p(a^{\textrm{in}}_{m})}{p(a^{\textrm{fin}}_{k})}\right),~~~~\text{and}~~~~
 \sigma_B(b^{\textrm{in}}_{h},b^{\textrm{fin}}_{l})=\ln\left(\frac{p(b^{\textrm{in}}_{h})}{p(b^{\textrm{fin}}_{l})}\right).
\end{equation}
If upon measurement the composite system is in a product state, the measurement outcomes for $A$ and $B$ are independent and the probabilities to obtain them factorize as
\begin{equation*}
\begin{cases}
p(c^{\textrm{in}}_{mh}) = p(a^{\textrm{in}}_{m})p(b^{\textrm{in}}_{h}) \\
p(c^{\textrm{fin}}_{kl}) = p(a^{\textrm{fin}}_{k})p(b^{\textrm{fin}}_{l})
\end{cases}.
\end{equation*}
As a direct consequence, the stochastic quantum entropy production becomes an additive quantity:
\begin{equation}\label{eq:sigma_sum}
\sigma_{A - B}(c^{\textrm{in}}_{mh},c^{\textrm{fin}}_{kl}) = \sigma_A(a^{\textrm{in}}_{m},a^{\textrm{fin}}_{k}) + \sigma_B(b^{\textrm{in}}_{h},b^{\textrm{fin}}_{l}) \equiv \sigma_{A + B}(c^{\textrm{in}}_{mh},c^{\textrm{fin}}_{kl}).
\end{equation}
In the more general case of correlated measurement outcomes, the probabilities do not factorize anymore. Instead, the mean value of the stochastic entropy production $\sigma_{A - B}(c^{\textrm{in}}_{mh},c^{\textrm{fin}}_{kl})$ becomes sub-additive. In other words
\begin{equation}\label{eq:sigma_sum}
\langle\sigma_{A - B}\rangle \leq \langle\sigma_{A}\rangle + \langle\sigma_{B}\rangle \equiv \langle\sigma_{A + B}\rangle,
\end{equation}
i.e. the mean value of the stochastic quantum entropy production $\sigma_{A - B}$ of the composite system $A - B$ is smaller than the sum of the mean values of the corresponding entropy production of its subsystems, when the latter are correlated. To see this, we recall the expression of the mean value of the stochastic entropy production in terms of the von Neumann entropies of the two post-measurement states (see appendix~\ref{chapter:appC}):
\begin{eqnarray*}
\langle\sigma_{A - B}\rangle &=& S(\rho_{\mathcal{T}})-S(\rho_{\textrm{in}})
=S(\rho_{\mathcal{T}})-S(\rho_{A,\textrm{in}})-S(\rho_{B,\textrm{in}})\\
&\leq& S(\rho_{A,\mathcal{T}})+S(\rho_{B,\mathcal{T}})-S(\rho_{A,\textrm{in}})-S(\rho_{B,\textrm{in}})\\
&=& \langle\sigma_{A}\rangle + \langle\sigma_{B}\rangle = \langle\sigma_{A + B}\rangle.
\end{eqnarray*}

\subsection{Probability distribution}

Depending on the values assumed by the measurement outcomes $c^{\textrm{in}}\in\{c^{\textrm{in}}_{mh}\}$ and $c^{\textrm{fin}}\in \{c^{\textrm{fin}}_{kl}\}$, $\sigma_{A - B}$ is a fluctuating variable as it is true also for the single subsystem contributions $\sigma_A\in\{\sigma_A(a^{\textrm{in}}_{m},a^{\textrm{fin}}_{k})\}$ and $\sigma_B\in\{\sigma_B(b^{\textrm{in}}_{h},b^{\textrm{fin}}_{l})\}$. We denote the probability distributions for the subsystems with $\textrm{Prob}(\sigma_A)$ and $\textrm{Prob}(\sigma_B)$ and $\textrm{Prob}(\sigma_{A-B})$ for the composite system. We will further compare this probability distribution for the composite system (containing the correlations of the local measurement outcomes) to the uncorrelated distribution of the sum of the single subsystems'contributions. We introduce the probability distribution $\textrm{Prob}(\sigma_{A+B})$ of the stochastic quantum entropy production $\sigma_{A+B}$ by applying the following discrete convolution sum:
\begin{equation}\label{eq:convolution}
\textrm{Prob}(\sigma_{A+B}) = \sum_{\{\xi_{B}\}}\textrm{Prob}((\sigma_{A+B} - \xi_{B})_{A})\textrm{Prob}(\xi_{B}),
\end{equation}
where $(\sigma_{A+B} - \xi_{B})_{A}$ and $\xi_{B}$ belong, respectively, to the sample space (i.e. the set of all possible outcomes) of the random variables $\sigma_A$ and $\sigma_B$.

The probability distribution for the single subsystem, {\em e.g.} the subsystem $A$, is fully determined by the knowledge of the measurement outcomes and the respective probabilities (relative frequencies). We obtain the measurement outcomes $(a^{\textrm{in}}_m,a^{\textrm{fin}}_k)$ with a certain probability $p_{a}(k,m)$, the joint probability for $a^{\textrm{in}}_m$ and $a^{\textrm{fin}}_k$, and this measurement outcome yields the stochastic entropy production $\sigma_A = \sigma_A(a^{\textrm{in}}_{m},a^{\textrm{fin}}_{k})$. Likewise, for system $B$ we introduce the joint probability $p_{b}(l,h)$ to obtain $(b^{\textrm{in}}_h,b^{\textrm{fin}}_l)$, which yields $\sigma_B = \sigma_B(b^{\textrm{in}}_{h},b^{\textrm{fin}}_{l})$.
Therefore, the probability distributions $\textrm{Prob}(\sigma_A)$ and $\textrm{Prob}(\sigma_B)$ are given by
\begin{equation}\label{prob_a}
\textrm{Prob}(\sigma_A) = \left\langle\delta\left[\sigma_A - \sigma_A(a^{\textrm{in}}_{m},a^{\textrm{fin}}_{k})\right]\right\rangle
= \sum_{k,m}\delta\left[\sigma_A - \sigma_A(a^{\textrm{in}}_{m},a^{\textrm{fin}}_{k})\right]p_{a}(k,m)
\end{equation}
and
\begin{equation}\label{prob_b}
\textrm{Prob}(\sigma_B) = \left\langle\delta\left[\sigma_B - \sigma_B(b^{\textrm{in}}_{h},b^{\textrm{fin}}_{l})\right]\right\rangle
= \sum_{l,h}\delta\left[\sigma_B - \sigma_B(b^{\textrm{in}}_{h},b^{\textrm{fin}}_{l})\right]p_{b}(l,h),
\end{equation}
where $\delta[\cdot]$ is the Dirac-delta distribution. In (\ref{prob_a}) and (\ref{prob_b}), the joint probabilities $p_{a}(k,m)$ and $p_{b}(l,h)$ read
\begin{equation}\label{joint}
\begin{cases}
p_{a}(k,m) = \textrm{Tr}\left[(\Pi^{\textrm{fin}}_{A,k}\otimes\mathbbm{1}_{B})\Phi(\Pi^{\textrm{in}}_{A,m}\otimes\rho_{\textrm{B,in}})\right]p(a_{m}^{\textrm{in}}) \\
p_{b}(l,h) = \textrm{Tr}\left[(\mathbbm{1}_{A}\otimes\Pi^{\textrm{fin}}_{B,l})\Phi(\rho_{\textrm{A,in}}\otimes\Pi^{\textrm{in}}_{B,h})\right]p(b_{h}^{\textrm{in}}).
\end{cases}
\end{equation}

By definition, given the reconstructed probability distributions $\textrm{Prob}(\sigma_A)$ and $\textrm{Prob}(\sigma_B)$, the probability $\textrm{Prob}(\sigma_{A+B})$ can be calculated straightforwardly by calculating the convolution of $\textrm{Prob}(\sigma_A)$ and $\textrm{Prob}(\sigma_B)$ according to (\ref{eq:convolution}).
Equivalently, the probability distribution $\textrm{Prob}(\sigma_{A - B})$ of the stochastic quantum entropy production of the composite system (containing the correlations between the local measurement outcomes) is given by:
\begin{eqnarray}
\textrm{Prob}(\sigma_{A - B}) &=& \left\langle\delta\left[\sigma_{A - B} - \sigma_{A - B}(c^{\textrm{in}}_{mh},c^{\textrm{fin}}_{kl})\right]\right\rangle\nonumber \\
&=&\sum_{mh,kl}\delta\left[\sigma_{A - B} - \sigma_{A - B}(c^{\textrm{in}}_{mh},c^{\textrm{fin}}_{kl})\right]p_{c}(mh,kl),
\end{eqnarray}
where
\begin{equation}
p_{c}(mh,kl) = \textrm{Tr}\left[\Pi^{\textrm{fin}}_{A - B,kl}\Phi\left(\Pi^{\textrm{in}}_{A,m}\otimes\Pi^{\textrm{in}}_{B,h}\right)\right]p(c^{\textrm{in}}_{mh}),
\end{equation}
with $p(c^{\textrm{in}}_{mh}) = p(a^{\textrm{in}}_{m})p(b^{\textrm{in}}_{h})$. Now, the integral quantum fluctuation theorems for $\sigma_A$, $\sigma_B$ and $\sigma_{A - B}$ can be derived just by computing the characteristic functions of the corresponding probability distributions $\textrm{Prob}(\sigma_A)$, $\textrm{Prob}(\sigma_B)$ and $\textrm{Prob}(\sigma_{A - B})$.

\subsection{Characteristic function}

As shown in the previous sections, the characteristic function of a real-valued random variable is given by its Fourier transform and it completely defines the properties of the corresponding probability distribution in the frequency domain. Thus, the characteristic function $G_C(\lambda)$ of the probability distribution $\textrm{Prob}(\sigma_C)$ (for $C\in \{A,B,A-B\}$) is defined as
$G_C(\lambda) = \int \textrm{Prob}(\sigma_C)e^{i\lambda\sigma_C} d\sigma_C$, where $\lambda\in\mathbb{C}$ is a complex number. For the two subsystems, by inserting (\ref{prob_a})-(\ref{joint}) and exploiting the linearity of the CPTP quantum maps and of the trace (see appendix~\ref{chapter:appC}), the characteristic functions for $\textrm{Prob}(\sigma_A)$ and $\textrm{Prob}(\sigma_B)$ can be written as
\begin{small}
\begin{equation}\label{eq:G_A}
G_A(\lambda) = \textrm{Tr}\left[\left((\rho_{A,\mathcal{T}})^{-i\lambda}\otimes\mathbbm{1}_{B}\right)\Phi\left(
(\rho_{\textrm{A,in}})^{1 + i\lambda}\otimes\rho_{\textrm{B,in}}\right)\right]
\end{equation}
\end{small}
and
\begin{small}
\begin{equation}\label{eq2}
G_B(\lambda) = \textrm{Tr}\left[\left(\mathbbm{1}_{A}\otimes(\rho_{B,\mathcal{T}})^{-i\lambda}\right)\Phi\left(
\rho_{\textrm{A,in}}\otimes(\rho_{B,\textrm{in}})^{1 + i\lambda}\right)\right].
\end{equation}
\end{small}
In a similar way, we can derive the characteristic function $G_{A - B}(\lambda)$ of the stochastic entropy production of the composite system $A - B$:
\begin{eqnarray}\label{eq:G_A-B}
 G_{A - B}(\lambda)=\mathrm{Tr}\left[\rho_{\mathcal{T}}^{-i\lambda}\Phi(\rho_{\mathrm{in}}^{1+i\lambda})\right]\,.
\end{eqnarray}
Furthermore, if we choose $\lambda = i$, the integral quantum fluctuation theorems can be straightforwardly derived, namely for $\sigma_A$ and $\sigma_B$:
\begin{small}
\begin{equation}\label{eq3}
\big\langle e^{-\sigma_A}\big\rangle \equiv G_A(i) = \textrm{Tr}\left[\left(\rho_{A,\mathcal{T}}\otimes\mathbbm{1}_{B}\right)\Phi\left(
\mathbbm{1}_{A}\otimes\rho_{\textrm{B,in}}\right)\right]
\end{equation}
\end{small}
and
\begin{small}
\begin{equation}\label{eq4}
\big\langle e^{-\sigma_B}\big\rangle \equiv G_B(i) = \textrm{Tr}\left[\left(\mathbbm{1}_{A}\otimes\rho_{B,\mathcal{T}}\right)\Phi\left(
\rho_{\textrm{A,in}}\otimes\mathbbm{1}_{B}\right)\right],
\end{equation}
\end{small}%
as well as
\begin{small}
\begin{equation}\label{eq4}
\big\langle e^{-\sigma_{A-B}}\big\rangle \equiv G_{A-B}(i) = \textrm{Tr}\left[\rho_{\mathcal{T}}\Phi\left(
\mathbbm{1}_{A-B}\right)\right]=1
\end{equation}
\end{small}
for $\sigma_{A-B}$ (with $\Phi$ unital). \\ \\
\textbf{Remark:} It is worth noting observe that the characteristic functions (\ref{eq:G_A})-(\ref{eq:G_A-B}) depend exclusively on appropriate powers of the initial and final density operators of each subsystem. These density operators are diagonal in the basis of the observable eigenvectors and can be measured by means of standard state population measurements for each value of $\lambda$. As will be shown in the following, this result can lead to a significant reduction of the number of measurements that is required to reconstruct the probability distribution of the stochastic quantum entropy production, beyond the direct application of the definition according to (\ref{prob_a})-(\ref{joint}).

\section{Reconstruction algorithm}
\label{VVV}

In this section, we present a novel algorithm for the reconstruction of the probability distribution of a generic thermodynamical quantity such as work, internal energy or entropy. Such protocol is based on the determination of the corresponding characteristic function, which is built over the stochastic realizations of the thermodynamical quantity after the second measurement of the protocol. The characteristic functions, that are measured, are evaluated over a given set of (real) parameters, in order to collect an adequate information to infer a complete statistics. In this regard, let us observe that the principles behind this procedure can be framed within the least squares approach to estimation theory~\cite{RaoLS}.

Without loss of generality, we will introduce the algorithm to reconstruct the probability distribution $\textrm{Prob}(\sigma)$ of the stochastic quantum entropy production $\sigma$. Being the procedure based on a parametric version of the integral quantum fluctuation theorem (i.e. $\langle e^{-\varphi \sigma}\rangle$, with $\varphi\in\mathbb{R}$), we introduce the moment generating functions $\chi_{C}(\varphi)$ for $C\in\{A,B,A-B\}$:
$$
\langle e^{-\varphi\sigma_C}\rangle = G_C(i\varphi) \equiv \chi_{C}(\varphi).
$$
$\chi_{C}(\varphi)$ can be expanded into a Taylor series, so to obtain
\begin{equation}
\chi_{C}(\varphi) = \langle e^{-\varphi\sigma_{C}}\rangle = \left\langle\sum_{k}\frac{(-\varphi)^{k}}{k!}\sigma_{C}^{k}\right\rangle
= 1 - \varphi\langle\sigma_{C}\rangle + \frac{\varphi^{2}}{2}\langle\sigma_{C}^{2}\rangle - \ldots
\end{equation}
Accordingly, the statistical moments of the stochastic quantum entropy production $\sigma_{C}$, denoted by $\{\langle\sigma_{C}^{k}\rangle\}$ for $k = 1,\ldots,N-1$, can be expressed in terms of the $\chi_{C}(\varphi)$'s defined over the parameter vector $\underline{\varphi} \equiv [\varphi_{1}, \ldots, \varphi_{N}]'$, i.e.
\begin{equation}
\begin{pmatrix} \chi_{C}(\varphi_{1}) \\ \chi_{C}(\varphi_{2}) \\ \vdots \\ \chi_{C}(\varphi_{N}) \end{pmatrix} = \underbrace{\begin{pmatrix} 1  & -\varphi_{1} & +\frac{\varphi_{1}^{2}}{2} & \ldots & \frac{(-\varphi_{1})^{N-1}}{N-1!} \\
1  & -\varphi_{2} & +\frac{\varphi_{2}^{2}}{2} & \ldots & \frac{(-\varphi_{2})^{N-1}}{N-1!} \\
\vdots & \vdots & \vdots & \vdots & \vdots \\
1  & -\varphi_{N} & +\frac{\varphi_{N}^{2}}{2} & \ldots & \frac{(-\varphi_{N})^{N-1}}{N-1!} \end{pmatrix}}_{A(\underline{\varphi})}
\begin{pmatrix} 1 \\ \langle\sigma_{C}\rangle \\ \langle\sigma_{C}^{2}\rangle \\ \vdots \\ \langle\sigma_{C}^{N-1}\rangle \end{pmatrix},
\end{equation}
where the matrix $A(\underline{\varphi})$ can be written as a Vandermonde matrix, as detailed below. It is clear at this point that the solution to the problem of inferring the set $\{\langle\sigma_{C}^{k}\rangle\}$ can be related to the resolution of a polynomial interpolation problem, where the experimental data-set is given by $N$ evaluations of the parametric integral fluctuation theorem of $\sigma_{C}$ in terms of the $\varphi$'s. Let us observe that only by choosing real values for the parameters $\varphi$ is it possible to set up the proposed reconstruction procedure via the resolution of an interpolation problem. By construction, the dimension of the parameters vector $\underline{\varphi}$ is equal to the number of statistical moments of $\sigma_{C}$ that we want to infer, including the trivial zero-order moment. In this regard, we define the vectors
\begin{equation*}
\widetilde{\underline{m}}\equiv\left(1, \, \,
-\langle\sigma_{C}\rangle, \,\,\, \ldots, \,\, (-1)^{N-1}\frac{\langle\sigma_{C}^{N-1}\rangle}{N-1!}\right)',
\end{equation*}
with element ${\widetilde{\underline{m}}}_j=(-1)^{j}\frac{\langle\sigma_{C}^{j}\rangle}{j!}$, $j=0, \ldots, N-1$, and
$$
\underline{\chi}_{C}\equiv(\chi_{C}(\varphi_{1}), \ldots, \chi_{C}(\varphi_{N}))'.
$$
Then one has
\begin{equation}\label{vandermonde}
\underline{\chi}_{C} = V(\underline{\varphi})\widetilde{\underline{m}},
\end{equation}
where
\begin{equation}
V(\underline{\varphi}) = \begin{pmatrix} 1  & \varphi_{1} & \varphi_{1}^{2} & \ldots & \varphi_{1}^{N-1} \\
1  & \varphi_{2} & \varphi_{2}^{2} & \ldots & \varphi_{2}^{N-1} \\
\vdots & \vdots & \vdots & \vdots & \vdots \\
1  & \varphi_{N} & \varphi_{N}^{2} & \ldots & \varphi_{N}^{N-1} \end{pmatrix}
\end{equation}
is the Vandermonde matrix built on the parameters vector $\underline{\varphi}$. $V(\underline{\varphi})$ is a matrix whose rows (or columns) have elements in geometric progression, i.e. $v_{ij} = \varphi^{j-1}_{i}$, where $v_{ij}$ denotes the $ij-$ element of $V(\underline{\varphi})$. Eq.~(\ref{vandermonde}) constitutes the formula for the inference of the statistical moments $\{\langle\sigma_{C}^{k}\rangle\}$ by means of a finite number $N$ of evaluations of $\chi_{C}(\varphi)$. Note that the determinant of the Vandermonde matrix, i.e. $\textrm{det}\left[V(\underline{\varphi})\right]$, is given by the product of the differences between all the elements of the vector $\underline{\varphi}$, which are counted only once with their appropriate sign. As a result, $\textrm{det}\left[V(\underline{\varphi})\right] = 0$ if and only if $\underline{\varphi}$ has at least two identical elements. Only in that case, the inverse of $V(\underline{\varphi})$ does not exist and the polynomial interpolation problem cannot be longer solved. However, although the solution of a polynomial interpolation by means of the inversion of the Vandermonde matrix exists and is unique, $V(\underline{\varphi})$ is an ill-conditioned matrix~\cite{Meyer2000}. This means that the matrix is highly sensitive to small variations of the set of the input data (in our case the parameters $\varphi$'s), such that the condition number of the matrix may be large and the matrix becomes singular. As a consequence, the reconstruction procedure will be computationally inefficient, especially in the case the measurements are affected by environmental noise. Numerically stable solutions of a polynomial interpolation problem usually rely on the Newton polynomials~\cite{Trefethen2000}. The latter allow us to write the characteristic function $\chi_{C}(\varphi)$ in polynomial terms as a function of each value of $\underline{\varphi}$, which is denoted as $\chi^{\textrm{pol}}_{C}(\varphi)$.

Then, the natural question arises on what is an optimal choice for $\underline{\varphi}$. It is essential, indeed, to efficiently reconstruct the set $\{\langle\sigma_{C}^{k}\rangle\}$ of the statistical moments of $\sigma_{C}$. For this purpose, we can take into account the error $e_{C}(\varphi) \equiv \chi_{C}(\varphi) - \chi^{\textrm{pol}}_{C}(\varphi)$ in solving the polynomial interpolation problem in correspondence of a value of $\varphi$ different from the interpolating points within the parameter vector $\underline{\varphi}$. The error
$e_{C}(\varphi)$ depends on the regularity of the function $\chi_{C}(\varphi)$, and especially on the values assumed by the parameters $\varphi$. As shown in~\cite{Trefethen2000}, the choice of the $\varphi$'s for which the interpolation error is minimized is given by the real zeros of the Chebyshev polynomial of degree $N$ in the interval $\left[\varphi_{\textrm{min}},\varphi_{\textrm{max}}\right]$, where $\varphi_{\textrm{min}}$ and $\varphi_{\textrm{max}}$ are, respectively, the lower and upper bound of the parameters $\varphi$. Accordingly, the optimal choice for $\underline{\varphi}$ is given by
\begin{equation}\label{cheby}
\varphi_{k} = \frac{(\varphi_{\textrm{min}}+\varphi_{\textrm{max}})}{2} + \frac{\varphi_{\textrm{max}}-\varphi_{\textrm{min}}}{2}\cos\left(\frac{2k-1}{2N}\pi\right),
\end{equation}
with $k = 1,\ldots,N$. Let us observe that the value of $N$, i.e. the number of evaluations of the characteristic function $\chi_{C}(\varphi)$, is equal to the number of statistical moments of $\sigma_{C}$ we want to infer. Therefore, in principle, if the probability distribution of the stochastic quantum entropy production is a Gaussian function, then $N$ could be taken equal to $2$. Hence, once all the evaluations of the characteristic functions $\chi_{C}(\varphi)$ have been collected, we can derive the statistical moments of the quantum entropy production $\sigma_{C}$, and consequently reconstruct the probability distribution $\textrm{Prob}(\sigma_{C})$ as
\begin{equation}\label{Fourier}
\textrm{Prob}(\sigma_{C}) \approx \mathcal{F}^{-1}\left[\sum_{k = 0}^{N-1}\frac{\langle\sigma_{C}^{k}\rangle}{k!}(i\mu)^{k}\right]
\equiv\frac{1}{2\pi}\int^{\infty}_{-\infty}\left(\sum_{k = 0}^{N-1}\frac{\langle\sigma_{C}^{k}\rangle}{k!}(i\mu)^{k}\right)e^{-i\mu\sigma_{C}}d\mu,
\end{equation}
where $\mu\in\mathbb{R}$ and $\mathcal{F}^{-1}[\cdot]$ denotes the inverse Fourier transform~\cite{Mnatsakanov}, which is numerically performed~\cite{Athanassoulis}. To do that, we fix a-priori the integration step $d\mu$ and we vary the integration limits of the integral, in order to minimize the error
$\sum_{k}\left|\widetilde{\langle\sigma_{C}^{k}\rangle} - \overline{\langle\sigma_{C}^{k}\rangle}\right|^{2}$ between the statistical moments $\widetilde{\langle\sigma_{C}^{k}\rangle}$, obtained by measuring the characteristic functions $\chi_{C}(\varphi)$ (i.e. after the inversion of the Vandermonde matrix), and the ones calculated from the reconstructed probability distribution, $\overline{\langle\sigma_{C}^{k}\rangle}$, which we derive by numerically computing the inverse Fourier transform for each value of $\sigma_{C}$. This procedure has to be done separately for $C\in\{A,B,A-B\}$, while, as mentioned, the probability distribution $\textrm{Prob}(\sigma_{A+B})$ is obtained by a convolution of $\textrm{Prob}(\sigma_{A})$ and $\textrm{Prob}(\sigma_{B})$. Here, it is worth observing that Eq. (\ref{Fourier}) provides an approximate expression for the probability distribution $\textrm{Prob}(\sigma_{C})$. Ideally, given a generic unital quantum CPTP map modeling the dynamics of the system, an infinite number $N$ of statistical moment of $\sigma_{C}$ is required to reconstruct $\textrm{Prob}(\sigma_{C})$ if we use the inverse Fourier transform as in Eq. (\ref{Fourier}). While we can always calculate the Fourier transform to reconstruct the probability distribution from its moments, in the case of a distribution with discrete support (as in our case), there is a different method that can lead to higher precision, especially when the moment generating function is not approximated very well by the chosen number $N$ of extracted moments. As a matter of fact, each statistical moment $\widetilde{\langle\sigma_{C}^{k}\rangle}$, with $C\in\{A,B,A-B\}$, is the best approximation of the true statistical moments of $\sigma_{C}$ from the measurement of the corresponding characteristic functions $\chi_C(\varphi)$. Hence, apart from a numerical error coming from the inversion of the Vandermonde matrix $A$ or the use of the Newton polynomials $\chi^{\textrm{pol}}_{C}$, we can state that
\begin{equation}\label{expansion_sigma_C}
\widetilde{\langle\sigma_{C}^{k}\rangle} \simeq \sum_{i = 1}^{M_{C}}\sigma^{k}_{C,i}\textrm{Prob}(\sigma_{C,i})
= \sigma_{C,1}^{k}\textrm{Prob}(\sigma_{C,1}) + \ldots + \sigma_{C,M_{C}}^{k}\textrm{Prob}(\sigma_{C,M_{C}}),
\end{equation}
with $k = 1,\ldots,N$. In (\ref{expansion_sigma_C}), $M_{C}$ is equal to the number of values that can be assumed by $\sigma_{C}$, while $\sigma_{C,i}$ denotes the $i-$th possible value for the stochastic quantum entropy production of the (sub)system $C$. As a result, the probabilities $\textrm{Prob}(\sigma_{C,i})$, $i = 1,\ldots,M$, can be approximately expressed as a function of the statistical moments $\left\{\widetilde{\langle\sigma_{C}^{k}\rangle}\right\}$, i.e.
\begin{equation}
\begin{pmatrix} \widetilde{\langle\sigma_{C}\rangle} \\ \widetilde{\langle\sigma_{C}^{2}\rangle} \\ \vdots \\ \widetilde{\langle\sigma_{C}^{N}\rangle} \end{pmatrix} = \underbrace{\begin{pmatrix} \sigma_{C,1}  & \sigma_{C,2} & \ldots & \sigma_{C,M} \\
\sigma_{C,1}^{2}  & \sigma_{C,2}^{2} & \ldots & \sigma_{C,M}^{2} \\
\vdots & \vdots & \vdots & \vdots \\
\sigma_{C,1}^{N}  & \sigma_{C,2}^{N} &  \ldots & \sigma_{C,M}^{N} \end{pmatrix}}_{\Sigma_{C}}
\begin{pmatrix} \textrm{Prob}(\sigma_{C,1}) \\ \textrm{Prob}(\sigma_{C,2}) \\ \vdots \\ \textrm{Prob}(\sigma_{C,M}) \end{pmatrix},
\end{equation}
where $\Sigma_{C}\in\mathbb{R}^{N\times M}$. By construction $\Sigma_{C}$ is a rectangular matrix, that is computed by starting from the knowledge of the values assumed by the stochastic quantum entropy production $\sigma_{C,i}$. Finally, in order to obtain the probabilities $\textrm{Prob}(\sigma_{C,i})$, $i = 1,\ldots,M_C$, we have to adopt the Moore-Penrose pseudo-inverse of $\Sigma_{C}$, which is defined as
\begin{equation}
\Sigma^{+}_{C}\equiv (\Sigma_{C}'\Sigma_{C})^{-1}\Sigma_{C}'.
\end{equation}

\begin{figure}[h!]
	\centering
	\includegraphics[scale=3.35]{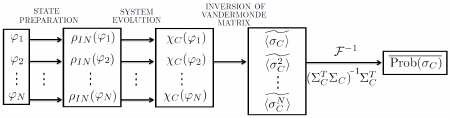}
	\caption{Pictorial representation of the reconstruction algorithm. The reconstruction algorithm starts by optimally choosing the parameters $\varphi\in\{\alpha,\beta,\gamma\}$ as the zeros of the Chebyshev polynomial of degree $N$ in the interval $[\varphi_{\textrm{min}},\varphi_{\textrm{max}}]$. Then, the moment generating functions $\chi_{C}(\varphi)$, with $C\in\{A,B,A-B\}$, are measured. Once the estimates $\widetilde{\langle \sigma_C^k\rangle}$ of the statistical moments of $\sigma_{C}$ are obtained, the inverse Fourier transform $\mathcal{F}^{-1}[\cdot]$ has to be numerically performed. Alternatively, the Moore-Penrose pseudo-inverse of $\Sigma_{C}$ can be adopted. As a result, an estimate $\overline{\textrm{Prob}(\sigma_{C})}$ for the probability distribution $\textrm{Prob}(\sigma_{C})$ is obtained.}
	\label{fig:procedure}
\end{figure}

A pictorial representation of the reconstruction protocol is shown in Fig.~\ref{fig:procedure}.
Let us observe, again, that the proposed algorithm is based on the expression of (\ref{sigma}) for the stochastic quantum entropy production, which has been obtained by assuming unital CPTP quantum maps for the system dynamics. It is expected that for a general open quantum system, not necessarily described by a unital CPTP map, one can extend the proposed reconstruction protocol, even though possibly at the price of a greater number of measurements. Notice that, since (\ref{equality_cond_prob}) is no longer valid in the general case, one has to use directly (\ref{general_sigma})-(\ref{general_mean_sigma}). However, we observe that, as shown in \cite{Manzano2015}, the ratio between the conditional probabilities may admit for a large family of CPTP maps the form $p(a^{\textrm{fin}}_{k}|a^{\textrm{in}}_{m}) / p(a^{\textrm{in}}_{m}|a^{\textrm{ref}}_{k})
\equiv e^{-\Delta V}$, where the quantity $\Delta V$ is related to the so-called non-equilibrium potential, so that $\sigma=\sigma_{unital}+V$ and $\sigma_{unital}$ again given by (\ref{sigma}).

\subsection*{Required number of measurements}

From an operational point of view, we need to measure (directly or indirectly) the quantities
\begin{equation}
\begin{cases}
\chi_{A}(\alpha) = \textrm{Tr}\left[\left((\rho_{A,\mathcal{T}})^{\alpha}\otimes\mathbbm{1}_{B}\right)\Phi\left(
(\rho_{\textrm{A,in}})^{1 - \alpha}\otimes\rho_{\textrm{B,in}}\right)\right] \\
\chi_{B}(\beta) = \textrm{Tr}\left[\left(\mathbbm{1}_{A}\otimes(\rho_{B,\mathcal{T}})^{\beta}\right)\Phi\left(
\rho_{\textrm{A,in}}\otimes(\rho_{\textrm{B,in}})^{1 - \beta}\right)\right]\\
\chi_{A-B}(\gamma) = \textrm{Tr}\left[(\rho_{\mathcal{T}})^{\gamma}\Phi\left(
(\rho_{\textrm{in}})^{1 - \gamma}\right)\right]
\end{cases},
\end{equation}
i.e. the moment generating functions of $\sigma_A$, $\sigma_B$ and $\sigma_{A-B}$, after a proper choice of the parameters $\alpha$, $\beta$ and $\gamma$, with $\alpha,\beta, \gamma \in\mathbb{R}$. For this purpose, as shown in appendix~\ref{chapter:appC}, it is worth mentioning that $\left(\rho_{C,\textrm{in}}\right)^{1-\varphi}\equiv\sum_{m}\Pi^{\textrm{in}}_{C,m}p(x_{m}^{\textrm{in}})^{1 - \varphi}$ and $\left(\rho_{C,\mathcal{T}}\right)^{\varphi}\equiv\sum_{k}\Pi^{\mathcal{T}}_{C,k}p(x^{\mathcal{T}}_{k})^{\varphi}$, where $C\in\{A,B,A - B\}$, $x\in\{a,b,c\}$ and $\varphi\in\{\alpha,\beta,\gamma\}$. A direct measurement of $\chi_{C}(\varphi)$, based for example on an interferometric setting as shown in \cite{MazzolaPRL2013} for the work distribution inference, is not trivial, especially for the general fully quantum case. For this reason, we propose a procedure, suitable for experimental implementation, requiring a limited number of measurements, based on the following steps:
\begin{enumerate}[(1)]
\item
Prepare the initial product state $\rho_{\textrm{in}}=\rho_{A,\textrm{in}}\otimes\rho_{B,\textrm{in}}$, as given in (\ref{rho_in}), with fixed probabilities $p(a^{\textrm{in}}_{m})$ and $p(b^{\textrm{in}}_{h})$. Then, after the composite system $A - B$ is evolved within the time interval $[0,\mathcal{T}]$, measure the occupation probabilities $p(a^{\textrm{fin}}_{k})$ and $p(b^{\textrm{fin}}_{l})$ via local measurements on $A$ and $B$. Then, compute the stochastic quantum entropy productions $\sigma_{A}(a_m^{\textrm{in}},a_k^{\textrm{fin}})$ and $\sigma_{B}(b_h^{\textrm{in}},b_l^{\textrm{fin}})$. Simultaneous measurements on $A$ and $B$ yield also the probabilities $p(c_{kl}^{\textrm{fin}})$ and thus $\sigma_{A-B}(c_{mh}^{\textrm{in}},c_{kl}^{\textrm{fin}})$.
\item
For every chosen value of $\alpha$, $\beta$ and $\gamma$, prepare, for instance by quantum optimal control tools~\cite{DoriaPRL2011}, the quantum subsystems in the states
\begin{equation*}
\begin{cases}
\displaystyle{\rho_{\textrm{IN}}(\alpha)\equiv\frac{(\rho_{A,\textrm{in}})^{1 - \alpha}\otimes\rho_{B,\textrm{in}}}{\textrm{Tr}\left[(\rho_{A,\textrm{in}})^{1 - \alpha}\otimes\rho_{B,\textrm{in}}\right]}} \\
\displaystyle{\rho_{\textrm{IN}}(\beta)\equiv\frac{\rho_{\textrm{A,in}}\otimes(\rho_{B,\textrm{in}})^{1 - \beta}}{\textrm{Tr}\left[\rho_{\textrm{A,in}}\otimes(\rho_{B,\textrm{in}})^{1 - \beta}\right]}}\\
\displaystyle{\rho_{\textrm{IN}}(\gamma)\equiv\frac{\left(\rho_{\textrm{A,in}}\otimes\rho_{B,\textrm{in}}\right)^{1 - \gamma}}{\textrm{Tr}\left[\left(\rho_{\textrm{A,in}}\otimes\rho_{B,\textrm{in}}\right)^{1 - \gamma}\right]}}
\end{cases},
\end{equation*}
and let the system evolve.
\item
Since the characteristic function $\chi_{C}(\varphi)$, with $C\in\{A,B,A-B\}$ and $\varphi\in\{\alpha,\beta,\gamma\}$, is given by performing a trace operation with respect to the composite system $A - B$, one can write the following simplified relation:
\begin{eqnarray}
\chi_{C}(\varphi) &=& \sum_{k}\sum_{m}\langle m|p(x^{\textrm{fin}}_{k})^{\varphi}|k\rangle\langle k|\rho_{\textrm{FIN}}(\varphi)|m\rangle\nonumber \\
&=& \sum_{m}p(x^{\textrm{fin}}_{m})^{\varphi}\langle m|\rho_{\textrm{FIN}}(\varphi)|m\rangle,
\end{eqnarray}
where $\{|l\rangle\}$, $l = m,k$, is the orthonormal basis of the composite system $A - B$, $x\in\{a,b,c\}$ and $\rho_{\textrm{FIN}}(\varphi)\equiv\Phi[\rho_{\textrm{IN}}(\varphi)]$ (with $p(x^{\textrm{fin}}_{m})$ measured in step 1 and $\rho_{\textrm{IN}}(\varphi)$ introduced in step 2). Thus, measure the occupation probabilities $\langle m|\rho_{\textrm{FIN}}(\varphi)|m\rangle$ in order to obtain all the characteristic functions $\chi_{C}(\varphi)$.
\end{enumerate}
It is observed that the measure of the characteristic functions $\chi_{C}(\varphi)$ relies only on the measure of occupation probabilities. Hence, the proposed procedure does \textit{not} require any tomographic measurement. Moreover, for the three steps of the protocol we can well quantify the required number of measurements to properly infer the statistics of the quantum entropy production regarding the composite quantum system. The required number of measurements, indeed, scales linearly with the number of possible measurement outcomes coming from each quantum subsystem at the initial and final stages of the protocol. Equivalently, if we define $d_{A}$ and $d_{B}$ as the dimension of the Hilbert space concerning the quantum subsystems $A$ and $B$, we can state that the number of measurements for both of the three steps scales linearly with $d_{A} + d_{B}$, i.e. with the number of values $M_{A} + M_{B}$ that can be assumed by $\sigma_{A}$ and $\sigma_{B}$, the stochastic quantum entropy production of the subsystems. It also
scales linearly with $M_A M_B$ for the reconstruction of the stochastic quantum entropy production $\sigma_{A-B}$ of the composite system. The reason is that the described procedure is able to reconstruct the distribution of the stochastic quantum entropy production, without directly measuring the joint probabilities $p_a(k,m)$ and $p_{b}(l,h)$ for the two subsystems and $p_c(mh,kl)$ for the composite system. Otherwise, the number of required measurements would scale, respectively, as $M_{A}^{2}$ and $M_{B}^{2}$ for the subsystems and as $(M_AM_B)^2$ for the composite system in order to realize all the combinatorics concerning the measurement outcomes.

\subsection{Illustrative example - M\o{}lmer-S\o{}rensen gate}

Here, in order to illustrate our theoretical results, we discuss an experimental implementation with trapped ions. Trapped ions have been demonstrated to be a versatile tool for quantum simulation~\cite{Friedenauer2008,Lanyon2011}, including simulation of quantum thermodynamics~\cite{Huber2008,AbahPRL2012,RossnagelPRL2014,An2015,Rossnagel325}. The application of our protocol on a physical example relies on the availability of experimental procedures for state preparation and readout, as well as an entangling operation.

We consider a system of two trapped ions, whose two internal states allow to encode the qubit states $|0\rangle$ and $|1\rangle$ of the standard computational basis. Then, the subsystems $A$ and $B$ are represented by the two qubits. The latter can interact by the common vibrational (trap) mode of the two ions, and external lasers allow to manipulate the ion states, generating arbitrary single qubit rotations through individual addressing or an entangling operation, as for example the M\o{}lmer-S\o{}rensen gate operation~\cite{SoerensenPRL82,RoosNJP2008,MonzPRL106,Nigg302}.
\begin{figure}[h!]
	\centering
	\includegraphics[scale=.38]{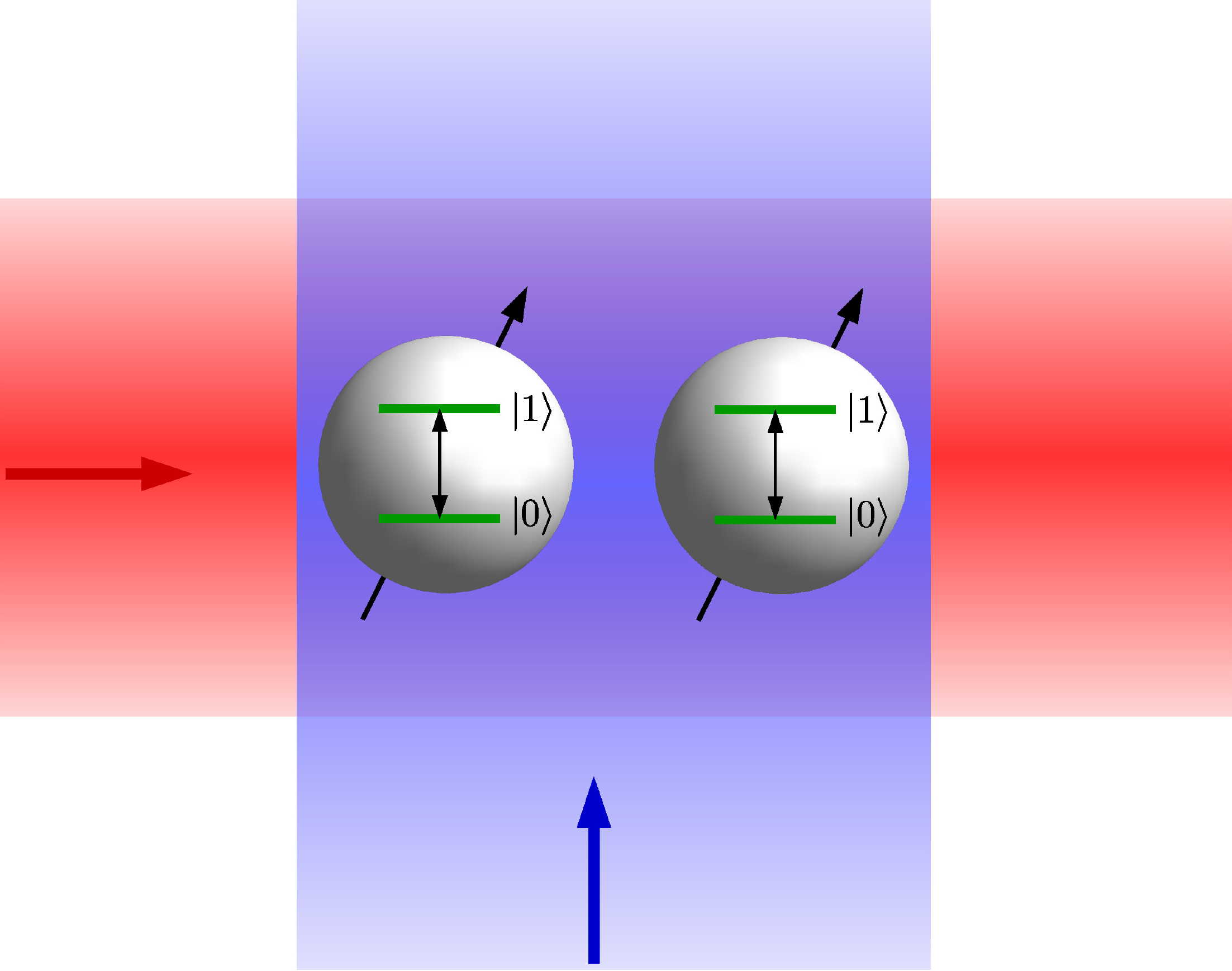}
	\caption{Pictorial representation of two trapped ions subjected to two laser fields. The internal levels of the ions allow to encode one qubit in each ion. The transition between these levels is driven by the lasers, where the driving depends on the state of the common vibrational (trap) mode of the two ions. The lasers can be focused to choose between single or global addressing. This allows to generate local gates as well as entangling gates.}
	\label{fig:fig2}
\end{figure}
Fig.~\ref{fig:fig2} shows a pictorial representation of the system. While usually universal state preparation for single qubits is supposed only for pure states, here we have to prepare mixed states. However, once we have prepared a pure state with the right amount of population in the two levels, we can reach the required mixed state by applying a random $Z$ rotation leading to a complete dephasing of the two levels, where $Z$ is the corresponding Pauli matrix. The two-qubit operation, that generates entanglement between $A$ and $B$, is chosen to be a partial M\o{}lmer-S\o{}rensen gate operation, given by the following unitary operation, depending on the phase $\phi$:
\begin{equation}\label{propagator}
\mathcal{U}(\phi)=e^{-i\phi\left(X^{A}\otimes X^{B}\right)},
\end{equation}
where $X^{A}$ and $X^{B}$ are equal, respectively, to the Pauli matrix $X$ for the quantum systems $A$ and $B$. In the following (and unless explicitly stated otherwise), we choose $\phi=\frac{\pi}{7}$, and start from the initial state $\rho_{0} = \textrm{diag}\left(\frac{6}{25},\frac{9}{25},\frac{4}{25},\frac{6}{25}\right)$ since this choice leads to a \textit{non-Gaussian} probability distribution $\textrm{Prob}(\sigma_{A - B})$ of the stochastic quantum entropy production. For the sake of simplicity, we remove the label $A$ and $B$ from the computational basis $\{|0\rangle,|1\rangle\}$ considered for the two subsystems. Thus, the corresponding projectors are $\Pi_{0}\equiv|0\rangle\langle 0|$ and $\Pi_{1}\equiv|1\rangle\langle 1|$, and each ion is characterized by $4$ different values of the stochastic quantum entropy production $\sigma_{C}$, with $C\in\{A,B\}$. As a consequence, the probability distribution $\textrm{Prob}(\sigma_{A - B})$ of the stochastic quantum entropy production for the composite system $A - B$ is defined over a discrete support given by $l$ samples, with $l \leq M_{A}M_{B} = 16$.

\subsubsection{Correlated measurement outcomes and correlations witness}

Generally, the outcomes of the second measurement of the protocol are correlated, as in our example, and the stochastic quantum entropy production of the composite system is sub-additive, i.e. $\langle\sigma_{A-B}\rangle\leq\langle\sigma_{A}\rangle+\langle\sigma_{B}\rangle$. Hence, by adopting the reconstruction algorithm proposed in Fig.~\ref{fig:procedure} we are able to effectively derive the upper bound of $\langle\sigma_{A - B}\rangle$, which defines the thermodynamic irreversibility for the quantum process. In the simulations we compare the fluctuation profile that we have derived by performing local measurements on the subsystems $A$ and $B$ with the ones that are obtained via a global measurement on the composite system $A - B$, in order to establish the amount of information which is carried by a set of local measurements. Furthermore, we discuss the changes of the fluctuation profile of the stochastic quantum entropy production both for unitary and noisy dynamics. The unitary operation describing the dynamics of the quantum system is given by (\ref{propagator}), while the noisy dynamics is given by the following differential Lindblad (Markovian) equation:
\begin{equation}\label{Lindblad}
\dot{\rho}(t) = -i\left[H,\rho\right] - \sum_{C\in\{A,B\}}\Gamma_{C}\left(\{\rho,L_{C}^{\dagger}L_{C}\} - 2L_{C}\rho L_{C}^{\dagger}\right).
\end{equation}
In (\ref{Lindblad}), $\rho(t)$ denotes the density matrix describing the composite quantum system $A - B$, $\{\cdot,\cdot\}$ is the anticommutator, $\Gamma_{A}$ and $\Gamma_{B}$ (rad/s) are dephasing rates corresponding to $L_{A}\equiv\Pi_{0}\otimes\mathbbm{1}_{B}$ and $L_{B}\equiv\mathbbm{1}_{A}\otimes\Pi_{0}$ are pure-dephasing Lindblad operators. The Hamiltonian of the composite system $A - B$ in (\ref{Lindblad}), instead, is given by $$H = \omega\left(X^{A}\otimes X^{B}\right),$$ where the interaction strength $\omega = \phi/\tau$ (rad/s) with $\tau$ kept fixed and chosen equal to $50$ s (leading to a largely relaxed system dynamics), consistently with the unitary operation (\ref{propagator}).
\begin{figure}[h!]
	\centering
	\includegraphics[scale=6.1]{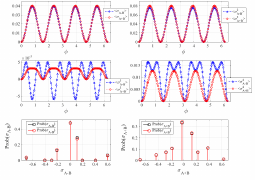}
	\caption{In the four top panels, we show the statistical moments $\langle\sigma^{k}_{A - B}\rangle$ and $\langle\sigma^{k}_{A + B}\rangle$, $k = 1,\ldots,4$, of the stochastic quantum entropy production $\sigma_{A - B}$ and $\sigma_{A + B}$ as a function of $\phi\in[0,2\pi]$, in the case where the dynamics of the composite quantum system $A - B$ is unitary. In the two bottom panels, moreover, we plot a comparison between the samples of the probability distributions $\textrm{Prob}(\sigma_{A - B})$, $\textrm{Prob}(\sigma_{A + B})$ (black squares) and the samples of the corresponding reconstructed distribution (red circles). The latter numerical simulations are performed by considering $\phi = \pi/7$, and $N$ is equal, respectively, to $20$ (for the fluctuation profile of $\sigma_{A - B}$) and $10$.}
	\label{fig:fig3}
\end{figure}

\begin{figure}[h!]
	\centering
	\includegraphics[scale=6.25]{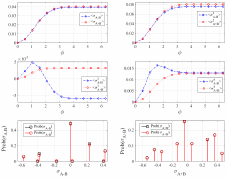}
	\caption{In the four top panels, the statistical moments $\langle\sigma^{k}_{A - B}\rangle$ and $\langle\sigma^{k}_{A + B}\rangle$, $k = 1,\ldots,4$, of the stochastic quantum entropy production $\sigma_{A - B}$ and $\sigma_{A + B}$ as a function of $\phi\in[0,2\pi]$ are shown, in the case where the dynamics of the composite quantum system $A - B$ is described by a Lindblad (Markovian) equation. In the two bottom panels, moreover, we plot a comparison between the samples of the probability distributions $\textrm{Prob}(\sigma_{A - B})$, $\textrm{Prob}(\sigma_{A + B})$ (black squares) and the samples of the corresponding reconstructed distribution (red circles). The latter numerical simulations are performed by considering $\phi = \frac{5\pi}{6}$, $\Gamma = \Gamma_{A} = \Gamma_{B} = 0.2$ rad/s, and $N$ is equal, respectively, to $20$ (for the fluctuation profile of $\sigma_{A - B}$) and $10$.}
	\label{fig:fig4}
\end{figure}
In Figs.~\ref{fig:fig3} and~\ref{fig:fig4}, we plot the first $4$ statistical moments of $\sigma_{A - B}$ and $\sigma_{A + B}$ as a function of the phase $\phi$, respectively, in case of unitary and noisy dynamics. Moreover, we show, for a given value of $\phi$, the probability distributions $\textrm{Prob}(\sigma_{A - B})$ and $\textrm{Prob}(\sigma_{A + B})$ for both unitary and noisy dynamics, compared with the corresponding reconstructed distributions obtained by applying the reconstruction algorithm, which we call $\overline{\textrm{Prob}(\sigma_{A - B})}$ and $\overline{\textrm{Prob}(\sigma_{A + B})}$, respectively. Let us recall that $\textrm{Prob}(\sigma_{A + B})$ is obtained by performing the two local measurements with observables $\mathcal{O}^{\textrm{fin}}_{A}$ and $\mathcal{O}^{\textrm{fin}}_{B}$ independently (disregarding the correlations of their outcomes) on the subsystems $A$, $B$, while the distribution $\textrm{Prob}(\sigma_{A + B})$ requires to measure $\mathcal{O}^{\textrm{fin}}_{A}$ and $\mathcal{O}^{\textrm{fin}}_{B}$ simultaneously, i.e. measuring the observable $\mathcal{O}^{\textrm{fin}}_{A - B}$, defined by (\ref{eq:ovservable_A-B}). For unitary dynamics, the statistical moments of the stochastic quantum entropy productions $\sigma_{A - B}$ and $\sigma_{A + B}$ follow the oscillations of the dynamics induced by changing the gate phase $\phi$. Conversely, for the noisy dynamics given by (\ref{Lindblad}), with $\Gamma = \Gamma_{A} = \Gamma_{B} > 0$, when $\phi$ increases the system approaches a fixed point of the dynamics. Consequently, the statistical moments of the stochastic quantum entropy production tend to the constant values corresponding to the fixed point, and the distribution of the stochastic entropy production becomes narrower. In both Figs.~\ref{fig:fig3} and~\ref{fig:fig4}, the first statistical moments (or mean values) $\langle\sigma_{A - B}\rangle$ and $\langle\sigma_{A + B}\rangle$ are almost overlapping, and the sub-additivity of $\sigma_{A - B}$ is confirmed by the numerical simulations. Furthermore, quite surprisingly, also the second statistical moments of $\sigma_{A - B}$ and $\sigma_{A + B}$ are very similar to each other. This means that the fluctuation profile of the stochastic entropy production $\sigma_{A + B}$ is able to well reproduce the probability distribution of $\sigma_{A - B}$ in its Gaussian approximation, i.e. according to the corresponding first and second statistical moments. In addition, we can state that the difference of the higher order moments of $\langle\sigma_{A+B}\rangle$ and $\langle\sigma_{A-B}\rangle$ reflects the presence of correlations between $A$ and $B$ created by the map, since for a product state $\sigma_{A - B} = \sigma_{A+B}$. Therefore, \textit{the difference between the fluctuation profiles of $\sigma_{A - B}$ and $\sigma_{A + B}$ constitutes a witness for classical and/or quantum correlations in the final state of the system before the second measurement}. As a consequence, if $\textrm{Prob}(\sigma_{A - B})$ and $\textrm{Prob}(\sigma_{A + B})$ are \textit{not} identical, then the final density matrix $\rho_{\textrm{fin}}$ is \textit{not} a product state, and (classical and/or quantum) correlations are surely present.
Notice that the converse statement is not necessarily true because the quantum correlations can be partially or fully destroyed by the second local measurements, while the classical ones are still preserved and thus detectable.

\begin{figure}[h!]
	\centering
	\includegraphics[scale=8.75]{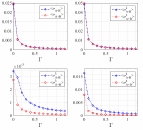}
	\caption{The statistical moments $\langle\sigma^{k}_{A - B}\rangle$ and $\langle\sigma^{k}_{A + B}\rangle$, $k = 1,\ldots,4$, of the stochastic quantum entropy production $\sigma_{A - B}$ and $\sigma_{A + B}$ as a function of $\Gamma\in[0,1.2]$ rad/s are shown, in the case the dynamics of the composite quantum system $A - B$ is described by a Lindblad (Markovian) equation, with $\phi = \pi/7$.}
	\label{fig:fig5}
\end{figure}

Furthermore, in Fig.~\ref{fig:fig5} we show the first $4$ statistical moments of $\sigma_{A - B}$ and $\sigma_{A + B}$ as a function of $\Gamma$ (rad/s). As before, we can observe a perfect correspondence between the two quantities when we consider only the first and second statistical moments of the stochastic quantum entropy productions, and, in addition, similar behaviour for the third and fourth statistical moments.
\\
Indeed, since the coherence terms of the density matrix describing the dynamics of the composite quantum system tend to zero for increasing $\Gamma$, the number of samples of $\sigma_{A - B}$ and $\sigma_{A + B}$ with an almost zero probability to occur is larger, and also the corresponding probability distribution approaches a Gaussian one, with zero mean and small variance. In accordance with Figs.~\ref{fig:fig3} and~\ref{fig:fig4}, this result confirms the dominance of decoherence in the quantum system dynamics (for large enough $\Gamma$), which coincides with no creation of correlations.

\subsubsection{Reconstruction for unitary dynamics}

Here, we show the performance of the reconstruction algorithm for the probability distribution of the stochastic quantum entropy production $\sigma_{A+B}$ via local measurements on the subsystems $A$ and $B$, when the dynamics of the quantum system is unitary. In particular, in the numerical simulations, we take the parameters $\alpha$ and $\beta$ of the algorithm, respectively, equal to the real zeros of the Chebyshev polynomial of degree $N$ in the intervals $\left[\alpha_{\textrm{min}},\alpha_{\textrm{max}}\right] = [0,N]$ and $\left[\beta_{\textrm{min}},\beta_{\textrm{max}}\right] = [0,N]$. This choice for the minimum and maximum values of the parameters $\alpha$ and $\beta$ ensures a very small numerical error (about $10^{-4}$) in the evaluation of each statistical moment of $\sigma_{A}$ and $\sigma_{B}$ via the inversion of the Vandermonde matrix, already for $N > 2$. Indeed, since all the elements of the vectors $\underline{\alpha}$ and $\underline{\beta}$ are different from each other, i.e. $\alpha_{i}\neq\alpha_{j}$ and $\beta_{i}\neq\beta_{j}$ $\forall i,j = 1,\ldots,N$, we can derive the statistical moments of $\sigma_{C}$, with $C\in\{A,B\}$, by inverting the corresponding Vandermonde matrix. The number $N$ of evaluations of the moment generating functions $\chi_{A}(\alpha)$ and $\chi_{B}(\beta)$, instead, has been taken as a free parameter in the numerics in order to analyze the performance of the reconstruction algorithm. The latter may be quantified in terms of the Root Mean Square Error~(RMSE)
defined as
\begin{equation}\label{RMSE_m}
\textrm{RMSE}\left(\{\langle\sigma^{k}_{A+B}\rangle\}_{k = 1}^{N_{\textrm{max}}}\right)\equiv\sqrt{\frac{\displaystyle{\sum_{k = 1}^{N_{\textrm{max}}}
\left|\langle\sigma^{k}_{A+B}\rangle - \overline{\langle\sigma^{k}_{A+B}\rangle}\right|^{2}}}{N_{\textrm{max}}}},
\end{equation}
where $\{\langle\sigma^{k}_{A+B}\rangle\}$ are the true statistical moments of the stochastic quantum entropy production $\sigma_{A+B}$, which have been numerically computed by directly using (\ref{eq:convolution})-(\ref{prob_b}), while $\overline{\langle\sigma^{k}_{A+B}\rangle}$ are the reconstructed statistical moments after the application of the inverse Fourier transform or the Moore-Penrose pseudo-inverse of $\Sigma_{C}$, $C\in\{A,B\}$. $N_{\textrm{max}}$, instead, is the largest value of $N$ considered for the computation of the $\textrm{RMSE}\left(\{\langle\sigma^{k}_{A+B}\rangle\}\right)$ in the numerical simulations (in this example $N_{\textrm{max}} = 16$).
\begin{figure}[h!]
	\centering
	\includegraphics[scale=6.25]{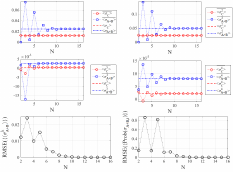}
	\caption{Reconstructed statistical moments of $\sigma_{A}$, $\sigma_{B}$ and $\sigma_{A+B}$ as a function of $N$ with unitary dynamics. In the four top panels we show the statistical moments $\langle\sigma^{k}_{C}\rangle$, $C = \{A,B\}$ (equal due to symmetry), and $\langle\sigma^{k}_{A+B}\rangle$, $k = 1,\ldots,4$, of the stochastic quantum entropy production $\sigma_{A}$, $\sigma_{B}$ and $\sigma_{A+B}$ as a function of $N$. As $N$ increases, the reconstructed statistical moments converge to the corresponding true value. The corresponding RMSEs $\textrm{RMSE}\left(\{\langle\sigma^{k}_{A+B}\rangle\}\right)$ and $\textrm{RMSE}\left(\{\textrm{Prob}(\sigma_{A+B,i})\}\right)$, instead, are plotted in the two bottom panels. All the numerical simulations in the figure are performed by considering unitary dynamics for the composite system $A - B$ with $\phi = \pi/7$.}
	\label{fig:fig6}
\end{figure}
Another measure for the evaluation of the algorithm performance, which will be used hereafter, is given by the RMSE
\begin{equation}\label{RMSE_p}
\textrm{RMSE}\left(\{\textrm{Prob}(\sigma_{A+B,i})\}_{i = 1}^{l}\right)\equiv\sqrt{\frac{\displaystyle{\sum_{i = 1}^{l}R_{i}^{2}}}{l}},
\end{equation}
where $R_{i} \equiv \left|\textrm{Prob}(\sigma_{A+B,i}) - \overline{\textrm{Prob}(\sigma_{A+B,i})}\right|$ is the reconstruction deviation, i.e. the discrepancy between the true and the reconstructed probability distribution $\textrm{Prob}(\sigma_{A+B})$. The $\textrm{RMSE}\left(\{\textrm{Prob}(\sigma_{A+B,i})\}\right)$ is computed with respect to the reconstructed values $\overline{\textrm{Prob}(\sigma_{A+B,i})}$ of the probabilities $\textrm{Prob}(\sigma_{A+B,i})$, $i = 1,\ldots,l$, for the stochastic quantum entropy production $\sigma_{A+B}$.

Fig.~\ref{fig:fig6} shows the performance of the reconstruction algorithm as a function of $N$ for the proposed experimental implementation with trapped ions in case the system dynamics undergoes a unitary evolution. In particular, we show the first $4$ statistical moments of $\sigma_{A}$, $\sigma_{B}$ and $\sigma_{A+B}$ as a function of $N$. In this regard, let us observe that the statistical moments of the stochastic quantum entropy production of the two subsystems $A$ and $B$ are equal due to the symmetric structure of the composite system. As expected, when $N$ increases, the reconstructed statistical moments converge to the corresponding true values, and also the reconstruction deviation tends to zero. This result is encoded in the RMSEs of (\ref{RMSE_m})-(\ref{RMSE_p}), which behave as monotonically decreasing functions. Both the $\textrm{RMSE}\left(\{\langle\sigma^{k}_{A+B}\rangle\}\right)$ and $\textrm{RMSE}\left(\{\textrm{Prob}(\sigma_{A+B,i})\}\right)$ sharply decrease for about $N \geq 6$, implying that the reconstructed probability distribution {\small $\overline{\textrm{Prob}(\sigma_{A+B})}$} overlaps with the true distribution $\textrm{Prob}(\sigma_{A+B})$ with very small reconstruction deviations $R_{i}$. Since the system of two trapped ions of this example is a small size system, we have chosen to derive the probabilities $\{\textrm{Prob}(\sigma_{A,i})\}$ and $\{\textrm{Prob}(\sigma_{B,i})\}$, $i = 1,\ldots,4$, without performing the inverse Fourier transform on the statistical moments $\{\widetilde{\langle\sigma_{C}^{k}\rangle}\}$, $C\in\{A,B\}$. Indeed, the computation of the inverse Fourier transform, which has to be performed numerically, can be a tricky step of the reconstruction procedure, because it can require the adoption of numerical methods with an adaptive step-size in order to solve the numerical integration. In this way, the only source of error in the reconstruction procedure is given by the expansion in Taylor series of the quantity $\chi_{C}(\varphi)$, with $C\in\{A,B\}$ and $\varphi\in\{\alpha,\beta\}$, around $\varphi = 0$ as a function of a \textit{finite} number of statistical moments $\langle\sigma_{C}^{k}\rangle$, $k = 1,\ldots,N-1$. As shown in Fig.~\ref{fig:fig6}, the choice of the value of $N$ is a degree of freedom of the algorithm, and it strictly depends on the physical implementation of the reconstruction protocol. In the experimental implementation above with two trapped ions, $N = 10$ ensures very good performance without making a larger number of measurements with respect to the number of values assumed by the stochastic quantum entropy production $\sigma_{A+B}$.

\begin{figure}[h!]
	\centering
	\includegraphics[scale = 6.25]{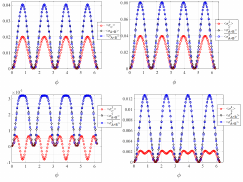}
	\caption{True and reconstructed statistical moments of $\sigma_{A}$, $\sigma_{B}$ and $\sigma_{A + B}$ as a function of the phase $\phi$ with unitary dynamics. We show the statistical moments $\langle\sigma^{k}_{C}\rangle$, $C = \{A,B\}$ (equal by symmetry), and $\langle\sigma^{k}_{A + B}\rangle$, $k = 1,\ldots,4$, of the stochastic quantum entropy production $\sigma_{A}$, $\sigma_{B}$ and $\sigma_{A + B}$ as a function of the phase $\phi$. All the numerical simulations are performed by considering unitary dynamics for the composite system $A - B$ with $N = 10$ and $\phi\in[0,2\pi]$.}
	\label{fig:Fig7}
\end{figure}
In Fig.~\ref{fig:Fig7}, moreover, we show for $N = 10$ the first $4$ true statistical moments of the stochastic quantum entropy productions $\sigma_{A}$ and $\sigma_{B}$ of the two subsystems, as well as the correlation-free convolution $\sigma_{A+B}$ as a function of $\phi\in[0,2\pi]$, along with the corresponding reconstructed counterpart $\overline{\langle\sigma^{k}_C\rangle}$, $k = 1,\ldots,4$, $C\in\{A,B,A+B\}$. As before, the reconstruction procedure yields values very close to the true statistical moments of $\sigma_{A}$, $\sigma_{B}$ and $\sigma_{A+B}$ for all values of the phase $\phi$.

\subsubsection{Reconstruction for noisy dynamics}

Let us consider, now, that the system dynamics is affected by pure-dephasing contributions, described via the differential Lindblad (Markovian) equation (\ref{Lindblad}), where the Hamiltonian of the composite system $A - B$ is defined as $H = \omega\left(X^{A}\otimes X^{B}\right)$.
\begin{figure}[h!]
	\centering
	\includegraphics[scale = 6.35]{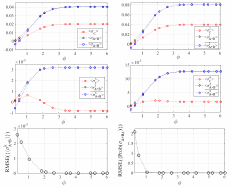}
	\caption{True and reconstructed statistical moments of $\sigma_{A}$, $\sigma_{B}$ and $\sigma_{A+B}$ as a function of the phase $\phi$ with noisy dynamics. In the first $4$ panels we show the statistical moments $\langle\sigma^{k}_{C}\rangle$, $C = \{A,B\}$ (equal by symmetry), and $\langle\sigma^{k}_{A+B}\rangle$, $k = 1,\ldots,4$, of the stochastic quantum entropy production $\sigma_{A}$, $\sigma_{B}$ and $\sigma_{A+B}$ as a function of the phase $\phi$. All the numerical simulations are performed by considering a Lindblad (Markovian) dynamics for the composite system $A - B$, given by (\ref{Lindblad}), with $N = 10$, $\Gamma = 0.2$ rad/s, and $\phi\in[0,2\pi]$. In the bottom panels of the figure, instead, we show the root mean square errors $\textrm{RMSE}\left(\{\langle\sigma^{k}_{A+B}\rangle\}\right)$ and $\textrm{RMSE}\left(\{\textrm{Prob}(\sigma_{A+B,i})\}\right)$.}
	\label{fig:Fig9}
\end{figure}
Since the fixed duration $\tau$ of the transformation has been chosen as before equal to $50$ s, we choose the desired phase $\phi$ by setting the interaction strength to $\omega \equiv \phi/\mathcal{T}$ (rad/s). Again, we evaluate the performance of the reconstruction algorithm also as a function of the phase $\phi = \omega\mathcal{T}$. As shown in Fig.~\ref{fig:Fig9}, when $\phi$ increases (with a fixed value of $\Gamma$, set to $0.2$) the statistical moments of $\sigma_{A+B}$ (but not necessarily the ones regarding the subsystems $A$ and $B$) increase as well, since when $\phi$ increases the system tends to a fixed point of the dynamics. Also the reconstruction procedure turns out to be more accurate for larger values of $\phi$, as shown in the two bottom panels of Fig.~\ref{fig:Fig9} (for this figure we use the Fourier transform). The reason is that when the dynamics approaches the fixed point, the distribution of the stochastic quantum entropy production becomes narrower and the convergence of the Fourier integral is ensured.

\begin{figure}[h!]
	\centering
	\includegraphics[scale = 9]{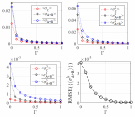}
	\caption{True and reconstructed statistical moments of $\sigma_{A}$, $\sigma_{B}$ and $\sigma_{A+B}$ as a function of the dephasing rate $\Gamma$. The first $3$ statistical moments of the stochastic quantum entropy productions for $A$, $B$ (equal by symmetry) and the composite system $A - B$ as a function of the dephasing rate $\Gamma = \Gamma_{A} = \Gamma_{B}$ (rad/s) are shown for the physical example of $2$ trapped ions. In particular, the statistical moments of $\sigma_{A+B}$ are put beside their reconstructed version, which have been obtained by choosing $N = 10$ and $\phi = \pi/7$. In the last panel, moreover, the corresponding root mean square error $\textrm{RMSE}\left(\{\langle\sigma^{k}_{A+B}\rangle\}\right)$ as a function of $\Gamma$ is shown.}
	\label{fig:Fig10}
\end{figure}
Finally, in Fig. \ref{fig:Fig10} we plot the behaviour of the first three statistical moments of $\sigma_{A}$, $\sigma_{B}$ and $\sigma_{A+B}$ as a function of the dephasing rate $\Gamma = \Gamma_{A} = \Gamma_{B}$, with $N = 10$ and $\phi = \pi/7$. As before, due to the symmetry of the bipartition, the statistical moments of $\sigma_{A}$ and $\sigma_{B}$ are identically equal. For increasing $\Gamma$ the dephasing becomes dominant over the interaction and all correlations between the subsystems are destroyed. As a consequence, the stochastic quantum entropy production tends to zero as is observed in the figure for all the investigated moments, both for the subsystems and the composite system.

\subsubsection{Probing irreversibility and dynamics correlations}

Once the fluctuation profile of the stochastic quantum entropy production (i.e. the corresponding probability distribution) is reconstructed, then the irreversibility properties of the composite system transformation can be successfully probed. The thermodynamic irreversibility is quantified by means of the mean value $\langle\sigma_{A - B}\rangle$, with $\langle \sigma_{A - B} \rangle = 0$ corresponding to thermodynamic reversibility. As previously shown in Figs.~\ref{fig:fig3},~\ref{fig:fig4} and \ref{fig:fig5}, the mean value $\langle\sigma_{A - B}\rangle$ can be well approximated by $\langle\sigma_{A + B}\rangle$ and from (\ref{eq:entropy-positivity}) and (\ref{eq:sigma_sum}) we have $0\leq\langle\sigma_{A-B}\rangle \leq \langle\sigma_{A+B}\rangle$. From Figs.~\ref{fig:fig5} and~\ref{fig:Fig10}, thus, we can observe that the implemented noisy transformation is more reversible with respect to the unitary one. Indeed, the statistical moments of the stochastic quantum entropy production, as well as the corresponding mean value, go to zero as $\Gamma$ increases. Since the dynamics originating from the Lindblad equation~(\ref{Lindblad}) admits as a fixed point the completely mixed state of the composite system $A - B$, if we increase the value of $\Gamma$ then the probability distribution of the quantum entropy production for the systems $A$, $B$ and $A - B$ tends to a Kronecker delta around zero, leading to a more-reversible system transformation with respect to the unitary case. For this reason, also the numerical simulations of Fig.~\ref{fig:Fig10} have been performed by using the inverse Fourier transform to reconstruct the probabilities $\{\textrm{Prob}(\sigma_{C,i})\}$, with $i = 1,\ldots,M_{C}$ and $C\in\{A,B\}$, instead of calculating the pseudo-inverse of the matrix $\Sigma_{C}$. As a matter of fact, as $\Gamma$ increases some values of $\sigma_{C}$ approach zero and $\Sigma_{C}$ becomes singular. Let us observe that, when the dynamics is unitary, the performance of the reconstruction algorithm adopting the inverse Fourier transform can be affected by a non-negligible error, as shown by the $\textrm{RMSE}\left(\{\langle\sigma^{k}_{A+B}\rangle\}\right)$ in the last panel of Fig. \ref{fig:Fig10}. For such case, the adoption of the pseudo-inverse of $\Sigma_{C}$ is to be preferred. Moreover, we expect that increasing the number of ions the thermodynamic irreversibility becomes more and more pronounced.

In conclusion, \textit{a system transformation on a multipartite quantum system involves stochastic quantum entropy production whenever correlations between the subsystems of the multipartite system is first created by the dynamics of the composite system and then destroyed by the second measurement}. This result, indeed, can be easily deduced from Figs.~\ref{fig:Fig9} and~\ref{fig:Fig10}, in which, for a fixed value of $\Gamma$ $(\Gamma = 0.2)$ and $\phi$ $(\phi = \pi/7)$ respectively, the behaviour of the statistical moments of the stochastic quantum entropy production as a function of $\phi$ ($\Gamma$) is monotonically increasing (decreasing). Indeed, the stronger is the interaction between the two ions, the larger is the corresponding production of correlations between them. On the other side, instead, the production of correlations within a multipartite system is inhibited due to the presence of strong decoherent processes.

\section{Conclusions and contributions}

To summarize, this chapter provides the following contributions:
\begin{itemize}
  \item We have studied the statistics of quantum-heat in a quantum system subjected to a sequence of projective measurements of a generic observable $\mathcal O$. At variance with previous works, we have investigated the case when the waiting time between consecutive measurements is a random variable. Previous works imply that when the waiting times are predetermined the quantum-heat obeys a integral fluctuation theorem which reads like the Jarzynski equality where quantum-heat replaces work. Here, we have shown that this continues to hold when the waiting times are random, and this can be understood by noticing that the corresponding quantum dynamics is unital.
  \item We have illustrated the theory with a two-level system, for which we have provided the explicit expressions of the characteristic function of quantum-heat. In particular, we have investigated both the case when the noise in the waiting times is annealed or quenched, and have noticed that, interestingly, in the annealed case more quantum heat is transferred by the two-level system as compared to the quenched noise case. Accordingly, our results reflects the intuition that a greater amount of noise in the waiting times between consecutive measurements of a given protocol is accompanied by higher quantum-heat transfer. Finally, we have found the existence of a discontinuity in the characteristic function $G(u)$ when the protocol relies on the application of many measurements of the Hamiltonian, i.e. $m\rightarrow\infty$ and $|a|^2\rightarrow 0,1$. This means that even an infinitesimal amount of noise in the measurement process will result in a finite amount of quantum-heat, also when measurements of the Hamiltonian are performed.
  \item We have characterized the stochastic quantum entropy production of an open quantum system by starting from a quantum fluctuation theorem (generalization of the Tasaki-Crooks theorem) based on the use of a two-time measurement protocol. In particular, the mean value $\langle\sigma\rangle$ of the stochastic quantum entropy production quantifies the amount of thermodynamic irreversibility of the system, with $\langle \sigma \rangle=0$ ($\langle \sigma \rangle>0$) corresponding to thermodynamic reversibility (irreversibility). At variance, we proved that $\langle \sigma \rangle<0$ -- violating the second law of the thermodynamics -- do not occur due to the non-negativity of the Kullback-Leibler divergence.
  \item Under the hypothesis that the open quantum system (at most described by an unital CPTP quantum map) is composed by mutually interacting subsystems, we have investigated the stochastic quantum entropy production both for the subsystems and for the composite system, showing that the mean values of the entropy production for the subsystems are sub-additive. However, their fluctuation profiles coincide if the composite system is described by a product state. As a consequence, by analyzing these fluctuation profiles one can witness classical and quantum correlations between each subsystem.
  \item We have proposed a suitable algorithm for the reconstruction of the fluctuation profiles of an arbitrary thermodynamical quantity. Without loss of generality, we have applied the procedure to infer the statistics of the stochastic quantum entropy production for each subsystem of a multipartite system. The algorithm is designed over a parametric version of the integral quantum fluctuation theorem, and provides the first $N$ statistical moments of the chosen thermodynamical quantity through the inversion of a Vandermonde matrix, which encodes the experimental evaluation of the corresponding characteristic function.
  \item By adopting the proposed reconstruction algorithm, we have proved that the required number of measurements to achieve the reconstruction scales linearly with the number of the values that can be assumed by the stochastic variable, and not quadratically as one would expect by a direct application of the definition of the corresponding probability distributions.
  \item We have observe that the fluctuation properties of the stochastic quantum entropy production strongly depend on the presence of decoherent channels between an arbitrary quantum system and the environment, which does not necessarily have to be a thermal bath. As a consequence, one could effectively determine not only the influence of the external noise sources on the system dynamics, but also characterize the environment structure and properties via quantum sensing procedures.
  \item Finally, we have proposed (in the form of an illustrative example) an experimental implementation of the reconstruction algorithm with trapped ions for the characterization of the thermodynamic irreversibility of a given system with Hamiltonian $H$.
\end{itemize}

\appendix
\appendixpage
\chapter{Proofs of chapter 1}
\label{chapter:appA}
\fancyhead{}
\fancyhead[LEH]{APPENDIX A}
\fancyhead[RO]{APPENDIX A}

\section{Closed-form solution of Problem $E_{t}^{A}$}

Let us consider the cost function (\ref{9}). Under the assumption $\bf {A}1$, (\ref{9}) can be written as the following quadratic form:
\begin{equation}\label{A1}
\begin{split}
J_{t}^{A} &=\hat{Y}_{t-N|t}'M_{t-N}\hat{Y}_{t-N|t}-\hat{Y}_{t-N|t}'D_{t-N}-D_{t-N}'\hat{Y}_{t-N|t}+r_{t-N}= \\
& =\hat{Y}_{t-N|t}'M_{t-N}\hat{Y}_{t-N|t}+2\hat{Y}_{t-N|t}'U_{t-N}+r_{t-N},
\end{split}
\end{equation}
where $\hat{Y}_{t-N|t}={\rm col}(\hat{x}_{t-N+i|t})_{i=0}^{N}\in\mathbb{R}^{nN}$, $D_{t-N}=-U_{t-N}\in\mathbb{R}^{nN}$ and the matrices $U_{t-N}\in\mathbb{R}^{nN}$, $M_{t-N}\in\mathbb{R}^{nN\times nN}$ are defined as
\begin{small}
\begin{equation*}
M_{t-N}=\begin{bmatrix} P+A'QA+\zeta_{j,1} & -A'Q & 0 & \cdots & 0 \\ -QA & Q+A'QA+\zeta_{j,2} & -A'Q & \cdots & 0 \\ \vdots & \vdots & \vdots & \cdots & \vdots \\ 0 & 0 & 0 & \cdots & Q+A'QA+\zeta_{j,N} \end{bmatrix}
\end{equation*}
\end{small}
and
\begin{equation*}
U_{t-N}=\begin{bmatrix} A'QBu_{0}-P\overline{x}_{t-N}-\pi_{j,1} \\ A'Q Bu_{1}-QBu_{0}-\pi_{j,2} \\ \vdots \\ A'Q Bu_{N-1}-QBu_{N-2}-\pi_{j,N-1} \\ A'Q Bu_{N}-QBu_{N-1}-\pi_{j,N}\end{bmatrix},
\end{equation*}
with
\begin{equation*}
\delta_{j,h}^{i}=\begin{cases} 1,\hspace{3mm}\text{if}\hspace{2mm}\exists j\in\mathfrak{I}^{i}_t:j=h \\ 0,\hspace{2mm}\text{else} \end{cases},\hspace{1.5mm}h=1,\ldots,N,
\end{equation*}
and
\begin{equation*}
\begin{split}
&\pi_{j,h}=\sum_{i=1}^{p}\delta_{j,h}^{i}C^{i'}R^{i}\tau^{i},\hspace{3mm}h=1,\ldots,N,\hspace{4mm}
\displaystyle{\zeta_{j,k}=\sum_{i=1}^{p}\delta_{j,h}^{i}C^{i'}R^{i}C^{i}}, \\
&r_{t-N} =\bar{x}_{t-N}'P\bar{x}_{t-N}+\sum_{k=t-N}^{t}u_{k}'B'QBu_{k}+\sum_{i=1}^{p}h_{i}R^{i}\tau^{i^{2}}\in\mathbb{R}, \\
&h_{i} =\dim(\mathfrak{I}^{i}_t).
\end{split}
\end{equation*}
Necessary condition for the minimum of the cost function (\ref{A1}) is
\begin{equation}\label{A2}
\nabla_{\hat{Y}_{t-N|t}}J_{t}^{A}(\hat{Y}_{t-N|t})=2M_{t-N}\hat{Y}_{t-N|t}+2U_{t-N}=0,
\end{equation}
for any $t =N,N+1,\ldots$ Solving (\ref{A2}) as a function of $\hat{x}_{t-N|t}$, we obtain the optimal estimates $\hat{x}^{\circ}_{t-N|t}$, $t=N,N+1,\ldots$ that minimize the cost function (\ref{9}), namely
\begin{equation}\label{A3}
\hat{x}^{\circ}_{t-N|t}=
\begin{bmatrix}\mathbbm{1}_{n} \underbrace{0 \ldots 0}_{\in\mathbb{R}^{(N-1)n\times n}} \end{bmatrix}M_{t-N}^{-1}D_{t-N},\hspace{3mm}t=N,N+1,\ldots
\end{equation}
Choosing the weighting matrices $P$ and $Q$ as positive semi-definite matrices and $R^{i}>0$, the solution (\ref{A3}) corresponds to a global minimum, since the Hessian matrix $M_{t-N}$ of the cost function is strictly positive definite. As a final remark, notice that there are many equivalent ways of writing the solution of Problem $E_{t}^{A}$ and the particular form presented here is a consequence of the fact
that we consider as optimization variables the state estimates $\hat x_{t-N+i|t}$ for $i = 0, \ldots, N$. An alternative would be to consider as optimization variables the state estimate
$\hat x_{t-N|t}$ at the beginning of the observation interval together with the estimates of the process disturbance $\hat w_{t-N+i|t} = \hat x_{t-N+i+1|t}  - A \hat x_{t-N+i|t} - B u_{t-N+i} $ for
$i = 0, \ldots, N-1$. In this case, each $\hat x_{t-N+i|t}$  would be written as a function of $\hat x_{t-N|t}$ and the observability matrix would explicitly appear in the solution.

\section{Proof of Proposition 1.1}

For each $k=t-N,\ldots,t-1$, we initially introduce the constraints for the $i-$th measurement equation, $i=1,\ldots,p$:
\begin{equation}\label{B1}
\begin{cases}
C^{i}\hat{x}_{k|t}<\tau^{i}+\rho_{V}^{i},\hspace{3mm}\text{if}\hspace{3mm}y_{k}^{i}=-1 \\
C^{i}\hat{x}_{k|t}>\tau^{i}-\rho_{V}^{i},\hspace{3mm}\text{if}\hspace{3mm}y_{k}^{i}=1
\end{cases}
\end{equation}
The system (\ref{B1}) is equivalent to the inequality
\begin{equation}\label{B2}
y_{k}^{i}(C^{i}\hat{x}_{k|t}+y_{k}^{i}\rho_{V}^{i})>y_{k}^{i}\tau^{i}.
\end{equation}
Observing that $(y_{k}^i)^{2}=1$, $\forall k=t-N,\ldots,t-1$, we obtain
\begin{equation*}\label{B3}
y_{k}^{i}(C^{i}\hat{x}_{k|t}-\tau^{i})+\rho_{V}^{i}>0,\hspace{3mm}k=t-N,\ldots,t-1.
\end{equation*}
If we define $\phi_{k}\equiv{\rm diag}(y_{k}^{i})\in\mathbb{R}^{p\times p}$, $i=1,\ldots,p$, $\tau_{p}={\rm col}(\tau^{i})_{i=1}^{p}\in\mathbb{R}^{p}$ and $\nu={\rm col}(\rho_{V}^{i})_{i=1}^{p}\in\mathbb{R}^{p}$, then we can write
\begin{equation*}
\phi_{k}(C\hat{x}_{k|t}-\tau_{p})+\nu>0,
\end{equation*}
since $\phi_{k}'\phi_{k}=\mathbbm{1}_{p}$. Moreover, introducing the matrices $\Phi_{t}$, $\mathcal{T}$ and $\mathcal{V}$ as in (\ref{16}), the constraints (\ref{B1}) can be written in matrix form, namely
\begin{equation}\label{B4}
\Phi_{t}~{\rm vec}\left[(C\hat{X}_{t}-\mathcal{T})'\right]<{\rm vec}(\mathcal{V})
\end{equation}
where $\hat{X}_{t}\equiv\left[\hat{x}_{t-N|t},\ldots,\hat{x}_{t|t}\right]'$. Observing that
\begin{equation}
{\rm vec}\left[(C\hat{X}_{t})'\right]\equiv(C\otimes I_{n}){\rm vec}\left(\hat{X}_{t}'\right),
\end{equation}
(\ref{B4}) is equal to (\ref{15}), so that the proposition is proved.

\section{Proof of Theorem 1.1}

Some preliminary definitions are needed. Notice first that, while the function $\omega(C^{i}x,y)\|C^{i}x-\tau^{i}\|$ is not differentiable for $C^{i}x=\tau^{i}$, for $C^{i}x \ne \tau^{i}$ one has
\begin{equation*}
\frac{\partial}{\partial x}  \omega(C^{i}x,y)\|C^{i}x-\tau^{i}\| = \left \{ \begin{array}{ll} 0, & \mbox{ if } y(C^{i}x-\tau^{i}) > 0 \, , \\ - y \, C^i , & \mbox{ if } y(C^{i}x-\tau^{i}) < 0 \, .\end{array} \right \}.
\end{equation*}
Hence $\omega(C^{i}x,y)\|C^{i}x-\tau^{i}\|$ is globally Lipschitz with Lipschitz constant $L^{i} = \| C^i \|$, for $i=1,\ldots,p$. Further, consider for each sensor $i$ and each sliding window $\mathfrak W_t$, the vector $\tilde{z}^{i}_{t|t}={\rm col}(C^{i}\hat{x}_{k|t}^{\circ})_{k\in\mathfrak{I}_{t}^{i}}$. Then, we can write
\begin{equation*}
\tilde{z}^{i}_{t|t}=\Theta_{t}^{i}\hat{x}_{t-N|t}^{\circ}+H_{t}^{i}\tilde{u}_{t}+D_{t}^{i}\tilde{w}_{t}^{\circ},
\end{equation*}
where
\begin{equation*}
\begin{split}
&\tilde{u}_{t}={\rm col}(u_{k})_{k\in[t-N,t]}, \\
&w_{k|t}^{\circ}=\hat{x}_{k+1|t}^{\circ}-A\hat{x}_{k|t}^{\circ}-Bu_{k}, \\
&\tilde{w}_{t}^{\circ}={\rm col}(w_{k|t}^{\circ})_{k\in[t-N,t]},
\end{split}
\end{equation*}
and $H_{t}^{i}$ and $D_{t}^{i}$ are suitable matrices. Moreover, let $\phi^i$ be defined as $\sup_{t \ge N} \overline{\lambda}({D}_{t}^{i\hspace{0.5mm}'} {D}^i_{t})^{1/2}$. Clearly, $\phi^i$ is finite since  $D^i_t$ can assume only a finite number of configurations in the estimation window.

Let us now consider the estimation error as $e_{t-N}= x_{t-N}-\hat{x}_{t-N}^{\circ}$; the aim is to find a lower and an upper bound for the optimal cost
\begin{eqnarray}
J_{t}^{\circ}
&=&\|\hat{x}_{t-N|t}^{\circ}-\overline{x}_{t-N}\|^{2}_{P}+\sum_{k=t-N}^{t-1}\|\hat{x}_{k+1|t}^{\circ}-A\hat{x}_{k|t}^{\circ}-Bu_{k}\|^{2}_{Q}\nonumber \\
&+&\sum_{i=1}^{p}\sum_{k=t-N}^{t}\omega(z_{k}^{i},y_{k}^{i})\|C^{i}\hat{x}_{k|t}^{\circ}-\tau^{i}\|^{2}_{R^{i}}.
\end{eqnarray}
to derive a bounding sequence on the norm of the estimation error. \\ \\ \\
\textit{-- Upper bound on the optimal cost $J_{t}^{\circ}$:} \\ \\
For the optimality of the cost function $J^{\circ}_{t}$, we have $\left.J^{\circ}_{t}\leq J_{t}^{B}\right|_{\hat{x}_{k|t}=x_{k},\hspace{1mm}k\in\mathfrak W_{t}}$ and hence
\begin{equation}\label{eq:upper_bound_cost}
J_{t}^{\circ}\leqslant\|x_{t-N}-\overline{x}_{t-N}\|^{2}_{P}+\sum_{k=t-N}^{t-1}\|w_{k}\|^{2}_{Q}+\sum_{i=1}^{p}
\sum_{k=t-N}^{t}\omega(z_{k}^{i},y_{k}^{i})\|z_{k}^{i}-\tau^{i}\|^{2}_{R^{i}}.
\end{equation}
The discontinuous function $\omega(z_{k}^{i},y_{k}^{i})$ is non zero if and only if $z_{k}^{i}-\tau^{i} \in V^{i}$, i.e. if the system output is close to the $i-$th sensor threshold and the measurement noise makes the sensor detection incoherent with the system evolution. Thus, the upper bound (\ref{eq:upper_bound_cost}) can be rewritten as
\begin{equation}\label{eq:upper_bound_cost_2}
J_{t}^{\circ}\leq\|x_{t-N}-\overline{x}_{t-N}\|^{2}_{P}+N\overline{\lambda}(Q)\rho^{2}_{W}+p(N+1)\overline{R}\overline{\rho}^{2}_{V}.
\end{equation}
\textit{-- Lower bound on the optimal cost $J_{t}^{\circ}$:} \\ \\
Let us consider a time instant $k\in\mathfrak{I}_{t}^{i}$ and suppose, for the sake of notational simplicity, that $y^{i}_{k}=1$ and $y^{i}_{k+1}=-1$
(up-down threshold crossing). Note that the dual case can be analysed in a similar way.
Thus, in the cost function $J_{t}^{\circ}$ the following contribution is present:
\begin{eqnarray*}
&&\iota(\hat{x}_{k|t}^{\circ},\hat{x}_{k+1|t}^{\circ})\equiv\omega(z_{k}^{i},1)\|C^{i}\hat{x}_{k|t}^{\circ}-\tau^{i}\|^{2}_{R^{i}}
+\omega(z_{k+1}^{i},-1)\|C^{i}\hat{x}_{k+1|t}^{\circ}-\tau^{i}\|^{2}_{R^{i}} \\
&&=\underbrace{\left[\omega(z_{k}^{i},1)+\omega(z_{k}^{i},-1)\right]}_{=1,\text{ by definition}}\|C^{i}\hat{x}_{k|t}^{\circ}-\tau^{i}\|^{2}_{R^{i}}
+\omega(z_{k+1}^{i},-1)\|C^{i}\hat{x}_{k+1|t}^{\circ}-\tau^{i}\|^{2}_{R^{i}} \\
&&-\omega(z_{k}^{i},-1)\|C^{i}\hat{x}_{k|t}^{\circ}-\tau^{i}\|^{2}_{R^{i}},
\end{eqnarray*}
where
\begin{eqnarray*}
\omega(z_{k+1}^{i},-1)\|C^{i}\hat{x}_{k+1|t}^{\circ}-\tau^{i}\|^{2}_{R^{i}}&-&\omega(z_{k}^{i},-1)\|C^{i}\hat{x}_{k|t}^{\circ}-\tau^{i}\|^{2}_{R^{i}}\nonumber \\
&\leq&(L^{i})^{2}\|(A-\mathbbm{1})\hat{x}_{k|t}^{\circ}+Bu_{k}+w^{\circ}_{k|t}\|^{2}.
\end{eqnarray*}
Since $\hat{x}^{\circ}_{k|t}\in\mathcal X$ for $k=t-N,\ldots,t$, it can be stated that each term $\iota(\hat{x}_{k|t}^{\circ},\hat{x}_{k+1|t}^{\circ})$ has a lower bound, such that
\begin{equation*}
\iota(\hat{x}_{k|t}^{\circ},\hat{x}_{k+1|t}^{\circ})\geq\|C^{i}\hat{x}_{k|t}^{\circ}-\tau^{i}\|^{2}_{R^{i}}-3(L^{i})^{2}\left(\|A-I\|^{2}\rho_{\mathcal X}^{2}+\|B\|^{2}\rho_{U}^{2}+\|w^{\circ}_{k|t}\|^{2}\right).
\end{equation*}
Since $k$ is a switching instant,
\begin{equation*}
y^{i}_{k}=h^{i}(C^{i}x_{k}+v^{i}_{k})=1
\end{equation*}
and
\begin{equation*}
y^{i}_{k+1}=h^{i}(C^{i}Ax_{k}+C^{i}Bu_{k}+C^{i}w_{k}+v^{i}_{k+1})=-1,
\end{equation*}
i.e. there exists $\alpha\in[0,1]$ such that $\alpha z^{i}_{k}+(1-\alpha)z^{i}_{k+1}=\tau^{i}$, from which
\begin{equation*}
\tau^{i}=C^{i}x_{k}+\zeta^{i}_{k},
\end{equation*}
where $\zeta^{i}_{k}=\delta^{i}_{k}+\eta^{i}_{k}$. Then,
\begin{equation*}
\|C^{i}\hat{x}_{k|t}^{\circ}-\tau^{i}\|^{2}_{R^{i}}=\|C^{i}\hat{x}_{k|t}^{\circ}-C^{i}x_{k}-\zeta^{i}_{k}\|^{2}_{R^{i}}
\geq\frac{1}{2}\|C^{i}\hat{x}_{k|t}^{\circ}-C^{i}x_{k}\|^{2}_{R^{i}}-\|\zeta^{i}_{k}\|^{2}_{R^{i}},
\end{equation*}
where
\begin{equation*}
\|\zeta^{i}_{k}\|^{2}_{R^{i}}\leq4R^{i}\left(\|C^{i}\|^{2}\|A-\mathbbm{1}\|^{2}\rho_{\mathcal X}^{2}+\|C^{i}\|^{2}\|B\|^{2}\rho_{U}^{2}+\|C^{i}\|^{2}\rho_{W}^{2}+(\rho_{V}^{i})^{2}\right).
\end{equation*}
Summarizing the previous results, if we consider $\forall i$ only the instants $k\in\mathfrak{I}_{t}^{i}$, we obtain
\begin{equation*}
J_{t}^{\circ}\geq\|\hat{x}_{t-N|t}^{\circ}-\overline{x}_{t-N}\|^{2}_{P}+\sum_{i=1}^{p}\sum_{k\in\mathfrak{I}_{t}^{i}}
\left(\|C^{i}\hat{x}_{k|t}^{\circ}-C^{i}x_{k}\|^{2}_{R^{i}}\right)-\beta_{t}-\sigma_{t},
\end{equation*}
where
\begin{eqnarray*}
\beta_{t}&=&\sum_{i=1}^{p}\sum_{k\in\mathfrak{I}_{t}^{i}}\left[4R^{i}\left(\|C^{j}\|^{2}\|A-\mathbbm{1}\|^{2}\rho_{\mathcal X}^{2}+\|C^{i}\|^{2}\|B\|^{2}\rho_{U}^{2}\right.\right. \\
&+&\left.\left.\|C^{i}\|^{2}\rho_{W}^{2}+(\rho_{V}^{i})^{2}\right)+3(L^{i})^{2}(\|A-\mathbbm{1}\|^{2}\rho_{\mathcal X}^{2}+
\|B\|^{2}\rho_{U}^{2})\right]
\end{eqnarray*}
and $\displaystyle{\sigma_{t}=\sum_{i=1}^{p}\sum_{k\in\mathfrak{I}_{t}^{i}}3(L^{i})^{2}\|w_{k|t}^{\circ}\|^{2}}$ are quantities with an upper bound. Indeed, it can be stated that:
\begin{eqnarray*}
\beta_{t}&\leq&4p(N+1)\overline{R}\left(\overline{C}^{2}\|A-\mathbbm{1}\|^{2}\rho_{\mathcal X}^{2}+\overline{C}^{2}\|B\|^{2}\rho_{U}^{2}+\overline{C}^{2}\rho_{W}^{2}+\overline{\rho}_{V}^{2}\right) \\
&+&3p(N+1)\overline{L}^{2}\left(\|A-\mathbbm{1}\|^{2}\rho_{\mathcal X}^{2}+\|B\|^{2}\rho_{U}^{2}\right)=\breve{\beta}_{t}
\end{eqnarray*}
and
\begin{eqnarray*}
\sigma_{t}&\leq&3p\left(\max_{i}L^{i}\right)^{2}\sum_{k=t-N}^{t-1}\|w_{k|t}^{\circ}\|^{2} \\
&\leq&\frac{3p\overline{L}^{2}}{\underline{\lambda}(Q)}\left[\|\hat{x}_{t-N|t}-
\overline{x}_{t-N}\|^{2}_{P}+N\overline{\lambda}(Q)\rho_{W}^{2}+p(N+1)\overline{R}\overline{\rho}_{V}^{2}\right]=\breve{\sigma}_{t}.
\end{eqnarray*}
To conclude the calculation of the lower bound, let us define
$\tilde{z}^{i}_{t}\equiv {\rm col}(z_{k})_{k\in\mathfrak{I}_{t}^{i}}$ and $\tilde{R}^{i}\equiv R^{i}I_{|\mathfrak{I}_{t}^{i}|}$ and write
\begin{eqnarray*}
\psi_{t}&\equiv&\sum_{i=1}^{p}\sum_{k\in\mathfrak{I}_{t}^{i}}\left(\|C^{i}\hat{x}_{k|t}^{\circ}-C^{i}x_{k}\|^{2}_{R^{i}}\right)=
\sum_{i=1}^{p}\|\tilde{z}^{i}_{t|t}-\tilde{z}^{i}_{t}\|^{2}_{\tilde{R}^{i}}= \\
&=&\sum_{i=1}^{p}\|\Theta_{t}^{i}\hat{x}_{t-N|t}^{\circ}+H^{i}_{t}\tilde{u}_{t}+D_{t}^{i}\tilde{w}_{t}^{\circ}-\Theta_{t}^{i}x_{t-N}-
H^{i}_{t}\tilde{u}_{t}-D^{i}_{t}\tilde{w}_{t}-\tilde{v}^{i}_{t}\|^{2}_{\tilde{R}^{i}},
\end{eqnarray*}
with $\tilde{w}_{t}\equiv {\rm col}(w_{k})_{k\in[t-N,t]}$ and $\tilde{v}^{i}_{t}\equiv {\rm col}(v_{k}^{i})_{k\in\mathfrak{I}_{t}^{i}}$. Hence,
\begin{eqnarray*}
\psi_{t}&\geq&\sum_{i=1}^{p}\left(\frac{1}{4}\|\Theta_{t}^{i}(\hat{x}_{t-N|t}^{\circ}-x_{t-N})\|^{2}_{\tilde{R}^{i}}-
\|D_{t}^{i}\tilde{w}_{t}^{\circ}\|^{2}_{\tilde{R}^{i}}-\|D_{t}^{i}\tilde{w}_{t}\|^{2}_{\tilde{R}^{i}}-\|\tilde{v}^{i}_{t}\|^{2}_{\tilde{R}^{i}}\right) \\
&\geq&\frac{1}{4}\|\Theta_{t}(\hat{x}_{t-N|t}^{\circ}-x_{t-N})\|^{2}_{\tilde{R}}-\breve{\mu}_{t},
\end{eqnarray*}
where
\begin{eqnarray*}
&&\mu_{t}=\sum_{i=1}^{p}R^{i}\left[\|D_{t}^{i}\|^{2}\left(\|\tilde{w}_{t}^{\circ}\|^{2}+\rho_{W}^{2}\right)+(\rho_{V}^{i})^{2}\right] \\
&&\leq p\overline{R}\left[\frac{\overline{\phi}^{2}}{\underline{\lambda}(Q)}\left(\|x_{t-N}-\overline{x}_{t-N}\|^{2}_{P}+N\overline{\lambda}
(Q)\rho_{W}^{2}+p(N+1)\overline{R}\overline{\rho}_{V}^{2}\right)+\overline{\phi}^{2}\rho_{W}^{2}+\overline{\rho}_{V}^{2}\right] \\
&&=\breve{\mu}_{t},
\end{eqnarray*}
i.e.
\begin{equation*}
\psi_{t}\geq\frac{\delta^{2}\underline{R}}{4\overline{\lambda}(P)}\|\hat{x}_{t-N|t}^{\circ}-x_{t-N}\|^{2}_{P}-\breve{\mu}_{t}=
\frac{\delta^{2}\underline{R}}{4\overline{\lambda}(P)}\|e_{t-N}\|_{P}^{2}-\breve{\mu}_{t}.
\end{equation*}
In conclusion
\begin{equation}\label{eq:lower_bound_cost}
J_{t}^{\circ}\geq\|\hat{x}_{t-N|t}^{\circ}-\overline{x}_{t-N}\|^{2}_{P}+\frac{\delta^{2}\underline{R}}{4\overline{\lambda}(P)}\|e_{t-N}\|_{P}^{2}
-\breve{\beta}_{t}-\breve{\sigma}_{t}-\breve{\mu}_{t}.
\end{equation}
Now we can exploit the bounds on the optimal cost $J_{t}^{\circ}$ in order to obtain a bounding sequence on the norm of the estimation error. More specifically, combining (\ref{eq:upper_bound_cost_2}) and (\ref{eq:lower_bound_cost}), we derive the following inequality:
\begin{eqnarray}\label{eq:inequality_proof}
&&\|\hat{x}_{t-N|t}^{\circ}-\overline{x}_{t-N}\|^{2}_{P}+\frac{\delta^{2}\underline{R}}{4\overline{\lambda}(P)}\|e_{t-N}\|_{P}^{2}
\leq\breve{\beta}_{t}+\breve{\sigma}_{t}+\breve{\mu}_{t}\nonumber \\
&&+\|x_{t-N}-\overline{x}_{t-N}\|^{2}_{P}+N\overline{\lambda}
(Q)\rho_{W}^{2}+p(N+1)\overline{R}\overline{\rho}_{V}^{2}.
\end{eqnarray}
But, noting that
\begin{equation*}
\|\hat{x}_{t-N|t}^{\circ}-\overline{x}_{t-N}\|^{2}_{P}\geq\frac{1}{2}\|e_{t-N}\|_{P}^{2}-\|x_{t-N}-\overline{x}_{t-N}\|^{2}_{P}
\end{equation*}
and
\begin{equation*}
x_{t-N}-\overline{x}_{t-N} =Ae_{t-N-1}+w_{t-N-1},
\end{equation*}
namely
\begin{equation*}
\|x_{t-N}-\overline{x}_{t-N}\|^{2}_{P}\leq 2\left(\|A\|^{2}_{P}\|e_{t-N-1}\|^{2}_{P}+\overline{\lambda}(P)\rho_{W}^{2}\right),
\end{equation*}
inequality (\ref{eq:inequality_proof}) can be rewritten as
\begin{equation*}
\|e_{t-N}\|^{2}_{P}\leq a_{1}\|e_{t-N-1}\|^{2}_{P}+a_{2},
\end{equation*}
where the coefficients $a_{1}$ and $a_{2}$ are defined as in formula~(\ref{20}) and
\begin{equation*}
\begin{split}
&d_{1}=2p\overline{\phi}^{2},\hspace{4mm}d_{2}=3\overline{L}^{2}\bar{\phi}^{-2},\hspace{4mm}c_{1}=c_{2}=p(N+1)\left(4\overline{R}~\overline{C}^{2}+3\overline{L}^{2}\right), \\
&c_{3}=b_{1}+N\overline{\lambda}(Q)\left(\frac{b_{1}}{2\overline{\lambda}(P)}-1\right)+p\overline{R}\left[4(N+1)\overline{C}^{2}+\overline{\phi}^{2}\right], \\
&c_{4}=p(N+1)\overline{R}\left(\frac{b_{1}}{2\overline{\lambda}(P)}-1\right)+p\overline{R}(4N+5).
\end{split}
\end{equation*}
Since $a_{2}$ is a positive scalar, if we further impose that $a_{1}<1$, the asymptotic upper bound $e^{\circ}_{\infty}$ can be easily derived, in that
\begin{equation*}
\|e_{t}\|^{2}_{P}<a_{1}^{t}\|e_{0}\|^{2}_{P}+a_{2}\sum^{t-1}_{j=0}a_{1}^{j},
\end{equation*}
which tends to $a_{2}/(1-a_{1})$ as $t\rightarrow\infty$.

\section{Proof of Proposition 1.2}

Notice first that the stability condition $a_1 < 1$ can be rewritten as
\[
\frac{\overline{\lambda}(P)}{\underline{\lambda}(P)} \left[4+\frac{d_1}{\underline{\lambda}(Q)}\left(d_2+\overline{R} \right)\right] \| A \|^2
\le \left(\frac{1}{2}+\frac{\delta^{2}\underline{R}}{4\overline{\lambda}(P)}\right) \, .
\]
By letting $P = \varepsilon \overline{P}$, with $\overline{P}$ any positive definite matrix, the above inequality becomes
\[
\frac{\overline{\lambda}(\overline{P})}{\underline{\lambda}(\overline{P})} \left[4+\frac{d_1}{\underline{\lambda}(Q)}\left(d_2+\overline{R} \right)\right] \| A \|^2
\le \left(\frac{1}{2}+\frac{\delta^{2}\underline{R}}{4\, \varepsilon \, \overline{\lambda}(\overline{P})}\right) \, .
\]
It can be seen that the left-hand side of such an inequality does not depend on $\varepsilon$, whereas the right-hand side goes to infinity as $\varepsilon$ goes to $0$, provided that $\delta^2 >0$. Hence, when $\delta^2 >0$, it is always possible to ensure that the stability condition $a_1 < 1$ holds by taking any $Q$, $R_i$, $i = 1, \ldots,p$, $\overline{P}$, and then choosing $\varepsilon$
suitably small.

\section{Proof of Proposition 1.3}

If assumption {\bf A3} holds, the dynamical system is linear and the noise are distributed as a Gaussian probability density function, then the cost function (\ref{37}) is convex if and only if $F^{i}(\tau^{i}-C^{i}x_{t})$ and $\Phi^{i}(\tau^{i}-C^{i}x_{t})$ are log-concave functions, $\forall i=1,\ldots,p$.

A function $f:\mathbb{R}^{n}\rightarrow\mathbb{R}$ is log-concave if $f(x)>0$ for all $x$ in its domain and $\ln~f(x)$ is concave~\cite{Boyd}, namely
\begin{equation}\label{eq:logconc}
\nabla^{2}ln~f(x)=\frac{1}{f^{2}(x)}\left[\frac{\partial^{2}f(x)}{\partial x^{2}}f(x)-\left(\frac{\partial f(x)}{\partial x}\right)'\left(\frac{\partial f(x)}{\partial x}\right)\right]<0.
\end{equation}
Now, let us consider the CDF $\Phi^{i}(\tau^{i}-C^{i}x_{t})$ and its complementary function $F^{i}(\tau^{i}-C^{i}x_{t})$, that are positive functions for all $\chi^{i}_{t}\equiv\tau^{i}-C^{i}x_{t}$, $i=1,\ldots,p$. From the fundamental theorem of calculus, namely
\begin{equation}
\frac{\partial}{\partial x}\left(\int^{a(x)}_{b(x)}f(x)dx\right)=f(a(x))\frac{\partial a(x)}{\partial x}-f(b(x))\frac{\partial b(x)}{\partial x},
\end{equation}
where $a(x)$ and $b(x)$ are arbitrary functions of $x$, the first and the second derivatives of the function $F^{i}(\tau^{i}-C^{i}x_{t})$ with respect to $x_{t}$ are, respectively, equal to
\begin{equation}
\frac{\partial F^{i}(\tau^{i}-C^{i}x_{t})}{\partial x_{t}}=
\frac{C^{i}}{\sqrt{2\pi r_{i}}}\exp\left(-\frac{(\tau^{i}-C^{i}x_{t})^{2}}{2r_{i}}\right)
\end{equation}
and
\begin{equation}
\frac{\partial^{2} F^{i}(\tau^{i}-C^{i}x_{t})}{\partial x_{t}^{2}}=\frac{(C^{i} )' C^{i}}{r_{i}\sqrt{2\pi r_{i}}}(\tau^{i}-C^{i}x_{t})\exp\left(-\frac{(\tau^{i}-C^{i}x_{t})^{2}}{2r_{i}}\right).
\end{equation}
If $\tau^{i}-C^{i}x_{t}\leq 0$, then $\frac{\partial^{2} F^{i}(\tau^{i}-C^{i}x_{t})}{\partial x_{t}^{2}}\leq 0$. Hence $\frac{\partial^{2}F^{i}}{\partial x^{2}}F^{i}\leq 0$ and, from (\ref{eq:logconc}), it follows that the Q-function $F^{i}$ is log-concave. Conversely, if $\tau^{i}-C^{i}x_{t}>0$, the log-concavity of $F^{i}$ depends on the sign of the term
\begin{eqnarray}
&&\frac{\partial^{2}F^{(i)}}{\partial x^{2}}F^{i}-\left(\frac{\partial F^{i}}{\partial x}\right)'\left(\frac{\partial F^{i}}{\partial x}\right) =\frac{(C^{i})'C^{i}}{2\pi r_{i}}\exp\left(-\frac{(\tau^{i}-C^{i}x_{t})^{2}}{2r_{i}}\right)\nonumber \\
&&\times\left[\frac{\tau^{i}-C^{i}x_{t}}{r_{i}}
\left(\displaystyle{\int^{\infty}_{\tau^{i}-C^{i}x_{t}}}\exp\left(-\frac{u^{2}}{2r_{i}}\right)du\right)- \exp\left(-\frac{(\tau^{i}-C^{i}x_{t})^{2}}{2r_{i}}\right)\right].\nonumber \\
&&
\end{eqnarray}
From the convexity properties of the function $f(x)=x^{2}/2$, it can be easily verified for any variable $s,k$ that $s^{2}/2\geq -k^{2}/2+sk$, and hence $\exp\left(-s^{2}/2\right)\leq\exp\left(-sk+k^{2}/2\right)$ (see e.g. \cite{Boyd}). Then, if $k>0$, it holds that
\begin{equation}
\int_{k}^{\infty}\exp\left(-\frac{s^{2}}{2}\right)ds\leq\int_{k}^{\infty}\exp\left(-sk+\frac{k^{2}}{2}\right)ds=
\frac{\exp\left(-\frac{k^{2}}{2}\right)}{k}.
\end{equation}
Since $\tau^{i}-C^{i}x_{t}>0$, with a simple change of variable, it can be stated that
\begin{equation}
\frac{\tau^{i}-C^{i}x_{t}}{r_{i}}\left(\displaystyle{\int^{\infty}_{\tau^{i}-C^{i}x_{t}}}\exp\left(-\frac{u^{2}}{2r_{i}}\right)du\right)
\leq\exp\left(-\frac{(\tau^{i}-C^{i}x_{t})^{2}}{2r_{i}}\right),
\end{equation}
proving, as a consequence, the log-concavity of the Q-function $F^{i}(\tau^{i}-C^{i}x_{t})$.

By using the complement rule, the cumulative distribution function can be written as
\begin{equation}
\Phi^{i}(\tau^{i}-C^{i}x_{t})=1-F^{i}(\tau^{i}-C^{i}x_{t})\geq 0
\end{equation}
and
\begin{equation}
\frac{\partial^{2} \Phi^{i}(\tau^{i}-C^{i}x_{t})}{\partial x_{t}^{2}}=-\frac{\partial^{2} F^{i}(\tau^{i}-C^{i}x_{t})}{\partial x_{t}^{2}}.
\end{equation}
If $\tau^{i}-C^{i}x_{t}>0$, then $\frac{\partial^{2}\Phi^{i}}{\partial x^{2}}\Phi^{i}<0$ such that $\Phi^{i}$ is log-concave. In the remaining case, i.e. $\tau^{i}-C^{i}x_{t}\leq 0$, noting that
\begin{equation}
\Phi^{i}=\frac{1}{\sqrt{2\pi r_{i}}}\int^{\tau^{i}-C^{i}(x_{t})}_{-\infty}\exp\left(-\frac{u^{2}}{2r_{i}}\right)du=\frac{1}{\sqrt{2\pi r_{i}}}\int_{-(\tau^{i}-C^{i}x_{t})}^{\infty}\exp\left(-\frac{u^{2}}{2r_{i}}\right)du,
\end{equation}
it can be observed that the sign of the term
\begin{eqnarray}
&\frac{\partial^{2}\Phi^{i}}{\partial x^{2}}\Phi^{i}-\left(\frac{\partial \Phi^{i}}{\partial x}\right)'\left(\frac{\partial \Phi^{i}}{\partial x}\right) = \frac{(C^{i})'C^{i}}{2\pi r_{i}}\exp\left(-\frac{(\tau^{i}-C^{i}x_{t})^{2}}{2r_{i}}\right) &\nonumber \\
&{} \times \left[\frac{-(\tau^{i}-C^{i}x_{t})}{r_{i}}   \left(\displaystyle{\int^{\infty}_{-(\tau^{i}-C^{i}x_{t})}}e^{-\frac{u^{2}}{2r_{i}}}du\right)
-\exp\left(-\frac{(\tau^{i}-C^{i}x_{t})^{2}}{2r_{i}}\right)\right]&\nonumber
\end{eqnarray}
is negative, thus proving the log-concavity of the CDF $\Phi^{i}(\tau^{i}-C^{i}x_{t})$ and the convexity of the whole cost function.

\chapter{Proofs of chapter 3}
\label{chapter:appB}
\fancyhead{}
\fancyhead[LEH]{APPENDIX B}
\fancyhead[RO]{APPENDIX B}

\section{Log-survival-probability distribution}\label{app_log_survP}

Here, we derive the distribution of the log-survival-probability, as given by (\ref{prob_log_surv}) in chapter~3. From the constraints (\ref{constraints}), we get
\begin{equation}\label{appBeq1}
m \ln q\left(\tau^{(d)}\right)-\mathcal{L}=\sum_{k=1}^{d-1}\tilde{n}_{k}\lambda\left(\mu^{(\alpha)}\right),
\end{equation}
with
\begin{equation}
\lambda(\tau^{(k)})\equiv\ln q(\tau^{(d)})- \ln q(\tau^{(k)}).
\end{equation}
Eq.~(\ref{appBeq1}) is solved with
\begin{equation}\label{sol1}
\tilde{n}_{\alpha} = \frac{m\ln q\left(\tau^{(d)}\right)-\mathcal{L}}
{(d-1)\lambda\left(\tau^{(k)}\right)};~k = 1,2,\ldots,d-1,
\end{equation}
while $\tilde{n}_d$ is given by
\begin{equation}\label{app0-eq1}
\tilde{n}_{d} = m - \sum_{k = 1}^{d-1}\tilde{n}_{k}.
\end{equation}
Then, (\ref{prob_L}) gives
%\begin{small}
\begin{eqnarray}\label{app0-eq2}
&&\textrm{Prob}(\mathcal{L})=\exp\left(\ln m!-\sum_{k = 1}^{d}\ln \tilde{n}_{k}!+
\sum_{k=1}^{d}\tilde{n}_{k}\ln p^{(k)}\right) \nonumber \\
&&\approx\exp\left(m\ln m -m-\sum_{k=1}^{d}\tilde{n}_{k}\ln \tilde{n}_{k}+\sum_{k = 1}^{d}\tilde{n}_{k}+
\sum_{k = 1}^{d}\tilde{n}_{k}\ln p^{(k)}\right) \nonumber \\
&&=\exp\left(m\ln m -m\sum_{k=1}^{d-1}\frac{\ln q\left(\tau^{(d)}\right)-\frac{\mathcal{L}}{m}}{(d-1)\lambda\left(\tau^{(k)}\right)}\ln\left(m\frac{\ln q\left(\tau^{(d)}\right)-\frac{\mathcal{L}}{m}}{(d-1)\lambda\left(\tau^{(k)}\right)}\right)\right.\nonumber \\
&&\left.-m\left(1-\sum_{k=1}^{d-1}
\frac{\ln q\left(\tau^{(d)}\right)-\frac{\mathcal{L}}{m}}{(d-1)\lambda\left(\tau^{(k)}\right)}\right)\ln\left[m\left(1-\sum_{k=1}^{d-1}\frac{\ln q\left(\tau^{(d)}\right)-\frac{\mathcal{L}}{m}}{(d-1)\lambda\left(\tau^{(k)}\right)}\right)\right]\right.\nonumber \\
&&\left.+m\sum_{k=1}^{d-1}\frac{\ln q\left(\tau^{(d)}\right)-\frac{\mathcal{L}}{m}}{(d-1)\lambda\left(\tau^{(k)}\right)}\ln p^{(k)}+m\left(1-\sum_{k=1}^{d-1}
\frac{\ln q\left(\tau^{(d)}\right)-\frac{\mathcal{L}}{m}}{(d-1)\lambda\left(\tau^{(k)}\right)}\right)\ln\left(1-\sum_{k=1}^{d-1}p^{(k)}\right)\right)\nonumber\\
&&\approx e^{-mI\left(\frac{\mathcal{L}}{m}\right)},
\end{eqnarray}
%\end{small}
where, in the second step, we have used the Stirling's approximation, while in the third step (\ref{sol1}) and (\ref{app0-eq1}) have been used. Being, from (\ref{app0-eq2}),
\begin{eqnarray}
&&I\left(\xi\right)=\sum_{k=1}^{d}f(\tau^{(k)})\ln\left(\frac{f(\tau^{(k)})}{p^{(k)}}\right),\\
&&f(\tau^{(k)})=\frac{\ln q(\tau^{(d)})-\xi}{(d-1)\lambda(\tau^{(k)})}\label{appBeq2};~k=1,\ldots,(d-1), \\
&&f(\tau^{(d)})=1-\sum_{k=1}^{d-1}f(\tau^{(k)}),
\end{eqnarray}
the distribution of the log-survival-probability, as given by (\ref{prob_log_surv}), has been derived.

\section{Most probable value of the log-survival-probability}

Here, we provide more details on the derivation of (\ref{L_star}). From (\ref{rate_function}), the condition $\partial
I\left(\mathcal{L}/m\right)/\partial \ln q(\tau^{(k)})|_{\mathcal{L}=\mathcal{L}^\star}=0$ gives for $k=1,\ldots,d-1$ the relation
\begin{equation}\label{appBeq3}
p^{(d)}f(\tau^{(k)}) = p^{(k)}\left(1 - \sum_{k = 1}^{d-1}f(\tau^{(k)})\right).
\end{equation}
Summing both sides over $k = 1,2,\ldots,d-1$, we get
\begin{equation}
p^{(d)}\sum_{k = 1}^{d-1}f(\tau^{(k)}) = \left(1-\sum_{k = 1}^{d-1}f(\tau^{(k)})\right)\sum_{k = 1}^{d-1}p^{(k)},
\end{equation}
which, by using $\sum_{k = 1}^{d}p^{(k)} = 1$, gives
\begin{equation}
\sum_{k = 1}^{d-1}f(\tau^{(k)}) = \sum_{k = 1}^{d-1}p^{(k)}.
\end{equation}
Using the above equation, and combining (\ref{appBeq2}) and (\ref{appBeq3}), one has
\begin{equation}
\left(\ln q(\tau^{(d)})-\frac{\mathcal{L}^\star}{m}\right)=(d-1)\frac{\left(1-\displaystyle{\sum_{k=1}^{d-1}}p^{(k)}\right)}{p^{(d)}}p^{(k)}\lambda(\tau^{(k)}),
\end{equation}
i.e.
\begin{equation}
\frac{\mathcal{L}^\star}{m} = \ln q(\tau^{(d)})-\left(1-\sum_{k=1}^{d-1}p^{(k)}\right)\left(\sum_{k=1}^{d-1}
\lambda(\tau^{(k)})\frac{p^{(k)}}{p^{(d)}}\right),
\end{equation}
which yields, finally, to the expression for the most probable value of the log-survival-probability, i.e.
\begin{equation}
\mathcal{L}^\star = m\sum_{k = 1}^{d} p^{(k)}\ln q(\tau^{(k)}).
\end{equation}

\section{LD form for the joint probability distribution $\textrm{Prob}(\mathcal{L},\mathcal{T})$}

To derive (\ref{LD-PLT}), we use (\ref{const_2}) and (\ref{T-defn1}) to get $\sum_{k = 1}^{d} \tilde{n}_{k} \tau^{(k)} = m\overline{\tau}$, which, by rewriting $\sum_{k=1}^{d}\tilde{n}_{k} = m$ as
\begin{equation}
\left(m-\sum_{k=1}^{d-1}\tilde{n}_{k}\right)\tau^{(d)} + \sum_{k=1}^{d-1}\tilde{n}_{k}\tau^{(k)} = m\overline{\tau},
\end{equation}
leads to
\begin{equation}\label{app-eq31}
m = \frac{\displaystyle{\sum_{k = 1}^{d-1}\tilde{n}_{k}\left(\tau^{(d)}-\tau^{(k)}\right)}}{(\tau^{(d)}-\overline{\tau})}.
\end{equation}
Then, from equation $\displaystyle{\sum_{k = 1}^{d}\tilde{n}_{k}\ln q(\tau^{(k)}) = \mathcal{L}}$, one obtain (similarly to the derivation of (\ref{sol1}))
\begin{equation}\label{app-eq32}
\tilde{n}_{k} = \frac{m\ln q\left(\tau^{(d)}\right)-\mathcal{L}}{(d-1)\lambda\left(\tau^{(k)}\right)};~k = 1,2,\ldots,d-1,
\end{equation}
while $\tilde{n}_{d}$ is
\begin{equation}
\tilde{n}_{d} = m - \sum_{k=1}^{d-1}\tilde{n}_{k}.
\end{equation}
Combining (\ref{app-eq31}) and (\ref{app-eq32}), and noting that $m\ne 0$, we get
\begin{equation}\label{d12}
\sum_{k = 1}^{d-1}\frac{\left(\ln q\left(\tau^{(d)}\right) - \mathcal{L}/m\right)(\tau^{(d)}-\tau^{(k)})}{(d-1)\lambda\left(\tau^{(k)}\right)(\tau^{(d)}-\overline{\tau})}=1,
\end{equation}
which is satisfied with
\begin{equation}
\left(\ln
q(\tau^{(d)})-\mathcal{L}/m\right)(\tau^{(d)}-\tau^{(k)})=(d-1)\lambda(\tau^{(k)})(\tau^{(d)}-\overline{\tau}).
\end{equation}
$\forall~k=1,\ldots,(d-1)$. Hence, from (\ref{app-eq32}), we get for $k = 1,2,\ldots,d-1$ that
\begin{eqnarray}
\tilde{n}_{k}&=&\frac{\tau^{(d)}(m\ln q\left(\tau^{(d)}\right)-\mathcal{L})}
{(d-1)\lambda\left(\tau^{(k)}\right)(\tau^{(d)}-\overline{\tau}+\overline{\tau})}\nonumber \\
&=&\frac{\tau^{(d)}(m\ln q\left(\tau^{(d)}\right)-\mathcal{L})}
{\left(\ln
q\left(\tau^{(d)}\right)-\mathcal{L}/m\right)(\tau^{(d)}-\tau^{(k)})+(\mathcal{T}/m)(d-1)\lambda\left(\tau^{(k)}\right)}.\nonumber \\
&&
\end{eqnarray}
Using the last expression, and proceeding in a way similar to Appendix~\ref{app_log_survP}, we get
\begin{equation}\label{app3-eq1}
\textrm{Prob}(\mathcal{L},\mathcal{T})\approx\exp\Big(-m\mathcal{I}\left(\frac{\mathcal{L}}{m},\frac{\mathcal{T}}{m}\right)\Big),
\end{equation}
where the expressions of $\mathcal{I}\left(\xi,y\right)$, $g(\tau^{(k)})$ and $g(\tau^{(d)})$ are given by (\ref{rate_function_2}),~(\ref{g_tau_k}) and (\ref{g_tau_d}) in chapter~3.

\section{Normalized discrepancy $\mathcal{D}$ in the Zeno regime}

Here, we show that the normalized discrepancy
\begin{eqnarray}
\mathcal{D}&\equiv&\frac{\overline{\mathcal{P}}_{a}-\overline{\mathcal{P}}_{g}}{\overline{\mathcal{P}}_{a}} = 1-e^{-\Delta q(\tau,m)}\nonumber \\
&\approx& \Delta q(\tau,m) = \ln\langle q(\tau)^{m}\rangle - \langle\ln q(\tau)^{m}\rangle
\end{eqnarray}
is not vanishing only at the fourth order in $\tau$. To this end, let us consider the series expansion of $q^{m}$ and its logarithm up to fourth order, namely
\begin{equation}\label{sviluppoP_1}
q^{m} = 1 - m\Delta^{2}H\tau^{2} + \frac{m}{12}\left[\gamma_{H}+3(2m-1)(\Delta^{2}H)^{2}\right]\tau^{4}+\mathcal{O}(\tau^{6})
\end{equation}
and
\begin{equation}\label{sviluppoP_2}
\ln q^{m}=-m\Delta^{2}H\tau^{2}+\frac{m}{12}\left[\gamma_{H}-3(\Delta^{2}H)^{2}\right]\tau^{4}+\mathcal{O}(\tau^{6}).
\end{equation}
In (\ref{sviluppoP_1}) and (\ref{sviluppoP_2}) $\gamma_{H}\equiv\overline{H^{4}}-4\overline{H^{3}}\overline{H}+6\overline{H^{2}}\overline{H}^{2}-3\overline{H}^{4}$ is the kurtosis of the system Hamiltonian. As a result, under this fourth order approximation, $\Delta q(\tau,m)\approx\mathcal{D}$ is identically equal to
\begin{eqnarray}
\Delta q &\approx& m\Delta^{2}H\langle\tau^{2}\rangle-\frac{m}{12}\left[\gamma_{H}-3(\Delta^{2}H)^{2}\right]\langle\tau^{4}\rangle\nonumber \\
&+&\ln\left[1-m\Delta^{2}H\langle\tau^{2}\rangle+\frac{m}{12}\left[\gamma_{H}+3(2m-1)(\Delta^{2}H)^{2}\right]\langle\tau^{4}\rangle\right],
\end{eqnarray}
i.e.
\begin{equation}
\Delta q \approx \frac{m^{2}}{2}(\Delta^{2}H)^{2}\langle\tau^{4}\rangle-\frac{m^{2}}{2}(\Delta^{2}H)^{2}\langle\tau^{2}\rangle^{2} = \frac{m^{2}}{2}(\Delta^{2}H)^{2}\left(\langle\tau^{4}\rangle - \langle\tau^{2}\rangle^2\right),
\end{equation}
where $\langle\tau^{2}\rangle\equiv\int_{\tau}d\tau p(\tau)\tau^{2}$ and $\langle\tau^{4}\rangle\equiv\int_{\tau}d\tau p(\tau)\tau^{4}$.

\section{Derivation of $\mathrm{Prob}(\mathcal{P})$ for a bimodal $p(\tau)$}

We analytically derive the expression for $\mathrm{Prob}(\mathcal{P})$ when the probability density function $p(\tau)$ is bimodal, with values $\tau^{(1)}$ and $\tau^{(2)}$ and probabilities $p_{1}$ and $p_{2} = 1 - p_{1}$. To this end, let us write the survival probability $\mathcal{P}(\{\tau_{j}\})$ as
\begin{equation}
\mathcal{P} = q(\tau^{(1)})^{k(\mathcal{P})}q(\tau^{(2)})^{m - k(\mathcal{P})},
\end{equation}
where $k(\mathcal{P})$ is the frequency of the event $\tau^{(1)}$. By taking the logarithm of $\mathcal{P}$ and resolving for $k(\mathcal{P})$, one has
\begin{equation}\label{k_P}
k(\mathcal{P}) = \frac{\ln \mathcal{P}-m\ln q(\tau^{(2)})}{\ln q(\tau^{(1)})-\ln q(\tau^{(2)})}.
\end{equation}
Moreover, being the frequency $k(\mathcal{P})$ binomially distributed, it can be stated that
\begin{equation}
\mathrm{Prob}(k(\mathcal{P})) = \frac{m!}{k(\mathcal{P})!(m - k(\mathcal{P}))!}p_{1}^{k(\mathcal{P})}p_{2}^{m - k(\mathcal{P})}.
\end{equation}
Then, assuming that for each value of $k(\mathcal{P})$ there exists a single solution $\mathcal{P}$ of (\ref{k_P}), $\mathrm{Prob}(\mathcal{P})$ is univocally determined from $\mathrm{Prob}(k(\mathcal{P}))$. Since by using the Stirling approximation the binomial distribution $\mathrm{Prob}(k(\mathcal{P}))$ is approximately equal (for $m$ sufficiently large) to a Gaussian distribution, we get
\begin{equation}
\mathrm{Prob}(\mathcal{P}) \approx \frac{1}{\sqrt{2\pi m p_1 p_2}}\exp\left(-\frac{(k(\mathcal{P})-m p_1)^2}{2m p_1 p_2}\right).
\end{equation}

\section{Fisher information operator in terms of the statistical moments of $p(\tau)$}

Here, we show how to transform the Fisher Information Operator (FIO)
\begin{equation}\label{FIO_appendix}
F(p) = m^2\frac{\mathcal{P}^\star}{1-\mathcal{P}^\star}|\ln q\rangle\langle \ln q|
\end{equation}
into the corresponding Fisher Information Matrix (FIM) expressed in terms of the statistical moments of the probability density function $p(\tau)$, defined as
\begin{equation}
 \langle\tau^{k}\rangle \equiv \int_{\tau}d\tau p(\tau)\tau^{k}.
\end{equation}
To this end, let us express the FIO (\ref{FIO_appendix}) in the the generic basis $\{|f_i\rangle\}$. We get the following relation:
\begin{equation}
 F_{ij} = m^{2}\frac{\mathcal{P}^\star}{1-\mathcal{P}^\star}\langle f_i|\ln q\rangle\langle \ln q|f_j\rangle.
\end{equation}
Then, by introducing the basis functions
\begin{equation}\label{stat_mom_basis}
f_k(\tau) = 2\frac{(-1)^k}{k!}\frac{\partial^k\delta(\tau)}{\partial\tau^{k}},
\end{equation}
with $\delta(\cdot)$ equal to the Dirac-delta distribution, we can express the FIO in terms of the statistical moments $\langle\tau^k\rangle$'s. Indeed, a small change of the probability density function $p(\tau)$ in the direction of $f_k(\tau)$ will only change its $k-$th moment $\langle\tau^{k}\rangle$, but not affect the other moments. Now, by means of a Taylor expansion around zero, we can write $\ln(q(\tau))$ as
\begin{equation}
\ln(q(\tau)) = \sum_{k = 1}^{\infty}\left.\frac{\partial^{k}\ln(q(\tau))}{\partial\tau^{k}}\right|_{\tau = 0}\frac{\tau^{k}}{k!},
\end{equation}
and, by defining
\begin{equation}
\beta_{k} \equiv \left.\frac{\partial^{k}\ln(q(\tau))}{\partial\tau^{k}}\right|_{\tau =0},
\end{equation}
we obtain
\begin{equation}
\mathcal{P}^{\star} = \exp\left(m\sum_{k = 1}^{\infty}\frac{\beta_{k}\langle\tau^{k}\rangle}{k!}\right),
\end{equation}
as well as
\begin{equation}
\langle f_i|\ln q\rangle = \frac{\beta_i}{i!}.
\end{equation}
This means that $\beta_{i}/i!$ is the effect of the system dynamics $\ln(q(\tau))$ in the direction of the basis function $|f_{i}\rangle$. In conclusion, the resulting FIM given by representing the FIO in the basis (\ref{stat_mom_basis}) is equal to
\begin{equation}\label{FIM_basis_stat}
\widetilde{F}_{ij} = m^{2}\frac{\mathcal{P}^{\star}}{(1-\mathcal{P}^{\star})}\frac{\beta_{i}\beta_{j}}{i!j!}.
\end{equation}
It is worth noting that, since
\begin{equation}
\frac{\partial \mathcal{P}^{\star}}{\partial\langle\tau^{h}\rangle} = m\mathcal{P}^{\star}\frac{\beta_{h}}{h!},
\end{equation}
(\ref{FIM_basis_stat}) is compatible with the standard definition of the FIM, i.e.
\begin{equation}
\widetilde{F}_{ij} = \frac{1}{\mathcal{P}^{\star}(1-\mathcal{P}^{\star})}\frac{\partial \mathcal{P}^{\star}}{\partial\langle\tau^{i}\rangle}\frac{\partial \mathcal{P}^{\star}}{\partial\langle\tau^{j}\rangle}.
\end{equation}
Finally, as observed for the FIO, also the rank of the FIM is equal to one. Indeed, the determinant of a generic $2\times 2$ minor of the Fisher matrix $\widetilde{F}_{ij}$ is equal to $0$:
\begin{equation}
\left(m^{2}\frac{\mathcal{P}^{\star}}{(1-\mathcal{P}^{\star})}\right)^2
\begin{vmatrix}
 \frac{\beta_{i}\beta_{j}}{i!j!} &\frac{\beta_{i}\beta_{j+1}}{i!(j+1)!}\\
 \frac{\beta_{i+1}\beta_{j}}{(i+1)!j!}&\frac{\beta_{i+1}\beta_{j+1}}{(i+1)!(j+1)!}
\end{vmatrix}
=0.
\end{equation}

\section{Proof of Theorem~3.1}

To prove Theorem~3.1, let us consider the survival probability's most probable value
\begin{equation}
\mathcal{P}^\star = \exp\left(m\int_{\tau}d\tau p(\tau)\ln(q(\tau))\right).
\end{equation}
Then, if we perform a Taylor expansion of $\ln q(\tau)$ as a function of the time intervals, then we can write
\begin{equation}\label{App_eq:Pstar-expansion}
\mathcal{P}^{\star} = \exp\left(m\sum_{k = 1}^{\infty}\frac{\alpha_{k}}{k!}\int_{\tau}d\tau p(\tau)\tau^{k}\right)
= \exp\left(m\sum_{k = 1}^{h/2}\frac{\alpha_{2k}}{2k!}\int_{\tau}d\tau p(\tau)\tau^{2k} + R_{h}(\xi)\right),
\end{equation}
where
\begin{equation}
\alpha_{k}\equiv\left.\frac{\partial^{k}\ln(q(\tau))}{\partial\tau^{k}}\right|_{\tau = 0}
\end{equation}
and $R_{h}(\xi)$ is the remainder of the Taylor expansion of $\ln(q(\tau))$ up to the $h-$th order, with $\xi\in[0,\mu]$ real number. For odd $k$, due to the symmetry of $q(\tau)$, we find $\alpha_{k} = 0$. Thus $h$ is assumed to be an even number greater than zero. For $h = 2$, namely by considering a second order approximation of the Taylor expansion (only the first term of the summation in (\ref{App_eq:Pstar-expansion}) is considered), the survival probability's most probable value is equal to
\begin{equation}
\mathcal{P}^{\star} = \exp\left(m\frac{\alpha_{2}}{2}(1+\kappa)\overline{\tau}^2\right)\exp\left(m\langle R_{2}(\xi)\rangle\right),
\end{equation}
where
\begin{equation}
\alpha_{2} = -2\Delta^{2}_{\rho_{0}}H_{\Pi}
\end{equation}
and
\begin{equation}
\kappa \equiv \frac{\Delta^{2}\tau}{\overline{\tau}^2}.
\end{equation}
Note that $\overline{\tau}$ and $\Delta^{2}\tau$ are, respectively, the expectation value and the variance of the probability density function $p(\tau)$, while the $2$nd order remainder of the Taylor expansion in the Lagrange form is
\begin{equation}
R_{2}(\xi)\equiv\frac{\partial^{3}\ln(q(\tau))}{\partial\tau^{3}}\big|_{\tau = \xi}\frac{\tau^{3}}{6},
\end{equation}
where
\begin{equation}
\left|\frac{1}{6}\frac{\partial^{3}\ln(q(\tau))}{\partial\tau^{3}}\big|_{\tau = \xi}\right|\leq C
\end{equation}
for some positive constant $C$ depending on the form of the specific system Hamiltonian $H$ and the initial state $\rho_{0}$. Hence
\begin{equation}
\langle R_{2}(\xi)\rangle \equiv \int_{\tau}d\tau p(\tau)R_{2}(\xi)
\end{equation}
is bounded by $C\tau^3$ and, if
\begin{equation}\label{app_weak_zeno}
\int_{\tau}d\tau p(\tau)\tau^{3}\ll\frac{1}{mC},
\end{equation}
then the term $\langle R_{2}(\xi)\rangle$ is negligible. Accordingly, we can now approximate the survival probability as
\begin{equation}\label{app:Zeno_approx}
\mathcal{P}^\star\approx\exp\left(-m\Delta^2_{\rho_{0}}H_{\Pi}(1+\kappa)\overline{\tau}^2 \right).
\end{equation}
(\ref{app:Zeno_approx}) generalizes the expression for the probability that the quantum system belongs to the measurement subspace after $m$ random projective measurements beyond the standard Zeno regime. We denote the inequality in (\ref{app_weak_zeno}) as the \textit{weak Zeno limit}, while the condition
\begin{equation}
m\Delta^2 H_{\rho_0}(1+\kappa)\overline{\tau}^2\ll 1
\end{equation}
is the \textit{strong Zeno limit}, which leads to
\begin{equation}
\mathcal{P}^\star\approx
1 - m\Delta^{2}_{\rho_{0}}H_{\Pi}(1+\kappa)\overline{\tau}^{2}.
\end{equation}

\section{Time-continuous stochastic Schr{\"o}dinger equation}

Let us consider an arbitrary quantum system that is coupled to a bath, whose effects on the system are encoded in the time fluctuating classical field $\Omega(t)$. Thus, the corresponding Hamiltonian is given by
\begin{equation}
H_{{\rm tot}}(t) = H_{0} + \Omega(t)H_{{\rm noise}},
\end{equation}
where $H_{0}$ is the coherent part of the Hamiltonian, while $H_{noise}$ describes the coupling of the environment with the system. Then, the system dynamics is governed by
following stochastic Schr\"{o}dinger equation:
\begin{equation}
\dot{\rho} = -i[H_{{\rm tot}}(t),\rho] = -i[H_{0},\rho] - i\Omega(t)[H_{{\rm noise}},\rho].
\end{equation}
The integral form of the initial value problem states that
\begin{equation}\label{eq:integral_form}
\rho(t) = \rho(0) - i\int_{0}^{t}[H_{{\rm tot}}(t'),\rho(t')] dt',
\end{equation}
so that
\begin{equation}\label{eq:general-2nd-order}
\dot{\rho}(t) = -i\bigg[H_{{\rm tot}}(t),\left(\rho(0) - i\int_{0}^{t}[H_{{\rm tot}}(t'),\rho(t')]dt'\right)\bigg].
\end{equation}
The random field $\Omega(t)$ is sampled from the probability density function $p(\Omega)$, such that
\begin{equation}
\langle\Omega(t)\rangle \equiv \int_{\Omega} p(\Omega)\Omega d\Omega,
\end{equation}
denotes its expectation value and
\begin{equation}
\langle\Omega(t)\Omega'(t')\rangle = \int_{\Omega}\int_{\Omega'} p(\Omega)p(\Omega')\Omega\Omega' d\Omega d\Omega'
\end{equation}
is the corresponding second-order time correlation function. Now, if we average (\ref{eq:general-2nd-order}) over the realizations of the noise term, we get
\begin{eqnarray}
\langle\dot{\rho}(t)\rangle &=& \int_{\Omega}\int_{\Omega'} p(\Omega)p(\Omega')\dot{\rho}(t)d\Omega d\Omega'
= -i\left[\left(H_0(t) + \int_{\Omega}p(\Omega)\Omega H_{{\rm noise}}d\Omega\right),\right.\nonumber \\
&&\left.\left(\rho(0) - i\int_0^t[(H_0(t')+\int_{\Omega'} p(\Omega')\Omega'H_{{\rm noise}} d\Omega'),\langle\rho(t')\rangle]dt'\right)\right],\nonumber \\
&&
\end{eqnarray}
and by using (\ref{eq:integral_form}), the general expression for $\langle\dot{\rho}(t)\rangle$ can be straightforwardly obtained:
\begin{small}
\begin{equation}
\langle\dot{\rho}(t)\rangle = -i\left[\left(H_{0} + \langle\Omega(t)H_{{\rm noise}}\rangle_{\Omega(t)}\right),
\left(\rho(t) - i\int_{0}^{t}\left[\langle\Omega(t')H_{{\rm noise}}\rangle_{\Omega(t')},\rho(t')\right]dt'\right)\right],
\end{equation}
\end{small}
where $\langle\xi(t)X\rangle_{\xi(t)}$ is equal to $\int_{\xi}d\xi p(\xi)\xi X$. Note that
\begin{eqnarray}
\left[\langle\Omega(t)H_{{\rm noise}}\rangle_{\Omega(t)},-i\int^{t}_{0}\left[\langle\Omega(t')H_{{\rm noise}}\rangle_{\Omega(t')},\rho(t')\right]dt'\right]\equiv\nonumber \\
-i\left[H_{{\rm noise}},\int^{t}_{0}\int_{\Omega}\int_{\Omega'}p(\Omega)p(\Omega')\Omega\Omega'd\Omega d\Omega'\big[H_{{\rm noise}},\rho(t')\big]dt'\right].
\end{eqnarray}
Hence, if we separate the contributions of the coherent term and the noise term, we get
\begin{eqnarray}\label{eq:Master-equation-correlation}
\langle\dot{\rho}(t)\rangle &=& -i\left[H_0 + \langle\Omega\rangle H_{{\rm noise}},\rho(t)\right]\nonumber \\
&-& \left[H_{{\rm noise}},\int^{t}_{0}\int_{\omega}\int_{\omega'}d\omega d\omega'dt' p(\omega)p(\omega')\omega\omega'\big[H_{{\rm noise}},\rho(t')\big]\right],
\end{eqnarray}
i.e.
\begin{eqnarray}
\langle\dot{\rho}(t)\rangle &=&
-i\left[H_0 + \langle\Omega\rangle H_{{\rm noise}},\rho(t)\right]-\left[H_{{\rm noise}},\left[H_{{\rm noise}},\int_{0}^{t}dt'\big\langle \omega(t)\omega(t')\big\rangle \rho(t')\right]\right]\nonumber \\
&=& -i\left[H_0 + \langle\Omega\rangle H_{{\rm noise}},\rho(t)\right] - \int_{0}^{t}dt'\langle\omega(t)\omega(t')\rangle [H_{{\rm noise}},[H_{{\rm noise}},\rho(t')]],\nonumber \\
&&
\end{eqnarray}
being assumed that $\Omega(t) = \langle\Omega\rangle + \omega(t)$, $\omega(t)$ is the fluctuating part of $\Omega(t)$ with vanishing expectation value.

\chapter{Proofs of chapter 4}
\label{chapter:appC}

\section*{Derivation of the characteristic function $G(u)$}

Here, we derive the expression for the characteristic function
\begin{equation}\label{G_u_app}
G(u) = \int P(Q_{q})e^{iuQ_{q}}dQ_{q}
\end{equation}
by taking into account, respectively, quenched and annealed disorder for the waiting times between measurements. In Eq. (\ref{G_u_app}) the quantum-heat probability distribution is defined as
\begin{equation}
P(Q_{q}) = \sum_{n,l}\delta(Q_{q} - E_{l} + E_{n})p_{l|n}~p_{n},
\end{equation}
where $p_{l|n}$ is the transition probability to get the final energy $E_{l}$ conditioned to have measured $E_{n}$ after the first energy measurement.

\subsubsection*{Quenched disorder}

Plugging the expression of the joint distribution
%\begin{equation}
$
p(\vec{\tau}) = p(\tau_1) \prod_{i=2}^m \delta(\tau_i-\tau_1)
%\end{equation}
$
into Eq. (\ref{conditional_prob_an}) in the main text, we obtain for the transition probability $p_{l|n}$ the expression
\begin{equation}
p_{l|n} = \sum_{\vec{k}}\int d\tau p(\tau){\rm Tr}\left[\Pi_{l}\mathcal{V}(\vec{k},\tau)\Pi_{n}\mathcal{V}^{\dagger}(\vec{k},\tau)\Pi_{l}\right],
\end{equation}
where
$$
\mathcal{V}(\vec{k},\tau) = \Pi_{k_{m}}\mathcal{U}(\tau)\cdots\Pi_{k_1}\mathcal{U}(\tau).
$$
Accordingly, the corresponding quantum-heat probability distribution is equal to
\begin{equation}
P(Q_{q}) = \int \sum_{n,\vec{k},l}{\rm Tr}\left[\Pi_{m}\mathcal{V}(\vec{k},\tau)\Pi_{n}\mathcal{V}^{\dagger}(\vec{k},\tau)\Pi_{l}\right]p_{n}p(\tau)d\tau,
\end{equation}
so that the characteristic function $G(u)$ reads
\begin{equation}
G(u) = \int\sum_{n,\vec{k},l}\bra{E_{l}}\mathcal{V}(\vec{k},\tau)\ket{E_{n}}\bra{E_{n}}\mathcal{V}^{\dagger}(\vec{k},\tau)\ket{E_{l}}e^{iu(E_{l} - E_{n})}p_{n}p(\tau)d\tau,
\end{equation}
where we used the relation
\begin{equation}
{\rm Tr}\left[\Pi_{l}\mathcal{V}\Pi_{n}\mathcal{V}^{\dagger}\Pi_{l}\right]=\bra{E_{m}}\mathcal{V}\ket{E_{n}}\bra{E_{n}}\mathcal{V}^{\dagger}\ket{E_{l}}.
\end{equation}
Finally, using
\begin{equation}\label{eq:rel_eigenvalue_eq}
\begin{cases}
e^{iuE_{l}}\ket{E_{l}} = e^{iuH}\ket{E_{l}} \\
\bra{E_{n}}e^{-iuE_{n}} = \bra{E_{n}}e^{-iuH}
\end{cases},
\end{equation}
we obtain
\begin{eqnarray}
G(u) &=& \sum_{\vec{k}}\int\sum_{n,l}
\bra{E_{l}}\mathcal{V}\ket{E_{n}}\bra{E_{n}}\rho_{0}\ket{E_{n}}\bra{E_{n}}e^{-iuH}\mathcal{V}^{\dagger}e^{iuH}\ket{E_{l}}p(\tau)d\tau\nonumber \\
&=& \sum_{\vec{k}}\int{\rm Tr}\left[\mathcal{V}e^{-iuH}\rho_{0}\mathcal{V}^{\dagger}e^{iuH}\right]p(\tau)d\tau,
\end{eqnarray}
i.e. Eq. (\ref{G_u_finale}) in the main text for the quenched disorder case.

\subsubsection*{Annealed disorder}

In case the stochasticity between consecutive projective measurements is distributed as an annealed disorder, the joint distribution of the waiting times is
$p(\vec{\tau}) = \prod_{j=1}^{m}p(\tau_{j})$, so that the transition probability $p_{l|n}$ is equal to
\begin{equation}
p_{l|n} = \sum_{\vec{k}}\int d^{m}\vec{\tau}p(\vec{\tau}){\rm Tr}\left[\Pi_{l}\mathcal{V}(\vec{k},\vec{\tau})\Pi_{n}\mathcal{V}^{\dagger}(\vec{k},\vec{\tau})\Pi_{l}\right].
\end{equation}
The latter corresponds to a multiple integral defined over the waiting times $\vec{\tau}$, where $\mathcal{V}(\vec{k},\vec{\tau}) = \Pi_{k_{m}}\mathcal{U}(\tau_{m})\cdots\Pi_{k_1}\mathcal{U}(\tau_{1})$. As a result, the quantum-heat probability distribution $P(Q_{q})$ and the corresponding characteristic function $G(u)$ can be written, respectively, as
\begin{equation}
P(Q_{q}) = \sum_{\vec{k}}\int\sum_{n,l}{\rm Tr}\left[\Pi_{l}\mathcal{V}(\vec{k},\vec{\tau})\Pi_{n}\mathcal{V}^{\dagger}(\vec{k},\vec{\tau})\Pi_{l}\right]p_{n}p(\vec{\tau})d^{m}\vec{\tau}
\end{equation}
and
\begin{equation}
G(u) = \sum_{\vec{k}}\int\sum_{n,l}
\bra{E_{l}}\mathcal{V}\ket{E_{n}}\bra{E_{n}}\rho_{0}\ket{E_{n}}\bra{E_{n}}e^{-iuH}\mathcal{V}^{\dagger}e^{iuH}\ket{E_{l}}p(\vec{\tau})d^{m}\vec{\tau}.
\end{equation}
Accordingly, by using again the relations of Eq.~(\ref{eq:rel_eigenvalue_eq}), we can derive the expression of $G(u)$, i.e.
\begin{equation}\label{G_u_finale_appendice}
G(u) = \overline{\left\langle {\rm Tr}\left[e^{iuH}\mathcal{V}(\vec{k},\vec{\tau})e^{-iuH}\rho_{0}\mathcal{V}^{\dagger}(\vec{k},\vec{\tau})\right] \right\rangle},
\end{equation}
i.e. Eq. (\ref{G_u_finale}) in the main text, where the angular bracket denote quantum-mechanical expectation, while the overline stands for the average over the noise realizations.

\subsubsection*{Fluctuation relation}

To derive $G(i\beta_{T}) = 1$, let us substitute the initial thermal state $\rho_0 = e^{-\beta_{T} H}/Z$ and $u = i\beta_{T}$ in the characteristic function of Eq. (\ref{G_u_finale_appendice}). We get
\begin{equation}
\begin{split}
G(i\beta_{T}) &= \overline{\left\langle {\rm Tr}\left[e^{-\beta_{T} H}\mathcal{V}(\vec{k},\vec{\tau})e^{\beta H}\frac{e^{-\beta_{T} H}}{Z}\mathcal{V}^{\dagger}(\vec{k},\vec{\tau})\right] \right\rangle}\\
&=\overline{{\rm Tr}\left[\frac{e^{-\beta_{T} H}}{Z}\sum_{\vec{k}}\mathcal{V}(\vec{k},\vec{\tau})\mathcal{V}^{\dagger}(\vec{k},\vec{\tau})\right]}=\overline{\frac{{\rm Tr}\left[e^{-\beta_{T} H}\right]}{Z}}=1,
\end{split}
\end{equation}
where we have exploited the unitality of the system dynamics, i.e. $\sum_{\vec{k}}\mathcal{V}(\vec{k},\vec{\tau})\mathcal{V}^{\dagger}(\vec{k},\vec{\tau})=\mathbb{I}$, and the normalisation $\int d^{m}\vec{\tau}p(\vec{\tau})=1$.

\section*{Analytical $G(u)$ for a two-level system}

\subsubsection*{Fixed waiting times sequence}

Let us consider a sequence of projective measurements applied to a $n-$level quantum system at fixed waiting times; we denote with $\overline{\tau}$ the (fixed) time between consecutive measurements. Then, the characteristic function of the quantum-heat is given by Eq.~(\ref{eq:G(u)_regular}), which can be rewritten as:
\begin{equation}
G(u) = f(u)L^{m-1}g(u).
\end{equation}
For a two-level system an explicit expression for $G(u)$ can be derived. To this end, we assume, without loss of generality, that the system energy values $E_{\pm}$ are equal to $\pm E$ and, then, we make use of the energy eigenvalue equation, i.e. $H|E_{\pm}\rangle = E_{\pm}|E_{\pm}\rangle$, so as to obtain
\begin{equation}\label{B2}
f(u)' = \begin{pmatrix}
\langle\alpha_{1}|e^{iuH}|\alpha_{1}\rangle \\
\langle\alpha_{2}|e^{iuH}|\alpha_{2}\rangle
\end{pmatrix} =
\begin{pmatrix}
|a|^{2}e^{iuE} + |b|^{2}e^{-iuE} \\
|a|^{2}e^{-iuE} + |b|^{2}e^{iuE}
\end{pmatrix},
\end{equation}
where $\{|\alpha_{j}\rangle\}$, $j = 1,2$, is the basis, defining the projective measurements of the protocol. As shown in the main text, the elements of the basis $\{|\alpha_{j}\rangle\}$ are chosen as linear combinations of the energy eigenstates $|E_{\pm}\rangle$ (see Eq.~(\ref{eq:abasis})). Instead, the transition matrix L turns out to be
\begin{equation}\label{B3}
L = \begin{pmatrix}
\left||a|^{2}e^{-iEt} + |b|^{2}e^{iEt} \right|^{2} &
\left|a^{\ast}be^{-iEt} - ab^{\ast}e^{iEt} \right|^{2} \\ \left|b^{\ast}ae^{-iEt} - ba^{\ast}e^{iEt} \right|^{2} & \left||b|^{2}e^{-iEt} + |a|^{2}e^{iEt} \right|^{2}
\end{pmatrix}
=\begin{pmatrix}
1-\overline{\nu} & \overline{\nu} \\ \overline{\nu} & 1-\overline{\nu}
\end{pmatrix},
\end{equation}
where
\begin{equation}\label{eq:nu}
\overline{\nu} \equiv 2|a|^{2}|b|^{2}\sin^{2}(\overline{\tau}\Delta E),
\end{equation}
and $\Delta E \equiv(E_{+} - E_{-}) = 2E$. Then, by using the decomposition of the initial density matrix $\rho_{0}$ in the energy basis - Eq.~(\ref{eq:rho_0_2_level_system})) - and again Eq.~(\ref{eq:abasis}), it holds that
\begin{equation}\label{eq:g(u)}
g(u) = \begin{pmatrix}
\langle\alpha_{1}|e^{-iuH}\rho_{0}|\alpha_{1}\rangle \\
\langle\alpha_{2}|e^{-iuH}\rho_{0}|\alpha_{2}\rangle
\end{pmatrix}
= \begin{pmatrix}
|a|^{2}c_{1}e^{-iuE}+|b|^{2}c_{2}e^{iuE}\\
|a|^{2}c_{2}e^{iuE} + |b|^{2}c_{1}e^{-iuE}
\end{pmatrix}.
\end{equation}
In conclusion, the explicit dependence of $G(u)$ from the set of parameters $(a,b,c_{1},c_{2},\overline{\tau})$ is given by the following equation:
\begin{equation}\label{eq:G_ordered_2LS}
G(u)=\begin{pmatrix}
|a|^{2}e^{iuE}+|b|^{2}e^{-iuE}\\
|a|^{2}e^{-iuE} + |b|^{2}e^{iuE}
\end{pmatrix}'\begin{pmatrix}
1-\overline{\nu} & \overline{\nu} \\ \overline{\nu} & 1-\overline{\nu}
\end{pmatrix}^{m-1}
\cdot\begin{pmatrix}
|a|^{2}c_{1}e^{-iuE}+|b|^{2}c_{2}e^{iuE}\\
|a|^{2}c_{2}e^{iuE} + |b|^{2}c_{1}e^{-iuE}
\end{pmatrix}.
\end{equation}
It is worth noting that the characteristic function $G(u)$ admits a discontinuity point in correspondence of $|a|^2\rightarrow 0,1$ and $m\rightarrow\infty$. In particular, when $|a|^2\rightarrow 0,1$ and a finite number $m$ of measurements is performed, $G(u)$ is identically equal to $1$. Conversely, under the asymptotic limit $m\rightarrow\infty$, the characteristic function does not longer depend on $a$ and it equals to
\begin{equation}
G(u) = \frac{(1+e^{2iuE})}{2} - c_{1}\sinh(2iuE),
\label{eq:B9}
\end{equation}
so that
$G(i\beta_{T}) = (1+e^{-2\beta_{T} E})/2 + c_{1}\sinh(2\beta_{T} E)$. The transition matrix $L$, indeed, admits as eigenvalues the values $1$ and $(1 - 2\overline{\nu})<1$, and, thus, after the eigendecomposition of the transition matrix, for $m\rightarrow\infty$ it holds that
\begin{equation}
L^{m-1} \longrightarrow V\begin{pmatrix} 0 & 0 \\ 0 & 1 \end{pmatrix}V^{T} = \begin{pmatrix} \frac{1}{2} & \frac{1}{2} \\ \frac{1}{2} & \frac{1}{2} \end{pmatrix},
\end{equation}
with
\begin{equation}
V = \begin{pmatrix}
     -\frac{1}{\sqrt{2}} & \frac{1}{\sqrt{2}} \\
     \frac{1}{\sqrt{2}} & \frac{1}{\sqrt{2}}
   \end{pmatrix}.
\end{equation}

\subsubsection*{Stochastic waiting times sequence}

Here, we take into account a sequence of projective measurements with stochastic waiting times $\tau_{k}$, $k = 1,\ldots,M$, which sampled by a bimodal probability density function $p(\tau)$, as shown in the main text.

The explicit expression of the characteristic function in the presence of quenched disorder can be derived from Eqs. (\ref{G_u_finale}) and (\ref{eq:G_ordered_2LS}). We obtain
\begin{small}
\begin{eqnarray}\label{B7}
G(u)&=&\sum_{j = 1}^{d_{\tau}}
\begin{pmatrix}
|a|^{2}e^{iuE}+|b|^{2}e^{-iuE}\\
|a|^{2}e^{-iuE} + |b|^{2}e^{iuE}
\end{pmatrix}'
\begin{pmatrix}
1-\nu_{j} & \nu_{j} \\
\nu_{j} & 1-\nu_{j}
\end{pmatrix}^{m-1}\nonumber \\
&\cdot&\begin{pmatrix}
|a|^{2}c_{1}e^{-iuE}+|b|^{2}c_{2}e^{iuE}\\
|a|^{2}c_{2}e^{iuE} + |b|^{2}c_{1}e^{-iuE}
\end{pmatrix}p_j,
\end{eqnarray}
\end{small}
where
\begin{equation}\label{eq:eta}
\nu_{j}\equiv\nu(\tau^{(j)})=
2|a|^{2}|b|^{2}\sin^{2}(2\tau^{(j)}E),
\end{equation}
and $d_{\tau} = 2$. As discussed in the main text, also in this case the characteristic function admits a discontinuity point in correspondence of $|a|^2\rightarrow 0,1$ and $m\rightarrow\infty$. As before, when $|a|^2\rightarrow 0,1$ and a finite number $m$ of measurements is performed, $G(u)$ is identically equal to $1$; while for $m\rightarrow\infty$ the characteristic function does not longer depend on $a$ and it equals again to
\begin{equation}
G(u) = \frac{(1+e^{2iuE})}{2} - c_{1}\sinh(2iuE),
\label{G_u_qu}
\end{equation}
as we obtained in the non-stochastic case. Indeed, the transition matrix $L\left(\tau^{(j)}\right)$ admits as eigenvalues the values $1$ and $(1 - 2\nu_{j})<1$, so that for $m\rightarrow\infty$
\begin{equation}
L\left(\tau^{(j)}\right)^{m-1} \longrightarrow \begin{pmatrix} \frac{1}{2} & \frac{1}{2} \\ \frac{1}{2} & \frac{1}{2} \end{pmatrix}~~\text{with}~~j=1,\ldots,d_{\tau}.
\end{equation}

Finally, we repeat the latter derivation when the stochasticity between measurements is distributed as annealed disorder. In this regard, the characteristic function
\begin{equation}
G(u) = \sum_{k = 0}^{m-1}\binom{m-1}{k}f(u)L(\tau^{(1)})^{k}L(\tau^{(2)})^{m-k-1}g(u)p_1^{k}p_2^{m-k-1}.
\label{eq:B12}
\end{equation}
and, by substituting the expressions of $f(u)$, $L$ and $g(u)$ as given in Eqs.~(\ref{B2}), \\
~(\ref{B3}),~(\ref{eq:nu}) and (\ref{eq:g(u)}), we obtain the following relation:
\begin{small}
\begin{eqnarray}
G(u)&=&\sum_{k = 0}^{m-1}\binom{m-1}{k}\begin{pmatrix}
|a|^2e^{iuE}+|b|^2e^{-iuE}\\
|a|^2e^{-iuE}+|b|^2e^{iuE}
\end{pmatrix}'\cdot\begin{pmatrix}
1-\nu_{1} & \nu_{1} \\ \nu_{1} & 1-\nu_{1}
\end{pmatrix}^{k}\nonumber \\
&\cdot&\begin{pmatrix}
1-\nu_{2} & \nu_{2} \\ \nu_{2} & 1-\nu_{2}
\end{pmatrix}^{m-k-1} \cdot \begin{pmatrix}
|a|^2c_{1}e^{-iuE}+|b|^2c_{2}e^{iuE}\\
|a|^2c_{2}e^{iuE}+|b|^2c_{1}e^{-iuE}
\end{pmatrix} p_1^{k}p_2^{m-k-1},\nonumber \\
&&
\end{eqnarray}
\end{small}
As for the other cases, we find the same discontinuity in $G(u)$ in the limits of $|a|^2\rightarrow 0,1$ and $m\rightarrow\infty$. Quite surprisingly, the discontinuity is exactly the same for both types of disorder. To observe this, let us take Eq.~(\ref{eq:B12}) with $a \neq 0$, and, then, use the binomial theorem, given by
$$
\displaystyle{(x+y)^{n}=\sum_{k=0}^{n} \binom{n}{k}x^{n-k}y^{k}}
$$
with $x$, $y$ arbitrary real variables. As a result, we obtain
\begin{equation}
G(u) = f(u)\left(L(\tau^{(1)})p_1+L(\tau^{(2)})p_2\right)^{m-1}g(u).
\end{equation}
By introducing the quantity $\zeta\equiv\nu_1p_1+\nu_2p_2$, the weighted sum (w.r.t. $p(\tau)$) of the transition matrices $L(\tau^{(1)})$ and $L(\tau^{(2)})$ can be simplified as
\begin{equation}
\left(L(\tau^{(1)})p_1+L(\tau^{(2)})p_2\right)=
\begin{pmatrix}
1-\zeta & \zeta \\ \zeta & 1-\zeta
\end{pmatrix},
\end{equation}
which admits eigenvalues 1 and $(1-2\zeta)\leq 1$. Thus, by performing the limit $m\rightarrow\infty$, the weighted sum of the transition matrices tends to a projector, so that $G(u)$ is effectively given by Eq.~(\ref{G_u_qu}).

\subsubsection*{$n-$th order derivative of $G(u)$}

Analytical expression for $\partial^{n}_{u}G(u)$ allows us to derive all the statistical moments of the quantum-heat, and, consequently, the its mean value $\langle Q_{q}\rangle$. In particular, the $n-$th order derivative of the quantum-heat characteristic function, when a protocol of projective measurements at fixed waiting times is considered, is
\begin{equation}\label{eq:partial_derivative}
\partial^{n}_{u}G(u)=
\sum_{k=0}^{n} A^k(u)'\cdot\begin{pmatrix}
1-\overline{\nu} & \overline{\nu} \\ \overline{\nu} & 1-\overline{\nu}
\end{pmatrix}^{m-1} \cdot B^{n-k}(u),
\end{equation}
where
\begin{equation}
A^l(u) \equiv
(i)^{l}
\begin{pmatrix}
\bra{\alpha_{1}}H^{l}e^{iuH}
\ket{\alpha_{1}} \\
\bra{\alpha_{2}}H^{l}e^{iuH}
\ket{\alpha_{2}}
\end{pmatrix}
\end{equation}
and
\begin{equation}
B^l(u)\equiv
(-i)^{l}
\begin{pmatrix}
\bra{\alpha_{1}}H^{l}e^{-iuH}\rho_{0}
\ket{\alpha_{1}} \\
\bra{\alpha_{2}}H^{l}e^{-iuH}\rho_{0}
\ket{\alpha_{2}}
\end{pmatrix}.
\end{equation}
Instead, in the quenched disorder case $\partial^{n}_{u}G(u)$ reads
\begin{equation}\label{eq:partial_derivative_quench}
\partial^{n}_{u}G(u) = \displaystyle{\sum_{j = 1}^{d_{\tau}}
\sum_{k=0}^{n}A^k(u)^{T}}\cdot\begin{pmatrix}
1-\nu(\tau^{(j)}) & \nu(\tau^{(j)})\\
\nu(\tau^{(j)}) & 1-\nu(\tau^{(j)})
\end{pmatrix}^{m-1}\cdot B^{n-k}(u)p_{j},
\end{equation}
while in the annealed case
\begin{eqnarray}\label{eq:partial_derivative_annealed}
\partial^{n}_{u}G(u) &=& \displaystyle{\sum_{k = 0}^{m-1}\sum_{l=0}^{n}A^{l}(u)'}
\cdot\begin{pmatrix}
1-\nu(\tau^{(1)}) & \nu(\tau^{(1)})\\
\nu(\tau^{(1)}) & 1-\nu(\tau^{(1)})
\end{pmatrix}^{k}\nonumber \\
&\cdot&\begin{pmatrix}
1-\nu(\tau^{(2)}) & \nu(\tau^{(2)})\\
\nu(\tau^{(2)}) & 1-\nu(\tau^{(2)})
\end{pmatrix}^{m-k-1}\cdot B^{n-l}(u)p_{1}^{k}p_{2}^{m-k-1}.
\end{eqnarray}

\section*{Proof of Theorem~4.1}

In this section, we prove the equality between the conditional probabilities $p(a^{\textrm{fin}}_{k}|a^{\textrm{in}}_{m})$ and $p(a^{\textrm{in}}_{m}|a^{\textrm{ref}}_{k})$, respectively, of the forward and backward processes of our two-time measurement scheme.
Let us recall the observables $\mathcal{O}_{\textrm{in}}\equiv\sum_{m}a^{\textrm{in}}_{m}\Pi^{\textrm{in}}_{m}$, $\mathcal{O}_{\textrm{fin}}\equiv\sum_{k}a^{\textrm{fin}}_{k}\Pi^{\textrm{fin}}_{k}$, $\widetilde{\mathcal{O}}_{\textrm{ref}}\equiv\sum_{k}a^{\textrm{ref}}_{k}\widetilde{\Pi}^{\textrm{ref}}_{k}$ and $\widetilde{\mathcal{O}}_{\textrm{in}} = \sum_{m}a^{\textrm{in}}_{m}\widetilde{\Pi}^{\textrm{in}}_{m}$. The dynamical evolution of the open quantum system between the two measurements is described by a unital CPTP map $\Phi(\cdot)$ (with $\Phi(\mathbbm{1}) = \mathbbm{1}$), whose Kraus operators $\{E_{u}\}$ are such that $\sum_{u}E_{u}^{\dagger}E_{u} = \mathbbm{1}$, where $\mathbbm{1}$ denotes the identity operator on the Hilbert space $\mathcal{H}$ of the quantum system.
Accordingly, $\Phi(\rho_{\textrm{in},m}) = \sum_{u}E_{u}\rho_{\textrm{in},m}E_{u}^{\dagger}$, where $\rho_{\textrm{in},m}\equiv\Pi^{\textrm{in}}_{m}\rho_{0}\Pi^{\textrm{in}}_{m}$, and thus the conditional probability $p(a^{\textrm{fin}}_{k}|a^{\textrm{in}}_{m})$ equals
\begin{eqnarray}\label{cond_prob_forward_appA}
&&p(a^{\textrm{fin}}_{k}|a^{\textrm{in}}_{m}) = \frac{\textrm{Tr}[\Pi^{\textrm{fin}}_{k}\Phi(\rho_{\textrm{in},m})]}
{\textrm{Tr}[\Pi^{\textrm{in}}_{m}\rho_{0}\Pi^{\textrm{in}}_{m}]} = \frac{\textrm{Tr}[\Pi^{\textrm{fin}}_{k}\sum_{u}E_{u}\rho_{\textrm{in},m}E_{u}^{\dagger}]}
{\textrm{Tr}[\Pi^{\textrm{in}}_{m}\rho_{0}\Pi^{\textrm{in}}_{m}]}\nonumber \\
&&= \sum_{u}\frac{\textrm{Tr}[\Pi^{\textrm{fin}}_{k}E_{u}\Pi^{\textrm{in}}_{m}\rho_{0}\Pi^{\textrm{in}}_{m}E_{u}^{\dagger}]}
{\textrm{Tr}[\Pi^{\textrm{in}}_{m}\rho_{0}\Pi^{\textrm{in}}_{m}]} = \sum_{u}|\langle\phi_{a_{k}}|E_{u}|\psi_{a_{m}}\rangle|^{2}.\nonumber \\
&&
\end{eqnarray}
Next, by inserting in (\ref{cond_prob_forward_appA}) the identity operator $\mathbbm{1} = \Theta\Theta^\dagger=\Theta^\dagger\Theta$, where $\Theta$ is the time-reversal operator, one has:
\begin{eqnarray}
|\langle\phi_{a_{k}}|E_{u}|\psi_{a_{m}}\rangle|^{2} &=& |\langle\phi_{a_{k}}|\Theta^{\dagger}\left(\Theta E_{u}\Theta^\dagger \right)\Theta|\psi_{a_{m}}\rangle|^{2}
= |\langle \widetilde{\phi}_{a_{k}}| \Theta E_{u} \Theta^\dagger |\widetilde{\psi}_{a_{m}}\rangle|^2 \nonumber\\
&=& |\langle\widetilde{\psi}_{a_{m}}|\Theta E_{u}^{\dagger}\Theta^\dagger|\widetilde{\phi}_{a_{k}}\rangle|^{2}.
\end{eqnarray}
where we have used complex conjugation and the modulus squared to flip the order of the operators. The time-reversal of a single Kraus operator is $\widetilde{E}_{u}\equiv\mathcal{A}\pi^{1/2}E^{\dagger}_{u}\pi^{-1/2}\mathcal{A}^{\dagger}$, where we choose $\mathcal{A} = \Theta$ and $\pi = \mathbbm{1}$ (as $\Phi$ is unital, such that $\Phi(\mathbbm{1}) = \mathbbm{1}$). We can now state that
\begin{equation}
|\langle\phi_{a_{k}}|E_{u}|\psi_{a_{m}}\rangle|^{2} = |\langle\widetilde{\psi}_{a_{m}}|\widetilde{E}_{u}|\widetilde{\phi}_{a_{k}}\rangle|^{2}.
\end{equation}
Moreover, by observing that
\begin{equation}
\sum_{u}|\langle\widetilde{\psi}_{a_{m}}|\widetilde{E}_{u}|\widetilde{\phi}_{a_{k}}\rangle|^{2} = \frac{\textrm{Tr}[\widetilde{\Pi}^{\textrm{in}}_{m}\widetilde{\Phi}(\rho_{\textrm{ref},k})]}
{\textrm{Tr}[\widetilde{\Pi}^{\textrm{ref}}_{k}\widetilde{\rho}_{\mathcal{T}}\widetilde{\Pi}^{\textrm{ref}}_{m}]}
= p(a^{\textrm{in}}_{m}|a^{\textrm{ref}}_{k}),
\end{equation}
where $\rho_{\textrm{ref},k}\equiv\widetilde{\Pi}^{\textrm{ref}}_{k}\widetilde{\rho}_{\mathcal{T}}\widetilde{\Pi}^{\textrm{ref}}_{m}$, the equality $p(a^{\textrm{fin}}_{k}|a^{\textrm{in}}_{m}) = p(a^{\textrm{in}}_{m}|a^{\textrm{ref}}_{k})$, as well as Theorem~4.1, follow straightforwardly.

\section*{Proof of Theorem~4.2}

Here, we prove Theorem~4.2, i.e. the inequality
\begin{equation*}
0\leq S(\rho_{\textrm{fin}}\parallel\rho_{\mathcal{T}})\leq\langle\sigma\rangle,
\end{equation*}
where $\rho_{\textrm{fin}}$ and $\rho_{\mathcal{T}}$ are the density operators of the open quantum system $\mathcal{S}$ before and after the second measurement of the forward process. $S(\rho_{\textrm{fin}}\parallel\rho_{\mathcal{T}})$ is the quantum relative entropy of $\rho_{\textrm{fin}}$ and $\rho_{\mathcal{T}}$ and $\langle\sigma\rangle$ is the average of the stochastic quantum entropy production. This inequality may be regarded as the quantum counterpart of the second law of thermodynamics for an open quantum system.

To this end, let us consider the stochastic entropy production $\sigma(a^{\textrm{fin}},a^{\textrm{in}})=\ln\left[\frac{p(a^{\textrm{in}})}{p(a^{\textrm{ref}})}\right]$ (as given in (\ref{sigma}) in chapter~\ref{chap:QTherno}) for the open quantum system $\mathcal{S}$, whose validity is subordinated to the assumptions of Theorem~4.1. Accordingly, the average value of $\sigma$ is
\begin{eqnarray}
\langle\sigma\rangle &=& \sum_{a^{\textrm{fin}},a^{\textrm{in}}}p(a^{\textrm{fin}},a^{\textrm{in}})\ln\left[\frac{p(a^{\textrm{in}})}{p(a^{\textrm{ref}})}\right]\nonumber \\
&=& \sum_{a^{\textrm{in}}}p(a^{\textrm{in}})\ln[p(a^{\textrm{in}})] - \sum_{a^{\textrm{fin}}}p(a^{\textrm{fin}})\ln[p(a^{\textrm{ref}})]\geq 0.
\end{eqnarray}
We observe that the mean quantum entropy production $\langle\sigma\rangle$ is a non-negative quantity due to the positivity of the classical relative entropy, or Kullback-Leibler divergence. Since $p(a^{\textrm{fin}})\equiv\langle\phi_{a}|\rho_{\textrm{fin}}|\phi_{a}\rangle$ and the reference state is diagonal in the basis $\{|\phi_a\rangle\}$, we have
\begin{eqnarray}
\sum_{a^{\textrm{fin}}}p(a^{\textrm{fin}})\ln[p(a^{\textrm{ref}})] &=& \sum_{a^{\textrm{fin}}}\langle\phi_{a}|\rho_{\textrm{fin}}|\phi_{a}\rangle\ln[p(a^{\textrm{ref}})]
= \sum_{a^{\textrm{fin}}}\langle\phi_{a}|\rho_{\textrm{fin}}\ln\rho_{\textrm{ref}}|\phi_{a}\rangle\nonumber \\
&=& \textrm{Tr}\left[\rho_{\textrm{fin}}\ln\rho_{\mathcal{T}}\right],
\end{eqnarray}
where the last identity is verified by assuming the equality between the reference state $\rho_{\textrm{ref}}$ and the density operator $\rho_{\mathcal{T}}$ after the second measurement of the protocol. One also has:
\begin{equation}
\sum_{a^{\textrm{in}}}p(a^{\textrm{in}})\ln[p(a^{\textrm{in}})] = \textrm{Tr}\left[\rho_{\textrm{in}}\ln\rho_{\textrm{in}}\right] = -S(\rho_{\textrm{in}}),
\end{equation}
where $S(\rho_{\textrm{in}})\equiv-\textrm{Tr}\left[\rho_{\textrm{in}}\ln\rho_{\textrm{in}}\right]$ is the von Neumann entropy for the initial density operator $\rho_{\textrm{in}}$ of the quantum system $\mathcal{S}$. The mean quantum entropy production $\langle\sigma\rangle$, thus, can be written in general as
\begin{equation}
\langle\sigma\rangle = -\textrm{Tr}\left[\rho_{\textrm{fin}}\ln\rho_{\mathcal{T}}\right] - S(\rho_{\textrm{in}}).
\end{equation}
The quantum relative entropy is defined as
$$
S(\rho_{\textrm{fin}}\parallel\rho_{\mathcal{T}}) = - \textrm{Tr}\left[\rho_{\textrm{fin}}\ln\rho_{\mathcal{T}}\right] - S(\rho_{\textrm{fin}})
$$
and trivially $S(\rho_{\textrm{fin}}\parallel\rho_{\mathcal{T}})\geq 0$. According to our protocol, the initial and the final states are connected by the unital CPTP map $\Phi$ as $\rho_{\textrm{fin}}=\Phi(\rho_{\textrm{in}})$. As a consequence of the unitality of $\Phi$, the von Neumann entropies obey the relation $S(\rho_{\textrm{in}})\leq S(\rho_{\textrm{fin}})$. Summarizing, we obtain
\begin{equation}
0\leq S(\rho_{\textrm{fin}}\parallel\rho_{\mathcal{T}}) = - \textrm{Tr}\left[\rho_{\textrm{fin}}\ln\rho_{\mathcal{T}}\right] - S(\rho_{\textrm{fin}})
\leq - \textrm{Tr}\left[\rho_{\textrm{fin}}\ln\rho_{\mathcal{T}}\right] - S(\rho_{\textrm{in}}) = \langle\sigma\rangle,
\end{equation}
proving the original inequality.

Note that if we perform the second measurement with a basis in which $\rho_{\textrm{fin}}$ is diagonal (i.e. vanishing commutator between measurement operator and final state, $[\mathcal{O}_{\textrm{fin}},\rho_{\textrm{fin}}]=0$), the state is unchanged by the second measurement and $\rho_{\textrm{fin}}=\rho_{\mathcal{T}}$. As a consequence
\begin{equation*}
0= S(\rho_{\textrm{fin}}\parallel\rho_{\mathcal{T}})\leq\langle\sigma\rangle=S(\rho_{\textrm{fin}})-S(\rho_{\textrm{in}}),
\end{equation*}
i.e. the quantum relative entropy vanishes, while the average of the stochastic entropy production equals the difference of final and initial von Neumann entropies, $\langle\sigma\rangle=S(\rho_{\textrm{fin}})-S(\rho_{\textrm{in}})$, and thus describes the irreversibility distribution of the map $\Phi$ only (and not of the measurement, as it would be in the general case). In the general case, i.e. if the condition $[\mathcal{O}_{\textrm{fin}},\rho_{\textrm{fin}}]=0$ does not hold, still the post-measurement state $\rho_{\mathcal{T}}$ is diagonal in the basis of the observable eigenstates and we obtain
\begin{equation}
\langle\sigma\rangle = -\textrm{Tr}\left[\rho_{\textrm{fin}}\ln\rho_{\mathcal{T}}\right] - S(\rho_{\textrm{in}}) = S(\rho_{\mathcal{T}})- S(\rho_{\textrm{in}}).
\end{equation}

\section*{Characteristic functions of the quantum entropy distribution}

We derive the expressions for the characteristic functions $G_A(\lambda)$ and $G_B(\lambda)$, respectively for the probability distributions $\textrm{Prob}(\sigma_A)$ and $\textrm{Prob}(\sigma_B)$, given by (\ref{eq3}) and (\ref{eq4}). We start with the definition
\begin{equation}
G_A(\lambda) = \int \textrm{Prob}_{A}(\sigma_A)e^{i\lambda\sigma_A}d\sigma_A,
\end{equation}
where
\begin{equation}
\textrm{Prob}(\sigma_A) = \sum_{k,m}\delta\left[\sigma_A - \sigma_A(a^{\textrm{in}}_{m},a^{\textrm{fin}}_{k})\right]p_a(k,m),
\end{equation}
as well as
\begin{equation}
p_{a}(k,m) = \textrm{Tr}\left[(\Pi^{\mathcal{T}}_{A,k}\otimes\mathbbm{1}_{B})\Phi(\Pi^{\textrm{in}}_{A,m}\otimes\rho_{\textrm{B,in}})\right]p(a_{m}^{\textrm{in}}),
\end{equation}
and
\begin{equation}
\sigma_A(a^{\textrm{in}}_{m},a^{\textrm{fin}}_{k})=\ln [p(a^{\textrm{in}}_{m})] - \ln[p(a^{\textrm{fin}}_{k})]
\end{equation}
Exploiting the linearity of $\Phi$ and the trace, we obtain
\begin{eqnarray}\label{eq:G-appendix}
&G_A(\lambda) = \sum_{k,m}p_{a}(k,m)e^{i\lambda\sigma_A(a^{\textrm{in}}_{m},a^{\textrm{fin}}_{k})}&\nonumber \\
&= \textrm{Tr}\left[\left(\sum_{k}\Pi^{\mathcal{T}}_{A,k}e^{-i\lambda \ln [p(a^{\textrm{fin}}_{k})]}\otimes\mathbbm{1}_{B}\right)\Phi\left(\sum_{m}\Pi^{\textrm{in}}_{A,m}e^{i\lambda \ln[p(a^{\textrm{in}}_{m})]}p(a_{m}^{\textrm{in}})\otimes\rho_{\textrm{B,in}}\right)\right].&\nonumber \\
&&
\end{eqnarray}
Recalling the spectral decompositions of the initial and final density operators, $\rho_{\textrm{A,in}}\equiv\sum_{m}\Pi^{\textrm{in}}_{A,m}p(a^{\textrm{in}}_{m})$ and $\rho_{A,\mathcal{T}}\equiv\sum_{k}\Pi^{\mathcal{T}}_{A,k}p(a^{\mathcal{T}}_{k})$, with eigenvalues $p(a^{\textrm{in}}_{m})$ and $p(a^{\mathcal{T}}_{k})=p(a^{\textrm{fin}}_{k})$, we get
\begin{equation}
\sum_{k}\Pi^{\mathcal{T}}_{A,k}e^{-i\lambda \ln[p(a^{\textrm{fin}}_{k})]} = \sum_{k}\Pi^{\mathcal{T}}_{A,k}e^{-i\lambda \ln[p(a^{\mathcal{T}}_{k})]} = \sum_{k}\Pi^{\mathcal{T}}_{A,k}p(a^{\mathcal{T}}_{k})^{-i\lambda} = \left(\rho_{A,\mathcal{T}}\right)^{-i\lambda},
\end{equation}
and
\begin{equation}
\sum_{m}\Pi^{\textrm{in}}_{A,m}e^{i\lambda \ln[p(a^{\textrm{in}}_{m})]}p(a_{m}^{\textrm{in}}) =
\sum_{m}\Pi^{\textrm{in}}_{A,m}p(a_{m}^{\textrm{in}})^{1 + i\lambda} = \left(\rho_{A,\textrm{in}}\right)^{1 + i\lambda}.
\end{equation}
If we insert these expressions into (\ref{eq:G-appendix}) we obtain the expression for the characteristic function $G_A(\lambda)$ given in (\ref{eq:G_A}). Analogously we can derive (\ref{eq2}) for $G_B(\lambda)$. In a similar way we can derive the characteristic function $G_{A - B}(\lambda)$ of the stochastic entropy production of the composite system $A - B$:
\begin{equation}
 G_{A - B}(\lambda)=\mathrm{Tr}\left[\rho_{\mathcal{T}}^{-i\lambda}\Phi(\rho_{\mathrm{in}}^{1+i\lambda})\right]\,.
\end{equation}

\chapter{Publications}
\label{mypublications}

This research activity has led to several publications in international journals and conferences. These are summarized below.\footnote{The author's bibliometric indices are the following: \textit{H}-index = 4, total number of citations = 35 (source: Google Scholar on Month 04, 2018).}

\begin{small}

\subsection*{International Journals}
\begin{enumerate}
\item \textbf{R. Mencucci, S. Matteoli, A. Corvi, L. Terracciano, E. Favuzza, S. Gherardini, F. Caruso, R. Bellucci}, ``Investigating the ocular temperature rise during femtosecond laser fragmentation: an in vitro study'', \textit{Graefe's Archive for Clinical and Experimental Ophthalmology}, \textbf{253(12)}, 2203-10 (2015).
\item \textbf{S. Gherardini, S. Gupta, F.S. Cataliotti, A. Smerzi, F. Caruso, S. Ruffo}, ``Stochastic quantum Zeno by large deviation theory'', \textit{New Journal of Physics} \textbf{18(1)}, 013048 (2016).
\item \textbf{M.M. M\"{u}ller, S. Gherardini, A. Smerzi, F. Caruso}, ``Fisher information from stochastic quantum measurements'', \textit{Physical Review A} \textbf{94}, 042322 (2016).
\item \textbf{S. Viciani, S. Gherardini, M. Lima, M. Bellini, F. Caruso}, ``Disorder and dephasing as control knobs for light transport in optical fiber cavity networks'', \textit{Scientific Reports} \textbf{6}, 37791 (2016).
\item \textbf{M.M. M\"{u}ller, S. Gherardini, F. Caruso}, ``Stochastic quantum Zeno-based detection of noise correlations'', \textit{Scientific Reports} \textbf{6}, 38650 (2016).
\item \textbf{S. Gherardini, C. Lovecchio, M.M. M\"{u}ller, P. Lombardi, F. Caruso, F.S. Cataliotti}, ``Ergodicity in randomly perturbed quantum systems'', \textit{Quantum Science and Technology} \textbf{2(1)}, 015007 (2017).
\item \textbf{M.M. M\"{u}ller, S. Gherardini, F. Caruso}, ``Quantum Zeno dynamics through stochastic protocols'', \textit{Annalen der Physik} \textbf{529(9)}, 1600206 (2017).
\item \textbf{G. Battistelli, L. Chisci, S. Gherardini}, ``Moving horizon estimation for discrete-time linear systems with binary sensors: algorithms and stability results'', \textit{Automatica} \textbf{85}, 374-385 (2017).
\item \textbf{S. Gherardini, S. Gupta, S. Ruffo}, ``Kuramoto models for synchronization: statistical mechanics of out-of-equilibrium globally coupled phase oscillations'', 2018, accepted in \textit{Contemporary Physics}.
\end{enumerate}

\subsubsection*{Submitted}
\begin{enumerate}
\item \textbf{S. Gherardini, M.M. M\"{u}ller, A. Trombettoni, S. Ruffo, F. Caruso}, ``Reconstruction of the stochastic quantum entropy production to probe irreversibility and correlations'', in \textit{Eprint arXiv:1706.02193}, 2017, submitted to \textit{Quantum Science and Technology - IOPscience}.
\item \textbf{M.M. M\"{u}ller, S. Gherardini, F. Caruso}, ``Noise-robust quantum sensing via optimal multi-probe spectroscopy'', in \textit{Eprint arXiv:1801.10220}, 2018, submitted to \textit{Scientific Reports}.
\item \textbf{S. Gherardini, L. Buffoni, M.M. M\"{u}ller, F. Caruso, M. Campisi, A. Trombettoni, S. Ruffo}, ``Non-equilibrium quantum-heat statistics under stochastic projective measurements'', in \textit{Eprint arXiv:1805.00773}, 2018, submitted to \textit{Physical Review E}.
\end{enumerate}

\subsubsection*{To be submitted}
\begin{enumerate}
\item \textbf{G. Battistelli, L. Chisci, N. Forti, S. Gherardini}, ``MAP moving horizon field estimation with threshold measurements for large-scale systems'', in preparation, 2018.
\end{enumerate}

\subsection*{International Conferences}
\begin{enumerate}
\item \textbf{G. Battistelli, L. Chisci, S. Gherardini}, ``Moving horizon state estimation for discrete-time linear systems with binary sensors'', in \textit{54th International Conference on Decision and Control (CDC)}, December 15-18, 2015, Osaka (Japan).
\item \textbf{G. Battistelli, L. Chisci, N. Forti, S. Gherardini}, ``MAP moving horizon state estimation with binary measurements'', in \textit{The 2016 American Control Conference (ACC)}, July 6-8, 2016, Boston (USA).
\end{enumerate}

\end{small}

%\nocite{*}
%\fancyhead[RO,LE]{\slshape BIBLIOGRAPHY}
%\fancyfoot[C]{\thepage}
%\bibliographystyle{plain}
%\bibliography{biblio}

\begin{thebibliography}{100}

\bibitem{AbahPRL2012}
O.~Abah, J.~Ro{\ss}nagel, G.~Jacob, S.~Deffner, F.~Schmidt-Kaler, K.~Singer,
  and E.~Lutz.
\newblock Single-ion heat engine at maximum power.
\newblock {\em Phys. Rev. Lett.}, 109:203006, 2012.

\bibitem{Aberg2016}
J.~Aberg.
\newblock Fully quantum fluctuation theorem.
\newblock {\em Eprint arXiv:1601.01302}, 2016.

\bibitem{Ahlswede1}
R.~Ahlswede and V.M. Blinovsky.
\newblock Large deviations in quantum information theory.
\newblock {\em Problems of Information Transmission}, 373:39, 2003.

\bibitem{Albash2013}
T.~Albash, D.A. Lidar, M.~Marvian, and P.~Zanardi.
\newblock Fluctuation theorems for quantum processes.
\newblock {\em Phys. Rev. E}, 88:032146, 2013.

\bibitem{AlBaBaTAC05}
A.~Alessandri, M.~Baglietto, and G.~Battistelli.
\newblock Receding-horizon estimation for switching discrete-time linear
  systems.
\newblock {\em IEEE Trans. on Automatic Control}, 50(11):1736--1748, 2005.

\bibitem{NLMHE}
A.~Alessandri, M.~Baglietto, and G.~Battistelli.
\newblock Moving horizon state estimation for nonlinear discrete-time systems:
  {N}ew stability results and approximation schemes.
\newblock {\em Automatica}, 44:1753--1765, 2008.

\bibitem{AlBaBaZavCDC10}
A.~Alessandri, M.~Baglietto, G.~Battistelli, and V.M. Zavala.
\newblock Advances in moving horizon estimation for nonlinear systems.
\newblock In {\em Proc. 49th IEEE Conference on Decision and Control}, pages
  5681--5688, Atlanta, GA, USA, 2010.

\bibitem{Alhambra2016}
A.M. Alhambra, L.~Masanes, J.~Oppenheim, and C.~Perry.
\newblock Fluctuating work: {F}rom quantum thermodynamical identities to a
  second law equality.
\newblock {\em Phys. Rev. X}, 6:041017, 2016.

\bibitem{Alicki1979}
R.~Alicki.
\newblock The quantum open system as a model of the heat engine.
\newblock {\em J. Phys. A: Math. Gen.}, 12:L103, 1979.

\bibitem{AlonsoPRL2016}
J.J. Alonso, E.~Lutz, and A.~Romito.
\newblock Thermodynamics of weakly measured quantum systems.
\newblock {\em Phys. Rev. Lett.}, 116:080403, 2016.

\bibitem{Amselem2009PRL103}
E.~Amselem, M.~Radmark, M.~Bourennane, and A.~Cabello.
\newblock State-independent quantum contextuality with single photons.
\newblock {\em Phys. Rev. Lett.}, 103:160405, 2009.

\bibitem{An2015}
S.~An, J-N. Zhang, M.~Um, D.~Lv, Y.~Lu, J.~Zhang, Z-Q. Yin, H.T. Quan, and
  K.~Kim.
\newblock Experimental test of the quantum {J}arzynski equality with a
  trapped-ion system.
\newblock {\em Nat. Phys.}, 11:193--199, 2014.

\bibitem{Aslam}
J.~Aslam, Z.~Butler, F.~Constantin, V.~Crespi, G.~Cybenko, and D.~Rus.
\newblock Tracking a moving object with a binary sensor network.
\newblock In {\em Proceedings 1st ACM Conf. on Embedded Networked Sensor
  Systems, Los Angeles, USA}, pages 150--161, 2003.

\bibitem{Athanassoulis}
G.A. Athanassoulis and P.N. Gavriliadis.
\newblock The truncated {H}ausdorff moment problem solved by using kernel
  density functions.
\newblock {\em Prob. Eng. Mech.}, 17:273--291, 2002.

\bibitem{Balasubramanian2008}
G.~Balasubramanian, I.Y. Chan, R.~Kolesov, M.~Al-Hmoud, J.~Tisler, C.~Shin,
  C.~Kim, A.~Wojcik, P.R. Hemmer, A.~Krueger, T.~Hanke, A.~Leitenstorfer,
  R.~Bratschitsch, F.~Jelezko, and J.~Wrachtrup.
\newblock Nanoscale imaging magnetometry with diamond spins under ambient
  conditions.
\newblock {\em Nature}, 455:648--651, 2008.

\bibitem{Bar-Shalom}
Y.~Bar-Shalom, X.~Rong~Li, and T.~Kirubarajan.
\newblock {\em Estimation with Applications to Tracking and Navigation}.
\newblock John Wiley \& Sons, 2001.

\bibitem{Barchielli1991}
A.~Barchielli and V.P. Belavkin.
\newblock Measurements continuous in time and a posteriori states in quantum.
\newblock {\em J. Phys. A: Math. Gen.}, 24:1495--1514, 1991.

\bibitem{BatalhaoPRL2015}
T.B. Batalhao, A.M. Souza, R.S. Sarthour, I.S. Oliveira, M.~Paternostro,
  E.~Lutz, and R.M. Serra.
\newblock Irreversibility and the arrow of time in a quenched quantum system.
\newblock {\em Phys. Rev. Lett.}, 115:190601, 2015.

\bibitem{BaBeCh}
G.~Battistelli, A.~Benavoli, and L.~Chisci.
\newblock Data-driven communication for state estimation with sensor networks.
\newblock {\em Automatica}, 48:926--935, 2012.

\bibitem{GherardiniMAP}
G.~Battistelli, L.~Chisci, N.~Forti, and S.~Gherardini.
\newblock {MAP} moving horizon state estimation with binary sensors.
\newblock In {\em Proc. 2016 American Control Conference (ACC)}, pages
  5413--5418, Boston, MA, USA, 2016.

\bibitem{GherardiniFIELDmap}
G.~Battistelli, L.~Chisci, N.~Forti, and S.~Gherardini.
\newblock {MAP} moving horizon field estimation with binary measurements for
  large-scale systems.
\newblock {\em in preparation}, 2018.

\bibitem{TACNick}
G.~Battistelli, L.~Chisci, N.~Forti, G.~Pelosi, and S.~Selleri.
\newblock Distributed finite-element {K}alman filter for field estimation.
\newblock {\em IEEE Transactions on Automatic Control}, 62(7):3309 -- 3322,
  2017.

\bibitem{GherardiniCDC}
G.~Battistelli, L.~Chisci, and S.~Gherardini.
\newblock Moving horizon state estimation for discrete-time linear systems with
  binary sensors.
\newblock In {\em Proc. 54th IEEE Conference on Decision and Control}, pages
  2414--2419, Osaka, Japan, 2015.

\bibitem{GherardiniAutomatica}
G.~Battistelli, L.~Chisci, and S.~Gherardini.
\newblock Moving horizon state estimation for discrete-time linear systems with
  binary sensors: algorithms and stability results.
\newblock {\em Automatica}, 85:374--385, 2017.

\bibitem{Baxter1}
R.J. Baxter.
\newblock {\em Exactly Solved Models in Statistical Mechanics}.
\newblock Academic Press, London, 1982.

\bibitem{Benedetti2016}
C.~Benedetti, F.~Buscemi, P.~Bordone, and M.G.A. Paris.
\newblock Non-{M}arkovian continuous-time quantum walks on lattices with
  dynamical noise.
\newblock {\em Phys. Rev. A}, 93:042313, 2016.

\bibitem{Benedetti2014}
C.~Benedetti, M.G.A. Paris, and S.~Maniscalco.
\newblock Non-{M}arkovianity of colored noisy channels.
\newblock {\em Phys. Rev. A}, 89:012114, 2014.

\bibitem{Biggerstaff2015}
D.N. Biggerstaff, R.~Heilmann, A.A. Zecevik, M.~Gr\"afe, M.A. Broome,
  A.~Fedrizzi, S.~Nolte, A.~Szameit, A.G. White, and I.~Kassal.
\newblock Enhancing coherent transport in a photonic network using controllable
  decoherence.
\newblock {\em Nat. Commun.}, 7:11282, 2016.

\bibitem{BlMi}
F.~Blanchini and S.~Miani.
\newblock Stabilization of {LPV} systems: {S}tate feedback, state estimation,
  and duality.
\newblock {\em SIAM Journal on Control and Optimization}, 42(1):76--97, 2003.

\bibitem{Boyd}
S.~Boyd and L.~Vandenderghe.
\newblock {\em Convex Optimization}.
\newblock Cambridge University Press, Cambridge, UK, 2004.

\bibitem{Brandao2015}
F.~Brand$\tilde{a}$o, M.~Horodecki, N.~Ng, J.~Oppenheim, and S.~Wehner.
\newblock The second laws of quantum thermodynamics.
\newblock {\em PNAS}, 112:3275--3279, 2015.

\bibitem{Braunstein1994}
S.L. Braunstein and C.M. Caves.
\newblock Statistical distance and the geometry of quantum states.
\newblock {\em Phys. Rev. Lett.}, 72:3439, 1994.

\bibitem{Brenner96}
S.C. Brenner and L.R. Scott.
\newblock {\em The mathematical theory of finite element methods}.
\newblock Springer--Verlag, New York, NY, 1996.

\bibitem{PetruccioneBook}
H.P. Breuer and F.~Petruccione.
\newblock {\em The Theory of Open Quantum Systems}.
\newblock Oxford University Press, 2003.

\bibitem{Brunelli2016}
M.~Brunelli and M.~Paternostro.
\newblock Irreversibility and correlations in coupled quantum oscillators.
\newblock {\em Eprint arXiv:1610.01172}, 2016.

\bibitem{Giovannetti2014}
D.K. Burgarth, P.~Facchi, V.~Giovannetti, H.~Nakazato, S.~Pascazio, and
  K.~Yuasa.
\newblock Exponential rise of dynamical complexity in quantum computing through
  projections.
\newblock {\em Nat. Comm.}, 5:6173, 2014.

\bibitem{Cai2012}
C.Y. Cai, Q.~Ai, H.T. Quan, and C.P. Sun.
\newblock Sensitive chemical compass assisted by quantum criticality.
\newblock {\em Phys. Rev. A}, 85:022315, 2012.

\bibitem{CamatiPRL2016}
P.A. Camati, J.P.S. Peterson, T.B. Batalhao, K.~Micadei, A.M. Souza, R.S.
  Sarthour, I.S. Oliveira, and R.M. Serra.
\newblock Experimental rectification of entropy production by {M}axwell's demon
  in a quantum system.
\newblock {\em Phys. Rev. Lett.}, 117:240502, 2016.

\bibitem{CampisiNJP2014}
M.~Campisi, R.~Blattmann, S.~Kohler, D.~Zueco, and P.~H\"{a}nggi.
\newblock Employing circuit {QED} to measure non-equilibrium work fluctuations.
\newblock {\em New J. Phys.}, 15:105028, 2014.

\bibitem{Campisi2011}
M.~Campisi, P.~Hanggi, and P.~Talkner.
\newblock {\it Colloquium}: {Q}uantum fluctuations relations: {F}oundations and
  applications.
\newblock {\em Rev. Mod. Phys.}, 83:771, 2011.

\bibitem{Campisi15NJP17}
M.~Campisi, J.~Pekola, and R.~Fazio.
\newblock Nonequilibrium fluctuations in quantum heat engines: theory, example,
  and possible solid state experiments.
\newblock {\em New J. Phys.}, 17:035012, 2015.

\bibitem{Campisi2009}
M.~Campisi, P.~Talkner, and P.~H\"{a}nggi.
\newblock Fluctuation theorem for arbitrary open quantum systems.
\newblock {\em Phys. Rev. Lett.}, 102:210401, 2009.

\bibitem{Campisi2010PRL}
M.~Campisi, P.~Talkner, and P.~H\"{a}nggi.
\newblock Fluctuation theorems for continuously monitored quantum fluxes.
\newblock {\em Phys. Rev. Lett.}, 105:140601, 2010.

\bibitem{Campisi2011PRE}
M.~Campisi, P.~Talkner, and P.~H\"{a}nggi.
\newblock Influence of measurements on the statistics of work performed on a
  quantum system.
\newblock {\em Phys. Rev. E}, 88:041114, 2013.

\bibitem{Capponi}
A.~Capponi, I.~Fatkullin, and L.~Shi.
\newblock Stochastic filtering for diffusion processes with level crossings.
\newblock {\em IEEE Transactions on Automatic Control}, 56(9):2201--2206, 2011.

\bibitem{FC2014}
F.~Caruso.
\newblock Universally optimal noisy quantum walks on complex networks.
\newblock {\em New J. Phys.}, 16:055015, 2014.

\bibitem{Caruso2009JChPh131}
F.~Caruso, A.W. Chin, A.~Datta, S.~F. Huelga, and M.B. Plenio.
\newblock Highly efficient energy excitation transfer in light-harvesting
  complexes: {T}he fundamental role of noise-assisted transport.
\newblock {\em J. Chem. Phys.}, 131:105106, 2009.

\bibitem{FC2016}
F.~Caruso, A.~Crespi, A.G. Ciriolo, F.~Sciarrino, and R.~Osellame.
\newblock Fast escape of a quantum walker from an integrated photonic maze.
\newblock {\em Nat. Commun.}, 7:11682, 2016.

\bibitem{CarusoRMP14}
F.~Caruso, V.~Giovannetti, C.~Lupo, and S.~Mancini.
\newblock Quantum channels and memory effects.
\newblock {\em Rev. Mod. Phys.}, 86:1203, 2014.

\bibitem{Caruso2010PRL105}
F.~Caruso, S.F. Huelga, and M.B. Plenio.
\newblock Noise-enhanced classical and quantum capacities in communication
  networks.
\newblock {\em Phys. Rev. Lett.}, 105:190501, 2010.

\bibitem{Cavities2012}
F.~Caruso, S.K. Saikin, E.~Solano, S.F. Huelga, A.~Aspuru-Guzik, and M.B.
  Plenio.
\newblock Probing biological light-harvesting phenomena by optical cavities.
\newblock {\em Phys. Rev. B}, 85:125424, 2012.

\bibitem{Hollenberg2009}
J.H. Cole and L.C.L. Hollenberg.
\newblock Scanning quantum decoherence microscopy.
\newblock {\em Nanotechnology}, 20:495401, 2009.

\bibitem{Strong2015}
D.M. Coles, Y.~Yang, Y.~Wang, R.T. Grant, R.A. Taylor, S.K. Saikin,
  A.~Aspuru-Guzik, D.G. Lidzey, J.K.-H. Tang, and J.M. Smith.
\newblock Strong coupling between chlorosomes of photosynthetic bacteria and a
  confined optical cavity mode.
\newblock {\em Nat. Commun.}, 5:5561, 2015.

\bibitem{CollinNAT2005}
D.~Collin, F.~Ritort, C.~Jarzynski, S.B. Smith, I.~Jr Tinoco, and
  C.~Bustamante.
\newblock Verification of the {C}rooks fluctuation theorem and recovery of rna
  folding free energies.
\newblock {\em Nature}, 437:231--234, 2005.

\bibitem{Collini2010NAT463}
E.~Collini, C.Y. Wong, K.E. Wilk, P.M.G. Curmi, P.~Brumer, and G.D. Scholes.
\newblock Coherently wired light-harvesting in photosynthetic marine algae at
  ambient temperature.
\newblock {\em Nature}, 463:644--647, 2010.

\bibitem{Cover2006}
T.M. Cover and J.A. Thomas.
\newblock {\em Elements of Information Theory}.
\newblock Wiley-Interscience New Jersey, 2006.

\bibitem{CrooksPRE1999}
G.~Crooks.
\newblock Entropy production fluctuation theorem and the nonequilibrium work
  relation for free energy differences.
\newblock {\em Phys. Rev. E}, 60:2721, 1999.

\bibitem{CrooksJSM2008}
G.E. Crooks.
\newblock On the {J}arzynski relation for dissipative quantum dynamics.
\newblock {\em J. Stat. Mech.}, page P10023, 2008.

\bibitem{CrooksPRA2008}
G.E. Crooks.
\newblock Quantum operation time reversal.
\newblock {\em Phys. Rev. A}, 77:034101, 2008.

\bibitem{davisson1928}
C.J. Davisson.
\newblock The diffraction of electrons by a crystal of nickel.
\newblock {\em Bell System Technical Journal}, 7:90 -- 105, 1928.

\bibitem{Groot1984}
S.R. de~Groot and P.~Mazur.
\newblock {\em Non-Equilibrium Thermodynamics}.
\newblock Dover Publications, 1984.

\bibitem{Deffner_tesi}
S.~Deffner.
\newblock {\em Nonequilibrium entropy production in open and closed quantum
  systems}.
\newblock PhD thesis. 2011.

\bibitem{DeffnerPRL2011}
S.~Deffner and E.~Lutz.
\newblock Nonequilibrium entropy production for open quantum systems.
\newblock {\em Phys. Rev. Lett.}, 107:140404, 2011.

\bibitem{Degen2017}
C.L. Degen, F.~Reinhard, and P.~Cappellaro.
\newblock Quantum sensing.
\newblock {\em Rev. Mod. Phys.}, 89:035002, 2017.

\bibitem{Delgado}
R.A. Delgado and G.C. Goodwin.
\newblock A combined {MAP} and {B}ayesian scheme for finite data and/or moving
  horizon estimation.
\newblock {\em Automatica}, 50(4):1116--1121, 2014.

\bibitem{Dembo1}
A.~Dembo and O.~Zeitouni.
\newblock {\em Large Deviations Techniques and Applications}.
\newblock Springer, Berlin, 2010.

\bibitem{Djuric_2}
P.M. Djuric, M.~Vemula, and M.F. Bugallo.
\newblock Target tracking by particle filtering in binary sensor networks.
\newblock {\em IEEE Trans. on Signal Processing}, 56:2229--2238, 2008.

\bibitem{DobekPRL2011}
K.~Dobek, M.~Karpinski, R.~Demkowicz-Dobrzanski, K.~Banaszek, and P.~Horodecki.
\newblock Experimental extraction of secure correlations from a noisy private
  state.
\newblock {\em Phys. Rev. Lett.}, 106:030501, 2011.

\bibitem{DoriaPRL2011}
P.~Doria, T.~Calarco, and S.~Montangero.
\newblock Optimal control technique for many-body quantum dynamics.
\newblock {\em Phys. Rev. Lett.}, 106:190501, 2011.

\bibitem{Dorner2013}
R.~Dorner, S.R. Clark, L.~Heaney, R.~Fazio, J.~Goold, and V.~Vedral.
\newblock Extracting work statistics and fluctuation theorems by single-qubit
  interferometry.
\newblock {\em Phys. Rev. Lett.}, 110:230601, 2013.

\bibitem{Edwards1975}
S.F. Edwards and P.W. Anderson.
\newblock Theory of spin glasses.
\newblock {\em J. Phys. F}, 5:965, 1975.

\bibitem{Ellis2006}
R.~Ellis.
\newblock {\em Entropy, Large Deviations, and Statistical Mechanics}.
\newblock Springer, New York, 2006.

\bibitem{Elouard2016}
C.~Elouard, D.A. Herrera-Mart\`i, M.~Clusel, and A.~Auff\'eves.
\newblock The role of quantum measurement in stochastic thermodynamics.
\newblock {\em Nature Quantum Information}, 3:9, 2017.

\bibitem{Engel2007NAT446}
G.S. Engel, T.R. Calhoun, E.L. Read, T.-K. Ahn, T.~Man\v{c}al, Y.-C. Cheng,
  R.E. Blankenship, and G.R. Fleming.
\newblock Evidence for wavelike energy transfer through quantum coherence in
  photosynthetic systems.
\newblock {\em Nature}, 446:782--786, 2007.

\bibitem{Kurizki2008}
N.~Erez, G.~Gordon, M.~Nest, and G.~Kurizki.
\newblock Thermodynamic control by frequent quantum measurements.
\newblock {\em Nature}, 452:724, 2008.

\bibitem{RTN2}
J.~Eroms, L.C. van Schaarenburg, E.F.C. Driessen, J.H. Plantenberg, C.M.
  Huizinga, R.N. Schouten, A.H. Verbruggen, C.J.P.M. Harmans, and J.E. Mooij.
\newblock Low-frequency noise in {J}osephson junctions for superconducting
  qubits.
\newblock {\em Appl. Phys. Lett.}, 89:122516, 2006.

\bibitem{Esposito2009}
M.~Esposito, U.~Harbola, and S.~Mukamel.
\newblock Nonequilibrium fluctuations, fluctuation theorems, and counting
  statistics in quantum systems.
\newblock {\em Rev. Mod. Phys.}, 81:1665, 2009.

\bibitem{Wim2006}
Wim C.~Van Etten.
\newblock {\em Introduction to Random Signals and Noise}.
\newblock John Wiley \& Sons, 2006.

\bibitem{FacchiPRA2004}
P.~Facchi, D.A. Lidar, and S.~Pascazio.
\newblock Unification of dynamical decoupling and the quantum {Z}eno effect.
\newblock {\em Phys. Rev. A}, 69:032314, 2004.

\bibitem{PascazioPRL2002}
P.~Facchi and S.~Pascazio.
\newblock Quantum {Z}eno subspaces.
\newblock {\em Phys. Rev. Lett.}, 89:080401, 2002.

\bibitem{PascazioJPA}
P.~Facchi and S.~Pascazio.
\newblock Quantum {Z}eno dynamics: mathematical and physical aspects.
\newblock {\em J. Phys. A}, 41:493001, 2008.

\bibitem{FacchiPRA2005}
P.~Facchi, S.~Tasaki, S.~Pascazio, H.~Nakazato, A.~Tokuse, and D.A. Lidar.
\newblock Control of decoherence: {A}nalysis and comparison of three different
  strategies.
\newblock {\em Phys. Rev. A}, 71:022302, 2005.

\bibitem{Farina1}
M.~Farina, G.~{Ferrari-Trecate}, and R.~Scattolini.
\newblock Distributed moving horizon estimation for linear constrained systems.
\newblock {\em IEEE Transactions on Automatic Control}, 55(11):2462--2475,
  2010.

\bibitem{FaFerrSca10}
M.~Farina, G.~{Ferrari-Trecate}, and R.~Scattolini.
\newblock Moving-horizon partition-based state estimation of large-scale
  systems.
\newblock {\em Automatica}, 46(5):910--918, 2010.

\bibitem{Farina2}
M.~Farina, G.~{Ferrari-Trecate}, and R.~Scattolini.
\newblock Distributed moving horizon estimation for nonlinear constrained
  systems.
\newblock {\em International Journal of Robust and Nonlinear Control},
  22(2):123--143, 2012.

\bibitem{Morari}
G.~Ferrari-Trecate, D.~Mignone, and M.~Morari.
\newblock Moving horizon estimation for hybrid systems.
\newblock {\em IEEE Transactions on Automatic Control}, 47(10):1663--1676,
  2002.

\bibitem{feynman93}
R.P. Feynman and A.R. Hibbs.
\newblock {\em Quantum Mechanics and Path Integrals, Emended edition}.
\newblock Science/Physics. Dover publications, 2005.

\bibitem{Fischer:2001}
M.C. Fischer, B.~Gutierrez-Medina, and M.G. Raizen.
\newblock Observation of the quantum {Z}eno and anti-{Z}eno effects in an
  unstable system.
\newblock {\em Phys. Rev. Lett.}, 87:040402, 2001.

\bibitem{Forti_tesi}
N.~Forti.
\newblock {\em Dynamic field estimation in complex environments}.
\newblock PhD thesis. 2016.

\bibitem{Friedenauer2008}
A.~Friedenauer, H.~Schmitz, J.T. Gl\"{u}ckert, D.~Porras, and T.~Sch\"{a}tz.
\newblock Simulating a quantum magnet with trapped ions.
\newblock {\em Nat. Phys.}, 4:757--761, 2008.

\bibitem{FuscoPRX2014}
L.~Fusco, S.~Pigeon, T.J.G. Apollaro, A.~Xuereb, L.~Mazzola, M.~Campisi,
  A.~Ferraro, M.~Paternostro, and G.~De~Chiara.
\newblock Assessing the nonequilibrium thermodynamics in a quenched quantum
  many-body system via single projective measurements.
\newblock {\em Phys. Rev. X}, 4:031029, 2014.

\bibitem{Gagliardi2010SCI330}
G.~Gagliardi, M.~Salza, S.~Avino, P.~Ferraro, and P.~DeNatale.
\newblock Probing the ultimate limit of fiber-optic strain sensing.
\newblock {\em Science}, 330:1081--1084, 2010.

\bibitem{Gallavotti1}
G.~Gallavotti, J.L. Lebowitz, and V.~Mastropietro.
\newblock Large deviation in rarefied quantum gases.
\newblock {\em J. Stat. Phys.}, 108:831, 2002.

\bibitem{Garrahan1}
J.P. Garrahan and I.~Lesanovsky.
\newblock Thermodynamics of quantum jump trajectories.
\newblock {\em Phys. Rev. Lett.}, 104:160601, 2010.

\bibitem{Gemmer2004}
J.~Gemmer, M.~Michel, and G.~Mahler.
\newblock {\em Quantum Thermodynamics}.
\newblock Springer, 2004.

\bibitem{GherardiniWORK}
S.~Gherardini, L.~Buffoni, M.M. M\"{u}ller, F.~Caruso, M.~Campisi,
  A.~Trombettoni, and S.~Ruffo.
\newblock Non-equilibrium quantum-heat statistics under stochastic projective
  measurements.
\newblock {\em Eprint arXiv:1805.00773}, 2018.

\bibitem{Gherardini2016NJP}
S.~Gherardini, S.~Gupta, F.S. Cataliotti, A.~Smerzi, F.~Caruso, and S.~Ruffo.
\newblock Stochastic quantum {Z}eno by large deviation theory.
\newblock {\em New J. Phys.}, 18:013048, 2016.

\bibitem{GherardiniQST}
S~Gherardini, C.~Lovecchio, M.M. M\"{u}ller, P.~Lombardi, F.~Caruso, and F.S.
  Cataliotti.
\newblock Ergodicity in randomly perturbed quantum systems.
\newblock {\em Quantum Science and Technology}, 2(1):015007, 2017.

\bibitem{GherardiniEntropy}
S.~Gherardini, M.M. M\"{u}ller, A.~Trombettoni, S.~Ruffo, and F.~Caruso.
\newblock Reconstruction of the stochastic quantum entropy production to probe
  irreversibility and correlations.
\newblock {\em Eprint arXiv:1706.02193}, 2017.

\bibitem{Gierling2011}
M.~Gierling, P.~Schneeweiss, G.~Visanescu, P.~Federsel, M.~H\''{a}ffner, D.P.
  Kern, T.E. Judd, A.~G\''{u}nther, and J.~Fort\'{a}gh.
\newblock Cold-atom scanning probe microscopy.
\newblock {\em Nat. Nanotech.}, 6:446--451, 2011.

\bibitem{Wiener2}
T.A. Glaria~L\'opez and D.~Sbarbaro.
\newblock Observer design for nonlinear processes with {Wiener} structure.
\newblock In {\em Proceedings 50th IEEE Conf. Decision and Control and European
  Control Conference, Orlando, FL, USA}, pages 2211--2316, 2011.

\bibitem{Gogolin2016}
C.~Gogolin and J.~Eisert.
\newblock Equilibration, thermalisation, and the emergence of statistical
  mechanics in closed quantum systems.
\newblock {\em Rep. Prog. Phys.}, 79:056001, 2016.

\bibitem{GooldJPA2016}
J.~Goold, M.~Huber, A.~Riera, L.~del Rio, and P.~Skrzypczyk.
\newblock The role of quantum information in thermodynamics - a topical review.
\newblock {\em J. Phys. A: Math. Theor.}, 49:143001, 2016.

\bibitem{GooldPRE2014}
J.~Goold, U.~Poschinger, and K.~Modi.
\newblock Measuring the heat exchange of a quantum process.
\newblock {\em Phys. Rev. E}, 90:020101, 2014.

\bibitem{GuoHuang13}
Y.~Guo and B.~Huang.
\newblock Moving horizon estimation for switching nonlinear systems.
\newblock {\em Automatica}, 49(11):3270--3281, 2013.

\bibitem{HabVerh13}
A.~Haber and M.~Verhaegen.
\newblock Moving horizon estimation for large-scale interconnected systems.
\newblock {\em IEEE Trans. on Automatic Control}, 58(11):2834--2847, 2013.

\bibitem{HatanoPRL2001}
T.~Hatano and S.~Sasa.
\newblock Steady-state thermodynamics of {L}angevin systems.
\newblock {\em Phys. Rev. Lett.}, 86:3463, 2001.

\bibitem{HekkingPRL2013}
F.W.J. Hekking and J.P. Pekola.
\newblock Quantum jump approach for work and dissipation in a two-level system.
\newblock {\em Phys. Rev. Lett.}, 111:093602, 2013.

\bibitem{Hildner2013SCI340}
R.~Hildner, D.~Brinks, J.B. Nieder, R.J. Cogdell, and N.F. van Hulst.
\newblock Quantum coherent energy transfer over varying pathways in single
  light-harvesting complexes.
\newblock {\em Science}, 340:1448--1451, 2013.

\bibitem{Hill1997JLT15}
K.O. Hill and G.~Meltz.
\newblock Fiber {B}ragg grating technology fundamentals and overview.
\newblock {\em J. Lightwave Technol.}, 15:1263--1276, 1997.

\bibitem{Hofferberth2008}
S.~Hofferberth, I.~Lesanovsky, T.~Schumm, A.~Imambekov, V.~Gritsev, E.~Demler,
  and J.~Schmiedmayer.
\newblock Probing quantum and thermal noise in an interacting many-body system.
\newblock {\em Nat. Phys.}, 4:489--495, 2008.

\bibitem{HorodeckiREVIEW}
M.~Horodecki and J.~J.~Oppenheim.
\newblock ({Q}uantumness in the context of) resource theory.
\newblock {\em Int. J. Mod. Phys. B}, 27:1345019, 2013.

\bibitem{Horodecki2013}
M.~Horodecki and J.~Oppenheim.
\newblock Fundamental limitations for quantum and nanoscale thermodynamics.
\newblock {\em Nat. Commun.}, 4:3059, 2013.

\bibitem{Horowitz2013}
J.M. Horowitz and J.M.R. Parrondo.
\newblock Entropy production along nonequilibrium quantum jump trajectories.
\newblock {\em New J. Phys.}, 15:085028, 2013.

\bibitem{Hoyer2014NJP16}
S.~Hoyer, F.~Caruso, S.~Montangero, M.~Sarovar, T.~Calarco, M.B. Plenio, and
  K.B. Whaley.
\newblock Realistic and verifiable coherent control of excitonic states in a
  light-harvesting complex.
\newblock {\em New J. Phys.}, 16:045007, 2014.

\bibitem{Huber2008}
G.~Huber, F.~Schmidt-Kaler, S.~Deffner, and E.~Lutz.
\newblock Employing trapped cold ions to verify the quantum {J}arzynski
  equality.
\newblock {\em Phys. Rev. Lett.}, 101:070403, 2008.

\bibitem{Itano1990}
W.M. Itano, D.J. Heinzen, J.J. Bollinger, and D.J. Wineland.
\newblock Quantum {Z}eno effect.
\newblock {\em Phys. Rev. A}, 41:2295, 1990.

\bibitem{JarzynskiPRL1997}
C.~Jarzynski.
\newblock Nonequilibrium equality for free energy differences.
\newblock {\em Phys. Rev. Lett.}, 78:2690, 1997.

\bibitem{Jazwinski}
A.H. Jazwinski.
\newblock Limited memory optimal filtering.
\newblock {\em IEEE Trans. on Automatic Control}, 13:558--563, 1968.

\bibitem{Jozsa}
J.~Jozsa.
\newblock Fidelity for mixed quantum states.
\newblock {\em J. Mod. Opt.}, 41:2315, 1994.

\bibitem{Kafri2012}
D.~Kafri and S.~Deffner.
\newblock Holevo's bound from a general quantum fluctuation theorem.
\newblock {\em Phys. Rev. A}, 86:044302, 2012.

\bibitem{Kammerlander2016}
P.~Kammerlander and J.~Anders.
\newblock Coherence and measurement in quantum thermodynamics.
\newblock {\em Sci. Rep.}, 6:22174, 2016.

\bibitem{Katz2006}
N.~Katz, M.~Ansmann, R.C. Bialczak, E.~Lucero, R.~McDermott, M.~Neeley,
  M.~Steffen, E.M. Weig, A.N. Cleland, J.M. Martinis, and A.N. Korotkov.
\newblock Coherent state evolution in a superconducting qubit from
  partial-collapse measurement.
\newblock {\em Science}, 312:1498--1500, 2006.

\bibitem{KimPRL2011}
S.W. Kim, T.~Sagawa, S.~De~Liberato, and M.~Ueda.
\newblock Quantum {S}zilard engine.
\newblock {\em Phys. Rev. Lett.}, 106:070401, 2011.

\bibitem{Kim2012}
Y.S. Kim, J.C. Lee, O.~Kwon, and Y.H. Kim.
\newblock Protecting entanglement from decoherence using weak measurement and
  quantum measurement reversal.
\newblock {\em Nat. Phys.}, 8:117, 2012.

\bibitem{Kliesch}
M.~Kliesch, C.~Gogolin, and J.~Eisert.
\newblock Lieb-{R}obinson bounds and the simulation of time evolution of local
  observables in lattice systems.
\newblock {\em {\em Many-Electron Approaches in Physics Chemistry and
  Mathematics, Bach V. \& Site L.D. (ed)} (Springer, Berlin)}, pages 301--318,
  2014.

\bibitem{kofman2000}
A.G. Kofman and G.~Kurizki.
\newblock Acceleration of quantum decay processes by frequent observations.
\newblock {\em Nature}, 405:546--550, 2000.

\bibitem{kofman2001}
A.G. Kofman and G.~Kurizki.
\newblock Universal dynamical control of quantum mechanical decay: {M}odulation
  of the coupling to the continuum.
\newblock {\em Phys. Rev. Lett.}, 87:270405, 2001.

\bibitem{Kominis2009}
I.K. Kominis.
\newblock Quantum {Z}eno effect explains magnetic-sensitive radical-ion-pair
  reactions.
\newblock {\em Phys. Rev. E}, 80:056115, 2009.

\bibitem{KorotkovPRB}
A.N. Korotkov.
\newblock Quantum efficiency of binary-outcomes detectors of solid-state
  qubits.
\newblock {\em Physical Review B}, 78:174512, 2008.

\bibitem{KorzekwaNJP2016}
K.~Korzekwa, M.~Lostaglio, J.~Oppenheim, and D.~Jennings.
\newblock The extraction of work from quantum coherence.
\newblock {\em New J. Phys.}, 18:023045, 2016.

\bibitem{Kosloff14ARPC65}
R.~Kosloff and A.~Levy.
\newblock Quantum heat engines and refrigerators: Continuous devices.
\newblock {\em Annual Review of Physical Chemistry}, 65:365--393, 2014.

\bibitem{Kurchan2001}
J.~Kurchan.
\newblock A quantum fluctuation theorem.
\newblock {\em Eprint arXiv:cond-mat/0007360}, 2001.

\bibitem{Kwiat:1995}
P.~Kwiat, H.~Weinfurter, T.~Herzog, A.~Zeilinger, and M.A. Kasevich.
\newblock Interaction-free measurement.
\newblock {\em Phys. Rev. Lett.}, 74:4763, 1995.

\bibitem{Lamperti1960}
J.~Lamperti.
\newblock Criteria for the recurrence or transience of stochastic process. {I}.
\newblock {\em J. Math. Analysis and App.}, 1:314--330, 1960.

\bibitem{Lanyon2011}
B.P. Lanyon, C.~Hempel, D.~Nigg, M.~M\"{u}ller, R.~Gerritsma, F.~Z\"{a}hringer,
  P.~Schindler, J.T. Barreiro, M.~Rambach, G.~Kirchmair, M.~Hennrich,
  P.~Zoller, R.~Blatt, and C.F. Roos.
\newblock Universal digital quantum simulation with trapped ions.
\newblock {\em Science}, 334:57--61, 2011.

\bibitem{Lee2007SCI316}
H.~Lee, Y.-C. Cheng, and G.R. Fleming.
\newblock Coherence dynamics in photosynthesis: Protein protection of excitonic
  coherence.
\newblock {\em Science}, 316:1462--1465, 2007.

\bibitem{Garrahan4}
I.~Lesanovsky, M.~van Horssen, M.~Guta, and J.P. Garrahan.
\newblock Characterization of dynamical phase transitions in quantum jump
  trajectories beyond the properties of the stationary state.
\newblock {\em Phys. Rev. Lett.}, 110:150401, 2013.

\bibitem{LR1972}
E.~Lieb and D.~Robinson.
\newblock The finite group velocity of quantum spin systems.
\newblock {\em Commun. Math. Phys.}, 28:251--257, 1972.

\bibitem{quantized_measurement}
A.~Liu, L.~Yu, W.-A. Zhang, and M.Z.Q. Chen.
\newblock Moving horizon estimation for networked systems with quantized
  measurements and packet dropouts.
\newblock {\em IEEE Trans. on Circuits and Systems I: Regular Papers},
  60:1823--1834, 2013.

\bibitem{LorenzoPRL2015}
S.~Lorenzo, R.~McCloskey, F.~Ciccarello, M.~Paternostro, and G.M. Palma.
\newblock Landauer's principle in multipartite open quantum system dynamics.
\newblock {\em Phys. Rev. Lett.}, 115:120403, 2015.

\bibitem{PWCP}
F.V. Louveaux.
\newblock Piecewise convex programs.
\newblock {\em Mathematical Programming}, 15(1):53--62, 1978.

\bibitem{LovecchioControl}
C.~Lovecchio, F.~Sch\"{a}fer, S.~Cherukattil, M.~Al\`{\i}~Khan, I.~Herrera,
  F.S. Cataliotti, T.~Calarco, S.~Montangero, and F.~Caruso.
\newblock Optimal preparation of quantum states on an atom-chip device.
\newblock {\em Phys. Rev. A}, 93:010304(R), 2016.

\bibitem{Maniscalco2008}
S.~Maniscalco, F.~Francica, R.L. Zaffino, N.L. Gullo, and F.~Plastina.
\newblock Protecting entanglement via the quantum {Z}eno effect.
\newblock {\em Phys. Rev. Lett.}, 100:090503, 2008.

\bibitem{Manzano2015}
G.~Manzano, J.M. Horowitz, and J.M.R. Parrondo.
\newblock Nonequilibrium potential and fluctuation theorems for quantum maps.
\newblock {\em Phys. Rev. E}, 92:032129, 2015.

\bibitem{Maze2008}
J.R. Maze, P.L. Stanwix, J.S. Hodges, S.~Hong, J.M. Taylor, P.~Cappellaro,
  L.~Jiang, M.V. Gurudev~Dutt, E.~Togan, A.S. Zibrov, A.~Yacoby, R.L.
  Walsworth, and M.D. Lukin.
\newblock Nanoscale magnetic sensing with an individual electronic spin in
  diamond.
\newblock {\em Nature}, 455:644--647, 2008.

\bibitem{MazzolaPRL2013}
L.~Mazzola, G.~De~Chiara, and M.~Paternostro.
\newblock Measuring the characteristic function of the work distribution.
\newblock {\em Phys. Rev. Lett.}, 110:230602, 2013.

\bibitem{McGuinness2015}
L.P. McGuinness and F.~Jelezko.
\newblock Quantum mechanics. {L}ook but don't touch the metals.
\newblock {\em Science}, 347:6226, 2015.

\bibitem{Meyer2000}
C.D. Meyer.
\newblock {\em Matrix Analysis and Applied Linear Algebra}.
\newblock SIAM, 2000.

\bibitem{Mezard1987}
M.~Mezard, G.~Parisi, and M.A. Virasoro.
\newblock {\em Spin glass theory and beyond}.
\newblock World Scientific (Singapore), 1987.

\bibitem{Misra1}
B.~Misra and E.C.G. Sudarshan.
\newblock The {Z}eno's paradox in quantum theory.
\newblock {\em J. Math. Phys.}, 18:756, 1977.

\bibitem{Mnatsakanov}
R.M. Mnatsakanov.
\newblock Hausdorff moment problem: {R}econstruction of probability density
  functions.
\newblock {\em Stat. Prob. Lett.}, 78:1869--1877, 2008.

\bibitem{qbiobook}
M.~Mohseni, Y.~Omar, G.S. Engel, and M.B. Plenio.
\newblock {\em Quantum effects in biology}.
\newblock Cambridge University Press, 2013.

\bibitem{Mohseni2008JCP129}
M.~Mohseni, P.~Rebentrost, S.~Lloyd, and A.~Aspuru-Guzik.
\newblock Environment-assisted quantum walks in photosynthetic energy transfer.
\newblock {\em J. Chem. Phys.}, 129:174106, 2008.

\bibitem{MonzPRL106}
T.~Monz, P.~Schindler, J.T. Barreiro, M.~Chwalla, D.~Nigg, W.A. Coish,
  M.~Harlander, W.~H\"{a}nsel, M.~Hennrich, and R.~Blatt.
\newblock 14-qubit entanglement: {C}reation and coherence.
\newblock {\em Phys. Rev. Lett.}, 106:130506, 2011.

\bibitem{Mukamel2003}
S.~Mukamel.
\newblock Quantum extension of the {J}arzynski relation: {A}nalogy with
  stochastic dephasing.
\newblock {\em Phys. Rev. Lett.}, 90:170604, 2003.

\bibitem{GherardiniSciRep}
M.M. M\"{u}ller, S.~Gherardini, and F.~Caruso.
\newblock Stochastic quantum {Z}eno-based detection of noise correlations.
\newblock {\em Scientific Reports}, 6:38650, 2016.

\bibitem{GherardiniAnnalen}
M.M. M\"{u}ller, S.~Gherardini, and F.~Caruso.
\newblock Quantum {Z}eno dynamics through stochastic protocols.
\newblock {\em Annalen der Physik}, 529(9):1600206, 2017.

\bibitem{MuellerSensing}
M.M. M\"{u}ller, S.~Gherardini, and F.~Caruso.
\newblock Noise-robust quantum sensing via optimal multi-probe spectroscopy.
\newblock {\em Eprint arXiv:1801.10220}, 2018.

\bibitem{GherardiniFisher}
M.M. M\"{u}ller, S.~Gherardini, A.~Smerzi, and F.~Caruso.
\newblock Fisher information from stochastic quantum measurements.
\newblock {\em Phys. Rev. A}, 94:042322, 2016.

\bibitem{NeilNatPhys}
C.~Neill, P.~Roushan, M.~Fang, Y.~Chen, M.~Kolodrubetz, Z.~Chen, A.~Megrant,
  R.~Barends, B.~Campbell, B.~Chiaro, A.~Dunsworth, E.~Jeffrey, J.~Kelly,
  J.~Mutus, P.J.J. O'Malley, C.~Quintana, D.~Sank, A.~Vainsencher, J.~Wenner,
  T.C. White, A.~Polkovnikov, and J.M. Martinis.
\newblock Ergodic dynamics and thermalization in an isolated quantum system.
\newblock {\em Nat. Phys.}, 12:1037--1041, 2016.

\bibitem{Netovcny1}
K.~Neto\v~cn\'y and F.~Redig.
\newblock Large deviation for quantum spin systems.
\newblock {\em J. Stat. Phys.}, 117:521, 2004.

\bibitem{Nigg302}
D.~Nigg, M.~M\"{u}ller, E.A. Martinez, P.~Schindler, M.~Hennrich, T.~Monz, M.A.
  Martin-Delgado, and R.~Blatt.
\newblock Quantum computations on a topologically encoded qubit.
\newblock {\em Science}, 345:302--305, 2014.

\bibitem{Norris2016}
L.M. Norris, G.A. Paz-Silva, and L.~Viola.
\newblock Qubit noise spectroscopy for non-{G}aussian dephasing environments.
\newblock {\em Phys. Rev. Lett.}, 116:150503, 2016.

\bibitem{Othons1997RSI68}
A.~Othonos.
\newblock Fiber {B}ragg gratings.
\newblock {\em Rev. Sci. Instrum.}, 68:4309--4341, 1997.

\bibitem{Panitchayangkoon10PNAS107}
G.~Panitchayangkoon, D.~Hayes, K.A. Fransted, J.R. Caram, E.~Harel, J.~Wen,
  R.E. Blankenship, and G.S. Engel.
\newblock Long-lived quantum coherence in photosynthetic complexes at
  physiological temperature.
\newblock {\em Proc. Natl. Acad. Sci. USA}, 107:12766, 2010.

\bibitem{Papoulis1984}
A.~Papoulis.
\newblock {\em Probability, Random Variables and Stochastic Processes}.
\newblock McGraw-Hill Inc., 1984.

\bibitem{ParisBook}
M.G.A. Paris and J.~Rehacek.
\newblock {\em Quantum State Estimation}.
\newblock Lectures Notes in Physics. Springer, 2004.

\bibitem{FC2016NATmat}
H.~Park and al.
\newblock Enhanced energy transport in genetically engineered excitonic
  networks.
\newblock {\em Nat. Mat.}, 15:211--216, 2016.

\bibitem{RTN1}
C.E. Parman, N.E. Israeloff, and J.~Kakalios.
\newblock Random telegraph-switching noise in coplanar current measurements of
  amorphous silicon.
\newblock {\em Phys. Rev. B}, 44:8391, 1991.

\bibitem{OPT2}
P.~Patrinos and H.~Sarimveis.
\newblock Convex parametric piecewise quadratic optimization: {T}heory and
  algorithms.
\newblock {\em Automatica}, 47(8):1770--1777, 2011.

\bibitem{Lidar2012}
G.A. Paz-Silva, A.~Rezakhani, J.M. Dominy, and D.A. Lidar.
\newblock {Z}eno effect for quantum computation and control.
\newblock {\em Phys. Rev. Lett.}, 108:080501, 2012.

\bibitem{Paz-Silva2014}
G.A. Paz-Silva and L.~Viola.
\newblock General transfer-function approach to noise filtering in open-loop
  quantum control.
\newblock {\em Phys. Rev. Lett.}, 113:250501, 2014.

\bibitem{PekolaNAT2015}
J.P. Pekola.
\newblock Towards quantum thermodynamics in electronic circuits.
\newblock {\em Nat. Phys.}, 11:118--123, 2015.

\bibitem{Perarnau-Llobet2015}
M.~Perarnau-Llobet, K.V. Hovhannisyan, M.~Huber, P.~Skrzypczyk, N.~Brunner, and
  A.~Ac\'{i}n.
\newblock Extractable work from correlations.
\newblock {\em Phys. Rev. X}, 5:041011, 2015.

\bibitem{Peres1}
A.~Peres.
\newblock Ergodicity and mixing in quantum theory. {I}.
\newblock {\em Phys. Rev. A}, 30:504, 1984.

\bibitem{Peterson2016}
J.P.S. Peterson, R.S. Sarthour, A.M. Souza, I.S. Oliveira, J.~Goold, K.~Modi,
  D.O. Soares-Pinto, and L.C. C\`eleri.
\newblock Experimental demonstration of information to energy conversion in a
  quantum system at the {L}andauer limit.
\newblock {\em Proc. R. Soc. A}, 472:20150813, 2016.

\bibitem{PetrovicCHIP}
J.~Petrovic, I.~Herrera, P.~Lombardi, F.~Sch\"afer, and F.S. Cataliotti.
\newblock A multi-state interferometer on an atom chip.
\newblock {\em New J. Phys.}, 15:043002, 2012.

\bibitem{Plenio2008NJP10}
M.B. Plenio and S.F. Huelga.
\newblock Dephasing-assisted transport: {Q}uantum networks and biomolecules.
\newblock {\em New J. Phys.}, 10:113019, 2008.

\bibitem{PlenioRMP}
M.B. Plenio and P.L. Knight.
\newblock The quantum-jump approach to dissipative dynamics in quantum optics.
\newblock {\em Rev. Mod. Phys.}, 70:101, 1998.

\bibitem{Polkovnikov2011}
A.~Polkovnikov, K.~Sengupta, A.~Silva, and M.~Vengalattore.
\newblock {\it Colloquium}: Nonequilibrium dynamics of closed interacting
  quantum systems.
\newblock {\em Rev. Mod. Phys.}, 83:863, 2011.

\bibitem{VenkateshNJP2015}
B.~Prasanna~Venkatesh, G.~Watanabe, and P.~Talkner.
\newblock Quantum fluctuation theorems and power measurements.
\newblock {\em New J. Phys.}, 17:075018, 2015.

\bibitem{RaoLS}
C.~R. Rao and H.~Toutenburg.
\newblock {\em Linear Models: {L}east Squares and Alternatives (3rd ed.)}.
\newblock Springer Series in Statistics. Springer, Berlin, 2008.

\bibitem{RaoRawLee01}
C.V. Rao, J.B. Rawlings, and J.~H. Lee.
\newblock Constrained linear estimation--a moving horizon approach.
\newblock {\em Automatica}, 37(10):1619--1628, 2001.

\bibitem{RaRaMa03}
C.V. Rao, J.B. Rawlings, and D.Q. Mayne.
\newblock Constrained state estimation for nonlinear discrete-time systems:
  {S}tability and moving horizon approximations.
\newblock {\em IEEE Trans. on Automatic Control}, 48(2):246--257, 2003.

\bibitem{RasteginJSM2013}
A.E. Rastegin.
\newblock Non-equilibrium equalities with unital quantum channels.
\newblock {\em J. Stat. Mech.}, page P06016, 2013.

\bibitem{Rebentrost2009NJP11}
P.~Rebentrost, M.~Mohseni, I.~Kassal, S.~Lloyd, and A.~Aspuru-Guzik.
\newblock Environment-assisted quantum transport.
\newblock {\em New J. Phys.}, 11:033003, 2009.

\bibitem{Reed1975}
M.C. Reed and B.~Simon.
\newblock {\em Methods of Modern Mathematical Physics, Volume II}.
\newblock Academic {P}ress, 1975.

\bibitem{Ribeiro1}
A.~Ribeiro and G.~B. Giannakis.
\newblock Bandwidth-constrained distributed estimation for wireless sensor
  networks - part {I}: {G}aussian case.
\newblock {\em IEEE Transactions on Signal Processing}, 54(3):1131--1143, 2006.

\bibitem{Ribeiro2}
A.~Ribeiro and G.~B. Giannakis.
\newblock Bandwidth-constrained distributed estimation for wireless sensor
  networks - part {II}: unknown probability density function.
\newblock {\em IEEE Transactions on Signal Processing}, 54(7):2784--2796, 2006.

\bibitem{Ristic}
B.~Ristic, A.~Gunatilaka, and R.~Gailis.
\newblock Achievable accuracy in {Gaussian} plume parameter estimation using a
  network of binary sensors.
\newblock {\em Information Fusion}, pages 42--48, 2015.

\bibitem{Rivas}
A.~Rivas, S.F. Huelga, and M.B. Plenio.
\newblock Quantum non-{M}arkovianity: {C}haracterization, quantification and
  detection.
\newblock {\em Rep. Prog. Phys.}, 77:094001, 2014.

\bibitem{RoncagliaPRL2014}
A.J. Roncaglia, F.~Cerisola, and J.P. Paz.
\newblock Work measurement as a generalized quantum measurement.
\newblock {\em Phys. Rev. Lett.}, 113:250601, 2014.

\bibitem{RoosNJP2008}
C.F. Roos.
\newblock Ion trap quantum gates with amplitude-modulated laser beams.
\newblock {\em New Journal of Physics}, 10(1):013002, 2008.

\bibitem{Rossi2015}
M.A.C. Rossi and M.G.A. Paris.
\newblock Entangled quantum probes for dynamical environmental noise.
\newblock {\em Phys. Rev. A}, 92:010302, 2015.

\bibitem{RossnagelPRL2014}
J.~Ro{\ss}nagel, O.~Abah, F.~Schmidt-Kaler, K.~Singer, and E.~Lutz.
\newblock Nanoscale heat engine beyond the {C}arnot limit.
\newblock {\em Phys. Rev. Lett.}, 112:030602, 2014.

\bibitem{Rossnagel325}
J.~Ro{\ss}nagel, S.T. Dawkins, K.N. Tolazzi, O.~Abah, E.~Lutz,
  F.~Schmidt-Kaler, and K.~Singer.
\newblock A single-atom heat engine.
\newblock {\em Science}, 352:325--329, 2016.

\bibitem{Sagawa2014}
T.~Sagawa.
\newblock {\em Lectures on Quantum Computing, Thermodynamics and Statistical
  Physics. Edited by Nakahara Mikio et al.}
\newblock World Scientific Publishing Co. Pte. Ltd., 2014.

\bibitem{SagawaPRL2010}
T.~Sakawa and M.~Ueda.
\newblock Generalized {J}arzynski equality under nonequilibrium feedback
  control.
\newblock {\em Phys. Rev. Lett.}, 104:090602, 2010.

\bibitem{Sakurai1994}
J.J. Sakurai.
\newblock {\em Modern Quantum Mechanics (2nd edition)}.
\newblock Addison-Wesley Publishing Company, 1994.

\bibitem{SchaferZeno}
F.~Sch\"afer, I.~Herrera, S.~Cherukattil, C.~Lovecchio, F.S. Cataliotti,
  F.~Caruso, and A.~Smerzi.
\newblock Experimental realization of quantum {Z}eno dynamics.
\newblock {\em Nat. Commun.}, 5:4194, 2014.

\bibitem{Schindler2013}
P.~Schindler, M.~M\"{u}ller, D.~Nigg, J.T. Barreiro, E.A. Martinez,
  M.~Hennrich, T.~Monz, S.~Diehl, P.~Zoller, and R.~Blatt.
\newblock Quantum simulation of dynamical maps with trapped ions.
\newblock {\em Nat. Phys.}, 9:361--367, 2013.

\bibitem{Schlosshauer2005}
M.~Schlosshauer.
\newblock Decoherence, the measurement problem, and interpretations of quantum
  mechanics.
\newblock {\em Reviews of Modern Physics}, 76(4):1267--1305, 2005.

\bibitem{SchnHannMarq15}
R.~Schneider, R.~Hannemann-Tamas, and W.~Marquardt.
\newblock An iterative partition-based moving horizon estimator with coupled
  inequality constraints.
\newblock {\em Automatica}, 61:302--307, 2015.

\bibitem{Scully03Science299}
M.O. Scully, M.S. Zubairy, G.S. Agarwal, and H.~Walther.
\newblock Extracting work from a single heat bath via vanishing quantum
  coherence.
\newblock {\em Science}, 299(5608):862--4, 2003.

\bibitem{Seifert2012Review}
U.~Seifert.
\newblock Stochastic thermodynamics, fluctuation theorems, and molecular
  machines.
\newblock {\em Rep. Prog. Phys.}, 75:126001, 2012.

\bibitem{likelihood}
D.~Shi, T.~Chen, and L.~Shi.
\newblock Event-triggered maximum likelihood state estimation.
\newblock {\em Automatica}, 50:247--254, 2014.

\bibitem{Shushin1}
A.I. Shushin.
\newblock The effect of measurements, randomly distributed in time, on quantum
  systems: stochastic quantum {Z}eno effect.
\newblock {\em J. Phys. A: Math. Theor.}, 44:055303, 2011.

\bibitem{SignolesZeno}
A.~Signoles, A.~Facon, D.~Grosso, I.~Dotsenko, S.~Haroche, J.-M. Raimond,
  M.~Brune, and S.~Gleyzes.
\newblock Confined quantum {Z}eno dynamics of a watched atomic arrow.
\newblock {\em Nat. Phys.}, 10:715, 2014.

\bibitem{Lazar}
J.~Sijs and M.~Lazar.
\newblock Event-based state estimation with time synchronous updates.
\newblock {\em IEEE Trans. on Automatic Control}, 57:2650--2655, 2012.

\bibitem{SkrzypczykNAT2014}
P.~Skrzypczyk, A.J. Shot, and S.~Popescu.
\newblock Work extraction and thermodynamics for individual quantum systems.
\newblock {\em Phys. Rev. X}, 5:5185, 2014.

\bibitem{SmerziPRL2012}
A.~Smerzi.
\newblock {Z}eno dynamics, indistinguishability of state, and entanglement.
\newblock {\em Phys. Rev. Lett.}, 109:150410, 2012.

\bibitem{Solinas2015}
P.~Solinas and S.~Gasparinetti.
\newblock Full distribution of work done on a quantum system for arbitrary
  initial states.
\newblock {\em Phys. Rev. E}, 92:042150, 2015.

\bibitem{SoerensenPRL82}
A.~S\o{}rensen and K.~M\o{}lmer.
\newblock Quantum computation with ions in thermal motion.
\newblock {\em Phys. Rev. Lett.}, 82:1971--1974, 1999.

\bibitem{Sozzi2008}
M.S. Sozzi.
\newblock {\em Discrete symmetries and CP violation}.
\newblock Oxford University Press, 2008.

\bibitem{Streed:2006}
E.W. Streed, J.~Mun, M.~Boyd, G.K. Campbell, P.~Medley, W.~Ketterle, and D.E.
  Pritchard.
\newblock Continuous and pulsed quantum {Z}eno effect.
\newblock {\em Phys. Rev. Lett.}, 97:260402, 2006.

\bibitem{Talkner2016}
P.~Talkner and P.~H\"{a}nggi.
\newblock Aspects of quantum work.
\newblock {\em Phys. Rev. E}, 93:022131, 2016.

\bibitem{Taylor2008}
J.M. Taylor, P.~Cappellaro, L.~Childress, L.~Jiang, D.~Budker, P.R. Hemmer,
  A.~Yacoby, R.~Walsworth, and M.D. Lukin.
\newblock High-sensitivity diamond magnetometer with nanoscale resolution.
\newblock {\em Nat. Phys.}, 4:810--816, 2008.

\bibitem{Tonomura}
A.~Tonomura, J.~Endo, T.~Matsuda, and T.~Kawasaki.
\newblock Demonstration of single-electron buildup of an interference pattern.
\newblock {\em Am. Journ. Phys.}, 57:117, 1989.

\bibitem{Touchette1}
H.~Touchette.
\newblock The large deviation approach to statistical mechanics.
\newblock {\em Phys. Rep.}, 478(1):1--69, 2009.

\bibitem{ToyabeNAT2010}
S.~Toyabe, T.~Sagawa, M.~Ueda, E.~Muneyuki, and M.~Sano.
\newblock Experimental demonstration of information to energy conversion and
  validation of the generalized {J}arzynski equality.
\newblock {\em Nat. Phys.}, 6:988--992, 2010.

\bibitem{Trefethen2000}
L.N. Trefethen.
\newblock {\em Spectral Methods in MATLAB}.
\newblock SIAM, 2000.

\bibitem{Uhlmann}
A.~Uhlmann.
\newblock The transition probability in the state space of a *-algebra.
\newblock {\em Rep. Math. Phys.}, 9:273, 1976.

\bibitem{Umegaki}
H.~Umegaki.
\newblock Conditional expectations in an operator algebra {IV} (entropy and
  information).
\newblock {\em Phys. Rev. Lett.}, 14:59--85, 1962.

\bibitem{Uzdin15PRX5}
R.~Uzdin, A.~Levy, and R.~Kosloff.
\newblock Equivalence of quantum heat machines, and quantum-thermodynamic
  signatures.
\newblock {\em Phys. Rev. X}, 5:031044, 2015.

\bibitem{Varadhan:1984}
S.R.S. Varadhan.
\newblock {\em Large Deviations and Applications}.
\newblock SIAM, Philadelphia, 1984.

\bibitem{Vedral2002}
V.~Vedral.
\newblock The role of relative entropy in quantum information theory.
\newblock {\em Rev. Mod. Phys.}, 74:197, 2002.

\bibitem{GherardiniNAT}
S.~Viciani, S.~Gherardini, M.~Lima, M.~Bellini, and F.~Caruso.
\newblock Disorder and dephasing as control knobs for light transport in
  optical fiber cavity networks.
\newblock {\em Scientific Reports}, 6:37791, 2016.

\bibitem{VicianiPRL2015}
S.~Viciani, M.~Lima, M.~Bellini, and F.~Caruso.
\newblock Observation of noise-assisted transport in an all-optical
  cavity-based network.
\newblock {\em Phys. Rev. Lett.}, 115:083601, 2015.

\bibitem{Wong2007}
S.~Vijayakumaran, Y.~Levinbook, and T.F. Wong.
\newblock Maximum likelihood localization of a diffusive point source using
  binary observations.
\newblock {\em IEEE Transactions on Signal Processing}, 55(2):665--676, 2007.

\bibitem{Viola1999}
L.~Viola, E.~Knill, and S.~Lloyd.
\newblock Dynamical decoupling of open quantum systems.
\newblock {\em Phys. Rev. Lett.}, 82:2417, 1999.

\bibitem{Neumann2}
J.~von Neumann.
\newblock Operatorenmethoden in der klassischen mechanik.
\newblock {\em Annals of Mathematics}, 33:587, 1932.

\bibitem{Neumann}
J.~(the English translation by R.~Tumulka) von Neumann.
\newblock Proof of the ergodic theorem and the {H}-theorem in quantum
  mechanics.
\newblock {\em Eur. Phys. J. H}, 35:201, 2010.

\bibitem{Irr-sampling}
L.Y. Wang, G.G. Li, L.~Guo, and C.-Z. Xu.
\newblock State observability and observers of linear-time-invariant systems
  under irregular sampling and sensor limitations.
\newblock {\em IEEE Trans. on Automatic Control}, 56:2639--2654, 2011.

\bibitem{state_reconstruction}
L.Y. Wang, G.~Xu, and G.G. Yin.
\newblock State reconstruction for linear time-invariant systems with
  binary-valued output observations.
\newblock {\em Systems and Control Letters}, 57:958--963, 2008.

\bibitem{Wang2}
L.Y. Wang, G.G. Yin, and J.F. Zhang.
\newblock Joint identification of plant rational models and noise distribution
  functions using binary-valued observations.
\newblock {\em Automatica}, 42:543--547, 2006.

\bibitem{Wang1}
L.Y. Wang, J.F. Zhang, and G.G. Yin.
\newblock System identification using binary sensors.
\newblock {\em IEEE Trans. on Automatic Control}, 48:1892--1907, 2003.

\bibitem{WatanabePRE2014}
G.~Watanabe, B.~Prasanna~Venkatesh, P.~Talkner, M.~Campisi, and P.~H\"{a}nggi.
\newblock Quantum fluctuation theorems and generalized measurements during the
  force protocol.
\newblock {\em Phys. Rev. E}, 89:032114, 2014.

\bibitem{Wiener1}
D.~Westwick and M.~Verhaegen.
\newblock Identifying {MIMO} {Wiener} systems using subspace model
  identification methods.
\newblock {\em Signal Processing}, 52(2):235--258, 1996.

\bibitem{Wootters1981}
W.K. Wootters.
\newblock Statistical distance and {H}ilbert space.
\newblock {\em Phys. Rev. D}, 23:357, 1981.

\bibitem{bounds}
L.~Wu.
\newblock Error bounds for piecewise convex quadratic programs and
  applications.
\newblock {\em SIAM Journal on Control and Optimization}, 33(5):1510--1529,
  1995.

\bibitem{Yi2013}
J.~Yi and Y.W. Kim.
\newblock Nonequilibirum work and entropy production by quantum projective
  measurements.
\newblock {\em Phys. Rev. E}, 88:032105, 2013.

\bibitem{Zhang2015}
Y.~Zhang and H.~Fan.
\newblock {Z}eno dynamics in quantum open systems.
\newblock {\em Sci. Rep.}, 5:11509, 2015.

\end{thebibliography}
%\addcontentsline{toc}{chapter}{Bibliography}

\nocite{*}
\fancyhead[RO,LE]{\slshape BIBLIOGRAPHY}
\fancyfoot[C]{\thepage}

\addcontentsline{toc}{chapter}{Bibliography}

\end{document}